\documentclass[aps,prl,reprint,notitlepage,nofootinbib,groupaddress]{revtex4-1}
\usepackage{pdfsync}
\usepackage{amsmath}    % need for subequations
\usepackage{amsfonts} % note how statements can be commented out
\usepackage{amssymb}
\usepackage{MnSymbol}
\usepackage{mathrsfs} % define the \mathscr font
\usepackage{textcomp} % to define \texttildelow font
\usepackage{bbm}
\usepackage{float}    % suppress float figures and tables
\usepackage[caption=false]{subfig}
\usepackage{accents} % put tilde under symbols using underaccent{}{}
\usepackage{mathtools}
\usepackage{graphics}
\usepackage{epsfig}
\usepackage{multirow}
\usepackage{bibentry}
\usepackage{braket}
\usepackage{algorithm}
\usepackage{algorithmic}
\usepackage{listings}
\usepackage{setspace}
\usepackage{fancyhdr}
\usepackage{slashed}
\usepackage{xfrac}
\usepackage{xcolor}
\usepackage{cases}
\usepackage[retainorgcmds]{IEEEtrantools}
\usepackage{hyperref}
\hypersetup{pdfborder= 0 0 0, citecolor=blue, urlcolor=blue, linkcolor=blue, colorlinks=true, bookmarksopen=true}

\allowdisplaybreaks

\newcommand{\Cc}{\mathbb{C}}
\newcommand{\Ss}{\mathbb{S}}
\newcommand{\Pp}{\mathbb{P}}

\newcommand{\Xx}{\mathbb{X}}
\newcommand{\pT}{\widehat{+}}
\newcommand{\mT}{\widehat{-}}
\newcommand{\muT}{\widehat{\mu}}
\newcommand{\nuT}{\widehat{\nu}}

\newcommand{\vT}{\hat{v}}

\newcommand{\itP}[1]{\widehat{#1}}

\begin{document}
\title{Interpolating Helicity Spinors Between the Instant Form and the Light-front Form }
\author{Ziyue Li}
%\affiliation{Department of Physics, North Carolina State University, Raleigh, North Carolina 27695-8202}
\author{Murat An}
%\affiliation{Department of Physics, North Carolina State University, Raleigh, North Carolina 27695-8202}
\author{Chueng-Ryong Ji}

\affiliation{Department of Physics, North Carolina State University, Raleigh, North Carolina 27695-8202}

\begin{abstract}
We discuss the helicity spinors interpolating between the instant form dynamics (IFD) and the front form dynamics, or the light-front dynamics (LFD), and
present the interpolating helicity amplitudes as well as their squares for the scattering of two fermions, and the annihilation of fermion and anti-fermion.
We parametrize the interpolation between the two dynamics, IFD and LFD, by an interpolation angle and derive not only the generalized helicity spinors in the
$(0,J)\oplus(J,0)$ chiral representation that links naturally the two typical IFD vs. LFD helicity spinors but also the generalized Melosh transformation
that relates these generalized helicity spinors to the usual Dirac spinors. Analyzing the directions of the particle momentum and spin with the variation of
the interpolation angle, we inspect the whole landscape of the generalized helicity intermediating between the usual Jacob-Wick helicity in the IFD
and the light-front helicity in the LFD. Our analysis clarifies the characteristic difference of the helicity amplitudes between the IFD and the LFD.
In particular, we find that the behavior of the angle between the momentum direction and the spin direction bifurcates at a critical interpolation angle
and the IFD and the LFD separately belong to the two different branches bifurcated at this critical interpolation angle.
This finding further clarifies any conceivable confusion in the prevailing notion of the equivalence between the infinite momentum frame and the LFD.
The existence of the universal J-curve found in our previous works of scalar field theory and the sQED theory is confirmed in the present work of
interpolating helicity amplitudes for the fermion scattering and annihilation processes. In conjunction with the bifurcation of branches,
the two boundaries appear in the interpolating helicity amplitudes and interestingly the J-curve persists within these two boundaries.
\end{abstract}

\maketitle

\section{Introduction}
\label{sec:introduction}

Among the three different forms of relativistic dynamics proposed by Dirac \cite{Dirac1949} in 1949,
the instant form ($x^{0}=0$) and the front form ($x^{+} \equiv (x^{0}+x^{3})/\sqrt{2}$=0)
have been the most popular choices in studying hadron physics. While the quantization at the equal time $t=x^{0}$ (in $c=1$ unit) produces the instant form dynamics (IFD) of quantum field theory, the quantization at equal light-front time $\tau = x^{+} = (t+z)/\sqrt{2}$ yields the front form dynamics, now known as the light-front dynamics (LFD). In an effort to link these two popular but different forms of relativistic dynamics, IFD and LFD, we have discussed~\cite{Ji1996,Ji2001,Ji2013,Ji2015}
the interpolation between the two forms of dynamics by introducing a parameter $\delta$ called ``an interpolation angle".
With this parameter $\delta$, the interpolating space-time coordinates between IFD and LFD are defined by a transformation from the ordinary space-time coordinates, $x^{\muT}=\mathcal{R}^{\muT}_{\phantom{\mu}{\nu}}x^{\nu}$, i.e.
\begin{align}\label{eqn:interpolation_angle_definition}
  \begin{pmatrix}
    x^{\pT}\\
    x^{\itP{1}}\\
    x^{\itP{2}}\\
    x^{\mT}
  \end{pmatrix}=
  \begin{pmatrix}
    \cos\delta & 0  & 0  & \sin\delta \\
    0          & 1  & 0  & 0 \\
    0          & 0  & 1  & 0 \\
    \sin\delta & 0  & 0  & -\cos\delta
  \end{pmatrix}
  \begin{pmatrix}
    x^{0}\\
    x^{1}\\
    x^{2}\\
    x^{3}
  \end{pmatrix},
\end{align}
where we use ``wide-hat" (``\textasciicircum'') on the indices to denote the interpolating variables with the parameter
$0\leq\delta\leq\pi/4$.
As $\mathcal{R}^{\itP{i}}_{j} = \delta^{i}_{j}$ for $i,j =1,2$ in Eq.~(\ref{eqn:interpolation_angle_definition}),
we will however omit the ``\textasciicircum''  notation unless necessary for the perpendicular indices $j=1,2$ in a 4-vector.
In the limits $\delta\rightarrow0$ and $\delta\rightarrow\pi/4$, we recover the corresponding variables in the instant form and the front form, respectively.  For example, the interpolating coordinates $x^{\itP{\pm}}$ in the limit $\delta\rightarrow\pi/4$ become the light-front coordinates $x^{\pm}=(x^{0}\pm x^{3})/\sqrt{2}$ without
``wide-hat" (``\textasciicircum'').

Our interpolation between the IFD and the LFD provides the whole picture of landscape between the two and clarifies the issue, if any, in linking them to each other. The same method of interpolating hypersurfaces has been used by Hornbostel \cite{Hornbostel1992} to analyze various aspects of field theories including the issue of nontrivial vacuum. The same vein of application to study the axial anomaly in the Schwinger model has been presented \cite{Ji1996}, and other related works \cite{Chen1971, Elizalde1976, Frishman1977, Sawicki1991,JiSurya1992} can also be found in the literature.

We started out this effort by studying the Poincar\'e algebra for any arbitrary interpolation angle~\cite{Ji2001} and recently provided the physical meaning of the kinematic vs. dynamic operators by introducing the interpolating time-ordered scattering amplitudes \cite{Ji2013}.
In particular, we demonstrated that the longitudinal boost invariance of each individual $x^{\pT}$-ordered scattering amplitude is realized only at $\delta=\pi/4$ and the disappearance of the connected contributions to the current arising from the vacuum occurs
independent of the reference frame only when the interpolation angle is taken to yield the LFD. This affirms the well-known saga of
the longitudinal boost $K^3$ which maximizes the number of kinematic (i.e. interaction independent) generators in LFD
as seven out of ten Poincare generators. Dramatic character change of $K^3$ from ``dynamic" for $0\leq\delta\le\pi/4$ to ``kinematic"
in $\delta=\pi/4$ brings indeed a great benefit to use LFD for the study of hadron physics.
Our interpolation between IFD and LFD made it also clear that the disappearance of the connected contributions to the current arising from the vacuum in LFD doesn't require the infinite momentum frame (IMF). It thus resolves the confusion in the prevailing notion of equivalence between the LFD and the IMF. For the study of hadron physics in QCD, the built-in boost invariance together with the simpler vacuum property
in LFD is certainly an appealing feature as it may save substantial computational efforts in getting QCD solutions that respect the full Poincar\'e symmetries.

Although we want ultimately to obtain a general formulation for the QCD using the interpolation between the IFD and the LFD, we start from the simpler theory to discuss first the bare-bone structure that will persist even in the more complicated theories. Subsequent to our study of the simple scalar field theory~\cite{Ji2013} involving just the fundamental degrees of freedom such as the momenta of particles in scattering processes, we considered very recently interpolating the electromagnetic gauge degree of freedom between the IFD and the LFD and found that the light-front gauge in the LFD is naturally linked to the Coulomb gauge in the IFD through the interpolation angle~\cite{Ji2015}.
We also extended our interpolation of the scattering amplitude presented in the simple scalar field theory~\cite{Ji2013} to the case of the electromagnetic gauge field theory but still with the scalar fermion fields known as the sQED theory~\cite{Ji2015} and analyzed the lowest-order scattering processes in sQED such as the analogues of the well-known QED processes
$``e\mu \rightarrow e\mu"$ and $``e^+ e^- \rightarrow \mu^+ \mu^-"$.

To promote the sQED calculation to the QED calculation, we now start discussing the fermion degrees of freedom and
their interpolation between IFD and LFD. Due to a few different representations available for the spinors, we examine the relationships
among the available spinor representations with the interpolation angle parameter $0\leq\delta\leq\pi/4$ in this work. As the first step,
we limit our discussion here only for the on-mass-shell spinors and analyze the interpolating helicity amplitudes for
the processes involving fermions as the external particles. We discuss a fermion and another fermion scattering process
as well as a fermion and anti-fermion pair annihilation and creation process analogous to ``$e\mu \rightarrow e\mu$" and ``$e^+ e^- \rightarrow \mu^+ \mu^-$", respectively.
In this work, we focus on the effects from the initial and final fermion degrees of freedom
rather than from the intermediate gauge boson which we have already studied in our previous work~\cite{Ji2015}.

% $p {\bar p} \rightarrow P_{11}({\rm Roper}) {\bar P_{11}}({\rm AntiRoper})$,

\begin{figure}[!b]
  \centering
  \includegraphics[width=\columnwidth]{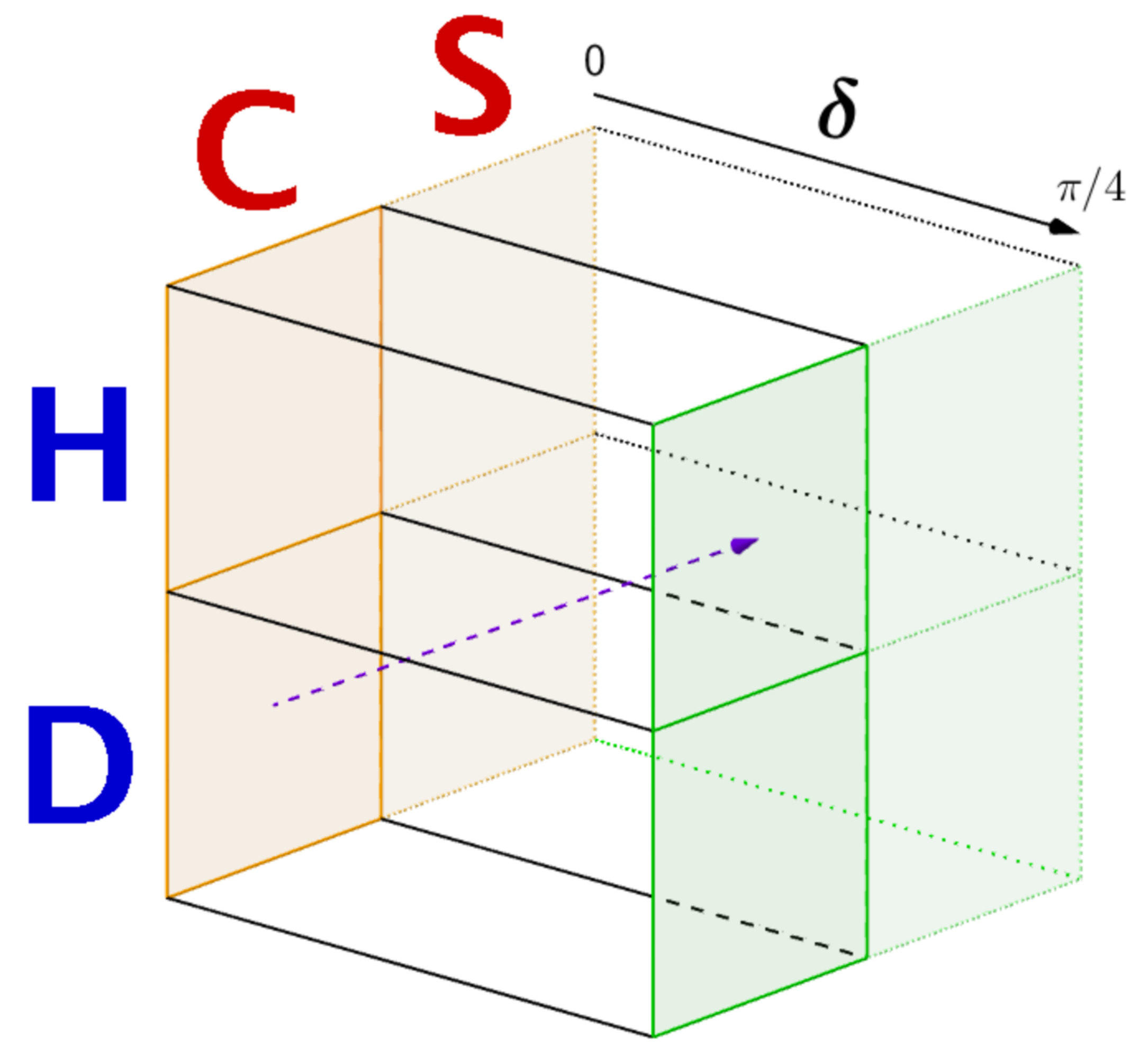}
  \caption{(color online) An illustration of the relations between different conventions and names we use in this paper.
			The red letters C and S stand for chiral and standard representations respectively.
			The blue letters H and D stand for helicity spinor and Dirac spinor respectively.
			The solid black arrow points from the instant form to the light front form, and the interpolation angle $\delta$ goes from $0$ to $\pi/4$.
			Both the helicity and Dirac spinors can be generated for arbitrary interpolation angle.
			The dashed purple arrow indicates the Melosh transformation, which relates the instant Dirac spinor to the light-front helicity spinor.
			\label{fig:spinors_overview}}
\end{figure}

For an overview of the available spinor representations, we may provide a schematic illustration as shown in Fig.~\ref{fig:spinors_overview}
and discuss the relationships among those representations. In Fig.~\ref{fig:spinors_overview}, we denote the so-called ``standard representation" and ``chiral representation" as ``S" and ``C". The transformation between the two representations ``S" and ``C" can be made by
the transformation matrix ${\it S}$ given by
\begin{align}
{\it S}={\it S^\dagger}=\dfrac{1}{\sqrt{2}}
	\begin{pmatrix}
	I & I\\
	I & -I\\
	\end{pmatrix},
	\label{eqn:SR_CR_fundamental_relation}
\end{align}
where $I$ is the $2 \times 2$ identity matrix.
As the Dirac matrices are related by
\begin{eqnarray}
\gamma^\mu_{\rm S} = {\it S} \gamma^\mu_{\rm C} {\it S^\dagger}, \nonumber \\
\gamma^\mu_{\rm C} = {\it S^\dagger} \gamma^\mu_{\rm S} {\it S},
\label{eagn:gammaMatrices}
\end{eqnarray}
the spinors are related by
\begin{eqnarray}
u_{\rm S}(p) = {\it S} u_{\rm C}(p), \nonumber \\
u_{\rm C}(p) = {\it S^\dagger} u_{\rm S}(p),
\end{eqnarray}
where the spinors $u_{\rm S}(p)$ and $u_{\rm C}(p)$ will be expressed explicitly later as the four-component column-vectors.
The six Lorentz group generators of rotation ($\mathbf{J}$) and boost ($\mathbf{K}$) in S and C representations
are also related by the transformation matrix ${\it S}$ such as $\mathbf{J}_{\rm S} = {\it S} \mathbf{J}_{\rm C} {\it S^\dagger},
\mathbf{K}_{\rm S} = {\it S} \mathbf{K}_{\rm C} {\it S^\dagger}, \mathbf{J}_{\rm C} = {\it S^\dagger} \mathbf{J}_{\rm S} {\it S}, \mathbf{K}_{\rm C} = {\it S^\dagger} \mathbf{K}_{\rm S} {\it S}$, etc. as given by
\begin{align}
\mathbf{J}_{\rm S}=\mathbf{J}_{\rm C}=\dfrac{1}{2}
	\begin{pmatrix}
	\boldsymbol\sigma & 0\\
	0 & \boldsymbol\sigma\\
	\end{pmatrix}, \nonumber \\
\mathbf{K}_{\rm S} = \dfrac{i}{2}
\begin{pmatrix}
0 & \boldsymbol\sigma \\
\boldsymbol\sigma & 0 \\
\end{pmatrix},
\mathbf{K}_{\rm C} = \dfrac{i}{2}
\begin{pmatrix}
\boldsymbol\sigma & 0  \\
0 & -\boldsymbol\sigma \\
\end{pmatrix},
\label{eqn:LorentzGen}
\end{align}
where $\boldsymbol\sigma$ denotes the Pauli matrices.
It is clear that the combinations of rotation and boost given by $\mathbf{J}_{\rm C}+i\mathbf{K}_{\rm C}$ and $\mathbf{J}_{\rm C}-i\mathbf{K}_{\rm C}$ operate only on the corresponding block-diagonal $2 \times 2$ matrices as their explicit representations are
\begin{align}
\mathbf{J}_{\rm C}+i\mathbf{K}_{\rm C}=
	\begin{pmatrix}
	0 & 0\\
	0 & \boldsymbol\sigma\\
	\end{pmatrix}, \nonumber \\
\mathbf{J}_{\rm C}-i\mathbf{K}_{\rm C}=
	\begin{pmatrix}
	\boldsymbol\sigma & 0\\
	0 & 0 \\
	\end{pmatrix}. \nonumber \\
\label{chirality}
\end{align}
Such decoupling in chiral representation may be understood from the transformation of the Lorentz group algebra
with a single invariant $SU(2)$ sub-algebra of the rotation $\mathbf{J}$ into a two decoupled invariant $SU(2) \times SU(2)$
sub-algebra by defining a pair of specific combinations between rotation and boost given by
\begin{align}\label{eqn:A_B_generator}
  \mathbf{A}&=\frac{1}{2}(\mathbf{J}+i\mathbf{K}), \\
  \mathbf{B}&=\frac{1}{2}(\mathbf{J}-i\mathbf{K}),
\end{align}
which satisfy the following commutation relations:
\begin{align}\label{eqn:A_B_commutation}
  [A_{i},A_{j}]&=i\epsilon_{ijk}A_{k}, \nonumber\\
  [B_{i},B_{j}]&=i\epsilon_{ijk}B_{k}, \nonumber\\
  [A_{i},B_{j}]&=0, (i,j,k=1,2,3),
\end{align}
where $\mathbf{A}$ and $\mathbf{B}$ each generates a corresponding $SU(2)$ group algebra.
The above specific combinations between rotation and boost may suggest the two decoupled helical
motions of the particle that may be denoted as the right-handed vs. left-handed chirality.
Such idea motivates to consider the transformation ${\it S}$ given by Eq.~(\ref{eqn:SR_CR_fundamental_relation})
between the standard representation ``S" and the chiral representation ``C" discussed above.
The irreducible representation may then be labeled by two angular momenta $(j,j')$
in the decoupled Lorentz group given by $SU(2)\times SU(2)$, where
$j$ and $j'$ denote the quantum numbers corresponding to each individual $SU(2)$ sub-group
consisted of $\mathbf{A}$ and $\mathbf{B}$ generators, respectively.
In the case that one of the two angular momenta is absent (or zero), $(j,j')$ corresponds to
\begin{alignat}{2}
  (0,j)\rightarrow \mathbf{J}&=-i\mathbf{K} &\quad (\mathbf{A}=0), \label{eqn:A_0_J_K}  \\
  (j,0)\rightarrow \mathbf{J}&=i\mathbf{K}  &(\mathbf{B}=0), \label{eqn:B_0_J_K}
\end{alignat}
as easily recognized in the chiral representation given by Eq.~(\ref{chirality}).
Due to such a transparent decoupling, we will write all of our spinors for this work in the $(0,J)\oplus(J,0)$ chiral representation of the Lorentz group.
Corresponding $(0,J)\oplus(J,0)$ standard representations can be found straightforwardly using the transformation matrix ${\it S}$
given by Eq.~(\ref{eqn:SR_CR_fundamental_relation}):
\begin{align}
  \Psi_{\rm S}={\it S}\,\Psi_{\rm C}=\dfrac{1}{\sqrt{2}}
	\begin{pmatrix}
	1 & 1\\
	1 & -1\\
	\end{pmatrix}
	\begin{pmatrix}
			\phi_{R}\\
			\phi_{L}
	\end{pmatrix}
	=\dfrac{1}{\sqrt{2}}
	\begin{pmatrix}
			\phi_{R}+\phi_{L}\\
			\phi_{R}-\phi_{L}
	\end{pmatrix},\label{eqn:SR_CR_relation}
\end{align}
where $\phi_{R}$ and $\phi_{L}$ are the right-handed and left-handed components in the chiral representation.
As one can easily get the corresponding standard representation using the above relation,
we won't list them explicitly in this work.

In Fig.~\ref{fig:spinors_overview}, we also denote the so-called ``helicity spinors'' and ``Dirac spinors'' as ``H" and ``D".
They represent the spinors obtained by two different procedures.
Following the procedure laid out by Jacob and Wick~\cite{Jacob1959} that defines the helicity in the IFD
and using the kinematic transformations defined in our previous works~\cite{Ji2013,Ji2015},
we may now define the helicity applicable to any arbitrary interpolation angle $\delta$.
To define the helicity spinor in IFD, Jacob and Wick \cite{Jacob1959} started with a state at rest
having a spin projection along the $z$ direction equal to the desired helicity,
then boosted it in the $z$ direction to get the desired magnitude of momentum $|{\vec P}|$, and
then rotated it subsequently to get the momentum and spin projection in the desired direction.
We follow the same procedure in an arbitrary interpolation angle $\delta$, replacing
the kinematic generators $J^1$ and $J^2$ in IFD by the corresponding kinematic generators
$\mathcal{K}^{\itP{1}}$ and $\mathcal{K}^{\itP{2}}$  given by~\cite{Ji2013,Ji2015}:
\begin{subequations}
		\label{eqn:mcK_1 and mcK_2}
		\begin{align}
				\mathcal{K}^{\itP{1}}&=-K^{1}\sin\delta-J^{2}\cos\delta,\nonumber \\
				\mathcal{K}^{\itP{2}}&=J^{1}\cos\delta-K^{2}\sin\delta,
		\end{align}
\end{subequations}
where the interpolating kinematic operators $\mathcal{K}^{\itP{1}}$ and $\mathcal{K}^{\itP{2}}$  coincide with the usual $E^{1}$ and $E^{2}$ of LFD
modulo sign for $\delta = \pi/4$.  As extensively discussed in Ref.~\cite{Ji2001}, the transverse rotations ($J^1, J^2$) are kinematic in IFD ($\delta = 0$),
while the LF transverse boosts ($E^1, E^2$) are kinematic in LFD ($\delta = \frac{\pi}{4}$).
The procedure set by Jacob and Wick~\cite{Jacob1959} in IFD can thus be generalized to any interpolation angle $\delta$
by the transformation matrix $T$ given by
\begin{align}\label{eqn:T_transformation_for_any_interpolation_angle}
  T=T_{12}T_{3}=e^{i\beta_{1}\mathcal{K}^{\itP{1}}+i\beta_{2}\mathcal{K}^{\itP{2}}}e^{-i\beta_{3}K^{3}},
\end{align}
where the values of $\beta_1, \beta_2, \beta_3$ for an interpolation angle $\delta$ are related to the desired final momentum
${\vec P}=(P^1,P^2,P^3)$ of the particle with mass $M$. The detailed derivation of the relationship between $(\beta_1, \beta_2, \beta_3)$ for a given $\delta$ and the 4-momentum components $P^{\muT}$ has been worked out  in our previous work~\cite{Ji2015}
and may be summarized as
  \begin{align}
    P^{\pT}&=\left(\cos\delta\cosh\beta_{3}+\sin\delta\sinh\beta_{3}\right)M, \nonumber \\
    P^{\itP{1}}&=\beta_{1}\frac{\sin\alpha}{\alpha}\left(\sin\delta\cosh\beta_{3}+\cos\delta\sinh\beta_{3}\right)M, \nonumber \\
    P^{\itP{2}}&=\beta_{2}\frac{\sin\alpha}{\alpha}\left(\sin\delta\cosh\beta_{3}+\cos\delta\sinh\beta_{3}\right)M, \nonumber \\
    P^{\mT}&= \frac{\Ss P^{\pT} - P_{\mT}}{\Cc},
  \label{eqn:four_momentum_transformation_from_rest_frame_any_interpolation_angle}
 % \label{eqn:4_momentum_transformation_from_rest_d}
  \end{align}
where $\Ss = \sin2\delta$, $\Cc=\cos2\delta$ and $P_{\mT}=\cos\alpha\left(\sin\delta\cosh\beta_{3}+\cos\delta\sinh\beta_{3}\right)M$ with $\alpha\equiv\sqrt{\Cc(\beta_{1}^{2}+\beta_{2}^{2})}$.
Here, we kept the ``\textasciicircum'' notation for the transverse momentum $P^{\itP{j}} (j=1,2)$ due to the apparent $\delta$ dependence in this equation although we will usually omit it as mentioned earlier.

From this set of equations in Eq.~(\ref{eqn:four_momentum_transformation_from_rest_frame_any_interpolation_angle}),
one can get a few equivalent useful relations between parameters $\beta_{1}, \beta_{2}, \beta_{3}, \alpha$ and the momentum components
as shown in Eq.~(\ref{eqn:useful_beta_P_relation}).
It may be worth to note that $P_{\mT}$ plays the role of the longitudinal momentum and
the factor $\left(\sin\delta\cosh\beta_{3}+\cos\delta\sinh\beta_{3}\right)$
in the 3-momentum ($P^{\itP{1}}, P^{\itP{2}}$, $P_{\mT}$) is due to the first boost $T_{3}=e^{-i\beta_{3}K^{3}}$.
The ``helicity spinors'' denoted by ``H" are thus obtained by applying the transformation $T$ given by Eq.~(\ref{eqn:T_transformation_for_any_interpolation_angle}) to the initial state at rest that has a spin projection along the $z$ direction.
On the other hand, the Dirac spinors denoted by ``D" are obtained by directly boosting the initial state at rest that has a spin projection
along the $z$ direction to the state with the desired momentum ${\vec P}=(P^1,P^2,P^3)$. The Dirac spinors in IFD are
the familiar spinors that show up in many textbooks. In IFD, however, the spin direction of the Dirac spinor is in general not aligned
to the moving direction (i.e. the 3-momentum direction or in short the momentum direction) of the particle represented by the same Dirac spinor while the spin direction of the helicity spinor is always aligned
to the momentum direction of the particle represented by the same helicity spinor.  The reason for this alignment between the spin
and momentum direction of the particle represented by ``H" in IFD is because the dynamical operation $T_3$ in Eq.~(\ref{eqn:T_transformation_for_any_interpolation_angle}) is made in the same direction as the initial state spin direction, i.e. the $z$ direction, and followed up by the entirely kinematical operation $T_{12}$, i.e. the transverse rotation in IFD, under which the alignment of the spin and
momentum direction is intact.  One should note that
the boost operation to get the ``D" spinors is entirely dynamical in contrast to the ``H" spinors, which would be the case
even if the interpolation angle $\delta$ is away from the IFD. Although one may still consider such interpolating ``D" representation
for any $\delta$, the operation to get those interpolating ``D" spinors is too dynamical to lend any useful discussion for the present work.
Thus, we will refer the ``D" spinors in this work to the familiar Dirac spinors in IFD only.
% The spins of the Dirac spinors are always orientated in the $\pm z$ direction regardless of the momentum of the particle.
We exhibit them for spin $J=\frac{1}{2}$ in Appendix~\ref{sec:dirac_spinor_in_chiral_representation} using the $(0,J)\oplus(J,0)$
chiral representation.

To study the properties of a spinor in general, we need to understand that a spinor for a particle is characterized by two pieces of information: the momentum of the particle, and the spin orientation. As discussed above, the helicity spinor defined in the IFD has the spin of the particle either aligned or anti-aligned with its momentum direction.
However, the generalized helicity spinor has its spin oriented at some angle away from the momentum direction in general
because the interpolating kinematic operators $\mathcal{K}^{\itP{1}}$ and $\mathcal{K}^{\itP{2}}$ given by Eq.~(\ref{eqn:mcK_1 and mcK_2}) in
the $T_{12}$ transformation involves the transverse rotations as well as the transverse boosts.
In this work, we will derive how this spin orientation angle changes with the interpolation angle.
In particular, it is useful to find a transformation which relates the usual Dirac spinor in IFD, i.e. the ``D" spinor defined in this work,
with the generalized helicity spinor because the initial state operated by the $T$ transformation given by Eq.~(\ref{eqn:T_transformation_for_any_interpolation_angle}) to get the ``H" spinor is nothing but the same initial state operated by
the boost to get the ``D" spinor and thus the inverse boost on the ``D" spinor can be subsequently operated by $T$ to link
the ``D" spinor with the ``H" spinor. Such transformation for the LFD helicity spinor is the well-known Melosh transformation~\cite{Melosh1974},
often referred as the connection between the Dirac spinors and the light-front spinors. In Fig.~\ref{fig:spinors_overview},
this Melosh transformation is schematically denoted by the dashed purple arrow from ``D" at $\delta = 0$ to ``H" at $\delta = \pi/4$.
In this work, we will work out the detailed derivation of a ``generalized Melosh transformation" that links the usual
``D" spinor to any interpolating ``H" spinor.

For explicit demonstration of the whole landscape picture for the helicity amplitude-squares, i.e. helicity probabilities, depending on the reference frame as well as the interpolation angle, we compute all 16 helicity probabilities not only for a fermion-fermion scattering process analogous to ``$e \mu \rightarrow e \mu$" but also for a fermion and anti-fermion pair annihilation and creation process analogous to ``$e^+ e^- \rightarrow \mu^+ \mu^-$" focusing on the effects from the initial and final fermion degrees of freedom. From this picture, one can see clearly that the helicity probabilities in LFD is independent of the reference frames while
all other interpolating helicity probabilities for $0\leq\delta < \pi/4$ do depend on the reference frames. This can be understood as a direct consequence of the remarkable character change in the longitudinal boost $K^3$ from ``dynamic" for $0\leq\delta < \pi/4$
to ``kinematic" at $\delta=\pi/4$. A simple example of the helicity that depends on the reference frame may be the helicity of the particle in IFD:
e.g. the particle moving in the $z$ direction with the spin parallel to the $z$ direction for the observer at rest would be seen as moving in the $-z$ direction with the spin still parallel to the $z$ direction for the observer moving faster than the particle in the $z$ direction so that the helicity would flip
from +1 in the rest frame to -1 in the moving frame. If the helicity is kept as a fixed value, then the spin direction would flip depending on the reference frame. Unless $\delta=\pi/4$ or LFD, the interpolating helicity is in general dependent on the reference frames as gleaned from this
simple example.
Thus, the landscape of the helicity probabilities will exhibit the boundaries that depend on the reference frame and the interpolation angle,
indicating the change of spin configurations for the particles involved in the process for the helicity probabilities.
In fact, we find that the angle difference between the momentum and the spin directions as a function of the momentum direction
bifurcates at a critical interpolation angle as shown in Fig.~\ref{fig:ang_diff_bifurcate}.
%
%{\color{red}\st{The boundary of interpolation angle indicates the bifurcation of the helicity amplitudes and consequently helicity probabilities
%}} \textcolor{cyan}{These boundaries can also be understood as due to the bifurcation of how the spin orientation angle $\theta_{s}$ changes with the momentum direction $\theta$, as shown in Fig.~{\ref{fig:thetas_on_theta_delta}} and Fig.~{\ref{fig:ang_diff_bifurcate}.
This bifurcation at the critical interpolation angle separates the branch that the LFD belongs to from the branch that the IFD belongs to.
This further clarifies the confusion in the prevailing notion of the equivalence between the IMF in IFD and the LFD,
demanding a clear distinction between the IMF and the LFD.
Moreover, it is interesting to note that the particular correlation, coined as the J-curve, between the total momentum of the system and the interpolation angle found in the scattering amplitude of the scalar field theory~\cite{Ji2013} as well as the sQED theory~\cite{Ji2015} persists even in the helicity amplitudes analyzed in the present work. Remarkably, the J-curve appears always within the two boundaries that define the changes of the initial two particle spin configurations in each and every helicity amplitude. We discuss the singular behavior of this remarkable correlation in conjunction with the zero-mode issue in the LFD.

The rest of the paper is organized as follows.
% In Appendix ~\ref{sec:Review_of_Interpolation_Angle}, we give a very brief review of the interpolation angle method and the generalized Poincar\'e algebra for arbitrary interpolation angle.
In Sec.~\ref{sec:helicity_operator_and_spinor_for_any_interpolation_angle}, we derive the generalized helicity operator for any interpolation angle and the corresponding $(0,\frac{1}{2})\oplus(\frac{1}{2},0)$ helicity spinors.
The spin orientations of these spinors are discussed in Sec.~\ref{sec:spin_orientation_and_generalized_melosh_transformation},
considering the two subsequent operations $B(\boldsymbol\eta)\mathcal{D}(\hat{\mathbf{m}},\theta_{s})$ on a spin-up spinor in the rest frame with the first operation $\mathcal{D}(\hat{\mathbf{m}},\theta_{s})=e^{-i\hat{\mathbf{m}}\cdot\mathbf{J}\theta_{s}}$ that rotates the spin around axis $\hat{\mathbf{m}}=(-\sin\phi_{s},\cos\phi_{s},0)$ by angle $\theta_{s}$ and the second operation $B(\boldsymbol\eta)=e^{-i\boldsymbol\eta\cdot\mathbf{K}}$
that boosts the spinor to momentum $\mathbf{P}$.
In Sec.~\ref{sec:spin_orientation_and_generalized_melosh_transformation}, we also derive the generalized Melosh transformation that relates these generalized helicity spinors to the Dirac spinors.
In Sec.~\ref{sec:interpolating_helicity_probabilities_for_fermion_scattering}, we calculate the generalized helicity probabilities for the
fermion-fermion scattering process analogous to ``$e \mu \rightarrow e \mu$" and plot their results in terms of both the total momentum $P^{z}$ of the system and the interpolation angle $\delta$ to reveal the entire landscape of the frame dependence and interpolation angle dependence of these probabilities.
%and also for the fermion and anti-fermion pair annihilation and creation
%process such as $e^{+}e^{-}\rightarrow\mu^{+}\mu^{-}$}},
% We observe that the same J curve that appeared in the time-ordered diagrams in the sQED theory in our previous work also exists here in every helicity amplitude.
% A detailed analysis of two boundaries across which the helicity amplitudes would suddenly change can also be found in the same section.
The summary and conclusions follow in Sec.~\ref{sec:summary_and_conclusions}.

For the completeness and clarity, we added several Appendices.
% Since the interpolation method we are using here were developed in our previous work \cite{Ji2001, Ji2012, Ji2015}, we won't repeat them here.
Although
the readers may refer to our previous work \cite{Ji2001, Ji2013, Ji2015} for the detailed discussion of the interpolation method,
a brief review of the interpolation angle method with some useful formulae for this paper can be found in Appendix~\ref{Appendix:Review_of_Interpolation_Angle}. We exhibit the explicit matrix form of operators $T$ and $B(\boldsymbol\eta)\mathcal{D}(\hat{\mathbf{m}},\theta_{s})$ in Appendix~\ref{sec:matrix_representation_of_T_and_BR}, the $(0,\frac{1}{2})\oplus(\frac{1}{2},0)$ helicity spinors for anti-particles in Appendix~\ref{sec:helicity_spinors_for_anti_particles}, and the Dirac spinors in Appendix~\ref{sec:dirac_spinor_in_chiral_representation}.
In Appendices~\ref{sec:generalized_helicity_spinors_for_arbitrary_interpolation_angle_up_to_spin_2} and \ref{Appendix:generalized_melosh_transformation},
we list the $(0,J)\oplus(J,0)$ helicity spinors and corresponding generalized Melosh transformations for higher spins up to $J=2$.
In Appendix~\ref{sec:apparent-spin}, the ``apparent" spin orientation angle $\theta_a$ plotted in Fig.~\ref{fig:LF_profile1} is derived for the discussion presented in Sec.~\ref{sec:spin_orientation_and_generalized_melosh_transformation}.
The plots of the scattering amplitudes corresponding to the generalized helicity probabilities presented in Sec.~\ref{sec:interpolating_helicity_probabilities_for_fermion_scattering}
are shown in Appendix~\ref{sec:Interpolated_scattering_amplitudes}.
We also calculate and plot the generalized helicity amplitudes and probabilities for the fermion and anti-fermion pair annihilation and creation process analogous to ``$e^{+}e^{-}\rightarrow\mu^{+}\mu^{-}$" in Appendix~\ref{sec:Interpolated_annihilation_amplitudes_and_probabilities}.

% section introduction (end)
\section{Helicity Operator and Helicity Spinor for Any Interpolation Angle}
\label{sec:helicity_operator_and_spinor_for_any_interpolation_angle}

%\subsection{Helicity Operator}
%\label{sec:helicity_operator}

As introduced in Sec.~\ref{sec:introduction}, the helicity spinors denoted by ``H" are obtained by applying
the transformation $T$ given by Eq. (\ref{eqn:T_transformation_for_any_interpolation_angle}) to the initial state at rest
that has a spin projection along the $z$ direction. We may denote a generalized helicity spinor in a given interpolation angle $\delta$
as $| p; j, m\rangle_\delta$ for a particle of spin $j$ moving with momentum $p$ and helicity $m$. This state $| p; j, m\rangle_\delta$
is obtained by the transformation $T$ from the spin eigenstate $|0; j,m\rangle$ at rest, which has a spin projection along the $z$ direction
satisfying $J_3 |0; j,m\rangle = m |0; j,m\rangle$. Thus, we may specify $| p; j, m\rangle_\delta =T | 0; j,m \rangle$.

Following the procedure of Leutwyler and Stern~\cite{Leutwyler1978}, we may then define a new spin operator $\mathcal{J}_{i}$
for a moving particle as $\mathcal{J}_{i}=TJ_{i}T^{-1}$ to get
\begin{align}\label{eqn:J_for_nonzero_p}
  \mathcal{J}_{3}| p; j, m\rangle_\delta =   T{J}_{3}T^{-1}T | 0; j, m \rangle = m | p; j, m\rangle_\delta,
\end{align}
where $m$ is now not only the eigenvalue of the ordinary spin operator $J_3$ for the initial state at rest  $| 0; j, m \rangle$
but also the eigenvalue of the operator $\mathcal{J}_{3}$ for the generalized helicity spinor state $| p; j, m\rangle_\delta$.
It is straightforward to verify that $\mathcal{J}_{i}$ satisfies the SU(2) algebra as $J_i$ does:
\begin{eqnarray}
[{\cal J}_i,{\cal J}_j] &=& T{J_i}{J_j}T^{-1} - T{J_j}{J_i}T^{-1}\nonumber \\
&=& T[J_i,J_j]T^{-1} \nonumber \\
&=&
i\epsilon_{ijk}T{J_k}T^{-1} \nonumber \\
&=& i\epsilon_{ijk}{{\cal J}_k}.
\end{eqnarray}
As we have shown in the previous work~\cite{Ji2001}, the new spin operator $\mathcal{J}_{i}$ commutes with
the mass operator $M$ defined by $M^2=P^{\muT}P_{\muT}=P_{\pT}^{2}\Cc-P_{\mT}^{2}\Cc+2P_{\pT}P_{\mT}\Ss-\mathbf{P}_{\perp}^{2}$
for any generalized helicity spinor state, i.e. $[{\mathcal J}_i,M]|p; j, m\rangle_\delta = 0$.
The operator $\mathcal{J}_{3}$ intermediates between the usual Jacob-Wick helicity operator in IFD and the light-front helicity operator in LFD
and thus offers the role of general helicity operator in-between for any interpolation angle $\delta$ as we now discuss below.

Using the Poincar\'e algebra for any arbitrary interpolation angle~\cite{Ji2001}, we find that
 the new spin operator written in terms of the parameters $\beta_{1}$, $\beta_{2}$ and $\beta_{3}$ remains unchanged
 with or without including $T_3$~\cite{Ji2001} as $[J^{3},K^{3}]=0$ and is given by
\begin{align}\label{eqn:J_beta_1_2_3}
  \mathcal{J}_{3}=J_{3}\cos\alpha+(\beta_{1}\mathcal{K}^{\itP{2}}-\beta_{2}\mathcal{K}^{\itP{1}})\dfrac{\sin\alpha}{\alpha}.
\end{align}
We can then use the relations in Eq.~(\ref{eqn:useful_beta_P_relation}) to rewrite it in terms of the particle's momentum, and get
\begin{align}
  \mathcal{J}_{3}=\dfrac{1}{\Pp}(P_{\mT}J_{3}+P^{1}\mathcal{K}^{\itP{2}}-P^{2}\mathcal{K}^{\itP{1}}),\label{eqn:J3_P}
\end{align}
where $\Pp\equiv\sqrt{(P^{\pT})^{2}-M^{2}\Cc}=\sqrt{P_{\mT}^{2}+\mathbf{P}_{\perp}^{2}\Cc}$.
It is interesting to note that this operator $\mathcal{J}_{3}$ can also be written in terms of the Pauli-Lubanski operator \cite{Leutwyler1978}
$W^{\mu}=\frac{1}{2}\epsilon^{\mu\nu\alpha\beta}P_{\nu}M_{\alpha\beta}$ simply as $\mathcal{J}_{3}=W^{\pT}/\Pp$.
In the instant form limit ($\delta\rightarrow0$), $\mathcal{K}^{\itP{1}}\rightarrow-J^{2}$, $\mathcal{K}^{\itP{2}}\rightarrow J^{1}$, $P_{\mT}\rightarrow P^{3}$ and $\Pp\rightarrow \sqrt{(P^{0})^{2}-M^{2}}=|\mathbf{P}|$, and thus the operator $\mathcal{J}_{3}$ coincides with the familiar IFD helicity operator
$\mathbf{P}\cdot\mathbf{J}/|\mathbf{P}|$. In the light-front limit ($\delta\rightarrow \pi/4$),
$\mathcal{K}^{\itP{1}}\rightarrow -E_{1}$, $\mathcal{K}^{\itP{2}}\rightarrow -E_{2}$, $P_{\mT}\rightarrow P^{+}$ and $\Pp\rightarrow \sqrt{(P^{+})^{2}}=
P^+$, and thus the operator $\mathcal{J}_{3}$ coincides with the light-front helicity operator $J_3 + \frac{1}{P^+}(P^2 E_1 - P^1 E_2)$
as discussed in Ref.~\cite{CarlsonJi2003}. Thus, $\mathcal{J}_{3}$ intermediates between the usual Jacob-Wick helicity operator in IFD and the light-front helicity operator in LFD and it is reasonable to identify the operator $\mathcal{J}_{3}$ as
the general helicity operator for any interpolation angle $\delta$. The helicity eigenvalue for the state $| p; j, m\rangle_\delta$ is $m$
as previously given in Eq.~(\ref{eqn:J_for_nonzero_p}). One should note, however, that the generalized helicity defined
by the operator $\mathcal{J}_{3}$ agrees with the ordinary notion of helicity defined usually by the spin parallel or anti-parallel to the
particle momentum direction only in the IFD. For different interpolation angles, in general, there's a relative angle between the spin orientation and the momentum direction according to the generalized helicity designated by $\mathcal{J}_{3}$. In LFD, the transverse light-front boost operators $E_1$ and $E_2$ involve the rotations and they generate the angle between the spin orientation and the momentum direction.

If the particle is moving in $+z$ or $-z$ direction, so that $P^{1}=P^{2}=0$, then the generalized helicity operator given by Eq.~(\ref{eqn:J3_P}) becomes
\begin{align}
  \mathcal{J}_{3}=\dfrac{P_{\mT}J_{3}}{\Pp}=\dfrac{P_{\mT}}{|P_{\mT}|}J_{3}.\label{eqn:J3_P_z_direction}
\end{align}
Thus, for an arbitrary interpolation angle, the helicity sign of a particle moving in $\pm z$ direction depends on the sign of $P_{\mT}$.
In the light-front limit, $P_{\mT}\rightarrow P^{+}$ which is always positive, and thus the light-front helicity
of the particle is positive once the spin is parallel to the $+z$ direction regardless whether the particle is moving in the $+z$ direction
or the $-z$ direction. This is dramatically different from the ordinary helicity defined in the IFD where $P_{\mT}\rightarrow P^{3}$.
For a particle moving in the $-z$ direction, the light-front helicity and the ordinary Jacob-Wick helicity is therefore opposite to each other.
The swap of helicity amplitudes caused by such dramatic difference between the light-front helicity and the ordinary Jacob-Wick helicity
has been noticed previously in deeply virtual Compton scattering process~\cite{JiBakker2013}.
We will discuss the sign flip effect of $P_{\mT}$ factor in Eq.~(\ref{eqn:J3_P_z_direction}) for the more general directions of
the particle spin and momentum in the next section, Sec.~\ref{sec:spin_orientation_and_generalized_melosh_transformation}.
% subsection interpolation_angle_dependent_boost_on_four_momentum (end)

As mentioned in Sec.~\ref{sec:introduction}, we write all of our spinors in the $(0,J)\oplus(J,0)$ chiral representation of the Lorentz group
due to a clear decoupling between the right-handed and left-handed components in the chiral representation.
In this representation, the transformation $T$ given by Eq.~(\ref{eqn:T_transformation_for_any_interpolation_angle}) decouples
as a block diagonal matrix given by
\begin{align}\label{eqn:T12_in_CR}
  T=
  \begin{pmatrix}
  T_{R} & 0\\
  0 & T_{L}
  \end{pmatrix}.
\end{align}
For spin $1/2$ fermions in the chiral representation $(0,\frac{1}{2})\oplus(\frac{1}{2},0)$ of the Lorentz group,
the right-handed part $T_{R}$ corresponds to the $(0,\frac{1}{2})$ representation ($\mathbf{A}=0$): $\mathbf{K}=i\mathbf{J}=i\boldsymbol\sigma/2$.
Plugging $\mathbf{K}$ and $\mathbf{J}$ into Eq.~(\ref{eqn:T_transformation_for_any_interpolation_angle}), we get
\begin{align}\label{eqn:T_RighHanded}
  T_{R}=e^{\mathbf{b}_{\perp}\cdot\boldsymbol\sigma_{\perp}/2}e^{\beta_{3}\sigma_{3}/2},
\end{align}
where $\mathbf{b}_{\perp}=\left(\beta_{1}\sin\delta+i\beta_{2}\cos\delta, -i\beta_{1}\cos\delta+\beta_{2}\sin\delta\right)$.
This transforms the right-handed component of the spinor.
Using the relation $\left(\mathbf{b}_{\perp}\cdot\boldsymbol\sigma_{\perp}\right)^{2}=-\alpha^{2}\cdot I$ and $\sigma_{3}^{2}=I$ where
we recall $\alpha = \sqrt{\Cc(\beta_{1}^{2}+\beta_{2}^{2})}$, we obtain
\begin{align}\label{eqn:T_RightHanded_representation}
  T_{12R}
  &=e^{\mathbf{b}_{\perp}\cdot\boldsymbol\sigma_{\perp}/2}
  =\cos\left(\frac{\alpha}{2}\right)\cdot I + \frac{\mathbf{b}_{\perp}\cdot\boldsymbol\sigma_{\perp}}{\alpha}\sin\left(\frac{\alpha}{2}\right),\\[5pt]
  T_{3R}&=e^{\beta_{3}\sigma_{3}/2}=
  \begin{pmatrix}
    e^{\beta_{3}/2} & 0 \\
    0 & e^{-\beta_{3}/2}
  \end{pmatrix}.
\end{align}
% So
% \begin{align}\label{eqn:TR_TL_Chiral_Representation_Any_Interpolation_Angle}
%   T_{R}&=T_{12R}T_{3R}\nonumber\\
%   &=
%   \resizebox{\columnwidth-30pt}{!}{
%     \begin{pmatrix}
%     \cos\dfrac{\alpha}{2}e^{\beta_{3}/2} &  -\dfrac{\beta_{L}\left(\cos\delta-\sin\delta\right)}{\alpha}\sin\dfrac{\alpha}{2}e^{-\beta_{3}/2}\\[8pt]
%     \dfrac{\beta_{R}\left(\sin\delta+\cos\delta\right)}{\alpha}\sin\dfrac{\alpha}{2}e^{\beta_{3}/2} & \cos\dfrac{\alpha}{2}e^{-\beta_{3}/2}
%     \end{pmatrix}.
%   }
% \end{align}
Similarly, the left-handed part $T_{L}$ corresponds to the $(\frac{1}{2},0)$ representation ($\mathbf{B}=0$): $\mathbf{K}=-i\mathbf{J}=-i\boldsymbol\sigma/2$.
Thus, we get
\begin{align}\label{eqn:T_LeftHanded}
  T_{L}=e^{-\mathbf{b}^{\ast}_{\perp}\cdot\boldsymbol\sigma_{\perp}/2}e^{-\beta_{3}\sigma_{3}/2},
\end{align}
where  $\mathbf{b}^{\ast}_{\perp}=\left(\beta_{1}\sin\delta-i\beta_{2}\cos\delta, i\beta_{1}\cos\delta+\beta_{2}\sin\delta\right)$.
With this, we obtain
\begin{align}\label{eqn:T_LeftHanded_representation}
  T_{12L}
  &=e^{-\mathbf{b}^{\ast}_{\perp}\cdot\boldsymbol\sigma_{\perp}/2}
  =\cos\left(\frac{\alpha}{2}\right)\cdot I - \frac{\mathbf{b}^{*}_{\perp}\cdot\boldsymbol\sigma_{\perp}}{\alpha}\sin\left(\frac{\alpha}{2}\right),\\[5pt]
  T_{3L}&=e^{-\beta_{3}\sigma_{3}/2}=
  \begin{pmatrix}
    e^{-\beta_{3}/2} & 0 \\
    0 & e^{\beta_{3}/2}
  \end{pmatrix}.
\end{align}
The explicit matrix form of $T$ can be found in Appendix~\ref{sec:matrix_representation_of_T_and_BR}.
To obtain the helicity spinor with arbitrary momentum, we apply this $T$ transformation on a spin eigenstate in its rest frame.
The definition for helicity operator $\mathcal{J}_{3}$ in Eq.~(\ref{eqn:J_for_nonzero_p}) assures that if $| 0; j, m \rangle$ is an eigenstate of $J_{3}$, then $T| 0; j, m \rangle$ is an eigenstate of $\mathcal{J}_{3}$.
Because here we use the chiral representation, the spinors in the rest frame are given by
\begin{alignat}{2}
  &u^{(\sfrac{1}{2})}(0)=
  \begin{pmatrix}
    \sqrt{M}\\
    0\\
    \sqrt{M}\\
    0
  \end{pmatrix},
  &\quad &u^{(-\sfrac{1}{2})}(0)=
  \begin{pmatrix}
    0\\
    \sqrt{M}\\
    0\\
    \sqrt{M}
  \end{pmatrix}, \label{eqn:dirac_CR_spinor_basis_rest_frame}
\end{alignat}
with the normalization $\bar{u}^{(\lambda)}u^{(\lambda)}=2M$.
After the $T$ transformation given by Eq.~(\ref{eqn:T12_in_CR}), we obtain
\begin{subequations}
  \label{eqn:u1u2_beta}
  \begin{align}
    u^{(\sfrac{1}{2})}_{H}(\beta)&=\sqrt{M}
    \begin{pmatrix}
    \cos\dfrac{\alpha}{2}e^{\beta_{3}/2}\\[8pt]
    \dfrac{\beta_{R}\left(\sin\delta+\cos\delta\right)}{\alpha}\sin\dfrac{\alpha}{2}e^{\beta_{3}/2}\\[8pt]
    \cos\dfrac{\alpha}{2}e^{-\beta_{3}/2}\\[8pt]
    \dfrac{\beta_{R}\left(\cos\delta-\sin\delta\right)}{\alpha}\sin\dfrac{\alpha}{2}e^{-\beta_{3}/2}
    \end{pmatrix},\label{eqn:u1_beta}\\
    u^{(-\sfrac{1}{2})}_{H}(\beta)&=\sqrt{M}
    \begin{pmatrix}
    -\dfrac{\beta_{L}\left(\cos\delta-\sin\delta\right)}{\alpha}\sin\dfrac{\alpha}{2}e^{-\beta_{3}/2}\\[8pt]
    \cos\dfrac{\alpha}{2}e^{-\beta_{3}/2}\\[8pt]
    -\dfrac{\beta_{L}\left(\sin\delta+\cos\delta\right)}{\alpha}\sin\dfrac{\alpha}{2}e^{\beta_{3}/2}\\[8pt]
    \cos\dfrac{\alpha}{2}e^{\beta_{3}/2}
    \end{pmatrix},\label{eqn:u2_beta}
  \end{align}
\end{subequations}
where $\beta_{R}=\beta_{1}+i\beta_{2}$ and $\beta_{L}=\beta_{1}-i\beta_{2}$.
The subscript ``H" is used to denote that these are the generalized helicity spinors as mentioned in Sec.~\ref{sec:introduction}
to distinguish them from the Dirac spinors that we will discuss in the next section,
Sec.~\ref{sec:spin_orientation_and_generalized_melosh_transformation}.
With the relations listed in Eq.~(\ref{eqn:useful_beta_P_relation}), these spinors can be rewritten in terms of the particle's momentum:
\begin{subequations}
 \label{eqn:u1u2_P}
 \begin{align}
  u^{(\sfrac{1}{2})}_{H}(P)&=
  \begin{pmatrix}
  \sqrt{\dfrac{P_{\mT}+\Pp}{2 \Pp}}
  \sqrt{\dfrac{P^{\pT}+\Pp}{\left(\sin\delta+\cos\delta\right)}}\\[16pt]
  P^{R}\sqrt{\dfrac{\sin\delta+\cos\delta}{2 \Pp(\Pp+P_{\mT})}}
  \sqrt{P^{\pT}+\Pp}\\[16pt]
  \sqrt{\dfrac{P_{\mT}+\Pp}{2 \Pp}}
  \sqrt{\dfrac{P^{\pT}-\Pp}{\left(\cos\delta-\sin\delta\right)}}\\[16pt]
  P^{R}\sqrt{\dfrac{\cos\delta-\sin\delta}{2 \Pp(\Pp+P_{\mT})}}
  \sqrt{P^{\pT}-\Pp}
  \end{pmatrix}\label{eqn:Spinor_u1_P_Any_Interpolation_Angle}
\intertext{and}
  u^{(-\sfrac{1}{2})}_{H}(P)&=
  \begin{pmatrix}
  -P^{L}\sqrt{\dfrac{\cos\delta-\sin\delta}{2 \Pp(\Pp+P_{\mT})}}
  \sqrt{P^{\pT}-\Pp}\\[16pt]
  \sqrt{\dfrac{P_{\mT}+\Pp}{2 \Pp}}
  \sqrt{\dfrac{P^{\pT}-\Pp}{\left(\cos\delta-\sin\delta\right)}}\\[16pt]
  -P^{L}\sqrt{\dfrac{\sin\delta+\cos\delta}{2 \Pp(\Pp+P_{\mT})}}
  \sqrt{P^{\pT}+\Pp}\\[16pt]
  \sqrt{\dfrac{P_{\mT}+\Pp}{2 \Pp}}
  \sqrt{\dfrac{P^{\pT}+\Pp}{\left(\sin\delta+\cos\delta\right)}}
  \end{pmatrix},\label{eqn:Spinor_u2_P_Any_Interpolation_Angle}
 \end{align}
\end{subequations}
where $P^{R}=P^{1}+iP^{2}$ and $P^{L}=P^{1}-iP^{2}$.
From $\Pp^2= (P^{\pT})^{2}-M^{2}\Cc = P_{\mT}^{2}+\mathbf{P}_{\perp}^{2}\Cc$,
we get
\begin{align}
  \sqrt{(\Pp-P_{\mT})/(\mathbf{P}_{\perp}^{2}\Cc)}=1/\sqrt{\Pp+P_{\mT}}\label{eqn:spinor_two_forms}
\end{align}
and use this equation, i.e. Eq.~(\ref{eqn:spinor_two_forms}), to obtain the above results given by Eqs.~(\ref{eqn:Spinor_u1_P_Any_Interpolation_Angle}) and
(\ref{eqn:Spinor_u2_P_Any_Interpolation_Angle}).
One should note that there are two equivalent ways of writing the same factor as given by Eq.~(\ref{eqn:spinor_two_forms}), i.e.
the left hand side or the right hand side of Eq.~(\ref{eqn:spinor_two_forms}).
The form in the right hand side is convenient for obtaining the light-front limit of these spinors since both $\Pp \rightarrow P^+$ and
$P_{\mT} \rightarrow P^+$ are positive and they cannot cancel each other.
On the other hand, $P_{\mT}$ can be negative for other interpolation angle $\delta \neq \pi/4$ and may cancel $\Pp$ to end up with a treacherous singularity in the form of the right hand side. In such case, it would be more convenient to use the form in the left hand side
as $\Cc \neq 0$ and instead take care of the limit $\mathbf{P}_{\perp}\rightarrow0$ noting that the factor $1/|\mathbf{P}_{\perp}|$ comes with
$P_{R}$ or $P_{L}$ in Eqs.~(\ref{eqn:Spinor_u1_P_Any_Interpolation_Angle}) and
(\ref{eqn:Spinor_u2_P_Any_Interpolation_Angle}) and the factor
$P_{R}/|\mathbf{P}_{\perp}|$ or $P_{L}/|\mathbf{P}_{\perp}|$ may be taken to be the unity by choosing an appropriate coordinate system.
%If the form in the right hand side is used instead, a much more careful limit would be necessary in taking $\mathbf{P}_{\perp}\rightarrow0$.

The antiparticle spinors are generated through charge conjugation as $v=C \bar{u}=i\gamma^{0}\gamma^{2}\gamma^{0}u^{*}$.
Here, the gamma matrices are written on the chiral basis as (see e.g. Ref.~\cite{Dziembowski1988a}):
% For completeness, they are listed in Appendix~\ref{app:gamma_matrices}.
\begin{align}
  \gamma^{0}=
  \begin{pmatrix}
  0 & I\\
  I & 0
  \end{pmatrix},
  \quad
  \gamma^{i}=
  \begin{pmatrix}
  0 & -\sigma^{i}\\
  \sigma^{i} & 0
  \end{pmatrix},
  \quad
  \gamma^{5}=
  \begin{pmatrix}
  I & 0\\
  0 & -I
  \end{pmatrix}.
\end{align}
The results of antiparticle spinors are summarized in Appendix~(\ref{sec:helicity_spinors_for_anti_particles}).

In the instant form limit ($\delta\rightarrow0$), $\Cc\rightarrow1$, $P^{\pT}\rightarrow P^{0}=E$, $P_{\mT}\rightarrow P^{3}$ and $\Pp\rightarrow|\mathbf{P}|$.
Using the relation
\begin{align}
  \sqrt{(E\pm|\mathbf{P}|)}=(E+M\pm|\mathbf{P}|)/\sqrt{2(E+M)},\label{eqn:E_P_M_useful_relation}
\end{align}
one can show that spinors in  Eqs.~(\ref{eqn:Spinor_u1_P_Any_Interpolation_Angle}) and
(\ref{eqn:Spinor_u2_P_Any_Interpolation_Angle})
 reduce to their instant form in chiral representation as follows:
\begin{subequations}
 \label{eqn:u1_u2_spinor_IF_CR}
 \begin{align}
  u_{H}^{(\sfrac{1}{2})}&=\sqrt{\frac{E+M}{2}}
  \begin{pmatrix}
    \dfrac{E+M+|\mathbf{P}|}{E+M}\cos\dfrac{\theta}{2}\\[8pt]
    \dfrac{E+M+|\mathbf{P}|}{E+M}\sin\dfrac{\theta}{2}e^{i\phi}\\[8pt]
    \dfrac{E+M-|\mathbf{P}|}{E+M}\cos\dfrac{\theta}{2}\\[8pt]
    \dfrac{E+M-|\mathbf{P}|}{E+M}\sin\dfrac{\theta}{2}e^{i\phi}
  \end{pmatrix}, \\[10pt]
  u_{H}^{(-\sfrac{1}{2})}&=\sqrt{\frac{E+M}{2}}
  \begin{pmatrix}
    -\dfrac{E+M-|\mathbf{P}|}{E+M}\sin\dfrac{\theta}{2}e^{-i\phi}\\[8pt]
    \dfrac{E+M-|\mathbf{P}|}{E+M}\cos\dfrac{\theta}{2}\\[8pt]
    -\dfrac{E+M+|\mathbf{P}|}{E+M}\sin\dfrac{\theta}{2}e^{-i\phi}\\[8pt]
    \dfrac{E+M+|\mathbf{P}|}{E+M}\cos\dfrac{\theta}{2}
  \end{pmatrix},
 \end{align}
\end{subequations}
where $\theta$ and $\phi$ specify the direction of the momentum $\mathbf{P}$: $\sin\theta\cos\phi=P^{1}/|\mathbf{P}|$, $\sin\theta\sin\phi=P^{2}/|\mathbf{P}|$ and $\cos\theta=P^{3}/|\mathbf{P}|$, i.e.
\begin{subequations}
  \begin{align}
    \cos(\theta/2)&=\sqrt{(|\mathbf{P}|+P^{3})/(2|\mathbf{P}|)},\label{eqn:cos_theta/2_relation_with_P}\\
    \sin(\theta/2)&=\sqrt{(|\mathbf{P}|-P^{3})/(2|\mathbf{P}|)},\\
    \exp(i\phi)&=P_{R}/|\mathbf{P}_{\perp}|,\\
    \exp(-i\phi)&=P_{L}/|\mathbf{P}_{\perp}|.
  \end{align}
\end{subequations}

In the light-front limit ($\delta\rightarrow\pi/4$), $\Cc\rightarrow0$, $P^{\pT}\rightarrow P^{+}$, $P_{\mT}\rightarrow P^{+}$, $\Pp\rightarrow P^{+}$, and $(P^{\pT}-\Pp)/\Cc \xrightarrow{\delta\rightarrow 0} M^{2}/(2P^{+})$.
One can show that the spinors in  Eqs.~(\ref{eqn:Spinor_u1_P_Any_Interpolation_Angle}) and
(\ref{eqn:Spinor_u2_P_Any_Interpolation_Angle})
 become the light-front spinors in the chiral representation \cite{Dziembowski1988a}, or identically the Kogut-Soper spinors \cite{Brodsky1998a}:
\begin{align}\label{eqn:u1_u2_LF_spinor_CR}
  &u^{(\sfrac{1}{2})}_{H}=\frac{1}{\sqrt{\sqrt{2}P^{+}}}
  \begin{pmatrix}
    \sqrt{2}P^{+}\\
    P^{R}\\
    M\\
    0
  \end{pmatrix},
  &&u^{(-\sfrac{1}{2})}_{H}=\frac{1}{\sqrt{\sqrt{2}P^{+}}}
  \begin{pmatrix}
    0\\
    M\\
    -P^{L}\\
    \sqrt{2}P^{+}
  \end{pmatrix}.
\end{align}
The $\sqrt{2}$ factors in front of $P^{+}$ in Eq.~(\ref{eqn:u1_u2_LF_spinor_CR}) are due to our definition of $P^{+}$ with the factor $1/\sqrt{2}$, i.e. $P^{+}=(P^{0}+P^{3})/\sqrt{2}$.

It is straightforward to extend our discussion for spin $1/2$ fermions and derive the generalized helicity spinors for any spin $J$.
The only thing one needs to do is to change the $\mathbf{K}$ and $\mathbf{J}$ matrices as well as the spinors in Eq.~(\ref{eqn:dirac_CR_spinor_basis_rest_frame}) from spin $1/2$ representation to spin $J$ representation.
% $2\times2$ pauli matrices to the appropriate matrix representation for spin $J$, and  in Eq.~(\ref{eqn:dirac_CR_spinor_basis_rest_frame}) to the appropiate spin $J$ representation.
The results of the generalized helicity spinors for spin up to $J=2$ are summarized in Appendix~\ref{sec:generalized_helicity_spinors_for_arbitrary_interpolation_angle_up_to_spin_2}.
It is interesting to note that the spin $1$ bosons in the chiral representation $(0,1)\oplus(1,0)$ of the Lorentz group are represented by
the column vectors of six components as given by Eq.~(\ref{eqn:spin_1}) while the spin $1$ polarization vectors such as transverse
$\epsilon^\mu(\pm,p)$ and longitudinal $\epsilon^\mu(0,p)$ are typically given by four components. We note that the typical
four-component polarization vectors correspond to the chiral representation $(1/2,1/2)$ of the Lorentz group and the relationship
between the two representations $(0,1)\oplus(1,0)$ and $(1/2,1/2)$ is analogous to the relationship between the six-component
electromagnetic fields, i.e. electric field ${\vec E}$ and magnetic field ${\vec B}$ or en masse $F^{\mu\nu}$, and the four-component electromagnetic potential, i.e. scalar potential $\Phi$ and vector potential ${\vec A}$ or en masse $A^\mu$, respectively. The details of discussion
regarding this relationship will be presented somewhere else~\cite{An-Li-Ji}.

% From this interpolation method, we can clearly see the strong connection between the helicity spinor in the instant form and the light-front spinor.
% They turn into each other smoothly through this parameter $\delta$.
% This suggests that we should really see the light-front spinor as an analog of the helicity spinors in the instant form, instead of an analog of the Dirac spinors, which we will discuss in the next subsection.

% \end{widetext}

% subsection helicity_spinor_for_any_interpolation_angle (end)
% section helicity_operator_and_spinor_for_any_interpolation_angle (end)

\section{Spin Orientation and Generalized Melosh Transformation}
\label{sec:spin_orientation_and_generalized_melosh_transformation}
\subsection{Spin Orientation for Generalized Helicity Spinors}
\label{sub:spin_orientation}
A spinor carries two pieces of information: the momentum of the particle and its spin orientation.
We may denote the momentum as $\mathbf{P}$, and the direction of momentum $\mathbf{P}$
as $(\theta, \phi)$, where $\theta$ is the angle between the momentum direction and the $z$ axis, and $\phi$ is the azimuthal angle.
Likewise, we may denote spin as $\mathbf{S}$, and the direction of spin $\mathbf{S}$
as $(\theta_{s}, \phi_{s})$. Since the direction of $\mathbf{S}$ depends on the reference frame in general,
we take $(\theta_{s}, \phi_{s})$ in the rest frame for the discussion of this Section.

We consider the following transformation on a spin-up spinor in the rest frame:
\begin{align}
  B(\boldsymbol\eta)\mathcal{D}(\hat{\mathbf{m}},\theta_{s})
	=e^{-i\boldsymbol\eta\cdot\mathbf{K}}e^{-i\hat{\mathbf{m}}\cdot\mathbf{J}\theta_{s}},
	\label{eqn:BR}
\end{align}
where the first operation $\mathcal{D}(\hat{\mathbf{m}},\theta_{s})=e^{-i\hat{\mathbf{m}}\cdot\mathbf{J}\theta_{s}}$ rotates the spin around the axis given by a unit vector $\hat{\mathbf{m}}=(-\sin\phi_{s},\cos\phi_{s},0)$ by angle $\theta_{s}$ and
then the second operation $B(\boldsymbol\eta)=e^{-i\boldsymbol\eta\cdot\mathbf{K}}$ boosts the spinor to momentum $\mathbf{P}$.
By combining these two operations, all possible spinors can be reached.
Therefore, we should be able to rewrite the $T$ transformation that we defined in Sec.~\ref{sec:helicity_operator_and_spinor_for_any_interpolation_angle} as a combination of a rotation and a boost with a suitable set of $(\theta_{s}, \phi_{s})$:
% In this subsection, we will prove that we can write $T$ operator in this form:
\begin{align}
  T=B(\boldsymbol\eta)\mathcal{D}(\hat{\mathbf{m}},\theta_{s}).
	\label{eqn:T_BR_equivalence}
\end{align}
% then by definition, the $T$ transformation is equivalent to a rotation and a boost.
% boosting a spinor whose spin already points in the $(\theta_{s}, \phi_{s})$ direction in its rest frame to momentum $\mathbf{P}$.
% Once we have spinors written in this most general form, we can compare our interpolating helicity spinors to it and extract the corresponding spin orientations.

% With our normalization convention, in chiral representation, a spinor whose spin points in the $(\theta_{s}, \phi_{s})$ direction in the rest frame can be written as
% \begin{align}\label{eqn:u1_theta_CR_rest_frame}
%   \sqrt{M}
%   \begin{pmatrix}
%   \cos\theta_{s}/2\\
%   e^{i\phi_{s}}\sin\theta_{s}/2\\
%   \cos\theta_{s}/2\\
%   e^{i\phi_{s}}\sin\theta_{s}/2
%   \end{pmatrix}.
% \end{align}
% We then apply the boost operation $e^{-i {\boldsymbol\eta}\cdot\mathbf{K}}$ on this spinor,
In the $(0,\frac{1}{2})\oplus(\frac{1}{2},0)$ chiral representation, we have
\begin{subequations}
		\begin{align}
				\mathcal{D}(\hat{\mathbf{m}},\theta_{s})
				&=e^{-i\hat{\mathbf{m}}\cdot\mathbf{J}\theta_{s}}=
				\begin{pmatrix}
						e^{-i\hat{\mathbf{m}}\cdot\boldsymbol\sigma \theta_{s}/2} & 0\\
						0 & e^{-i\hat{\mathbf{m}}\cdot\boldsymbol\sigma \theta_{s}/2}
				\end{pmatrix},\label{eqn:R(m,theta_s)}\\
				B(\boldsymbol\eta)
				&=e^{-i {\boldsymbol\eta}\cdot\mathbf{K}}
				=
				\begin{pmatrix}
						e^{\boldsymbol\eta \cdot \boldsymbol\sigma/2} & 0 \\
						0 & e^{- \boldsymbol\eta \cdot \boldsymbol\sigma/2}
				\end{pmatrix},\label{eqn:single_boost_transformation}
		\end{align}
\end{subequations}
with
\begin{subequations}
 \begin{align}
  e^{- i\hat{\mathbf{m}}\cdot\boldsymbol\sigma \theta_{s}/2}&=
   \mathbf{I} \cos\dfrac{\theta_{s}}{2} - i \hat{\mathbf{m}}\cdot\boldsymbol\sigma \sin\dfrac{\theta_{s}}{2}
	 \label{eqn:e_-i_sigma_m/2},\\
  e^{\pm \boldsymbol\sigma \cdot \boldsymbol\eta/2}&=
  \mathbf{I} \cosh\dfrac{\eta}{2} \pm \hat{\mathbf{n}}\cdot\boldsymbol\sigma \sinh\dfrac{\eta}{2}.\label{eqn:e_+-_sigma_eta/2}
 \end{align}
\end{subequations}
% \begin{subequations}
%  \begin{align}
%   e^{\boldsymbol\sigma \cdot \boldsymbol\eta/2}&=
%   \begin{pmatrix}
%   \cosh\dfrac{\eta}{2}+n_{3}\sinh\dfrac{\eta}{2} & n_{L}\sinh\dfrac{\eta}{2}\\
%   n_{R}\sinh\dfrac{\eta}{2} & \cosh\dfrac{\eta}{2}-n_{3}\sinh\dfrac{\eta}{2}
%   \end{pmatrix},\label{eqn:e_+_sigma_eta/2}\\
%   e^{-1/2 \boldsymbol\sigma \cdot \boldsymbol\eta}&=
%   \begin{pmatrix}
%   \cosh\dfrac{\eta}{2}-n_{3}\sinh\dfrac{\eta}{2} & -n_{L}\sinh\dfrac{\eta}{2}\\
%   -n_{R}\sinh\dfrac{\eta}{2} & \cosh\dfrac{\eta}{2}+n_{3}\sinh\dfrac{\eta}{2}
%   \end{pmatrix},\label{eqn:e_-_sigma_eta/2}
%  \end{align}
% \end{subequations}
In the above expressions, $\hat{\mathbf{n}}=(\sin\theta\cos\phi,\sin\theta\sin\phi,\cos\theta)$ is the unit vector in the momentum $\mathbf{P}$ direction, and the magnitude of $\boldsymbol\eta$ is given by $\sinh(\eta)=|\mathbf{P}|/M$ and $\cosh(\eta)=E/M=\sqrt{\mathbf{P}^{2}+M^{2}}/M$.
The explicit matrix form of $B(\boldsymbol\eta)\mathcal{D}(\hat{\mathbf{m}},\theta_{s})$ is written in Appendix~\ref{sec:matrix_representation_of_T_and_BR}.

% Applying this transformation to Eq.~(\ref{eqn:u1_theta_CR_rest_frame}) with the relation $\cosh(\eta/2)=|\mathbf{P}|/\sqrt{2M(E+M)}$ and $\sinh(\eta/2)=(E+M)/\sqrt{2M(E+M)}$, we get
% \begin{align}\label{eqn:u1_theta_CR_moving}
%   \dfrac{1}{\sqrt{2\left(E+M\right)}}
%   \begin{pmatrix}
%   \left(\sqrt{2}P^{+}+M\right)\cos\dfrac{\theta_{s}}{2}+P^{L}e^{i\phi_{s}}\sin\dfrac{\theta_{s}}{2}\\[6pt]
%   P^{R}\cos\dfrac{\theta_{s}}{2}+\left(\sqrt{2}P^{-}+M\right)e^{i\phi_{s}}\sin\dfrac{\theta_{s}}{2}\\[6pt]
%   \left(\sqrt{2}P^{-}+M\right)\cos\dfrac{\theta_{s}}{2}-P^{L}e^{i\phi_{s}}\sin\dfrac{\theta_{s}}{2}\\[6pt]
%   -P^{R}\cos\dfrac{\theta_{s}}{2}+\left(\sqrt{2}P^{+}+M\right)e^{i\phi_{s}}\sin\dfrac{\theta_{s}}{2}
%   \end{pmatrix}.
% \end{align}
% Again, the $\sqrt{2}$ factors in front of $P^{\pm}$ are due to our conventions used in this work: $P^{\pm}=(P^{0}\pm P^{3})/\sqrt{2}$.

% \begin{align}
%   \dfrac{1}{\sqrt{2\left(E+M\right)}}
%   \begin{pmatrix}
%   \left(P^{+}+M\right)\cos\dfrac{\theta_{s}}{2}+P^{L}\sin\dfrac{\theta_{s}}{2}\\[6pt]
%   P^{R}\cos\dfrac{\theta_{s}}{2}+\left(P^{-}+M\right)\sin\dfrac{\theta_{s}}{2}\\[6pt]
%   \left(P^{-}+M\right)\cos\dfrac{\theta_{s}}{2}-P^{L}\sin\dfrac{\theta_{s}}{2}\\[6pt]
%   -P^{R}\cos\dfrac{\theta_{s}}{2}+\left(P^{+}+M\right)\sin\dfrac{\theta_{s}}{2}
%   \end{pmatrix}
% \end{align}

Comparing the explicit matrix representations of $T$ and $B(\boldsymbol\eta)\mathcal{D}(\hat{\mathbf{m}},\theta_{s})$ that result in
the identical spinor, i.e. the same momentum and spin, we find that indeed Eq.~(\ref{eqn:T_BR_equivalence}) holds, provided that the angles $(\theta_{s},\phi_{s})$ are given by
\begin{widetext}
\begin{subequations}
 \label{eqn:theta_s_phi_s_relation_with_alpha_and_beta}
 \begin{align}
  \cos\dfrac{\theta_{s}}{2}&=\sqrt{\dfrac{2M}{E+M}}\cos\dfrac{\alpha}{2}\cosh\dfrac{\beta_{3}}{2}=\dfrac{1}{2}\sqrt{\dfrac{2M}{E+M}}\sqrt{\dfrac{P_{\mT}+\Pp}{2\Pp}}\left(\sqrt{\dfrac{P^{\pT}+\Pp}{M(\sin\delta+\cos\delta)}}+\sqrt{\dfrac{M(\sin\delta+\cos\delta)}{P^{\pT}+\Pp}}\right), \label{eqn:theta_s_relation_with_alpha_and_beta}\\
  \cos\phi_{s}&=\dfrac{\beta_{1}}{\sqrt{\beta_{1}^{2}+\beta_{2}^{2}}}=\dfrac{P^{1}}{\sqrt{\mathbf{P}_{\perp}^{2}}}=\cos\phi, \quad \sin\phi_{s}=\dfrac{\beta_{2}}{\sqrt{\beta_{1}^{2}+\beta_{2}^{2}}}=\dfrac{P^{2}}{\sqrt{\mathbf{P}_{\perp}^{2}}}=\sin\phi, \label{eqn:phi_s_relation_with_alpha_and_beta}
  % \sin\dfrac{\theta_{s}}{2}\cos\phi_{s}&=\sqrt{\dfrac{2M}{E+M}}\beta_{1}\dfrac{\sin\dfrac{\alpha}{2}}{\alpha}\left(\cos\delta\cosh\dfrac{\beta_{3}}{2}+\sin\delta\sinh\dfrac{\beta_{3}}{2}\right), \label{eqn:sin_theta_{s}/2_cos_phi_{s}_relation_with_alpha_and_beta} \\
  % \sin\dfrac{\theta_{s}}{2}\sin\phi_{s}&=\sqrt{\dfrac{2M}{E+M}}\beta_{2}\dfrac{\sin\dfrac{\alpha}{2}}{\alpha}\left(\cos\delta\cosh\dfrac{\beta_{3}}{2}+\sin\delta\sinh\dfrac{\beta_{3}}{2}\right). \label{eqn:sin_theta_{s}/2_sin_phi_{s}_relation_with_alpha_and_beta}
 \end{align}
\end{subequations}
\end{widetext}
% $\beta_{1}$, $\beta_{2}$ and $\beta_{3}$ are related to the momentum components by Eq.~(\ref{eqn:useful_beta_P_relation}).
where we used $\sinh(\eta/2)=|\mathbf{P}|/\sqrt{2M(E+M)}$ and $\cosh(\eta/2)=(E+M)/\sqrt{2M(E+M)}$.

Because our generalized helicity spinors are obtained by applying the $T$ transformation on a rest spinor, and now we know that the $T$ transformation is the same as first a rotation around the axis $\hat{\mathbf{m}}$ and then a boost, we conclude that for the positive helicity spinors the spin points in the $(\theta_{s}, \phi_{s})$ direction, and for the negative helicity spinors, the spin will point in the exact opposite direction, namely $(\pi-\theta_{s},\pi+\phi_{s})$.
Since Eq.~(\ref{eqn:T_BR_equivalence}) is an operator equation, which doesn't depend on the total spin of the system,
this expression should also hold for higher spins although we obtained it using the $J=1/2$ representation.

We notice that the spin $\mathbf{S}$, momentum $\mathbf{P}$ and $z$ axis are all in the same plane due to the fact that $\phi_{s}$ is the same as $\phi$ as shown in Eq.~(\ref{eqn:phi_s_relation_with_alpha_and_beta}).
Therefore, for the following discussions, we may take $\phi=\phi_{s}=0$ without loss of any generality and focus on the relation between $\theta$ and $\theta_{s}$.

\begin{figure}[!b]
  \centering
  \includegraphics[width=\columnwidth]{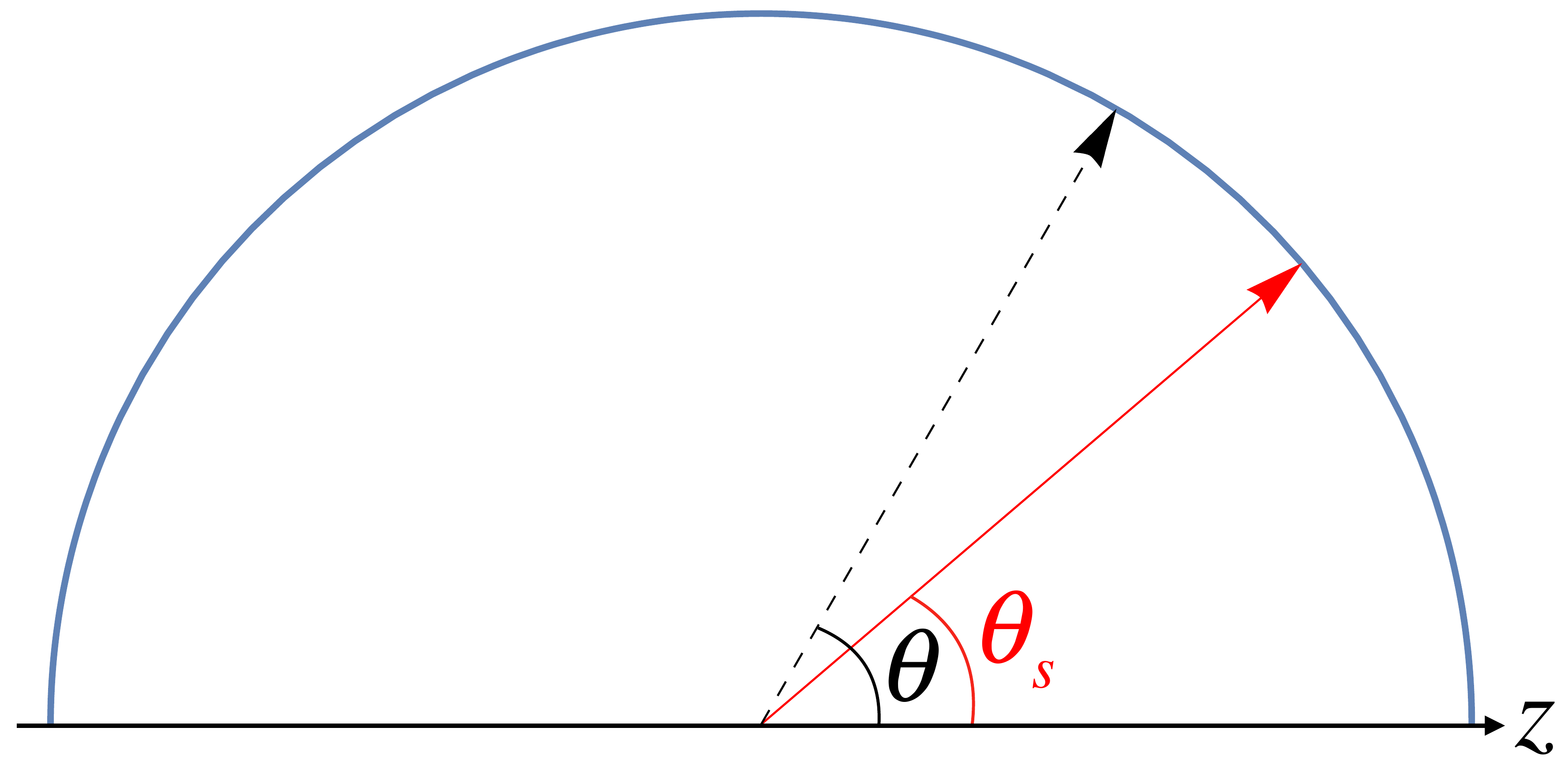}
  \caption{(color online) At interpolation angle $\delta=\pi/6$, a particle's momentum direction at $\theta=\pi/3$ (dashed black arrow) and the corresponding spin orientation (solid red arrow) of the positive helicity spinor.
			We take as an example a particle with mass $M=1\;\text{GeV}$ and total momentum $|\mathbf{P}|=1\;\text{GeV}$.\label{fig:spinor_orientation}}
\end{figure}

First of all, in the instant form limit ($\delta\rightarrow0$), $P^{\pT}\rightarrow P^{0}=E$, $P_{\mT}\rightarrow P^{3}$ and $\Pp\rightarrow|\mathbf{P}|$, and we get
\begin{align}
  \cos(\theta_{s}/2)=\sqrt{(|\mathbf{P}|+P^{3})/(2|\mathbf{P}|)}=\cos(\theta/2).\label{eqn:theta_{s}/2_IF_limit}
\end{align}
So $\theta=\theta_{s}$ when $\delta=0$, and the spin and the momentum directions are aligned, as expected
from the usual Jacob and Wick helicity~\cite{Jacob1959} defined in the IFD.
However, as $\delta$ gets different from zero, the two angles $\theta$ and $\theta_{s}$ differ from each other.
% The expressions of $\theta_{s}$ and $\phi_{s}$ given as functions of particle momentum are relegated to Appendix~\ref{sec:relation_between_spin_orientation_theta_phi_and_the_particle_momentum_for_helicity_spinor}.
To give an illustration of the angles, we plot in Fig.~\ref{fig:spinor_orientation} the spin orientation (red solid vector) and the momentum direction (black dashed vector) of the positive helicity spinor $u_{H}^{(\sfrac{1}{2})}$ for a particle with mass $M=1\;\text{GeV}$ and total momentum $|\mathbf{P}|=1\;\text{GeV}$ when $\theta=\pi/3$ and $\delta=\pi/6$.
It's clear from Eq.~(\ref{eqn:theta_s_relation_with_alpha_and_beta}) that the spin orientation depends on both the particle's momentum and the interpolation angle.
In particular, using Eq.~(\ref{eqn:theta_s_relation_with_alpha_and_beta}), one can analyze how the angle difference between the momentum direction and the spin direction, i.e.
$\theta-\theta_s$, changes as the momentum direction $\theta$ changes for each different interpolation angle $\delta$.

\begin{figure}[!b]
  \includegraphics[width=\columnwidth]{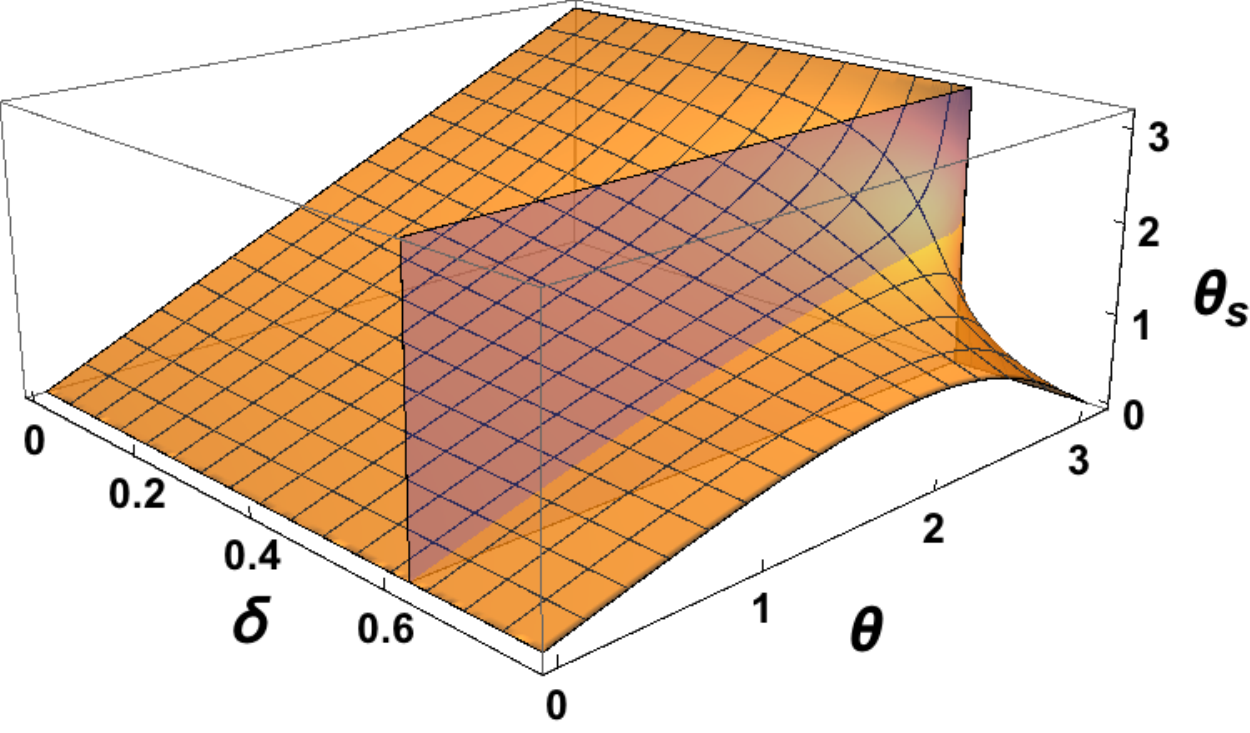}
  \caption{(color online) The dependence of $\theta_{s}$ of a positive helicity spinor on the $\theta$ angle and the interpolation angle $\delta$.
			Again we take a particle with mass $M=1\;\text{GeV}$ and total momentum $|\mathbf{P}|=1\;\text{GeV}$ as our illustrative example.
			\label{fig:thetas_on_theta_delta}}
\end{figure}

\begin{figure}[!b]
  \includegraphics[width=\columnwidth]{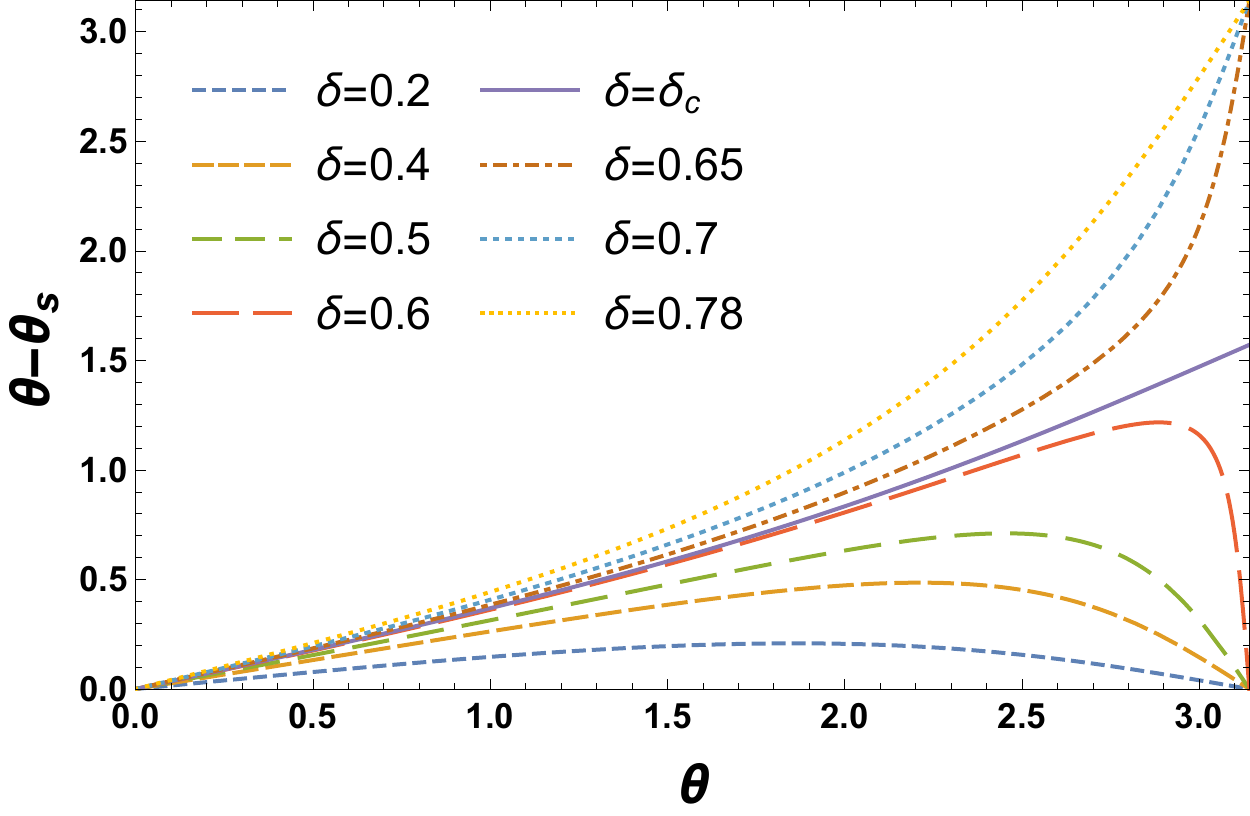}
  \caption{(color online) The angle between the momentum direction and the spin direction, i.e. $\theta - \theta_s$, as a function of $\theta$ with the variation of the interpolation angle $\delta$. The bifurcation point appears at the critical interpolation angle $\delta_{c}=\arctan \left( \frac{|\mathbf{P}|}{E} \right) \approx 0.61548$ as discussed in the text.
			\label{fig:ang_diff_bifurcate}}
\end{figure}

To show how the spin orientation changes with the interpolation angle and the momentum direction, we plot $\theta_{s}$ of a postive helicity spinor $u_{H}^{(\sfrac{1}{2})}$
given by Eq.~(\ref{eqn:theta_s_relation_with_alpha_and_beta}) in terms of both $\delta$ and $\theta$ in Fig.~\ref{fig:thetas_on_theta_delta}.
Here again, we take the particle mass $M=1\;\text{GeV}$ and total momentum $|\mathbf{P}|=1\;\text{GeV}$ for illustration.
For $\delta=0$, the equality $\theta=\theta_{s}$ from Eq.~(\ref{eqn:theta_{s}/2_IF_limit}) yields the straight line relation at $\delta=0$ as shown in Fig.~\ref{fig:thetas_on_theta_delta}.
As we vary the interpolation angle between $\delta=0$ and $\delta=\pi/4$, we see that for a fixed $\theta$, i.e. fixed momentum direction, $\theta_{s}$ decreases as $\delta$ increases in general, and correspondingly the angle between the spin and the momentum will increase with $\delta$.
In fact, the angle between the momentum and the spin directions, i.e. $\theta-\theta_{s}$, increases with the interpolation angle $\delta$ and becomes the largest at the light front.
This feature is shown in Fig.~\ref{fig:ang_diff_bifurcate} as what's shown in Fig.~\ref{fig:thetas_on_theta_delta} is now plotted in terms of the angle difference $\theta-\theta_s$ between the momentum and the spin directions for various interpolation angles listed in the legend.
Note here that the increment of the angle difference $\theta-\theta_s$ with the increment of the interpolation angle $\delta$ bifurcates
at a critical interpolation angle $\delta_{c}$. This bifurcation of two branches shown in Fig.~\ref{fig:ang_diff_bifurcate} is also
indicated by the transparent vertical plane in Fig.~\ref{fig:thetas_on_theta_delta}.
At a fixed interpolation angle $\delta<\delta_{c}$, the spin orientation $\theta_{s}$ increases with $\theta$, and when $\theta=\pi$, we also have $\theta_{s}=\pi$.
% Particularly, in the instant form ($\delta=0$), we see that the relation between $\theta$ and $\theta_{s}$ is a straight line.
% This is because in the instant form, the spin aligns with the momentum direction and $\theta=\theta_{s}$.
On the other hand, at a fixed interpolation angle $\delta>\delta_{c}$, $\theta_{s}$ doesn't follow $\theta$ all the way to $\pi$, but instead starts to decrease beyond a certain point and goes back to $0$ when $\theta=\pi$.
% This is true for the light front form, and as a result the spin of a positive light-front spinor points in the opposite direction of the momentum when the momentum is in the $-z$ direction.
This phenomenon is due to the change of sign for $P_{\mT}$ in Eq.~(\ref{eqn:J3_P_z_direction}) as we discussed
in Sec.~\ref{sec:helicity_operator_and_spinor_for_any_interpolation_angle}, and the critical interpolation angle $\delta_{c}$ is thus given by $P_{\mT}=0$ when
$\theta=\pi$:
\begin{align}
		\delta_{c}=\arctan \left( \frac{|\mathbf{P}|}{E} \right).\label{eqn:critical_delta}
\end{align}
The critical value for the case $M=|\mathbf{P}|=1\;\text{GeV}$ shown in Figs.~\ref{fig:thetas_on_theta_delta} and \ref{fig:ang_diff_bifurcate}
is $\delta_c \approx 0.61548$.
If we change the particle's mass and total momentum, the position of this critical plane in Fig.~\ref{fig:thetas_on_theta_delta} will change accordingly.
The swap of helicity amplitudes~\cite{JiBakker2013} mentioned in the previous section, Sec.~\ref{sec:helicity_operator_and_spinor_for_any_interpolation_angle},
is directly linked to the fact that the IFD and the LFD separately belong to the two different branches bifurcated and divided out at the critical interpolation angle $\delta_c$.
This bifurcation indicates the necessity of the distinction in the spin orientation between the IFD and the LFD and clarifies further
the characteristic of helicity amplitudes in the two distinguished forms of the relativistic dynamics. In particular, the discussion of the spin orientation and the helicity amplitudes clarifies any conceivable confusion in the prevailing notion of the equivalence between the IMF formulated in IFD and the LFD. We thus discuss below
the distinguished features of the light-front helicity.

%{\color{red} So, let's now look at the light-front limit of Eq.~(\ref{eqn:theta_s_relation_with_alpha_and_beta}).}
As we get to the light-front limit ($\delta\rightarrow\pi/4$), $P^{\pT}\rightarrow P^{+}$, $P_{\mT}\rightarrow P^{+}$, $\Pp\rightarrow P^{+}$,
Eq.~(\ref{eqn:theta_s_relation_with_alpha_and_beta}) becomes
\begin{align}
  \label{eqn:cos_theta_s/2_LF}
  \cos\dfrac{\theta_{s}}{2}=\dfrac{\sqrt{2}P^{+}+M}{\sqrt{2 \sqrt{2}P^{+}(E+M)}}.
\end{align}
For a particle that is moving in the $\pm z$ direction, the numerator becomes $E\pm|\mathbf{P}|+M$ and the denominator becomes $\sqrt{2(E\pm|\mathbf{P}|)(E+M)}$.
Using Eq.~(\ref{eqn:E_P_M_useful_relation}), we get $\cos(\theta_{s}/2)\rightarrow1$.
Therefore in the light-front limit, a positive helicity spinor, regardless of whether the particle is moving in the $+z$ or $-z$ direction, always has its spin pointed in the $+z$ direction. This produces the feature of $\theta_{s}=0$ at both $\theta=0$ and $\theta=\pi$ as shown in Fig.~\ref{fig:thetas_on_theta_delta} as well as in Fig.~\ref{fig:ang_diff_bifurcate}.
By the same token, a negative helicity spinor moving in $\pm z$ direction always has its spin pointed to the $-z$ direction in LFD, even when the particle is moving in the $-z$ direction.
This already reveals a distinct feature of the light-front helicity and provides the rationale for the swap of helicity amplitudes discussed
in deeply virtual Compton scattering process~\cite{JiBakker2013}.
Because the light-front helicity is obtained in general by the two kinds of boost operations, i.e. the light-front transverse boost $\mathbf{E_\perp}$ and the longitudinal boost $K^3$, as one can get $T=e^{-i\mathbf{\beta_{\perp}}\cdot\mathbf{E_{\perp}}}e^{-i\beta_{3}K^{3}}$ from Eq.~(\ref{eqn:T_transformation_for_any_interpolation_angle}) in the limit $\delta\rightarrow\pi/4$, we now discuss the salient features of the spin orientations due to each of these two kinds of light-front kinematic operations separately.

\subsubsection{Light-front Transverse Boost $\mathbf{E_\perp}$ Operation}
\label{subsub:E_perp}
Without loss of any generality, we may take the coordinate system where everything lies in the $x-z$ plane
and consider just the transverse boost $E^1=(K^1+J^2)/\sqrt{2}$ to discuss the effect of the light-front boost operation on the spin orientation.
Suppose that the particle at rest with the spin up in the positive $z$ direction is boosted by $E^1$ to gain the particle energy
$E=\sqrt{\mathbf{P}^{2}+M^{2}}$. As $E^1$ consists of not only the boost in the $x$ direction, $K^1$, but also the rotation around the $y$ axis, $J^2$,
the particle gains the momentum in the $-z$ direction as well as in the $x$ direction. Interestingly, the momentum gained in the $-z$ direction
takes the non-relativistic form $P^z = -\frac{\vec{\bf P}_\perp^2}{2M}$ (here of course $\vec{\bf P}_\perp=P^x \hat{x}$) in such a way that the energy of the particle becomes $E=P^0=M+\frac{\vec{\bf P}_\perp^2}{2M}$ while the light-front plus momentum $P^+=(P^0+P^z)/\sqrt{2}$ gets invariant~\cite{Ji2013}.
Of course, the relativistic energy-momentum dispersion relation still holds as one may expect:
\begin{equation}
\label{eqn:LF-dispersion}
\left(P^0\right)^2-\vec{\bf P}^2 = \left(M+\frac{\vec{\bf P}_\perp^2}{2M}\right)^2-\vec{\bf P}_\perp^2-\left(-\frac{\vec{\bf P}_\perp^2}{2M}\right)^2 = M^2 \, .
\end{equation}
In terms of the energy $E$ and the particle mass $M$, one may express the particle momentum components as $P^x=\sqrt{2M(E-M)}$ and $P^z=-\frac{(P^x)^2}{2M}=M-E$
as well as $\sqrt{2}P^+=M$. Thus, the resulted momentum direction is given by $\cos\theta_0 = \frac{P^z}{|\mathbf{P}|}=\frac{M-E}{\sqrt{E^2-M^2}}=-\sqrt{\frac{E-M}{E+M}}$.
Since the 4-vector of the particle spin $S^\mu$ is transverse to the 4-momentum $P^\mu$, i.e. $S\cdot P=0$, while its normalization is fixed by $S^2=-1$,
one can find the specific direction of the spin that the particle takes under the $E^1$ operation. Thus, we consider here effectively the $E^1$ operation that results in the final state with the four-momentum $P^\mu$ and the spin 4-vector $S^\mu$ starting from the initial state at rest  with the spin up in the $\hat z$ direction.
As the $E^1$ operation is the only source of the spin rotation here, the rotated spin angle gets maximum only when
the given energy $E=\sqrt{\mathbf{P}^{2}+M^{2}}$ is attained entirely by the $E^1$ operation.
If the particle is tempered with the $K^3$ operation prior to $E^1$ operation, then the amount of the $E^1$ operation required to achieve the same energy $E=\sqrt{\mathbf{P}^{2}+M^{2}}$ would be reduced.
We may discuss the ``apparent'' orientation of the spin in the moving frame by analyzing $S^{\mu}$ directly as we describe in Appendix~\ref{sec:apparent-spin}, where we define the ``apparent" spin orientation angle $\theta_a$ and derive the relationship between $\theta_a$ and $\theta$.
Using Eq.~(\ref{eqn:BR}), however, one may also take a look at the spin orientation in the particle's rest frame, i.e. the rotated $\theta_{s}$ angle prior to the boost operation $B(\boldsymbol\eta)=e^{-i\boldsymbol\eta\cdot\mathbf{K}}$ with which
the final 4-momentum $P^\mu$ and the spin 4-vector $S^\mu$ are obtained.
The maximum ``apparent'' spin orientation in the moving frame, i.e. the final state, must correspond to the maximum initial spin orientation $(\theta_s)_{\rm max}$ since the boost $B(\boldsymbol\eta)=e^{-i\boldsymbol\eta\cdot\mathbf{K}}$ tilts the initial spin orientation toward the direction of the rapidity $\boldsymbol\eta$ or the final momentum $\mathbf{P}$.
Therefore, for a certain total momentum $|\mathbf{P}|$ or energy $E$, $(\theta_s)_{\rm max}$ should also be obtained at $\theta=\theta_{0}$
that was attained entirely by the $E^1$ operation.

\begin{figure}[!b]
  \centering
  \includegraphics[width=\columnwidth]{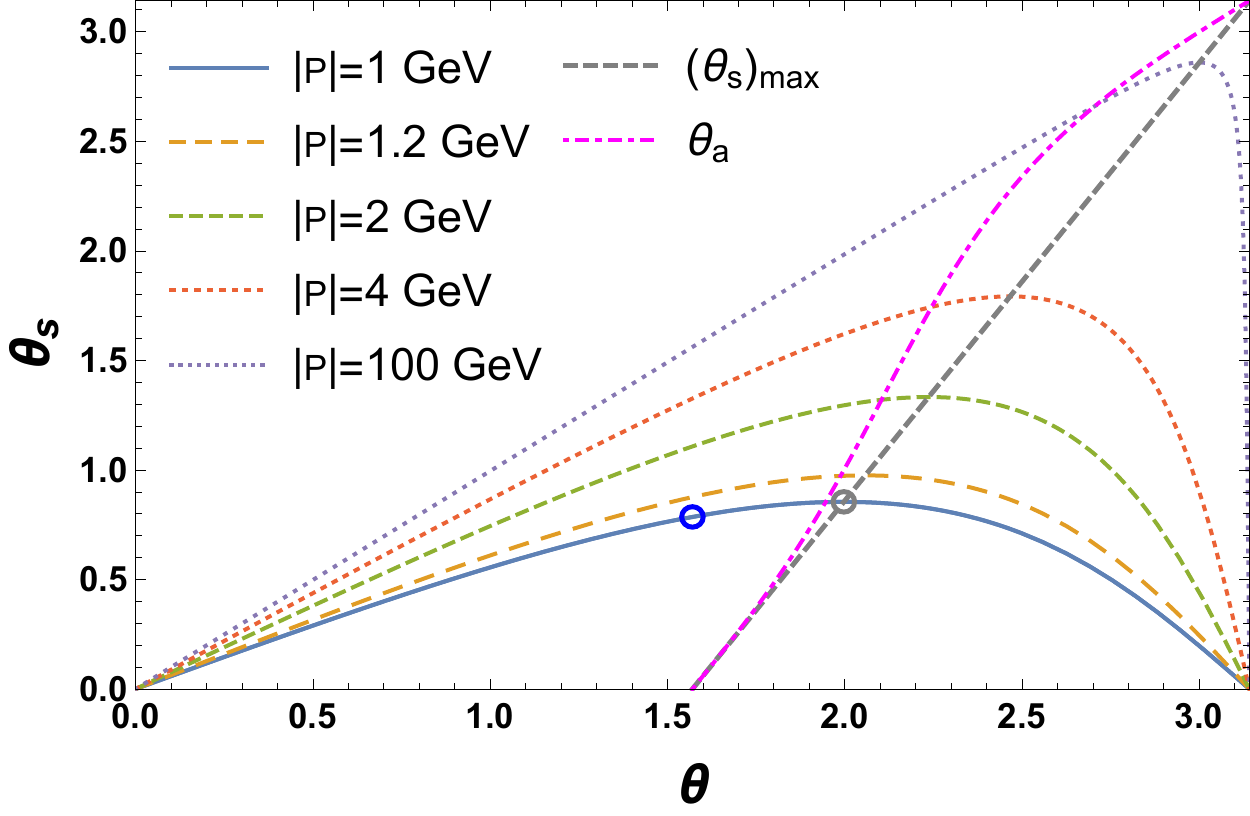}
  \caption{(color online) The light-front profiles of $\theta_{s}$ vs $\theta$ for a particle with mass $M=1\;\text{GeV}$ and the magnitude of the momentum $|\mathbf{P}|=1,\;1.2,\;2,\;4,\;100\;\text{GeV}$.
			The thick gray dashed line indicates the position of $\theta_0$ which provides the maximum $\theta_s$.
			As a comparison, the magenta dot-dashed line indicates how the ``apparent'' spin orientation angle $\theta_{a}$ changes with the momentum direction. The gray circle is marked for the intersecting point
between the line of $|\mathbf{P}|=1$ GeV (blue solid line) and the thick gray dashed line.
			The blue circle corresponds to the case when the spinor doesn't have any momentum in $z$ direction but moves perpendicular to the $z$ direction, e.g. $x$ direction.
			\label{fig:LF_profile1}}
\end{figure}

This initial spin orientation $\theta_s$ given by Eq.~(\ref{eqn:cos_theta_s/2_LF})
can be written as
\begin{align}
	\dfrac{\theta_{s}}{2}=\dfrac{\pi}{2}-\arctan\left( \dfrac{\sqrt{2}{P}^{+}+M}{{P}^{x}} \right),\label{eqn:spin_orientation_LFlimit}
\end{align}
since $\cot\dfrac{\theta_{s}}{2} = \left( \dfrac{\sqrt{2}{P}^{+}+M}{{P}^{x}} \right)$ or $\tan\dfrac{\theta_{s}}{2} = \left( \dfrac{{P}^{x}}{\sqrt{2}{P}^{+}+M} \right)$.
One may obtain $(\theta_s)_{\rm max}$ from Eq.~(\ref{eqn:spin_orientation_LFlimit})
by taking $P^x = |\mathbf{P}|\sin\theta$ and
$\sqrt{2}P^+ = \sqrt{|\mathbf{P}|^2+M^2}+|\mathbf{P}|\cos\theta$ and differentiating $\theta_{s}$ with respect to $\theta$
to gets zero. As expected, we find the angle $\theta_0$ precisely
given by $\cos\theta_0 = - \frac{|\mathbf{P}|}{E+M}=-\sqrt{\frac{E-M}{E+M}}$ with the corresponding $(\theta_s)_{\rm max}= \pi-2\arctan(\sqrt{\frac{2M}{E-M}})$ or equivalently $(\theta_s)_{\rm max} = 2 \arctan \sqrt{\frac{E-M}{2M}}$.
The negative sign here indicates that the $z$ component of the particle momentum must be negative.
As $|\mathbf{P}|$ (or $E$) gets larger, $\theta_0$ gets closer to $\pi$.
This behavior is shown in Fig.~\ref{fig:LF_profile1}.
We choose $M=1$ GeV for an illustration and first plot $\theta_s$ given by Eq.~(\ref{eqn:spin_orientation_LFlimit}) as a function of $\theta$ for fixed values
of $|\mathbf{P}|$, e.g. $|\mathbf{P}|=1,\;1.2,\;2,\;4,\;100\;\text{GeV}$, as indicated in the legend of Fig.~\ref{fig:LF_profile1}.
The values of $(\theta_s)_{\rm max}$ is traced out with the thick gray dashed line along the values of $\theta_0$  as $|\mathbf{P}|$ varies in Fig.~\ref{fig:LF_profile1}.
This thick gray dashed line is obtained entirely by the $E^1$ operation on the initial rest particle and thus
the value of $\sqrt{2}P^+$ is identical to the particle mass $M$ throughout this line. As expected, this line approaches to $\theta=\pi/2$
when $|\mathbf{P}|$ gets close to zero.
We also plot the ``apparent" spin orientation angle $\theta_a$ defined in Appendix~\ref{sec:apparent-spin} as the magenta dot-dashed line,
overlaying with the thick gray dashed line of $(\theta_s)_{\rm max}$.

As an example of the $(\theta_s)_{\rm max}$ state lying on the thick gray dashed line, a gray circle is marked for the intersecting point
between the line of $|\mathbf{P}|=1$ GeV (blue solid line) and the thick gray dashed line in Fig.~\ref{fig:LF_profile1}. Since $|\mathbf{P}|=1$ GeV for this state, the value of $\theta_0$ is given by $\arccos\left(-\frac{1}{1+\sqrt{2}} \right) \approx 1.99787$ and the corresponding value of $(\theta_s)_{\rm max}$ is given by $2\arctan\left(\frac{1}{\sqrt{2(1+\sqrt{2})}}\right) \approx 0.854157$.
%which corresponds to the intersecting point marked by a gray circle between the line of $|\mathbf{P}|=1$ GeV (blue solid line) and the thick gray dashed line in Fig.~\ref{fig:LF_profile1}.
As a comparison, we displayed the state with the final momentum entirely given by $P^x$ without any $P^z$ component, e.g. $P^x=|\mathbf{P}|=1$ GeV,
or $\theta = \pi/2$, marked by the blue circle in Fig.~\ref{fig:LF_profile1}.
To get this final momentum, the initial momentum prior to the $E^1$ operation should have had the positive $P^z$ momentum to compensate
the negative $P^z$ component, i.e. $P^z=-\frac{\vec{\bf P}_\perp^2}{2M}$, gained by the $E^1$ operation. The initial positive $P^z$ momentum
should have been attained by the $K^3$ operation prior to the $E^1$ operation and consequently the amount of $E^1$ operation required to achieve the state with $|\mathbf{P}|=1$ GeV gets reduced. Thus, the corresponding $\theta_s$ for the blue circle is smaller than $(\theta_s)_{\rm max}$
marked by gray circle. Numerically, the corresponding $\theta_s$ value $2 \arctan\left(\frac{1}{\sqrt{2}+1}\right) \approx 0.785398$
for $\theta=\pi/2$ (blue circle) may be compared with $(\theta_s)_{\rm max}=2\arctan\left(\frac{1}{\sqrt{2(1+\sqrt{2})}}\right) \approx 0.854157$
for $\theta_0=\arccos\left(-\frac{1}{1+\sqrt{2}} \right) \approx 1.99787$ (gray circle) on the line of $|\mathbf{P}|=1$ GeV.
When $E\rightarrow \infty$, both $\theta_0 \rightarrow \pi$ and $(\theta_s)_{\rm max}\rightarrow \pi$ as shown in Fig.~\ref{fig:LF_profile1}.

\subsubsection{Longitudinal Boost $K^3$ Operation}
\label{subsub:K^3}

\begin{figure}[!b]
  \centering
  \includegraphics[width=\columnwidth]{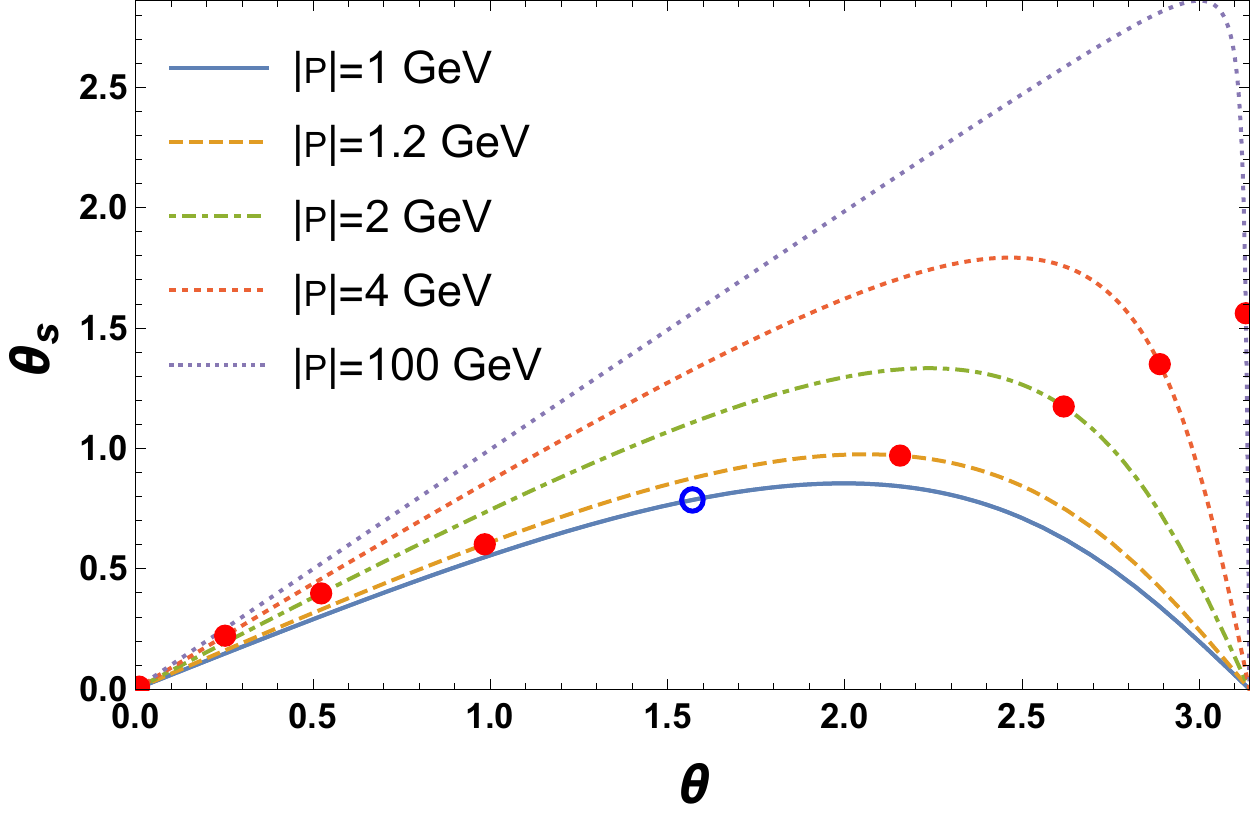}
  \caption{(color online) The light-front profiles of $\theta_{s}$ vs $\theta$ for a particle with mass $M=1\;\text{GeV}$ and the magnitude of the momentum $|\mathbf{P}|=1,\;1.2,\;2,\;4,\;100\;\text{GeV}$.
			The red solid dots indicates how the momentum direction $\theta$ and the spin direction $\theta_{s}$ changes as the spinor is boosted in $\pm z$ direction.
			The blue circle corresponds to the case when the spinor doesn't have any momentum in $z$ direction but moves perpendicular to the $z$ direction, e.g. $x$ direction.
			\label{fig:LF_profile}}
\end{figure}

As we have completed our discussion on the $\mathbf{E_\perp}$ operation, we now take a look at how the spin direction changes under
the longitudinal boost $K^3$ when the particle is not moving in the $\pm z$ direction.
The change in $|\mathbf{P}|$ will result in a change in the $\theta_{s}$ vs. $\theta$ profile as we have already discussed
and the changed momentum direction $\theta$ pinpoints the resulted spin direction $\theta_{s}$ on the corresponding profile.
For an illustration, we start with the case when the 3-momentum is perpendicular to the $z$ direction, e.g. $|\mathbf{P}|=P^{x}=1$ GeV,
which we again denote by the blue circle now in Fig.~\ref{fig:LF_profile}.
As discussed earlier, the corresponding spin direction is given by Eq.~(\ref{eqn:spin_orientation_LFlimit}), i.e. $\theta_s = 2 \arctan\left(\frac{1}{\sqrt{2}+1}\right) \approx 0.785398$.
As the particle is boosted in the $+z$ direction while $P^{x}$ is fixed as 1 GeV, the momentum angle $\theta$ is decreased but the profile is lifted higher as the magnitude of momentum $|\mathbf{P}|$ is increased.
The corresponding change of ($\theta$, $\theta_{s}$) values follows the red solid dots in the region where $\theta < \pi/2$ as shown in Fig.~\ref{fig:LF_profile}.
It is interesting to note that the angle $\theta$ decreases fast enough to get $\theta_{s}$ decreased despite the fact that the profiles are lifted.
When boosted in the $-z$ direction, the momentum angle $\theta$ is increased and in fact it quickly goes over to the $\theta>\theta_0$ side,
where $\cos\theta_0 = - \sqrt{\frac{E-M}{E+M}}$ as previously discussed.
As the whole profile moves upward, the resulted $\theta_{s}$ is increased following the red solid dots in the region where $\theta > \pi/2$.
As can be seen from the trend of the red dots as $\theta \rightarrow \pi$ in Fig.~\ref{fig:LF_profile}, the $\theta_{s}$ approaches to a value that's not $\pi$.
In other words, even under the maximum boost in the $-z$ direction, the spin does not align in parallel with the momentum direction.
While $P^x$ is fixed, the maximum $\theta_{s}$ value that can be achieved  in Eq.~(\ref{eqn:spin_orientation_LFlimit})
is $\pi-2\arctan(M/P^{x})$ as $P^{z} \rightarrow -\infty$. For $P^x=1$ GeV, $\pi-2\arctan(M/P^{x}) \approx 1.5708$.
As $P^{x}$ is fixed in this case, $P^{+}$ can approach to zero.
This is in constrast to the maximum on the $\theta_{s}$ vs. $\theta$ profile (i.e. $(\theta_s)_{max}$) discussed before, for which $P^{+}$ is fixed as $M/\sqrt{2}$ and $P^{x} = \sqrt{2M(E-M)}$,
so that $(\theta_s)_{max} \rightarrow \pi-2\arctan(2M/P^{x}) \approx \pi$ as the $|\mathbf{P}|$ or $E$ goes to $\infty$.
Effectively, in the fixed $P^x = 1$ GeV case, the red solid dots follows a monotonic relation between the $\theta_{s}$ and $\theta$ under the boost in $z$ direction going through the blue circle in Fig.~\ref{fig:LF_profile}.

This may also be understood from Eq.~(\ref{eqn:spin_orientation_LFlimit}) by noting that the
rotated angle of the spin after the boost in {$\pm z$} direction satisfies the following equation:
\begin{align}
  \label{eqn:wigner_rotation_angle}
-\dfrac{\Delta\theta_{s}}{2}=&\arctan\left(\dfrac{\sqrt{2}{P'}^{+}+M}{{P'}^{x}}\right)-\arctan\left(\dfrac{\sqrt{2}{P}^{+}+M}{P^{x}}\right),
\end{align}
where $'$ denotes the boosted quantities.
For boosting the particle in the $+z$ ($-z$) direction without changing its momentum in the perpendicular direction, i.e. $x$ direction,
the right-hand-side is positive (negative), because $P'^{+}>P^{+}$ ($P'^{+}<P^{+}$) and $P'^{x}=P^{x}$.
The negative sign in front of $\Delta\theta_{s}$ indicates that $\theta_{s}$ is decreased (increased) as the particle is boosted in the  $+z$ ($-z$) direction.
This illustrates that the rotation of the spin occurs around the $\mathbf{P}\times (\pm \mathbf{\hat{z}})$ axis as the particle is boosted in the
$\pm z$ direction, respectively. This is of course the well-known Wigner rotation and $\Delta\theta_{s}$ is exactly the corresponding Wigner rotation angle given by Eq.~(A10) in \cite{Osborn1968} when it is written using our coordinate system and notations.
% Thus, this is indeed the Wigner rotation that's needed \cite{Osborn1968}.
The fact that the Wigner rotation is not needed when boosting the light-front helicity spinors has been known for a long time, but our discussion using the spin orientation angle provides a clear understanding on this interesting feature of the light-front helicity.
 This is also related with the reason why the amplitudes computed
with the light-front helicity, i.e. the light-front helicity amplitudes, are independent of the reference frame as we shall demonstrate in
the scattering processes discussed in the next section, Sec.~\ref{sec:interpolating_helicity_probabilities_for_fermion_scattering}.
Since both $\mathbf{E_{\perp}}$ and $K^3$ in $T=e^{-i\mathbf{\beta_{\perp}}\cdot\mathbf{E_{\perp}}}e^{-i\beta_{3}K^{3}}$
are the kinematic operators in LFD, the light-front helicity formulation takes care of the Wigner rotation which is a rather complicate dynamic effect, and offers an effective computation of spin observables in hadron physics.

\subsection{Generalized Melosh Transformation for Any Interpolation Angle}
\label{sub:generalized_melosh_transformation}
On the light-front, people usually use the Melosh transformation \cite{Melosh1974} to connect between the Dirac spinors and the light-front spinors.
Now that we have our generalized helicity spinors, we can derive a generalized Melosh transformation that connects between the Dirac spinors and our generalized helicity spinors.

Dirac spinors $u^{\pm\sfrac{1}{2}}_{D}(P)$ can be obtained by applying the boost operator $e^{-i\boldsymbol\eta\cdot\mathbf{K}}$ directly on the rest frame spinors $u^{\pm\sfrac{1}{2}}(0)$ (see Appendix~\ref{sec:dirac_spinor_in_chiral_representation} for explicit expressions).
Using Eq.~(\ref{eqn:T_BR_equivalence}), this means that the Dirac spinors represent particles moving with momentum $\mathbf{P}$ while their spins are parallel or anti-parallel to the $z$ axis in their rest frame.
% As we discussed in the last subsection, a single boost from the rest frame does not change the spin orientation by definition.
% This means the Dirac spinors represent particles moving with momentum $\mathbf{P}$ while having their spins parallel (anti-parallel) to the $z$ axis.
Pictorially, when the red solid arrow in Fig.~\ref{fig:spinor_orientation} is horizontal while the dashed black arrow remains in its original position, it becomes a Dirac spinor defined for that momentum.
Therefore, it is clear that the helicity spinors are related to the Dirac spinors by a pure rotation of the spin.

One can also prove this mathematically.
Suppose that the Dirac spinors and the helicity spinors are related by the generalized Melosh transformation in the following way \cite{Dziembowski1988a} \cite{Ahluwalia1993}:
\begin{align}
  u_{D}^{(\lambda)}=\Omega_{\lambda\rho}^{[u_{D}u_{H}]}u_{H}^{(\rho)}+\Omega_{\lambda\rho}^{[u_{D}v_{H}]}v_{H}^{(\rho)},\\
  v_{D}^{(\lambda)}=\Omega_{\lambda\rho}^{[v_{D}u_{H}]}u_{H}^{(\rho)}+\Omega_{\lambda\rho}^{[v_{D}v_{H}]}v_{H}^{(\rho)},
	\label{eqn:Melsoh_transformation}
\end{align}
where $H$ and $D$ subscripts denote the generalized helicity spinors and the Dirac spinors respectively, and $\rho, \lambda=\sfrac{1}{2},-\sfrac{1}{2}$ denote positive or negative helicity in the case of our generalized helicity spinors and spin up or spin down in the case of Dirac spinors.
Using our orthonormal convention $\bar{u}_{H}^{(\rho)}u_{H}^{(\lambda)}=\bar{v}_{H}^{(\rho)}v_{H}^{(\lambda)}=2M\delta_{\rho\lambda}$, we then have
\begin{align}
		\Omega_{\lambda\rho}^{[u_{D}u_{H}]}
		&=\dfrac{1}{2M}\bar{u}^{(\rho)}_{H}(P)u^{(\lambda)}_{D}(P)\nonumber\\
    &=\dfrac{1}{2M}u^{(\rho)\dagger}(0)e^{i\hat{\mathbf{m}} \cdot \mathbf{J}\theta_{s}}e^{-i\boldsymbol\eta\cdot\mathbf{K}}\gamma^{0} e^{-i\boldsymbol\eta\cdot\mathbf{K}}u^{(\lambda)}(0)  \nonumber\\
    &=\dfrac{1}{2M}u^{(\rho)\dagger}(0)e^{i\hat{\mathbf{m}} \cdot \mathbf{J}\theta_{s}}\gamma^{0}u^{(\lambda)}(0)  \nonumber\\
		&=\xi^{(\rho)\dagger}e^{i\hat{\mathbf{m}} \cdot \mathbf{J}\theta_{s}}\xi^{(\lambda)}\nonumber\\
		&=\mathcal{D}(\hat{\mathbf{m}},-\theta_{s})_{\rho\lambda},
		\label{eqn:Melosh_matrix_elements_uu}
\end{align}
where $(e^{-i\boldsymbol\eta\cdot\mathbf{K}})^{\dagger}=e^{-i\boldsymbol\eta\cdot\mathbf{K}}$ because the boost operator $\mathbf{K}$ is antihermitian, and we've rewritten Eq.~(\ref{eqn:dirac_CR_spinor_basis_rest_frame}) as
\begin{align}
  u^{(i)}=\sqrt{M}
	\begin{pmatrix}
    \xi^{(i)}\\
		\xi^{(i)}
  \end{pmatrix}
	\text{ with }
	\xi^{(\sfrac{1}{2})}=
	\begin{pmatrix}
	  1\\
		0
	\end{pmatrix},\:
	\xi^{(-\sfrac{1}{2})}=
	\begin{pmatrix}
	  0\\
		1
	\end{pmatrix}.\label{eqn:u_restframe_xi}
\end{align}
Similarly, we have
\begin{align}
		\Omega_{\lambda\rho}^{[v_{D}v_{H}]}=\mathcal{D}(\hat{\mathbf{m}},-\theta_{s})_{\rho\lambda}.\label{eqn:Melosh_matrix_elements_vv}
\end{align}
Using the same method and the orthogonal condition $\bar{u}^{(\rho)}v^{(\lambda)}=\bar{v}^{(\rho)}u^{(\lambda)}=0$, one can easily check that
\begin{align}
	\Omega_{\lambda\rho}^{[u_{D}v_{H}]}=\Omega_{\lambda\rho}^{[v_{D}u_{H}]}=0.
	\label{eqn:Melosh_matrix_elements_uv_vu}
\end{align}
From Eq.~(\ref{eqn:Melosh_matrix_elements_uu}) and Eq.~(\ref{eqn:Melosh_matrix_elements_vv}), we see that the generalized Melosh transformation is nothing but the transpose of the Wigner rotation matrix $\mathcal{D}(\hat{\mathbf{m}},-\theta_{s})$.
Plugging $\hat{\mathbf{m}}=(-\sin\phi_{s},\cos\phi_{s},0)$ and $-\theta_{s}$ into Eq.~(\ref{eqn:e_-i_sigma_m/2}), we find the generalized Melosh transformation for spin $1/2$ as
\begin{align}
  \Omega\left(\dfrac{1}{2}\right)&=
	\begin{pmatrix}
	  \omega\left(\dfrac{1}{2}\right)& 0\\
		0 & \omega\left(\dfrac{1}{2}\right)
	\end{pmatrix},
\end{align}
where
\begin{align}
		\omega\left(\dfrac{1}{2}\right)&=
  \begin{pmatrix}
  \cos\dfrac{\theta_{s}}{2} & -e^{i\phi_{s}}\sin\dfrac{\theta_{s}}{2}\\
  e^{-i\phi_{s}}\sin\dfrac{\theta_{s}}{2} & \cos\dfrac{\theta_{s}}{2}
  \end{pmatrix},
	\label{eqn:}
\end{align}
with $\theta_{s}, \phi_{s}$ given by Eq.~(\ref{eqn:theta_s_phi_s_relation_with_alpha_and_beta}) in terms of the particle's momentum and ``$1/2$'' denotes the fact that this is for spin $J=1/2$.
% Note that this matrix does not act upon the individual components of the spinor (that would be rotating both and momentum and the spin orientation of the spinor), but rather, it
% This matrix gives the combinatorial factors for spinors in the sense that $u_{D}^{(\lambda)}=\itP{\mathcal{D}}_{\rho\lambda}u_{H}^{(\rho)}$, where $H$ and $D$ subscripts denote helicity and Dirac spinors respectively, and $\rho, \lambda=\sfrac{1}{2},-\sfrac{1}{2}$ denote positive or negative helicity in the case of our generalized helicity spinors and spin up or spin down in the case of Dirac spinor.
% % The two $\itP{R}$ blocks in $\itP{\mathcal{D}}$ are for particle and antiparticle spinors.
% % and $v_{D}^{(\lambda)}=\itP{\mathcal{D}}_{\rho\lambda}\vT_{H}^{(\rho)}$
% % and $v_{D}^{(\lambda)}=\itP{\Omega}_{\lambda\rho}\vT_{H}^{(\rho)}$
%
% On the other hand, Melosh transformation is defined as $u_{D}^{(\lambda)}=\itP{\Omega}_{\lambda\rho}u_{H}^{(\rho)}$ \cite{Dziembowski1988a} \cite{Ahluwalia1993}.
% So the Melosh transformation is nothing but the transpose of this Wigner rotation matrix: $\itP{\Omega}=\itP{\mathcal{D}}^{T}$.

In the light-front limit ($\delta\rightarrow\pi/4$), we have Eq.~(\ref{eqn:phi_s_relation_with_alpha_and_beta}) and Eq.~(\ref{eqn:cos_theta_s/2_LF}), so $\Omega(1/2)$ reduces to the well-known Melosh transformation that relates the light-front spinors in Eq.~(\ref{eqn:u1_u2_LF_spinor_CR}) to the Dirac spinors in Eq.~(\ref{eqn:Dirac_spinors}), i.e.
\begin{align}\label{eqn:Melosh_transformation_matrix}
  \omega\left(\dfrac{1}{2}\right)=\frac{1}{\sqrt{2 \sqrt{2}P^{+}(E+M)}}
  \begin{pmatrix}
   \sqrt{2}P^{+}+M & -P^{R}  \\
   P^{L}  & \sqrt{2}P^{+}+M
  \end{pmatrix}.
  % \begin{pmatrix}
  %   B & 0 \\
  %   0 & B
  % \end{pmatrix},
\end{align}
% where
% \begin{align}\label{eqn:Melosh_B}
%   B=
%   \begin{pmatrix}
%    P^{+}+M & -P^{R}  \\
%    P^{L}  & P^{+}+M
%   \end{pmatrix}.
% \end{align}
The $\sqrt{2}$ in front of $P^{+}$ is due to our convention.
This expression agrees with Eq.~(A8) in Ref.~\cite{Dziembowski1988a} and Eq.~(C1), Eq.~(C2) in Ref.~\cite{Ahluwalia1993}.

As Eq.~(\ref{eqn:Melosh_matrix_elements_uu}), Eq.~(\ref{eqn:Melosh_matrix_elements_vv}) and Eq.~(\ref{eqn:Melosh_matrix_elements_uv_vu}) are independent of the spin $J$, our derivation is also applicable to higher spins.
We can therefore quickly write down the generalized Melosh transformations for any spin $J$, as the transpose of the corresponding Wigner rotation matrix $\mathcal{D}(\hat{\mathbf{m}},-\theta_{s})$.
The generalized Melosh transformation matrices for spins up to $J=2$ are listed in Appendix~\ref{Appendix:generalized_melosh_transformation} for reference.
% subsection generalized_melosh_transformation (end)

\section{Interpolating Helicity Scattering Probabilities}
% \label{sec:interpolating_helicity_amplitudes_for_fermion_scattering}

\begin{figure}[!htp]
  \includegraphics[width=\columnwidth]{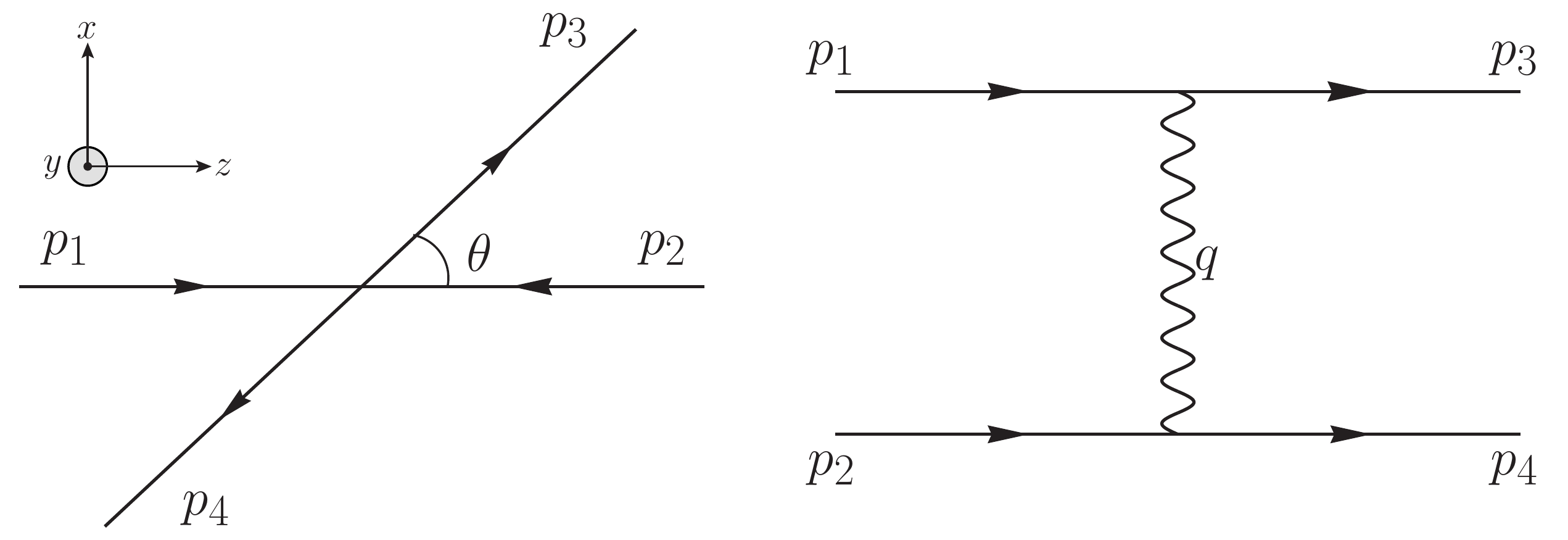}
  \caption{\label{fig:scattering} Scattering process at angle $\theta$, and its corresponding Feynman diagram at the lowest tree level.}
\end{figure}

% section interpolating_helicity_amplitudes_for_fermion_scattering (end)
\label{sec:interpolating_helicity_probabilities_for_fermion_scattering}
Using the generalized helicity spinors we can calculate the helicity-dependent amplitudes, for example, for a scattering process depicted in Fig.~\ref{fig:scattering}.
At the lowest order, the helicity amplitude $\mathcal{M}(\lambda_{1},\lambda_{2},\lambda_{3},\lambda_{4 })$ is given by
\begin{align}
  \bar{u}^{(\lambda_{3})}(p_{3})\gamma^{\mu}u^{(\lambda_{1})}(p_{1}) \bar{u}^{(\lambda_{4})}(p_{4})\gamma_{\mu}u^{(\lambda_{2})}(p_{2}),\label{eqn:scattering_M}
\end{align}
where $\lambda$ denotes the helicity of the particle.
We dropped the coupling constant factor $(-ie)^{2}$ and the Lorentz invariant part of the propagator $-1/q^{2}$, since they are irrelevant to our discussion.
In this section, we investigate the scattering probabilities which are the square of Eq.~(\ref{eqn:scattering_M}).

It needs to be emphasized that the ``helicity'' we discuss here is the ``generalized helicity'' for an arbitrary interpolation angle, as defined in Sec.~\ref{sec:helicity_operator_and_spinor_for_any_interpolation_angle}.
For a certain helicity, the spin orientation depends on both the interpolation angle $\delta$ and the momentum of the particle, as given by Eq.~(\ref{eqn:theta_s_relation_with_alpha_and_beta}).
Because a boost in the $z$ direction changes the 4-momentum, it will also change the spin orientation in general.
As a result, for a certain helicity configuration, the scattering probability will have both the interpolation angle dependence and the frame dependence.
To demonstrate both dependences, we plot the helicity probabilities in terms of both the interpolation angle $\delta$ and the total momentum of the system $P^{z}$, which reflects the boost effect in the $z$ direction.

As illustrated in Fig.~\ref{fig:scattering}, we set our coordinate system in such a way that the initial particles are lying on the $z$ axis, and $\theta$ is the scattering angle between the momentum $\mathbf{p}_{1}$ and momentum $\mathbf{p}_{3}$.
In the center of mass frame, the 4-momentum of these four particles are given by
\begin{subequations}
  \label{eqn:p_for_four_particles}
  \begin{align}
    p_{1}&=(\epsilon_{1},0,0,p_{\text{initial}}),\\
    p_{2}&=(\epsilon_{2},0,0,-p_{\text{initial}}),\\
    p_{3}&=(\epsilon_{3},p_{\text{final}} \sin\theta,0,p_{\text{final}} \cos\theta),\\
    p_{4}&=(\epsilon_{4},-p_{\text{final}} \sin\theta,0,-p_{\text{final}}\cos\theta),
  \end{align}
\end{subequations}
where $p_{\rm initial}=p_{\rm final}\equiv p$, $\epsilon_{1}=\epsilon_{3}=\sqrt{m_{1}^{2}+p^{2}}$ and $\epsilon_{2}=\epsilon_{4}=\sqrt{m_{2}^{2}+p^{2}}$, from the conservation of energy and momentum.
To incorporate the frame dependence for the probabilities, we boost the whole system to the total momentum $P^{z}$, and get
\begin{subequations}
  \label{eqn:p_for_four_particles_boosted}
 \begin{align}
		 {p'}_{i}^{0}&=\gamma p_{i}^{0}+\gamma\beta p_{i}^{z}=\dfrac{E}{M}p_{i}^{0}+\dfrac{P^{z}}{M}p_{i}^{z},\label{eqn:boosted_p0}\\
		 {p'}_{i}^{z}&=\gamma p_{i}^{z}+\gamma\beta p_{i}^{0}=\dfrac{E}{M}p_{i}^{z}+\dfrac{P^{z}}{M}p_{i}^{0},\label{eqn:boosted_pz}\\
		 {\mathbf{p'}}_{i}^{\perp}&=\mathbf{p}_{i}^{\perp},\label{eqn:boosted_pp}\quad (i=1,2,3,4)
 \end{align}
\end{subequations}
where $M=\epsilon_{1}+\epsilon_{2}$ is the center of mass energy and $E=\sqrt{(P^{z})^{2}+M^{2}}$ is the total energy in the boosted frame.
We use these frame-dependent 4-momentum in Eq.~(\ref{eqn:u1u2_P}) to get frame dependent helicity spinors for an arbitrary interpolation angle $\delta$.
The probabilities as the square of Eq.~(\ref{eqn:scattering_M}) are then calculated using these generalized helicity spinors.

\begin{figure*}[!htp]
 \begin{align}
    &\subfloat{
    \includegraphics[width=0.22\textwidth]{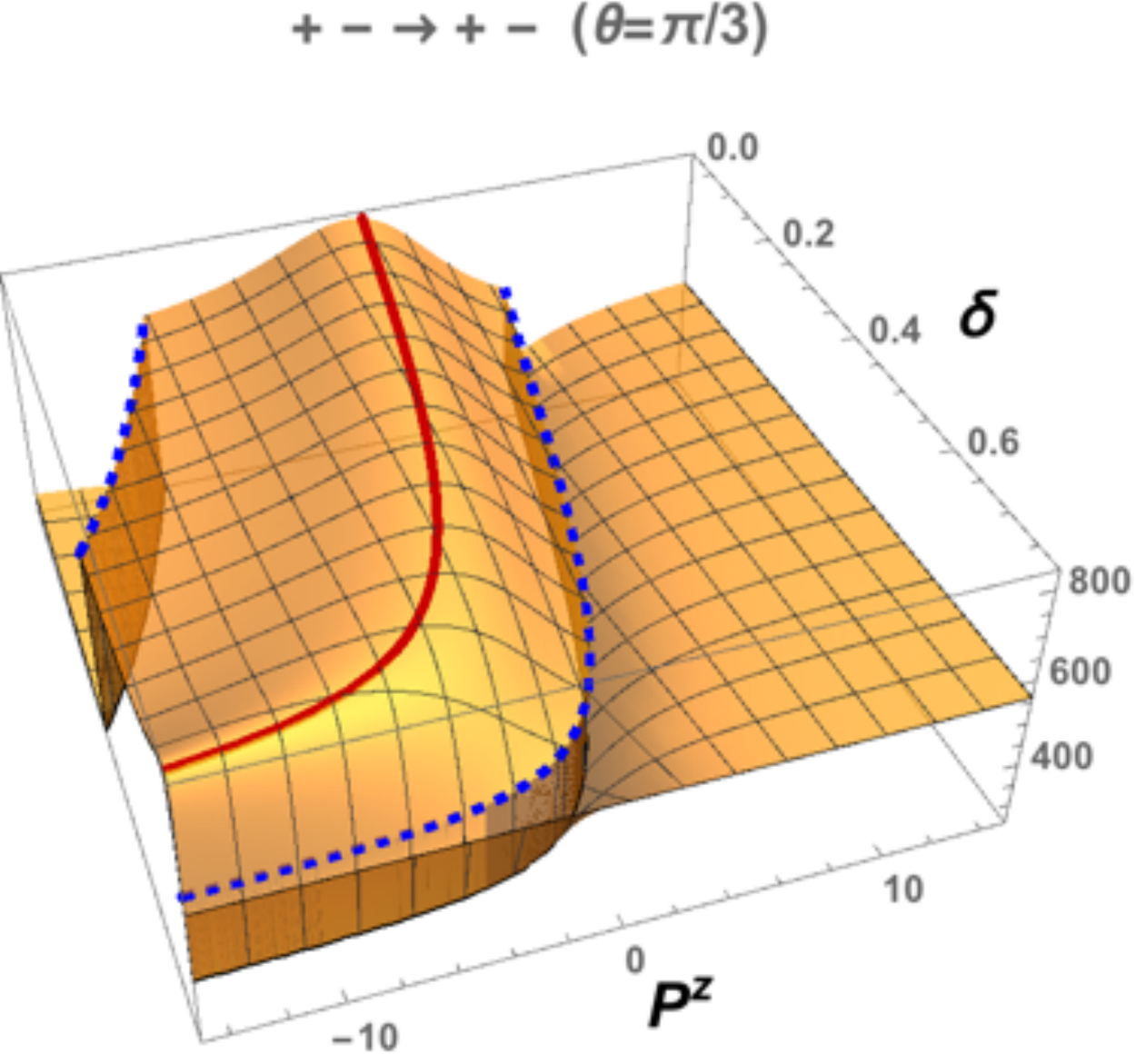}
    }
    %\quad
    \subfloat{
    \includegraphics[width=0.22\textwidth]{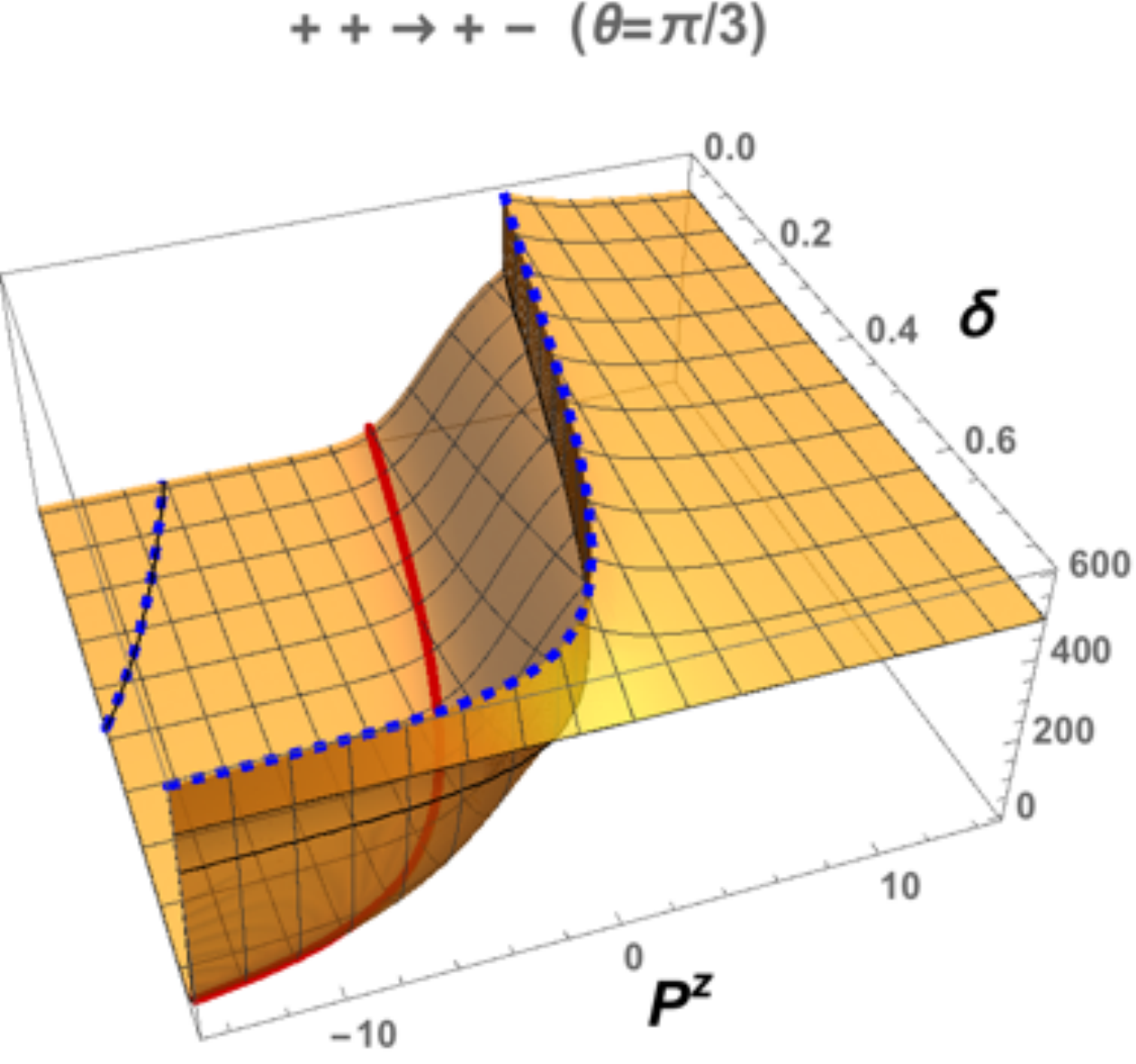}
    }
    %\quad
    \subfloat{
    \includegraphics[width=0.22\textwidth]{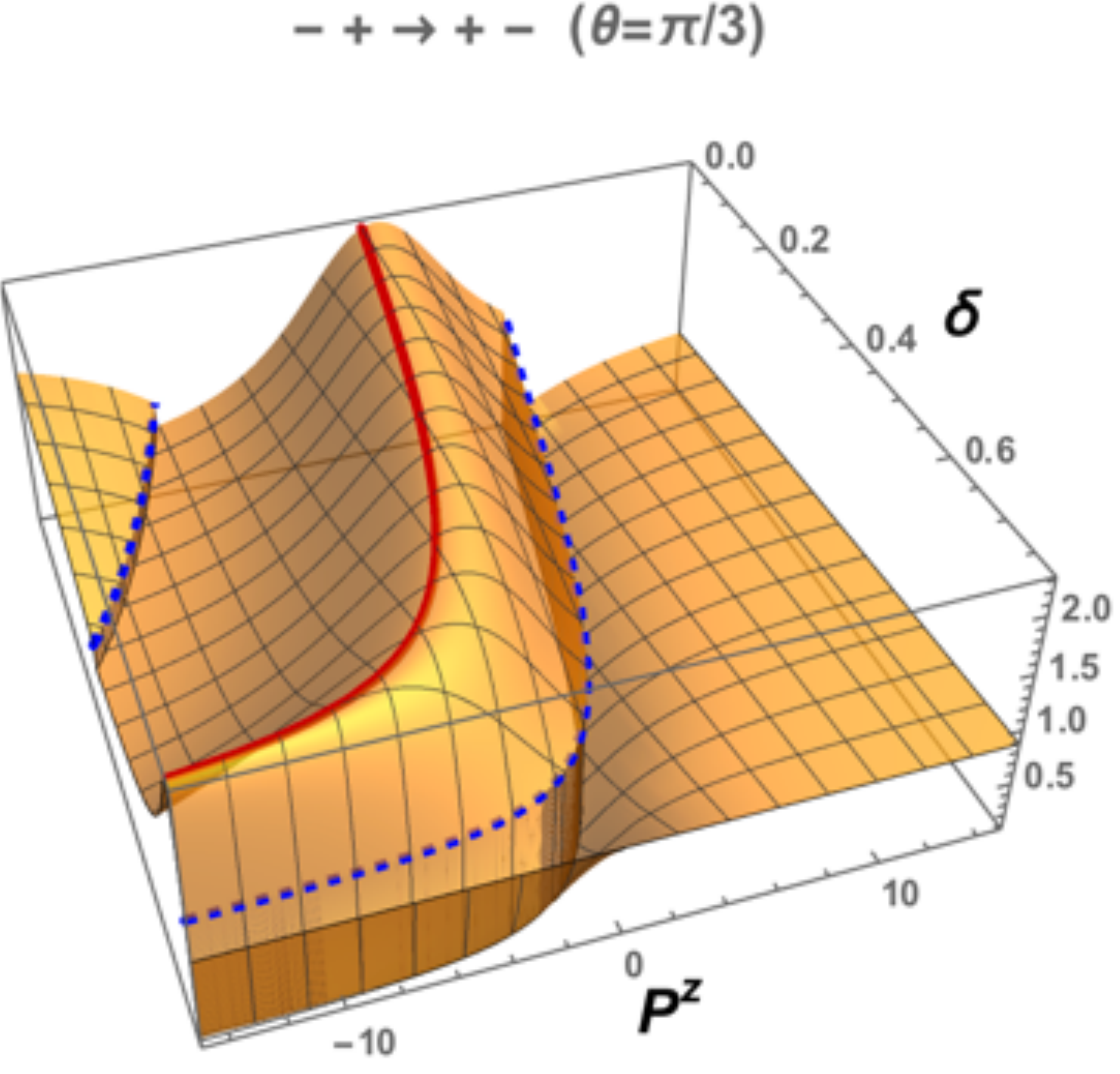}
    }
		%\quad
    \subfloat{
    \includegraphics[width=0.22\textwidth]{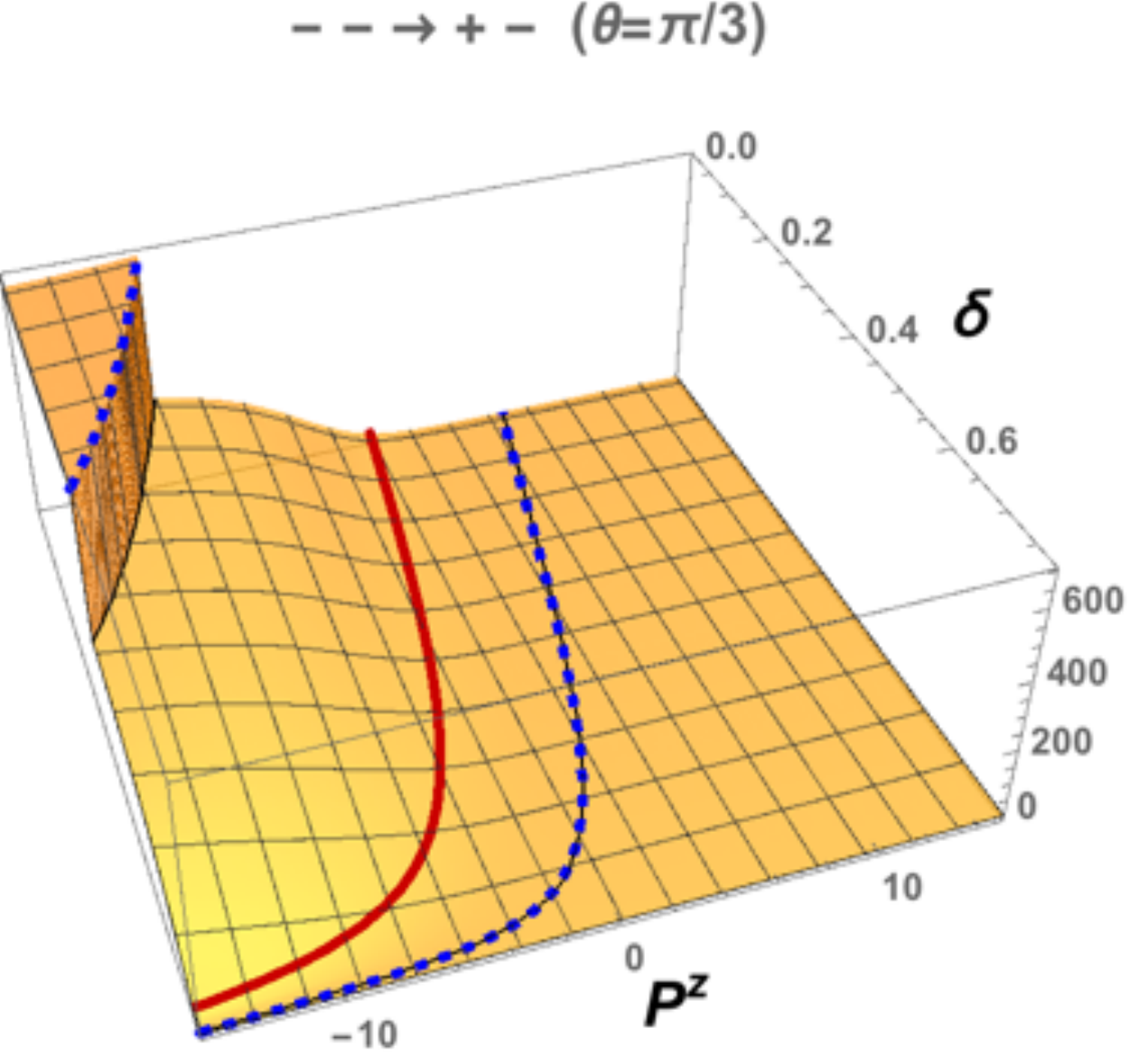}
    }
    \nonumber\\
    &\subfloat{
    \includegraphics[width=0.22\textwidth]{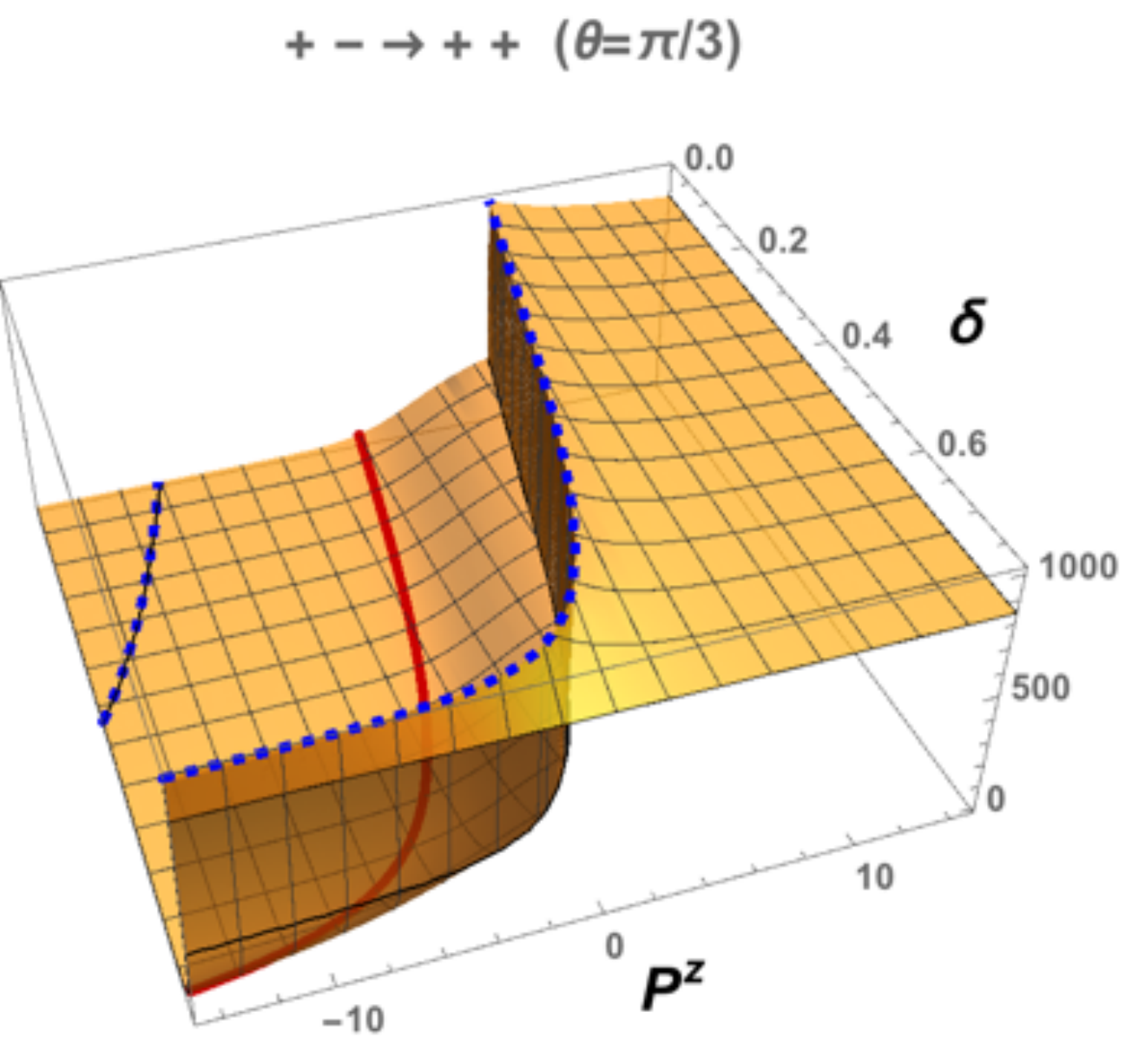}
    }
    %\quad
    \subfloat{
    \includegraphics[width=0.22\textwidth]{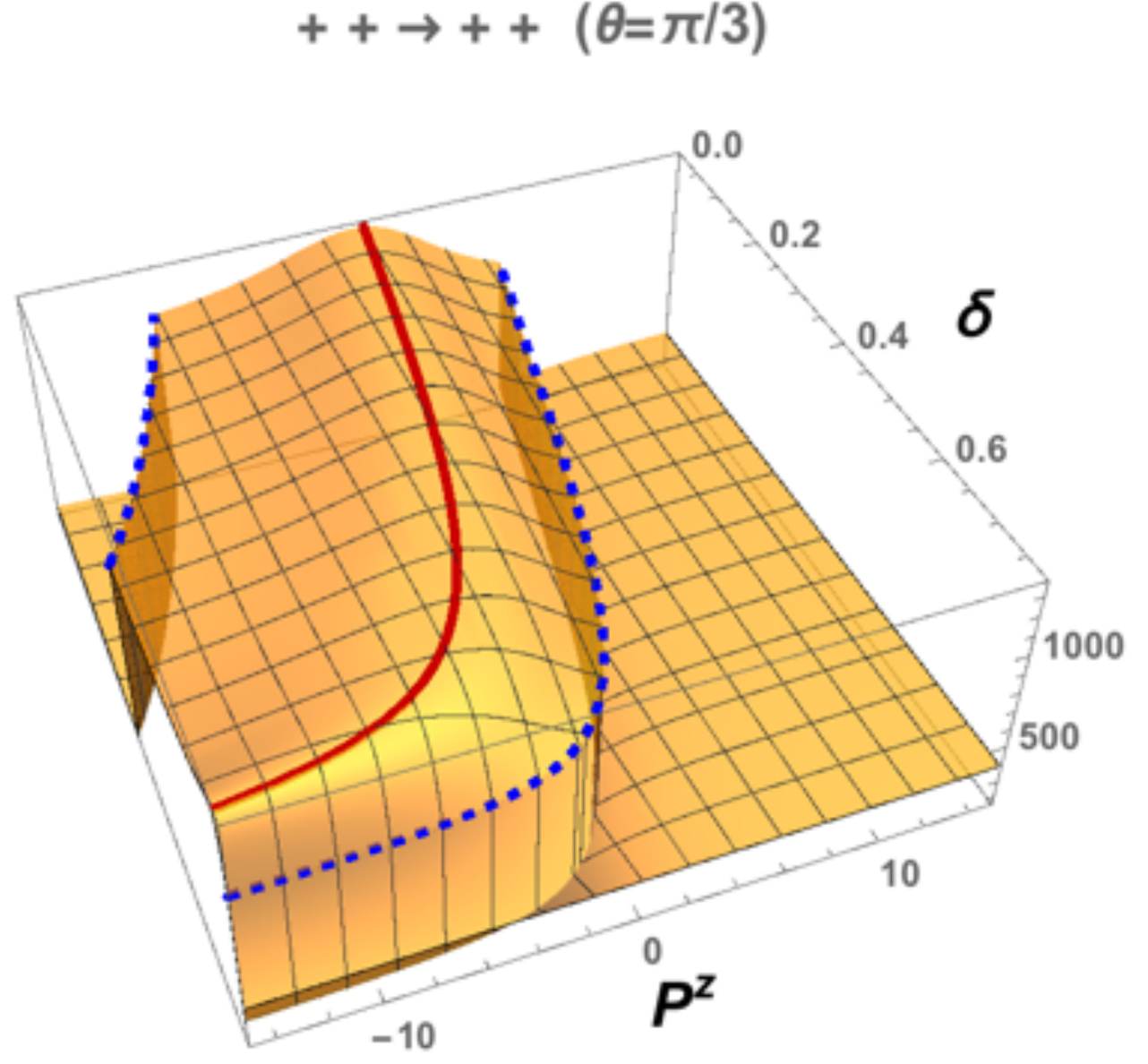}
    }
    %\quad
    \subfloat{
    \includegraphics[width=0.22\textwidth]{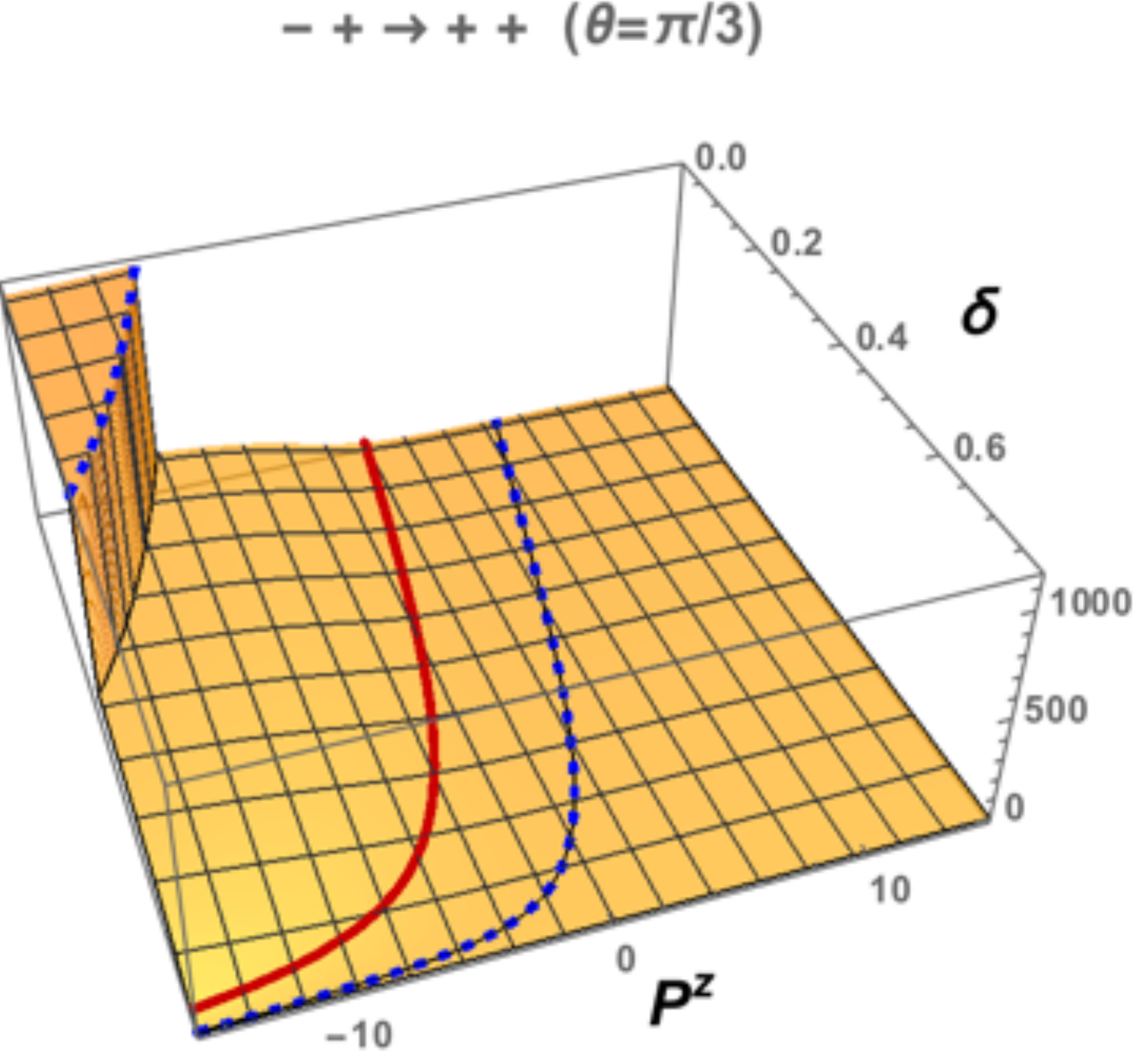}
    }
    %\quad
    \subfloat{
    \includegraphics[width=0.22\textwidth]{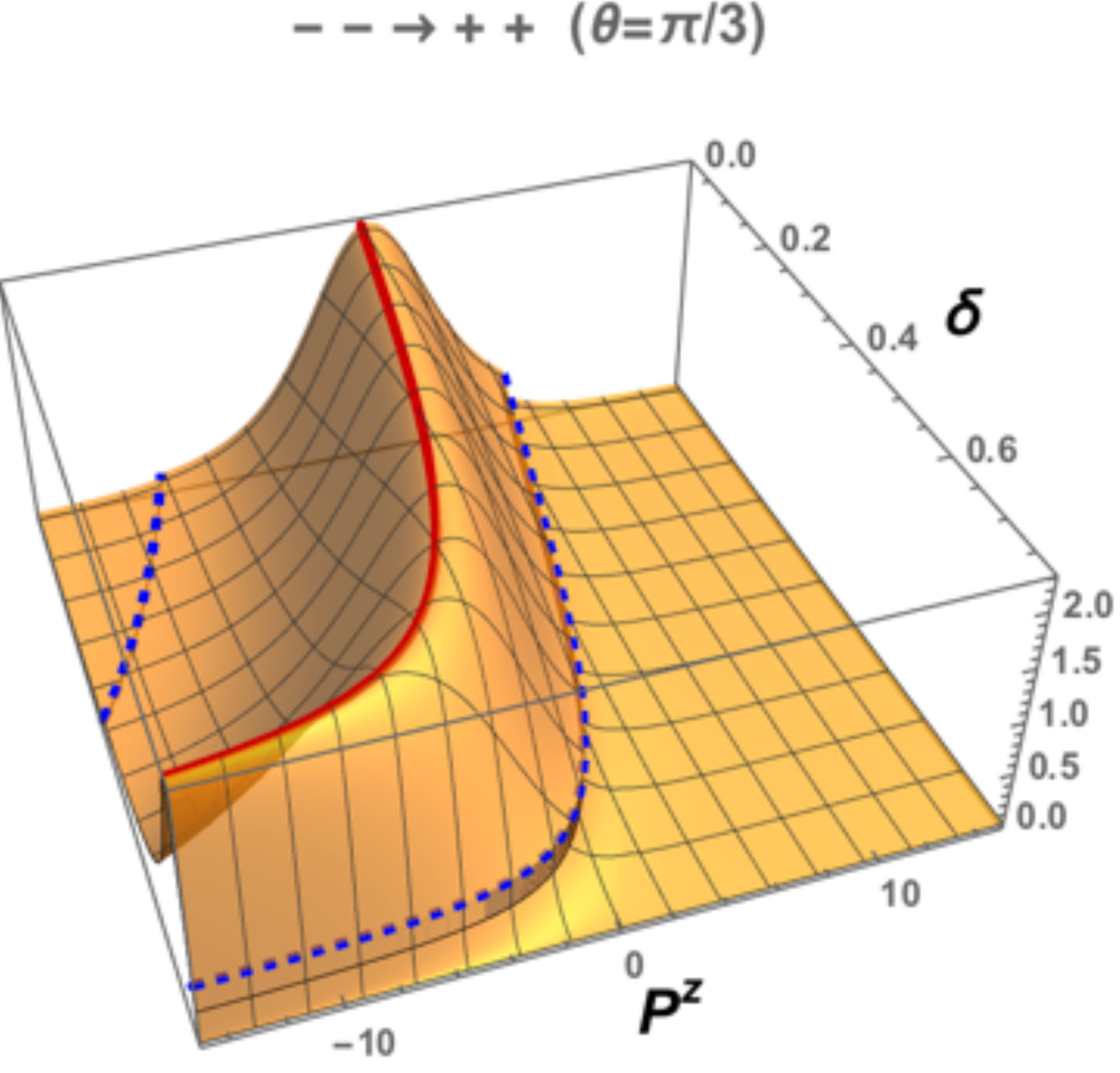}
    }
    \nonumber\\
    &\subfloat{
    \includegraphics[width=0.22\textwidth]{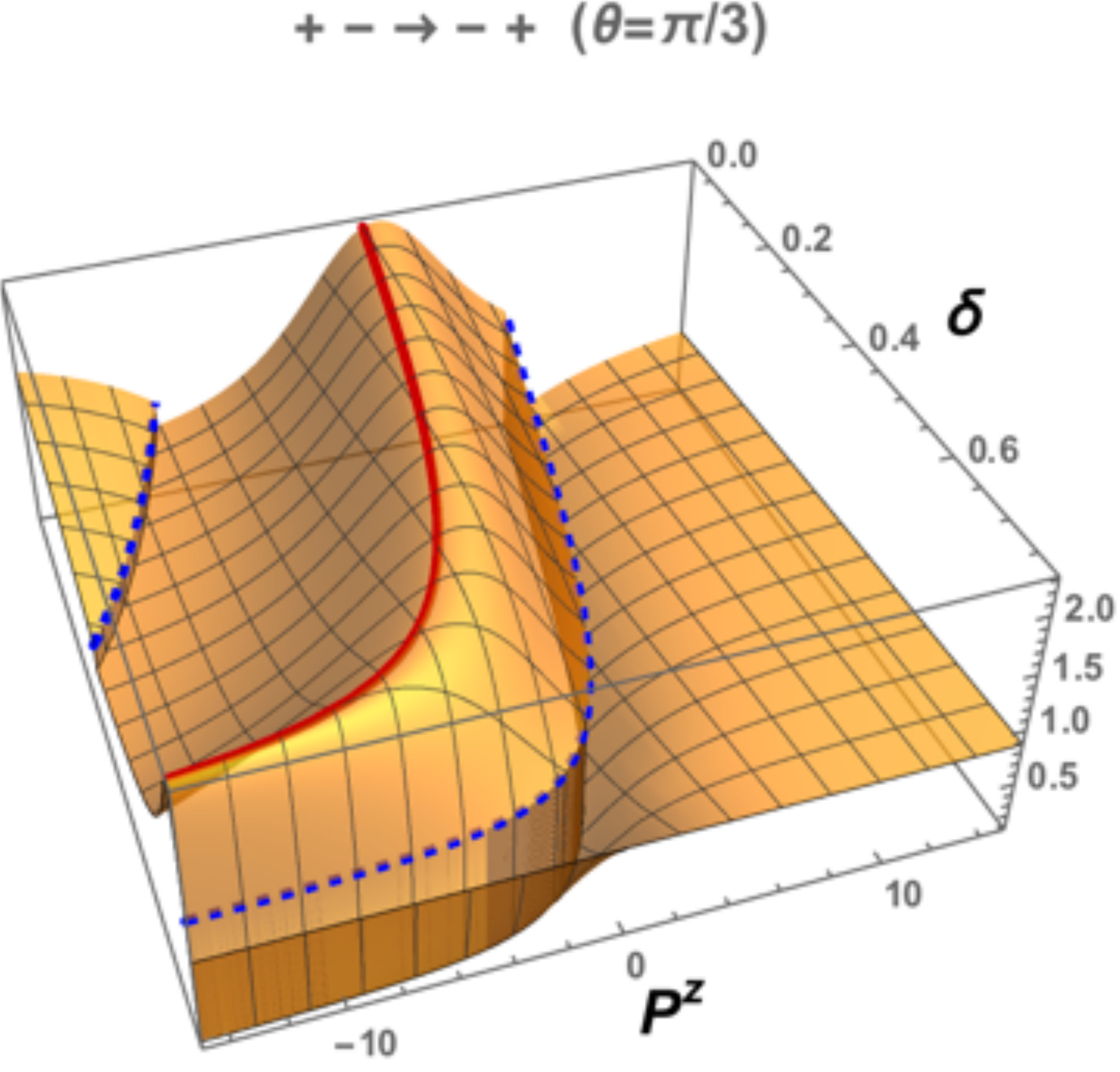}
    }
    %\quad
    \subfloat{
    \includegraphics[width=0.22\textwidth]{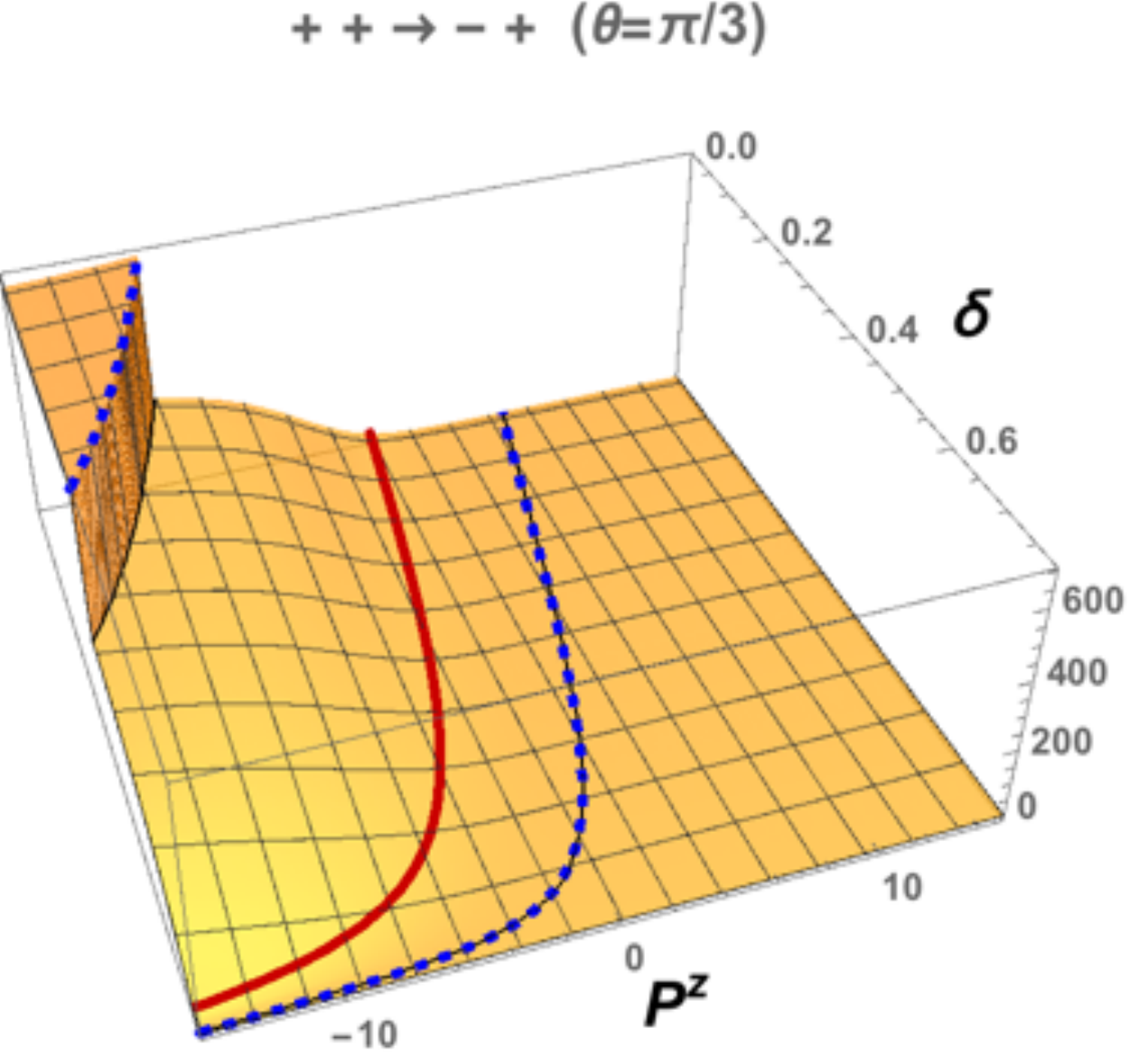}
    }
    %\quad
    \subfloat{
    \includegraphics[width=0.22\textwidth]{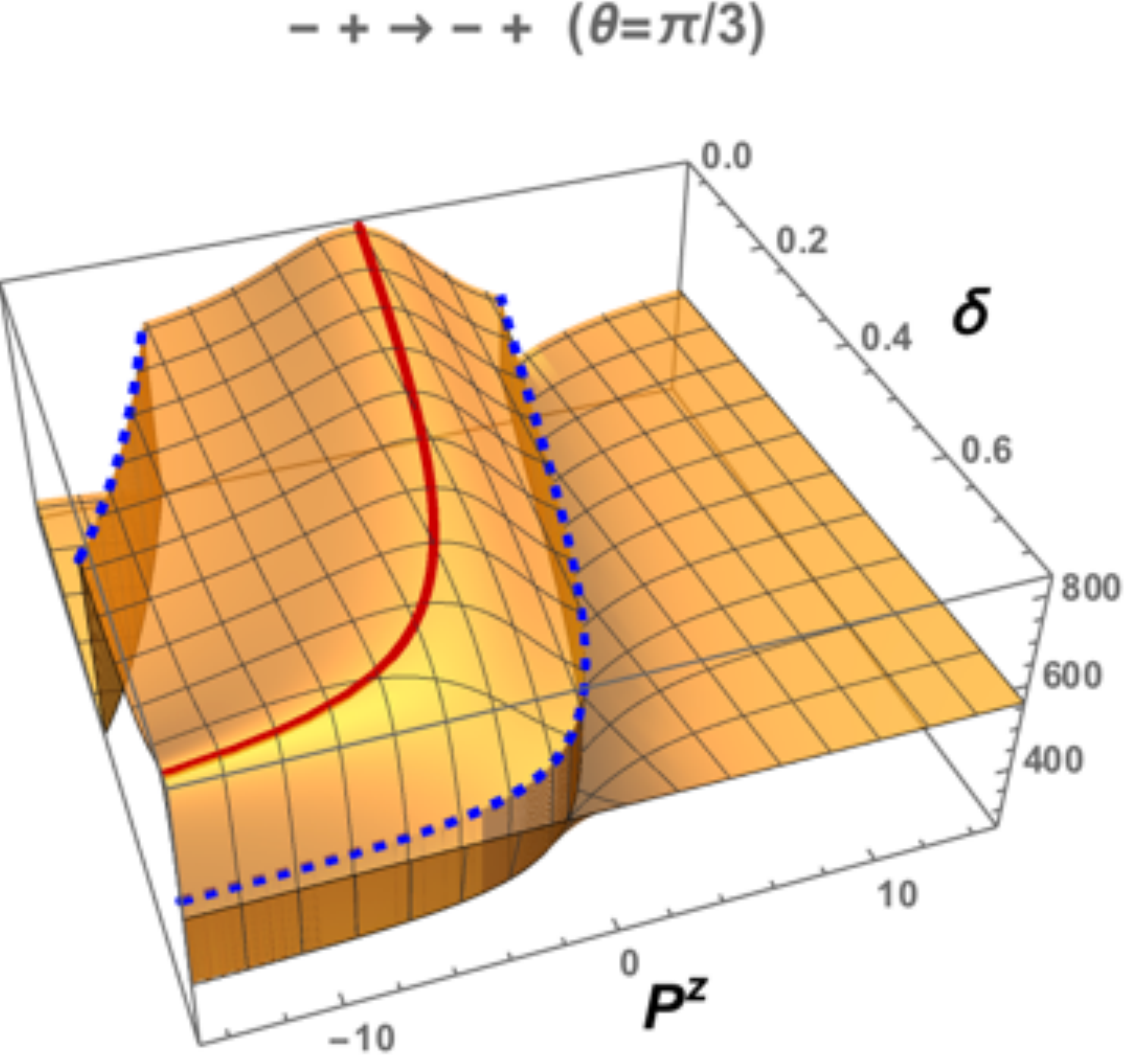}
    }
    %\quad
    \subfloat{
    \includegraphics[width=0.22\textwidth]{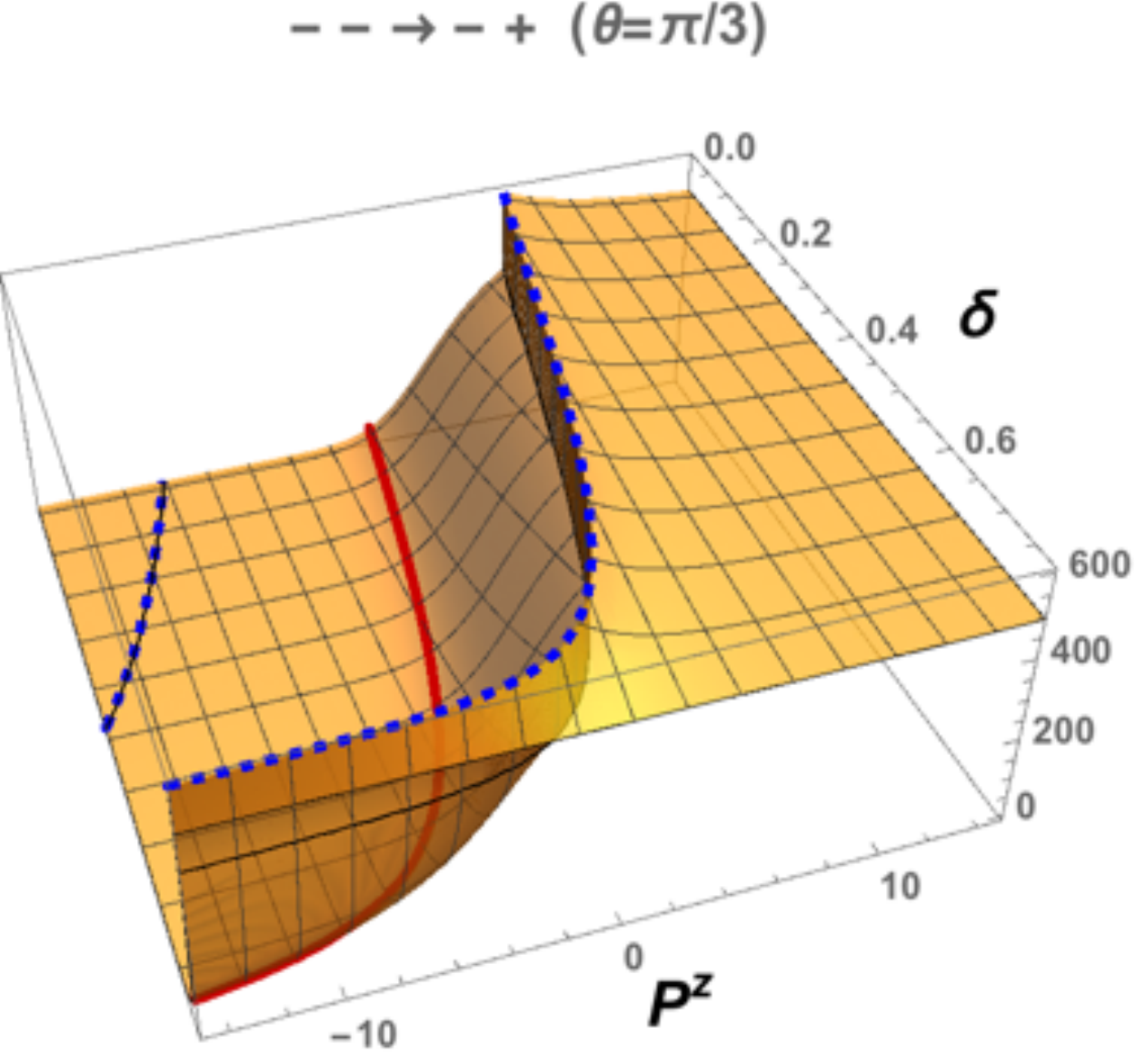}
    }
    \nonumber\\
    &\subfloat{
    \includegraphics[width=0.22\textwidth]{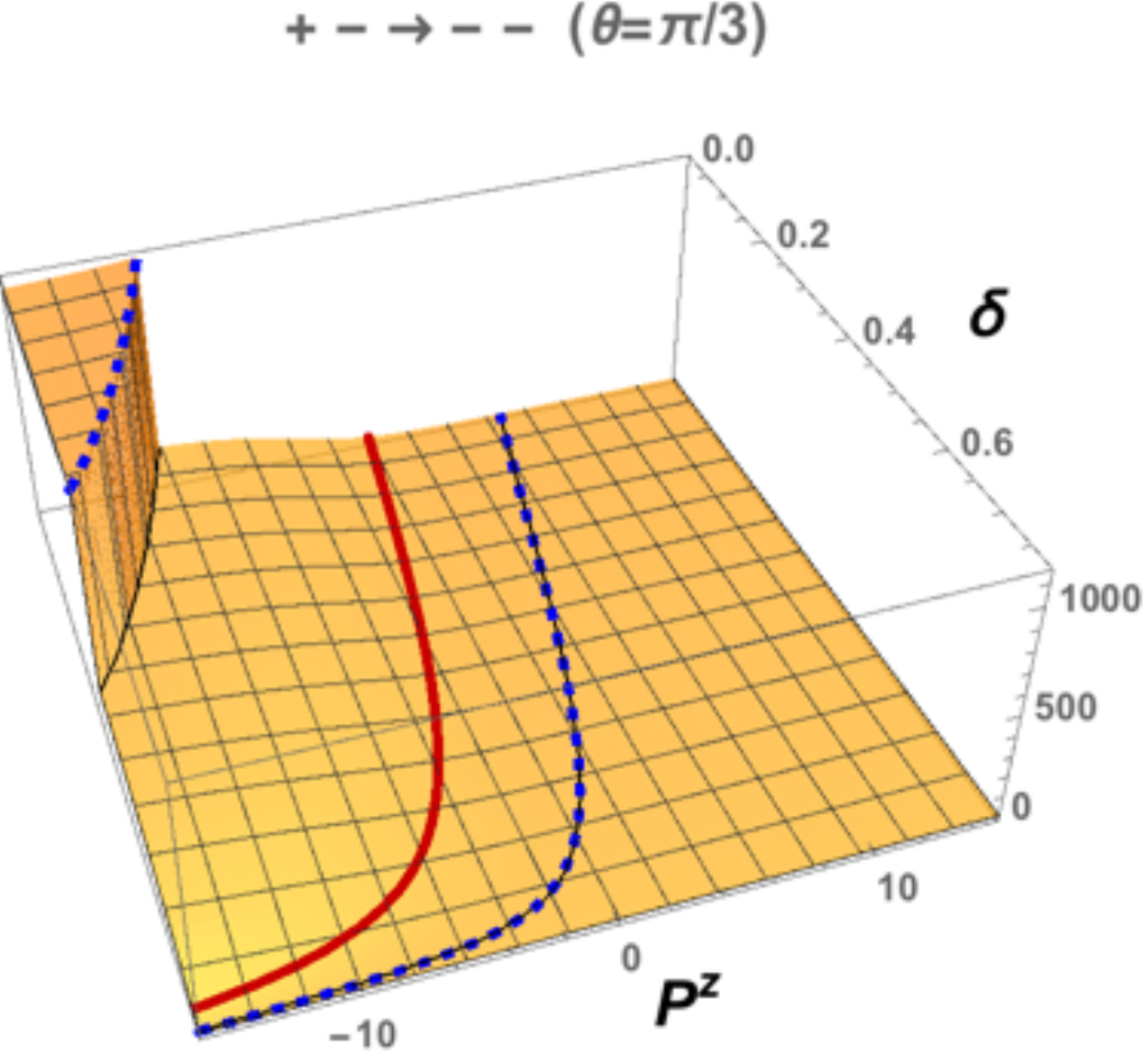}
    }
    %\quad
    \subfloat{
    \includegraphics[width=0.22\textwidth]{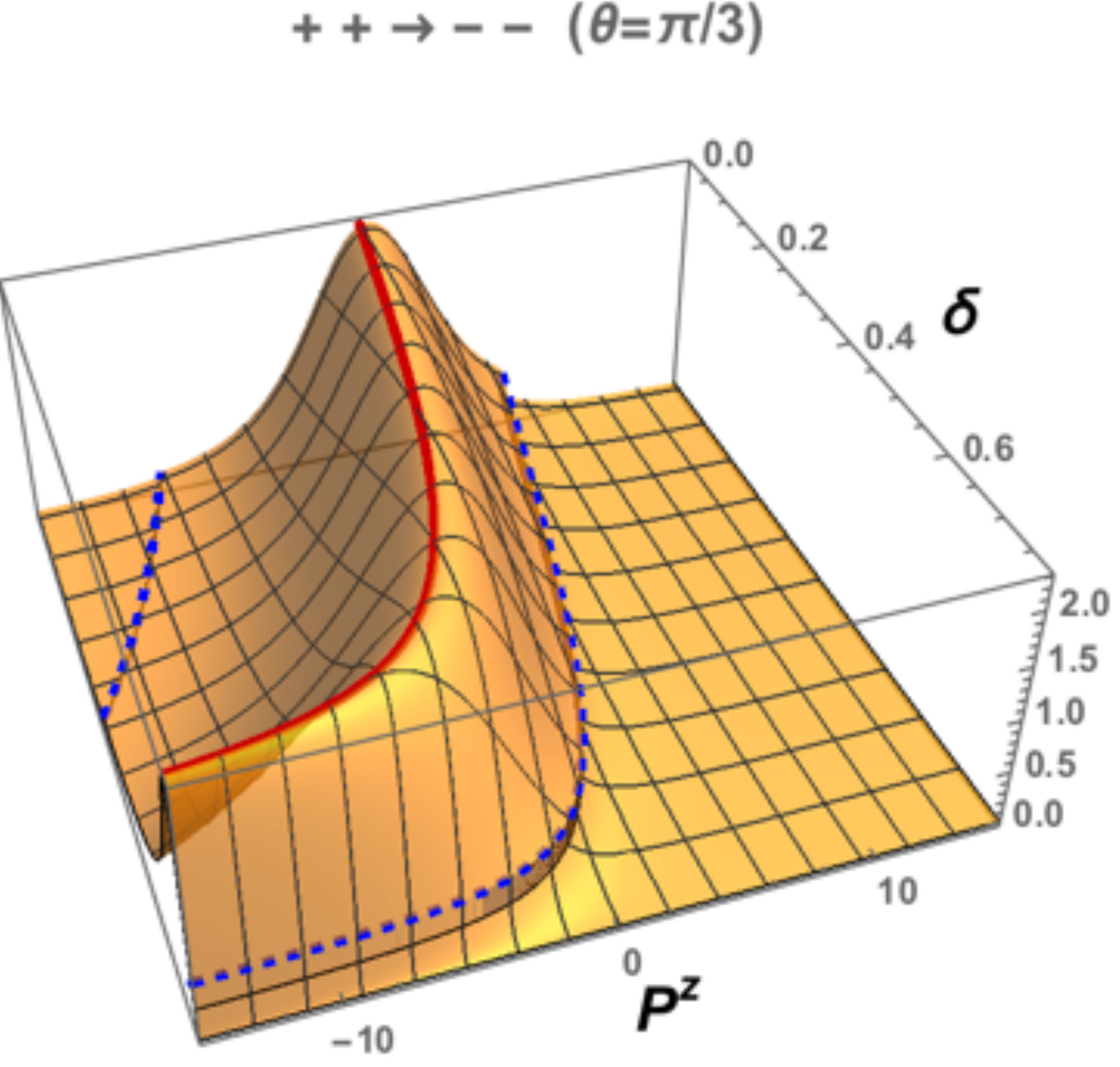}
    }
    %\quad
    \subfloat{
    \includegraphics[width=0.22\textwidth]{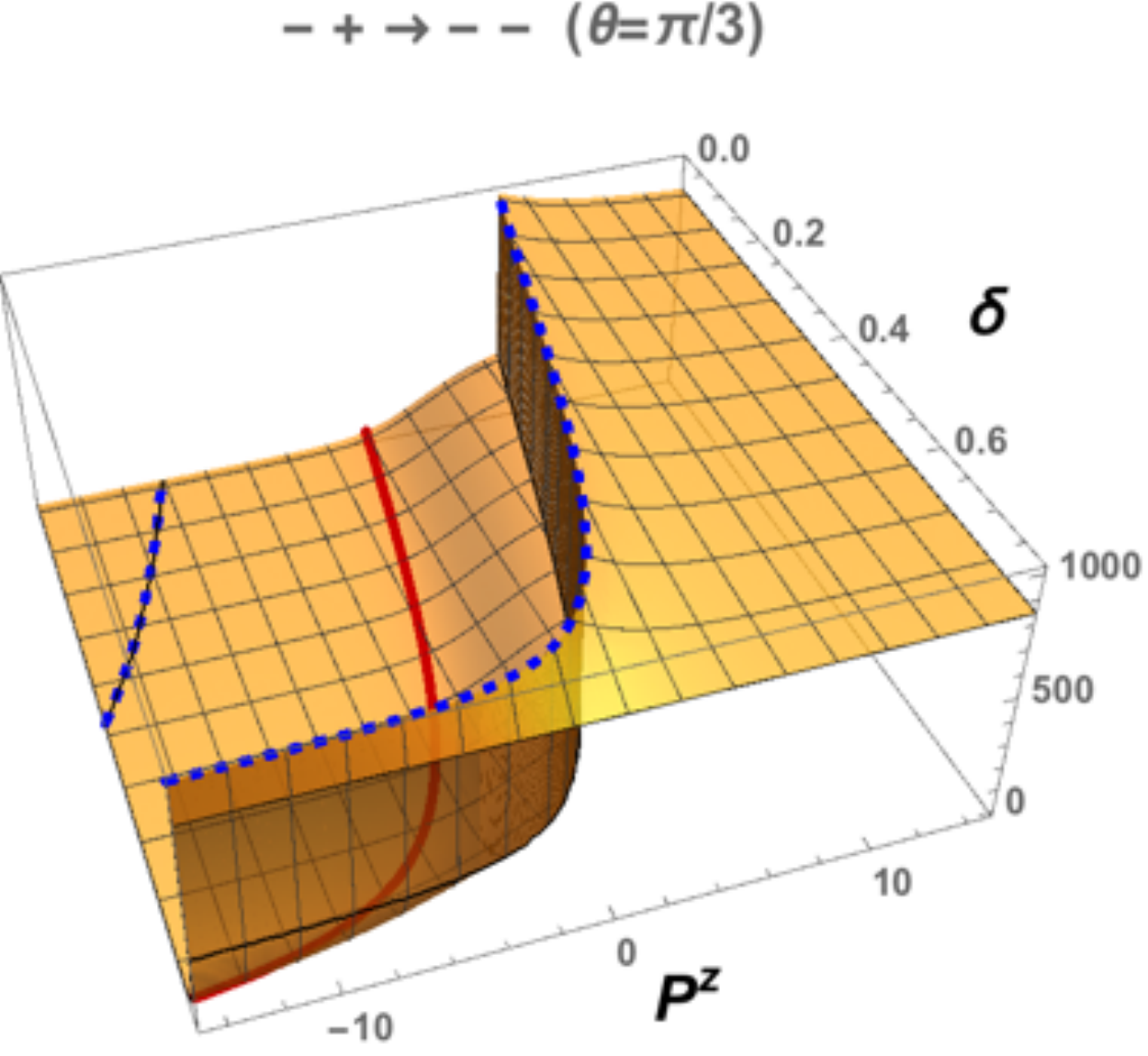}
    }
    %\quad
    \subfloat{
    \includegraphics[width=0.22\textwidth]{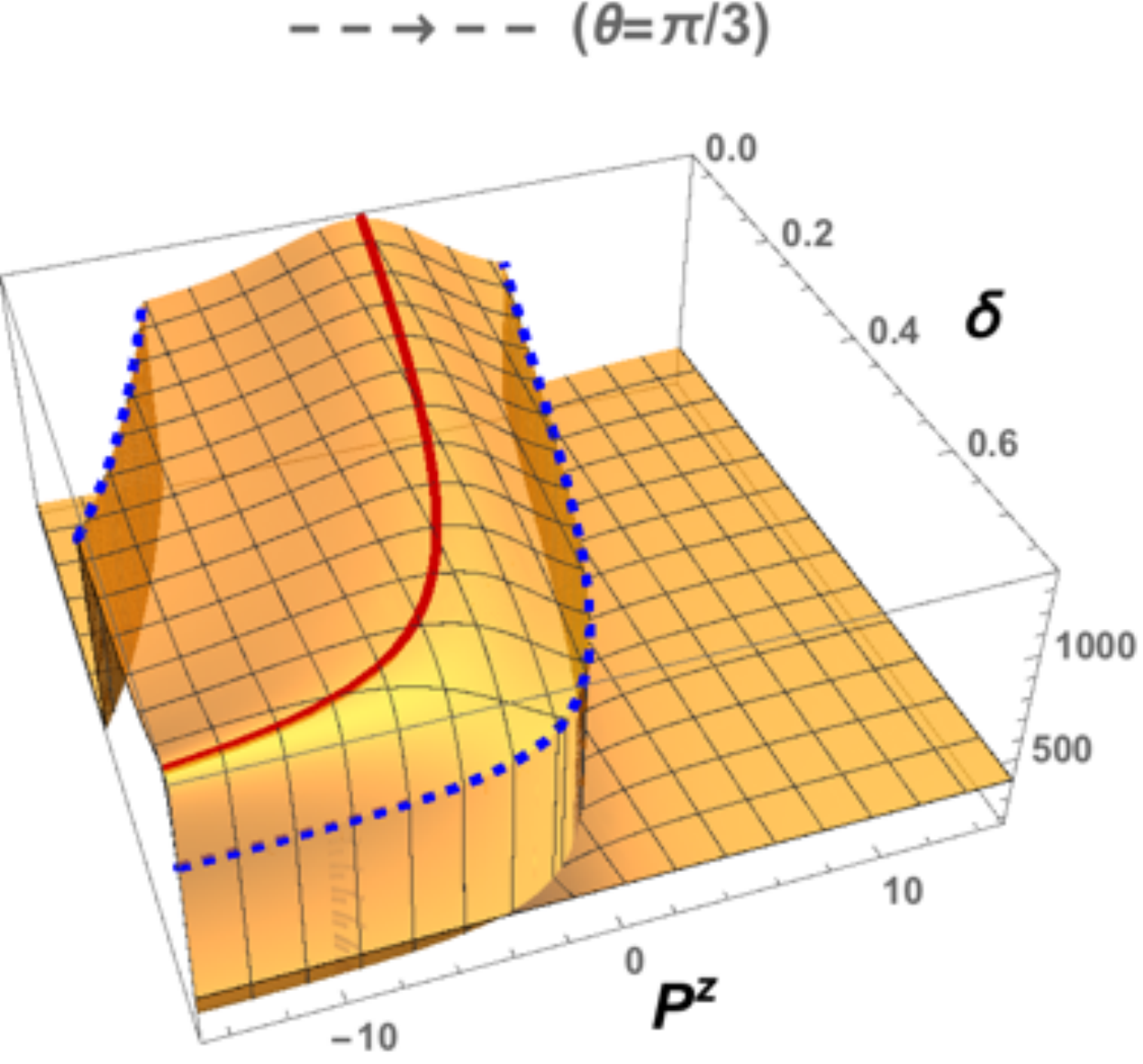}
    }
		\nonumber\\
    &\subfloat{
    \includegraphics[width=0.22\textwidth]{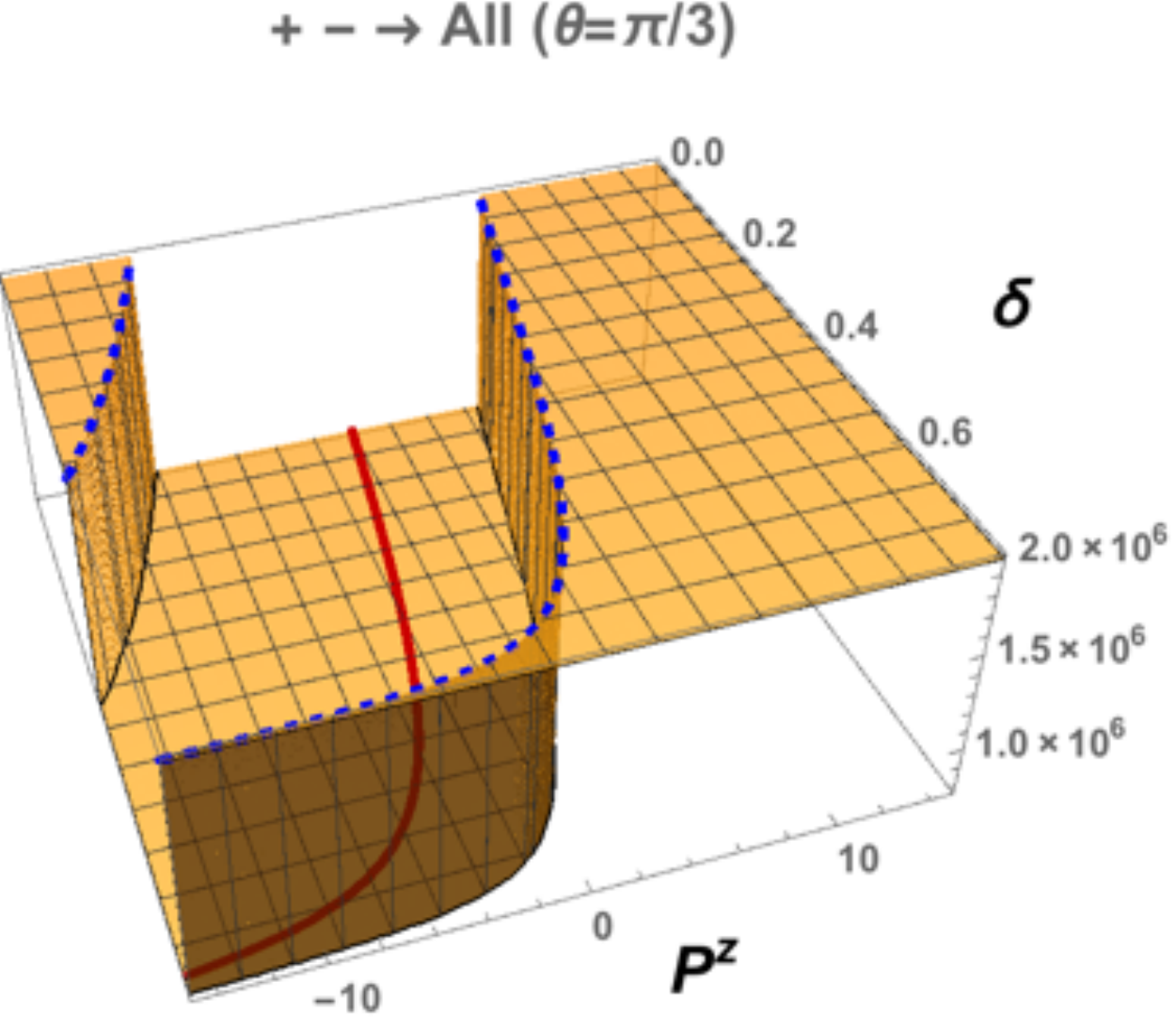}
    }
    %\quad
    \subfloat{
    \includegraphics[width=0.22\textwidth]{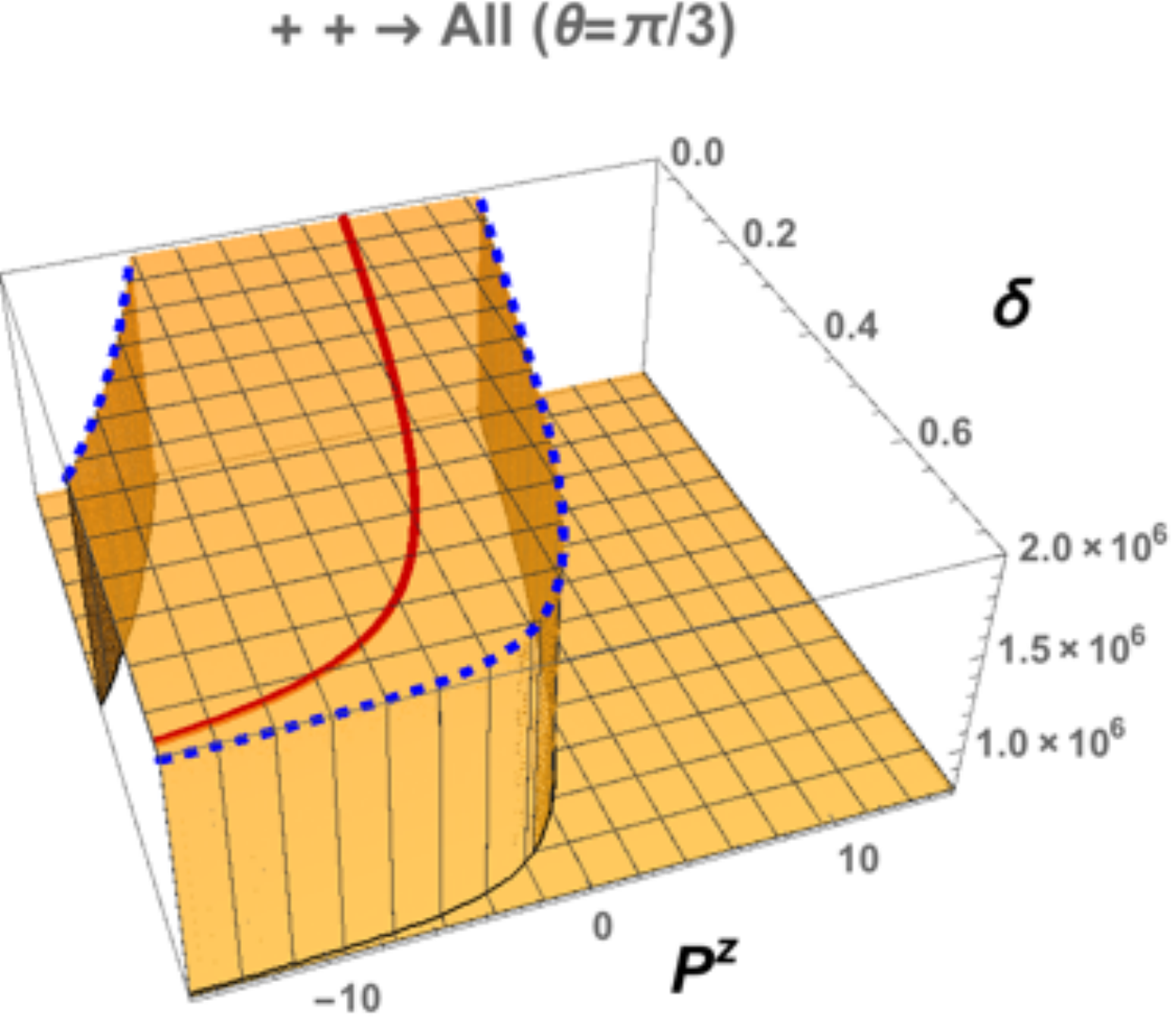}
    }
    %\quad
    \subfloat{
    \includegraphics[width=0.22\textwidth]{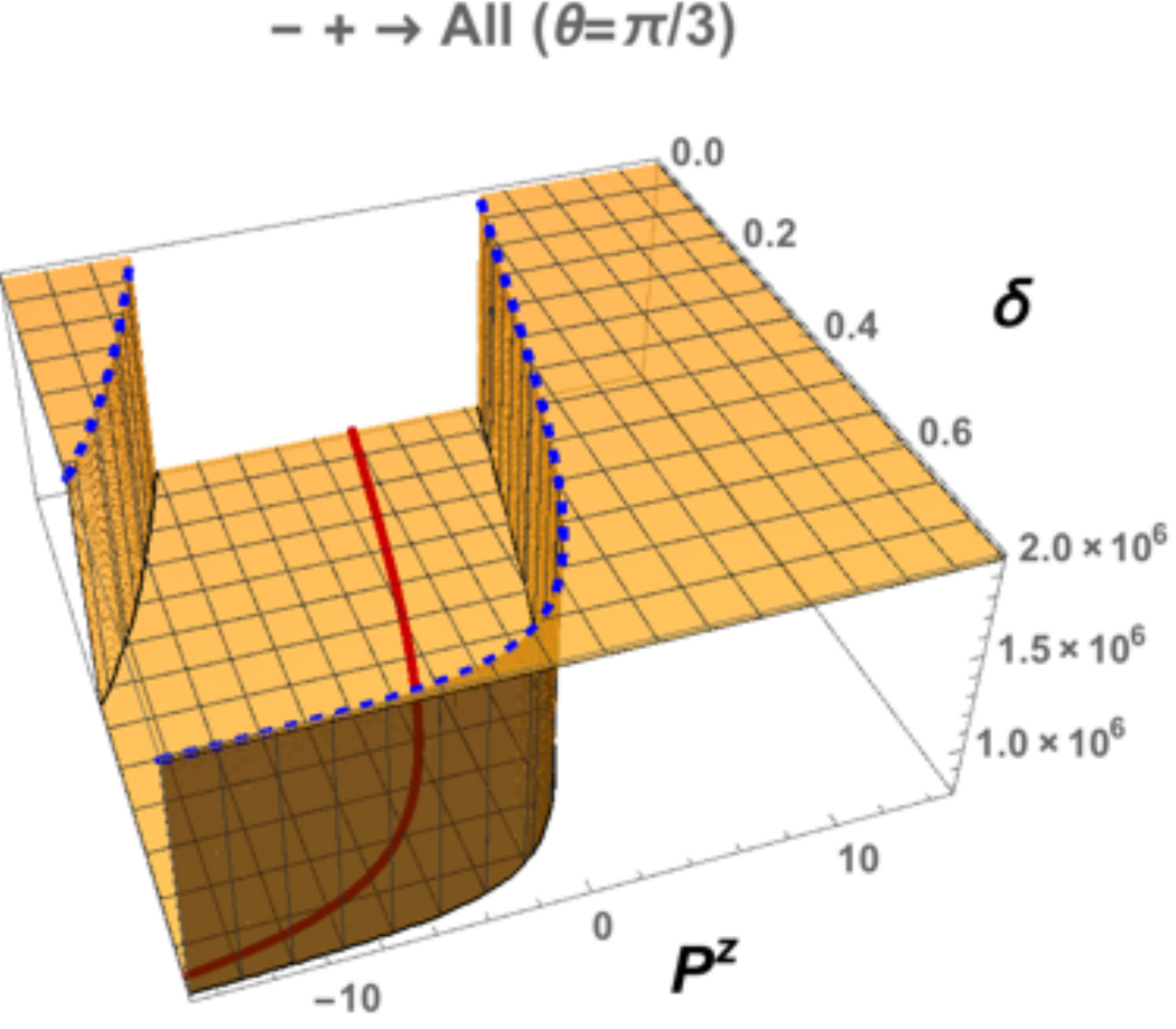}
    }
    %\quad
    \subfloat{
    \includegraphics[width=0.22\textwidth]{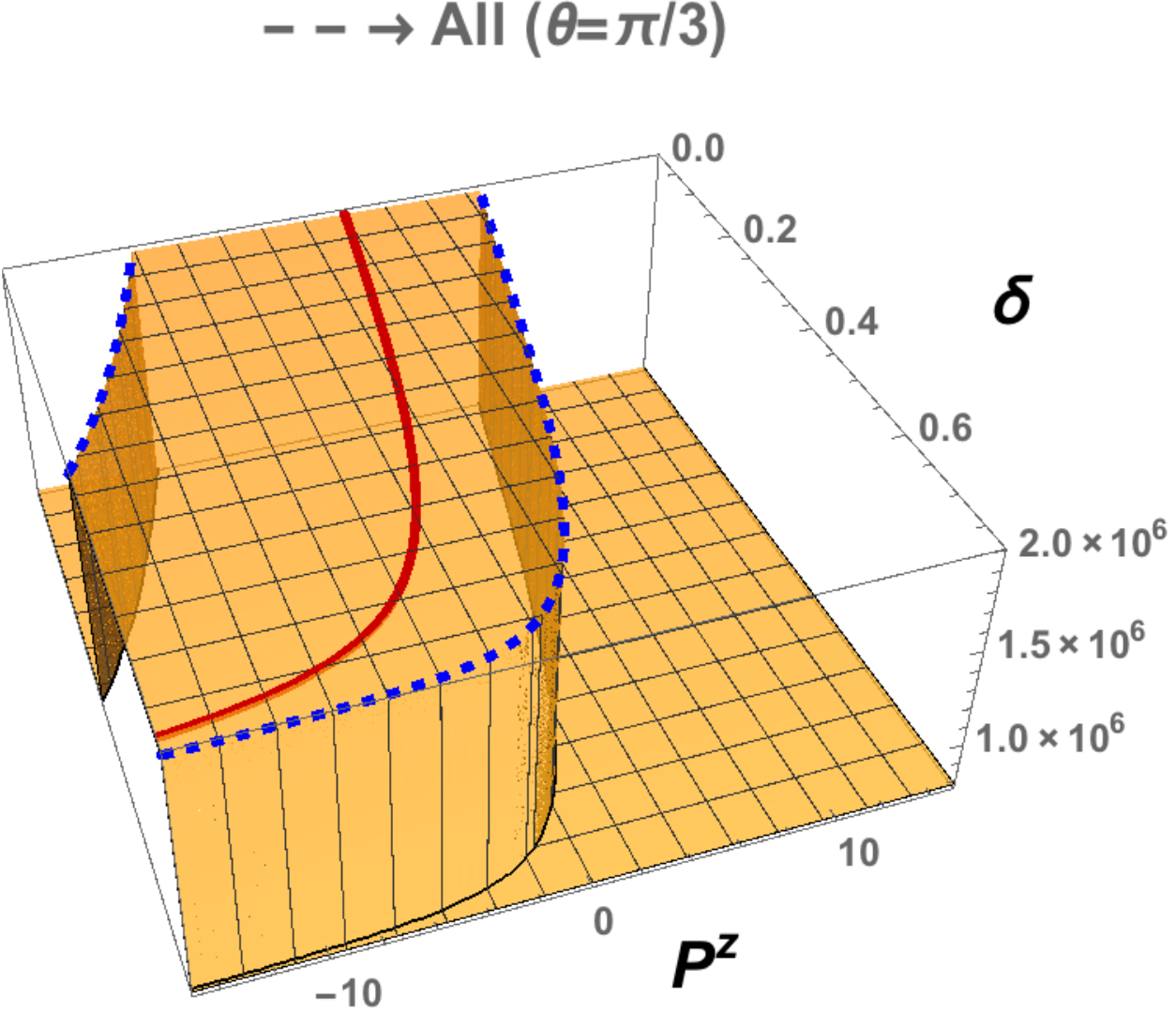}
    }
    \nonumber
 \end{align}
  \caption{\label{fig:Scattering_Helicity_Probabilities}(Color online) Scattering probabilities (with the factor $e^{4}/q^{4}$ dropped) for 16 different helicity configurations, where the masses of the two colliding particles are $m_{1}=1$ GeV and $m_{2}=1.5$ GeV, the center of mass momentum for each particle is $p=2$ GeV and the scattering angle $\theta=\pi/3$.
	The 4 plots in the last row are each a summation of the 4 plots above, representing the probability of a certain helicity configuration going into all possible final helicity states.}
\end{figure*}

In our calculation, we choose $\theta=\pi/3$ in the center of mass frame, and the center of mass momentum for each particle as $p=2$ GeV.
The masses of the colliding particles are chosen to be $m_{1}=m_{3}=1$ GeV and $m_{2}=m_{4}=1.5$ GeV as an illustrative example.
The scattering probabilities for all 16 different helicity configurations are plotted in terms of both the interpolation angle $\delta$ and the total momentum $P^{z}$ in the first 4 rows of Fig.~\ref{fig:Scattering_Helicity_Probabilities}, where we denote the positive and negative helicities with ``+'' and ``-''.
For example, ``$+-\rightarrow +-$'' represents a process where particle 1 with positive helicity and particle 2 with negative helicity scatter to particle 3 with positive helicity and particle 4 with negative helicity.

The first striking feature that these plots exhibit is the two boundaries indicated by the blue dashed lines in Fig.~\ref{fig:Scattering_Helicity_Probabilities}, across which the probabilities suddenly change their values. The appearance of these boundaries is due to the
the bifurcation of the branches that are distinguished by the critical interpolation angle $\delta_c$ as we have discussed
in Sec.~\ref{sub:spin_orientation}. Since $\delta_c$ is obtained from the condition $P_{\mT}=0$, the initial two particles that move only in the $\pm z$ direction yield the corresponding two boundaries upon the boost of the reference frame in the $\pm z$ direction.
The sudden change of the helicity amplitude-square values is thus caused by the spin flip of the incoming scattering particles 1 and 2 for the given helicity assignment in the amplitude.
When this spin flip happens, the probabilities for the same helicity configuration represent different spin configurations.
In other words, we are not calculating the same physical process any more but three different initial spin configurations
for the corresponding three different branches divided by the two boundaries although the given helicity configuration is
identical regardless of the boundaries. Thus, the helicity probability suddenly shifts to a different value crossing a boundary.
The only helicity amplitude immune from such bizarre behaviors is the light-front helicity amplitude without any border at $\delta=\pi/4$.
%\textcolor{red}{\st{ It may clarify the reason why the helicity amplitude defined only in LFD can be regarded as physical.}}

\begin{figure}[!t]
  \includegraphics[width=\columnwidth]{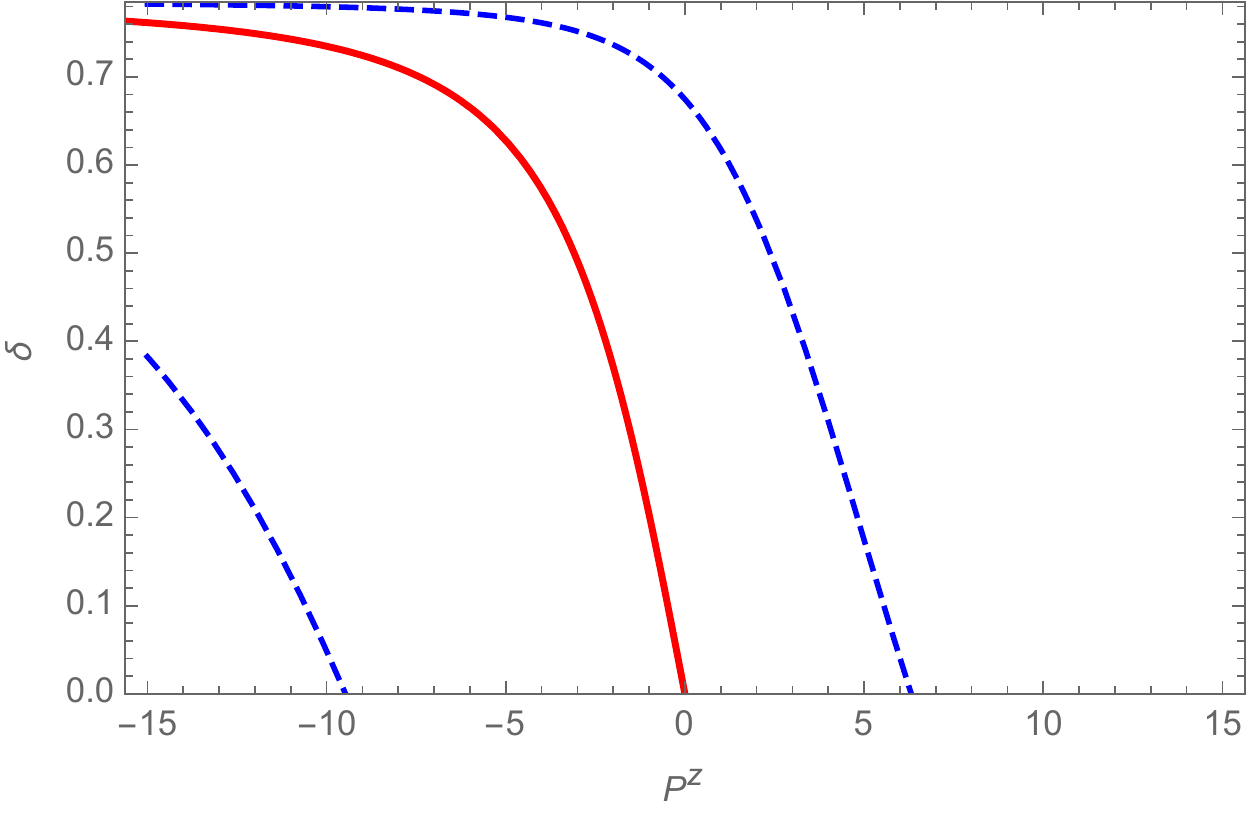}
  \caption{\label{fig:P_mt_region}(color online) The two dashed blue lines are the boundaries defined by Eq.~(\ref{eqn:Boundary_lines}), across which the spin of electron or positron will flip.
  The red solid line is the J-curve described by Eq.~(\ref{eqn:J_curve}).}
\end{figure}

% The argument that the spin does not flip for electron and positron with definite helicities only applies to the light front.
% This is easiest to understand in the instant form limit, where we all know that we can move to a frame where the momentum changes sign.
% When that happens the spin also needs to flip in order to have the same ``helicity'' as before.
To see why the spin flips for $\delta \neq \pi/4$, we can approach it in two different ways.
First, we can fix a certain value for the total $P^{z}$, i.e. we fix the inertial frame, and look at what happens when we change the interpolation angle $\delta$.
From the $\theta=\pi$ edge of Fig.~(\ref{fig:thetas_on_theta_delta}), it's clear that for a particle moving in the $-z$ direction, as we increase the $\delta$, at some critical angle $\delta_{c}$, the spin of ``positive helicity'' will flip from $\theta_{s}=\pi$ to $\theta_{s}=0$.
And because the spin direction of ``negative helicity'' is always exactly the opposite of that of the ``positive helicity'', the spin of the ``negative helicity'' will flip at this $\delta_{c}$ from $\theta_{s}=0$ to $\theta_{s}=\pi$.
Alternatively, and perhaps more conveniently, this spin flip can be understood by fixing a certain value for the interpolation angle $\delta$, and see what happens when we boost the system in the $z$ direction.
As we discussed in Sec.~\ref{sec:helicity_operator_and_spinor_for_any_interpolation_angle}, using Eq.~(\ref{eqn:J3_P_z_direction}), the ``helicity'' for particles moving in $\pm z$ direction is defined as whether the spin has the same sign as $P_{\mT}$.
So for the same helicity configuration, as we boost the system in $\pm z$ direction, the spin of particle 1 or 2 will flip if the sign of its $P_{\mT}$ changes.
Therefore, we can find these boundary lines by setting $P_{\mT}=0$ for particles 1 and 2, with their momenta given by Eqs.~(\ref{eqn:boosted_p0})-(\ref{eqn:boosted_pp}) and the form of $P_{\mT}$ given by Eq.~(\ref{eqn:P_interpolation_4}).
The equations that describe these two boundaries are respectively given by
\begin{subequations}
  \label{eqn:Boundary_lines}
  \begin{align}
    \tan\delta=-\dfrac{\epsilon_{1}P^{z}+ p E}{\epsilon_{1}E+p P^{z}}, \\
    \tan\delta=-\dfrac{\epsilon_{2}P^{z}- p E}{\epsilon_{2}E-p P^{z}}.
  \end{align}
\end{subequations}
% The reason we don't see these two boundaries in Fig.~\ref{fig:Helicity_amplitudes} is because the mass of electron (positron) is too small compared to its center of mass energy $\epsilon=120$ MeV.
% The amount of boost required to flip the sign of $P_{\mT}$ is therefore quite large, and these two boundaries are pushed far away to large $P^{z}$ regions.
Each of the 4 plots in the last row of Fig.~\ref{fig:Scattering_Helicity_Probabilities} is the summation of the 4 corresponding plots above it.
They are the probabilities of a certain helicity configuration going into all possible final helicity configurations.
By summing over all final states, we see clearly that the varying landscapes due to the helicity configurations of the scattered particles 3 and 4 disappears and the two boundaries due to the spin flips of the incoming particles 1 and 2 remain.
They divide the whole landscape into three regions.
To show the positions of these boundaries more clearly, we plot them in the $\delta$ vs $P^{z}$ plane as the two blue dashed lines in Fig.~\ref{fig:P_mt_region}.
% Across these two blue dashed lines, the helicity amplitudes shifts to a different value.
% And right at these two boundaries, the helicity and helicity amplitudes are not well defined, as $P_{\mT}=0$ in Eq.~(\ref{eqn:J3_P_z_direction}).

Another interesting feature exhibited by all 16 helicity probability plots is the J-curve indicated by the red solid line in both Fig.~\ref{fig:Scattering_Helicity_Probabilities} and Fig.~\ref{fig:P_mt_region}.
This is the same curve that appeared in the time-ordered amplitudes in the $\phi^{3}$ theory \cite{Ji2013} and in the sQED theory \cite{Ji2015} we studied previously.
It starts out in the center of mass frame in the $\delta=0$ limit and maintains the same probability value throughout the whole range of interpolation angle.
It follows exactly the same formula as before \cite{Ji2013, Ji2015}:
\begin{align}
  P^{z}=-\sqrt{\dfrac{M^{2}(1-\Cc)}{2\Cc}},\label{eqn:J_curve}
\end{align}
where $M^{2}$ is the same as the invariant Mandelstam variable ``$s$'' used in \cite{Ji2013}.
As one can see, this curve is independent of the specific kinematics of the scattering process, e.g. the $\theta$ angle, the masses of these four particles and so on.
It's only scaled by the center of mass energy $M$, and has a universal shape.
In our example, we chose $\theta=\pi/3$ in the center of mass frame.
As the $\theta$ angle changes, the landscape will continuously change, but the main features remain the same, and this J-curve is always present.
Incidentally, this J-curve as given by Eq.~(\ref{eqn:J_curve}) is exactly $P_{\mT}=0$ for the system, which can be verified by setting $P^{0}=\sqrt{M^{2}+(P^{z})^{2}}$ and $P^{3}=P^{z}$ in Eq.~(\ref{eqn:P_interpolation_4}).
As $\delta\rightarrow\pi/4$, $P^{z}\rightarrow-\infty$, and therefore if we take the limit to the light front along this J-curve, we will have the same probability value as in the center of mass frame in the IFD, which doesn't agree with the invariant light-front result in general, as can be seen from Fig.~\ref{fig:Scattering_Helicity_Probabilities}. This raises the issue of non-commuting orders in taking the two independent limits, $\delta \rightarrow \pi/4$ and $P_z \rightarrow -\infty$. Thus, it requires a great caution in taking the limit to $\delta = \pi/4$.
This is reminiscent of the zero-mode issue in LFD as we have previously discussed in great details~\cite{Ji2013, Ji2015}.
%Therefore, just like in Ref.~\cite{Ji2015}, this J-curve gives rise to a treacherous point (the zero mode issue) at $P^{z}=-\infty$ in the light-front limit.
%It thus requires great caution in taking the light-front limit from $\Cc\neq0$.
% It also appears in every one of these 16 different amplitudes with different spin configurations.
% We haven't been able to figure out why this universal shape appears not only in time-ordered amplitudes, but also in these helicity amplitudes.

%For interpolation angle $\delta\neq \pi/4$, the calculated helicity probabilities are frame dependent.
%This is because the same helicity configuration in different inertial frames actually represent different physical spin configurations.
%This gives rise to the varying landscapes we see in Fig.~\ref{fig:Scattering_Helicity_Probabilities}.
In LFD ($\delta=\pi/4$), all 16 helicity probabilities are frame independent as we have discussed.
We may relate this LFD invariance with our previous examination on the spin orientations of the light-front spinors.
% This has been understood before, but since we've just done our analysis of spin orientations, we are in a good position to make this clear with a physical argument.
First of all, as we discussed in Sec.~\ref{sub:spin_orientation}, for light-front spinors moving in the $\pm z$ directions, positive (negative) helicity always means the spin is orientated in the $+z$ ($-z$) direction.
Thus, for a certain helicity configuration, the spin orientations of particles 1 and 2 are fixed and there are no boundaries generated by spin flips.
% That means no matter which frame we are in, the helicity states of electron and positron represent the same physical spin configuration.
% Thus, no matter which frame you choose, you can be sure that for a certain helicity configuration, the electron and positron will always have their spin directions fixed.
% What about the spins of the muon and the antimuon?
If $\theta=0$ or $\pi$, then particles 3 and 4 are also moving in $\pm z$ directions, and their spin orientations will also be fixed.
In this case, the probabilities are clearly frame independent, since the helicity configuration in different inertial frames represent the same physical spin configuration.
What about the case where $0<\theta<\pi$ (the case plotted in Fig.~\ref{fig:Scattering_Helicity_Probabilities})?
As we discussed in Sec.~\ref{sub:spin_orientation} using Eq.~(\ref{eqn:wigner_rotation_angle}), the light-front spinors naturally generates the required Wigner rotations when boosted in $\pm z$ directions.
This means even for the scattered particles 3 and 4 not moving in the $\pm z$ direction, the light-front spinors in different inertial frames still represent the same physical spin states.
Since we are calculating the same physical process in different frames, we are guaranteed to get the same probability.

\begin{figure}[!thp]
  \includegraphics[width=\columnwidth]{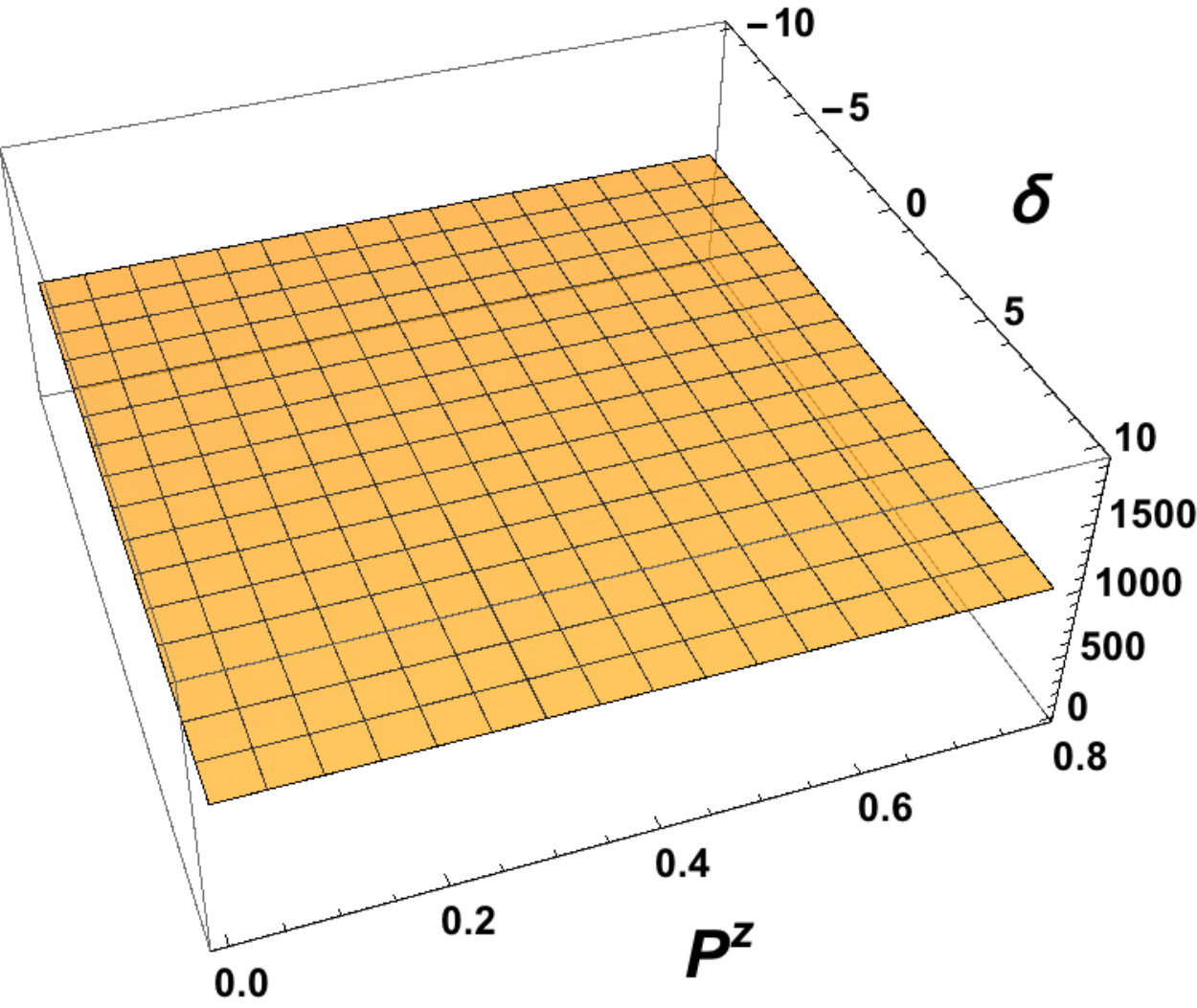}
  \caption{\label{fig:Ptot} The total scattering probability (with the factor $(e^{2}/q^{2})^{2}$ dropped) as a sum of all 16 spin contributions for the lowest tree diagram shown in Fig.~\ref{fig:Scattering_Helicity_Probabilities}.}
\end{figure}

Finally, the helicity probability plots in Fig.~\ref{fig:Scattering_Helicity_Probabilities} show a clear asymmetry between the $+P^{z}$ direction and the $-P^{z}$ direction unless $\delta = \pi/4$. The corresponding helicity amplitudes
that provide the probabilities shown in Fig.~\ref{fig:Scattering_Helicity_Probabilities} are plotted in Appendix~\ref{sec:Interpolated_scattering_amplitudes}.
The LFD result is smoothly connected only to the $P^{z}=+\infty$ result but not to the  $P^{z}=-\infty$ result.
This disparity is a good example of showing the difference between the IMF in IFD and the LFD.
Our plots obviously clear up the prevailing confusion between the two.
Regardless of all the discussion on individual helicity probabilities, however,
the sum of all 16 helicity probabilities plotted in Fig.~\ref{fig:Ptot} is both frame independent
and interpolation angle independent as it must be.

%\textcolor{cyan}{The corresponding helicity amplitudes for this scattering process are plotted in Appendix~\ref{sec:Interpolated_scattering_amplitudes}.}
For completeness, the helicity amplitudes and probabilities for annihilation to the lowest order are calculated and plotted in Appendix~\ref{sec:Interpolated_annihilation_amplitudes_and_probabilities}.

% The total amplitude as a sum of all possible helicity combinations should however be both frame independent and interpolation angle independent.
% This is verified by the plots in the last row of Fig.~\ref{fig:Helicity_amplitudes}, which show the total probabilities of all four initial spin configurations.

\section{Summary and Conclusions}
\label{sec:summary_and_conclusions}
As a continuation of our effort to interpolate between the IFD and the LFD, we generalized the helicity operator to any interpolation angle $\delta$
in this work and derived the generalized helicity spinor that links the instant form helicity spinor to the light-front helicity spinor.

For a given generalized helicity spinor, the spin direction does not coincide with the momentum direction in general.
Thus, we studied how the spin orientation angle changes in terms of both $\delta$ and the angle $\theta$ that defines the momentum direction
of the particle. Applying the transformation matrix $T$ given by Eq.~(\ref{eqn:T_transformation_for_any_interpolation_angle}) to an initial
spin state at rest, we obtained a generalized helicity spinor as we discussed in Sec. \ref{sec:helicity_operator_and_spinor_for_any_interpolation_angle}.
We then used the operator relation given by Eq.~(\ref{eqn:T_BR_equivalence}) to analyze the spin orientation angles ($\theta_{s}, \phi_{s}$),
where $\phi_{s}$ can be taken to be zero without loss of any generality because the spin $\mathbf{S}$, momentum $\mathbf{P}$ and $z$ axis are
all in the same plane. As shown in Figs. \ref{fig:thetas_on_theta_delta} and \ref{fig:ang_diff_bifurcate},  the angle between the momentum and the spin directions, i.e. $\theta-\theta_{s}$, increases with the interpolation angle $\delta$ and becomes the largest at the light front.
In particular, the increment of the angle difference $\theta-\theta_s$ with the increment of the interpolation angle $\delta$ bifurcates
at a critical interpolation angle $\delta_{c}$ as shown in Fig.~\ref{fig:ang_diff_bifurcate}. We found $\delta_{c}$ as given by Eq.~(\ref{eqn:critical_delta})
and noted that the IFD and the LFD separately belong to the two different branches bifurcated and divided out at the critical interpolation angle $\delta_c$.
This bifurcation indicates the necessity of the distinction in the spin orientation between the IFD and the LFD and clarifies any conceivable confusion in the prevailing notion of the equivalence between the IMF formulated in IFD and the LFD.

As the light-front helicity is obtained in general by the two kinds of boost operations, i.e. the light-front transverse boost $\mathbf{E_\perp}$ and the longitudinal boost $K^3$,  we discussed the salient features of the spin orientations
due to each of these two kinds of light-front kinematic operations separately in Sec. \ref{sub:spin_orientation}.
It is interesting to note that the $\mathbf{E_\perp}$ operation on a rest particle of mass $M$ provides the non-relativistic form of energy gain
$E=M+\frac{\vec{\bf P}_\perp^2}{2M}$, yet satisfying the relativistic energy-momentum dispersion relation given by Eq.~(\ref{eqn:LF-dispersion}).
We have analyzed the change of spin orientation under the $\mathbf{E_\perp}$ operation and obtained the ``apparent" spin orientation angle $\theta_a$
given by Eq.~(\ref{eqn:theta_apparent_in_terms_of_theta}). In Fig. \ref{fig:LF_profile1}, we plotted $\theta_a$ in comparison with the corresponding
initial spin orientation $(\theta_s)_{\rm max}$ discussed in Sec. \ref{sub:spin_orientation} along with various $\theta_{s}$ profiles depending on
the particle momentum $|\mathbf{P}|$ or the energy $E$. The change of the spin direction under the $K^3$ operation was also very interesting
and our result given by Eq.~(\ref{eqn:wigner_rotation_angle}) agrees with the well-known result of the Wigner rotation angle given by Eq.~(A10) in \cite{Osborn1968}.  Thus, the light-front helicity formulation takes care of the Wigner rotation which is a rather complicate dynamic effect and
offers an effective computation of spin observables in hadron physics. With the spin orientation analysis presented in Sec. \ref{sub:spin_orientation},
we have also derived the generalized Melosh transformation that connects between the Dirac spinors and our generalized helicity spinors as shown in Sec. \ref{sub:generalized_melosh_transformation}.

Using the generalized helicity spinors, we computed all 16 helicity amplitudes and their squares (or probabilities) for the scattering process analogous to the QED process ``$e \mu \rightarrow e \mu$" in Sec. \ref{sec:interpolating_helicity_probabilities_for_fermion_scattering}.
\footnote{For completeness of our discussion, we have also plotted the helicity amplitudes and probabilities for
the fermion and anti-fermion pair annihilation and creation process analogous to the QED process ``$e^+ e^- \rightarrow \mu^+ \mu^-$" in
in Appendix~\ref{sec:Interpolated_annihilation_amplitudes_and_probabilities}.}
This computation provided
an explicit demonstration of the whole landscape picture for the helicity amplitudes and amplitude-squares that depend on the reference frame as well as the interpolation angle. From this picture, one can see clearly that the helicity probabilities in LFD is independent of the reference frames while
all other interpolating helicity probabilities for $0\leq\delta < \pi/4$ do depend on the reference frames.
The striking feature of the interpolating helicity probabilities as well as amplitudes is the appearance of the two boundaries due to the
the bifurcation of the branches that are distinguished by the critical interpolation angle $\delta_c$.
As $\delta_c$ is obtained from the condition $P_{\mT}=0$, the initial two particles that move only in the $\pm z$ direction yield
the corresponding two boundaries upon the boost of the reference frame in the $\pm z$ direction.
For the given initial helicity states, the two boundaries indicate the spin flip of each of the two incoming scattering particles
as we have shown explicitly in Fig.~\ref{fig:Scattering_Helicity_Probabilities} by the blue dashed lines.

Another interesting feature exhibited by all 16 helicity probability plots in Fig.~\ref{fig:Scattering_Helicity_Probabilities}
is the J-shaped curved indicated by the red solid line which is the same curve that appeared in
our previous works for the $\phi^{3}$ theory \cite{Ji2013} and the sQED theory \cite{Ji2015}.
In Fig.~\ref{fig:P_mt_region}, we showed that the J-curve lies between the two boundaries discussed above.
The J-curve has a universal shape independent of the kinematics in the scattering process no matter what the
the underlying theory is. The scattering amplitudes maintain the same value along the J-curve
in the whole range of interpolation angle $0\leq\delta\leq\pi/4$. However, in taking the limit to $\delta = \pi/4$, it requires a great caution
due to the non-commutativity in the order of taking the two independent limits, $\delta \rightarrow \pi/4$ and $P_z \rightarrow -\infty$.
This non-commutativity brings up the zero-mode issue in LFD as we have discussed previously in great details~\cite{Ji2013, Ji2015}.
The disparity between $P_z \rightarrow \infty$ and $P_z \rightarrow -\infty$ also further clears up the confusion between the IMF in IFD and the LFD.

As we discussed in this work as well as in our previous works~\cite{Ji2013, Ji2015}, the interpolation analysis between IFD and LFD appears to
be beneficial in understanding the distinguished features of LFD advantageous ultimately for the study of hadron physics in QCD.
The extension of the present on-mass-shell fermion analysis into the off-mass-shell fermion analysis involving the intermediate fermion
propagator is underway.

% It gives rise to the trencherous point at $P^{+}=0$ at the light-front limit.

% section conclusion (end)

\appendix

\section{Useful formulae from interpolation angle method}
\label{Appendix:Review_of_Interpolation_Angle}
% \subsection{Interpolation Angle Dependent Poincar\'e Algebra}
% \label{sub:Interpolation_Angle_Dependent_Poincare_Algebra}
The interpolation angle $0\leq\delta\leq\pi/4$ is defined in Eq.~(\ref{eqn:interpolation_angle_definition}).
In the limit $\delta\rightarrow0$, we recover the space-time coordinates in the instant form.
In the other limit $\delta\rightarrow\pi/4$, $x^{\itP{\pm}}=(x^{0}\pm x^{3})/\sqrt{2}$, we get the light-front coordinates.
Notice that the speed of light is taken to be 1.
In this new coordinate system, the metric becomes
\begin{align}\label{eqn:g_munu_interpolation}
  g^{\muT\nuT}
	= g_{\muT\nuT}
  = \begin{pmatrix}
    \Cc & 0  & 0  & \Ss \\
    0          & -1 & 0  & 0 \\
    0          & 0  & -1 & 0 \\
    \Ss & 0  & 0  & -\Cc
  \end{pmatrix},
\end{align}
where $\Ss=\sin2\delta$ and $\Cc=\cos2\delta$.
The covariant and contravariant components are related by
\begin{alignat}{2}
  a_{\pT}=\Cc a^{\pT}+\Ss a^{\mT}; &\quad a^{\pT}=\Cc a_{\pT}+\Ss a_{\mT} \label{eqn:relation_between_covariant_and_contravariant_components_with_any_interpolation}\\
  a_{\mT}=\Ss a^{\pT}-\Cc a^{\mT}; &\quad a^{\mT}=\Ss a_{\pT}-\Cc a_{\mT} \nonumber\\
 a_{\itP{j}}= a_{j}=-a^{j}=-a^{\itP{j}},&\quad (j=1,2) \nonumber.
\end{alignat}
Since the perpendicular components remain the same ($a^{\itP{j}}=a^{j},a_{\itP{j}}=a_{j}, j=1,2$), we will omit the ``\textasciicircum''  notation unless necessary for the perpendicular indices $j=1,2$ in a four-vector.
And inner product can be written as
\begin{align}\label{eqn:inner_product_of_four_vectors_interpolation_angle}
  a^{\muT}b_{\muT}&=(a^{\pT}b^{\pT}-a^{\mT}b^{\mT})\Cc+(a^{\pT}b^{\mT}+a^{\mT}b^{\pT})\Ss-a^{1}b^{1}-a^{2}b^{2}\nonumber\\
  &=a^{\mu}b_{\mu}.
\end{align}

The same transformations also apply to momentum:
\begin{subequations}
  \label{eqn:P_interpolation}
  \begin{align}
    P^{\pT}&=P^{0}\cos\delta + P^{3}\sin\delta,\label{eqn:P_interpolation_1}\\
    P^{\mT}&=P^{0}\sin\delta - P^{3}\cos\delta,\label{eqn:P_interpolation_2}\\
    P_{\pT}&=P^{0}\cos\delta - P^{3}\sin\delta,\label{eqn:P_interpolation_3}\\
    P_{\mT}&=P^{0}\sin\delta + P^{3}\cos\delta.\label{eqn:P_interpolation_4}
  \end{align}
\end{subequations}
% The same transformations also apply to momentum: $P^{\pT}=P^{0}\cos\delta +P^{3}\sin\delta, P^{\mT}=P^{0}\sin\delta - P^{3}\cos\delta \text{ and } P_{\pT}=P^{0}\cos\delta - P^{3}\sin\delta , P_{\mT}=P^{0}\sin\delta  + P^{3}\cos\delta $.
% Since the perpendicular components remain the same ($a^{\itP{j}}=a^{j},a_{\itP{j}}=a_{j}, j=1,2$), from now on we will omit the ``\textasciicircum''  notation for the perpendicular indices $j=1,2$ in a four vector.

In this new basis, the Poincar\'e matrix becomes
\begin{align}\label{eqn:Poincare_Matrix_Interpolation_superscripts}
  M^{\muT\nuT} =
  \begin{pmatrix}
    0 & {E}^{\itP{1}} & {E}^{\itP{2}} & -{K}^{3}\\
    -{E}^{\itP{1}} & 0 & {J}^{3} & -{F}^{\itP{1}}\\
    -{E}^{\itP{2}} & -{J}^{3} & 0 & -{F}^{\itP{2}}\\
    {K}^{3} & {F}^{\itP{1}} & {F}^{\itP{2}} & 0
  \end{pmatrix}
\end{align}
and
\begin{align}\label{eqn:Poincare_Matrix_Interpolation_subscripts}
  M_{\muT\nuT}
  % =
  % g_{\muT\itP{\alpha}}M^{\itP{\alpha}\itP{\beta}}g_{\itP{\beta}\nuT}
  =
  \begin{pmatrix}
    0 & {\mathcal{D}}^{\itP{1}} & {\mathcal{D}}^{\itP{2}} & {K}^{3}\\
    -{\mathcal{D}}^{\itP{1}} & 0 & {J}^{3} & -{\mathcal{K}}^{\itP{1}}\\
    -{\mathcal{D}}^{\itP{2}} & -{J}^{3} & 0 & -{\mathcal{K}}^{\itP{2}}\\
    -{K}^{3} & {\mathcal{K}}^{\itP{1}} & {\mathcal{K}}^{\itP{2}} & 0
  \end{pmatrix},
\end{align}
where
\begin{align}\label{eqn:E_F_D_K_Definition_Interpolation_Angle}
  &E^{\itP{1}}=J^{2}\sin\delta+K^{1}\cos\delta,
  &&\mathcal{K}^{\itP{1}}=-K^{1}\sin\delta-J^{2}\cos\delta, \nonumber\\
  &E^{\itP{2}}=K^{2}\cos\delta-J^{1}\sin\delta,
  &&\mathcal{K}^{\itP{2}}=J^{1}\cos\delta-K^{2}\sin\delta, \nonumber\\
  &F^{\itP{1}}=K^{1}\sin\delta-J^{2}\cos\delta,
  &&\mathcal{D}^{\itP{1}}=-K^{1}\cos\delta+J^{2}\sin\delta, \nonumber\\
  &F^{\itP{2}}=K^{2}\sin\delta+J^{1}\cos\delta,
  &&\mathcal{D}^{\itP{2}}=-J^{1}\sin\delta-K^{2}\cos\delta.
\end{align}
The commutation relations between these operators constitute the generalized Poincar\`e algebra, which is listed in Table~\ref{tab:Poincare_Algebra_Interpolation_Angle}.

\begin{table}[b]
    \caption{\label{tab:Poincare_Algebra_Interpolation_Angle} Poincar\'e Algebra for any interpolation angle.
    The commutation relation reads [element in the first column, element in the first row] = element at the intersection of the corresponding row and column.}
    \resizebox{\columnwidth}{!}{
      \begin{tabular}{|c|cccccccccc|}
	\hline
	& $P_{\pT}$ & $\mathcal{D}^{1}$ & $\mathcal{D}^{2}$ & $K^{3}$ & $\mathcal{K}^{1}$ & $\mathcal{K}^{2}$ & $J^{3}$ & $P_{1}$ & $P_{2}$ & $P_{\mT}$ \\
	\hline
	\hline
	$P_{\pT}$ & 0 & $i\Cc P_{1}$ & $i\Cc P_{2}$ & -$i P^{\mT}$ & $i\Ss P_{1}$ & $i\Ss P_{2}$ & 0 & 0 & 0 & 0 \\
	$\mathcal{D}^{1}$ & -$i\Cc P_{1}$ & 0 & -$iJ^{3}\Cc$ & $i F^{1}$ & $i K^{3}$ & -$i j^{3} \Ss $ & -$i\mathcal{D}^{2}$ & -$i P_{\pT}$ & 0 & -$i\Ss P_{1}$ \\
	$\mathcal{D}^{2}$ & -$i\Cc P_{2}$ & $iJ^{3}\Cc$ & 0 & $i F^{2}$ & $i J^{3}\Ss$ & $i K^{3}$ & $i \mathcal{D}^{1}$ & 0 & -$i P_{\pT}$ & -$i\Ss P_{2}$ \\
	$K^{3}$ & $i P^{\mT}$ & -$i F^{1}$ & -$i F^{2}$ & 0 & $i E^{1}$ & $i E^{2}$ & 0 & 0 & 0 & -$i P^{\pT}$ \\
	$\mathcal{K}^{1}$ & -$i\Ss P_{1}$ & -$i K^{3}$ & -$i J^{3}\Ss$ & -$i E^{1}$ & 0 & $iJ^{3}\Cc$ & -$i \mathcal{K}^{2}$ & -$i P_{\mT}$ & 0 & $i\Cc P_{1}$ \\
	$\mathcal{K}^{2}$ & -$i\Ss P_{2}$ & $i J^{3}\Ss$ & -$i K^{3}$ & -$i E^{2}$ & -$iJ^{3}\Cc$ & 0 & $i \mathcal{K}^{1}$ & 0 & -$i P_{\mT}$ & $i\Cc P_{2}$ \\
	$J^{3}$ & 0 & $i \mathcal{D}^{2}$ & -$i \mathcal{D}^{1}$ & 0 & $i \mathcal{K}^{2}$ & -$i \mathcal{K}^{1}$ & 0 & $i P_{2}$ & -$i P_{1}$ & 0 \\
	$P_{1}$ & 0 & $i P_{\pT}$ & 0 & 0 & $i P_{\mT}$ & 0 & -$i P_{2}$ & 0 & 0 & 0 \\
	$P_{2}$ & 0 & 0 & $i P_{\pT}$ & 0 & 0 & $i P_{\mT}$ & $i P_{1}$ & 0 & 0 & 0 \\
	$P_{\mT}$ & 0 & $i\Ss P_{1}$ & $i\Ss P_{2}$ & $i P^{\pT}$ & -$i\Cc P_{1}$ & -$i\Cc P_{2}$ & 0 & 0 & 0 & 0 \\[1Pt]
	\hline
      \end{tabular}}
\end{table}

Using Eq.~(\ref{eqn:inner_product_of_four_vectors_interpolation_angle}), a useful relation for $P^{\pT}, P^{\perp}, P_{\mT}$ can be found:
\begin{align}
  (P^{\pT})^2-M^{2}\Cc=P_{\mT}^{2}+\mathbf{P}_{\perp}^{2}\Cc,
	\label{eqn:Pplus_Pminus_Pperp_relation}
\end{align}
% from which, we also get
% \begin{align}\label{eqn:P_square_in_P_ph_mh_pp}
%   P^{2}=M^{2}=\dfrac{P^{\pT2}-P_{\mT}^{2}-\mathbf{P}_{\perp}^{2}\Cc}{\Cc}.
% \end{align}
% The same result can be derived alternatively by using Eq.~(\ref{eqn:inner_product_of_four_vectors_interpolation_angle}).
Since this quantity in Eq.~(\ref{eqn:Pplus_Pminus_Pperp_relation}) appears so often in our calculations, we now give it a special symbol to simplify our notation
\begin{align}
  \Pp\equiv\sqrt{(P^{\pT})^{2}-M^{2}\Cc}=\sqrt{P_{\mT}^{2}+\mathbf{P}_{\perp}^{2}\Cc}. \label{eqn:Pp_definition}
\end{align}
% \begin{align}
%   \Pp\equiv\sqrt{(P^{\pT})^{2}-M^{2}\Cc}.\label{eqn:definition_of_Pp}
% \end{align}$
Solving Eq.~(\ref{eqn:four_momentum_transformation_from_rest_frame_any_interpolation_angle}), one get the following useful relations between parameters $\beta_{1}, \beta_{2}, \beta_{3}, \alpha$ and the momentum components:
% \begin{subequations}
%   \label{eqn:beta123_relation_with_P}
%   \begin{align}
%   \sin\delta\cosh\beta_{3}+\cos\delta\sinh\beta_{3}&=\dfrac{ \Pp }{M},\\
%   \cos\delta\cosh\beta_{3}+\sin\delta\sinh\beta_{3}&=\dfrac{ P^{\pT} }{M}.
%   \end{align}
% \end{subequations}
% and
\begin{subequations}
 \label{eqn:useful_beta_P_relation}
  \begin{align}
  \cos\alpha&=\dfrac{P_{\mT}}{\Pp},\\
  \sin\alpha&=\dfrac{\sqrt{\mathbf{P}_{\perp}^{2}\Cc}}{\Pp},\\
  e^{\beta_{3}}&=\dfrac{P^{\pT}+\Pp}{M\left(\sin\delta+\cos\delta\right)},\\
  e^{-\beta_{3}}&=\dfrac{P^{\pT}-\Pp}{M\left(\cos\delta-\sin\delta\right)},\\
  \dfrac{\beta_{j}}{\alpha}&=\dfrac{P^{j}}{\sqrt{\mathbf{P}_{\perp}^{2}\Cc}},(j=1,2).
  \end{align}
\end{subequations}

\begin{widetext}

\section{\texorpdfstring{$(0,\frac{1}{2})\oplus(\frac{1}{2},0)$}{(0,1/2)+(1/2,0)} chiral representation of $T$ and
\texorpdfstring{$B(\boldsymbol\eta)\mathcal{D}(\hat{\mathbf{m}},\theta_{s})$}{BR}}
\label{sec:matrix_representation_of_T_and_BR}
Written in the $(0,\frac{1}{2})\oplus(\frac{1}{2},0)$ chiral representation, $T$ and
 $B(\boldsymbol\eta)\mathcal{D}(\hat{\mathbf{m}},\theta_{s})$ are
\begin{align}
  T=
  \begin{pmatrix}
  T_{R} & 0\\
  0 & T_{L}
  \end{pmatrix},
	&\qquad\qquad
	 B(\boldsymbol\eta)\mathcal{D}(\hat{\mathbf{m}},\theta_{s})=
	\begin{pmatrix}
	(B(\boldsymbol\eta)\mathcal{D}(\hat{\mathbf{m}},\theta_{s}))_{R} & 0 \\
	0 & (B(\boldsymbol\eta)\mathcal{D}(\hat{\mathbf{m}},\theta_{s}))_{L}
	\end{pmatrix},
\end{align}
where
\begin{subequations}
		\begin{align}
				T_{R}&=
				\begin{pmatrix}
						\cos\dfrac{\alpha}{2}e^{\frac{\beta_{3}}{2}} & -\dfrac{\beta_{L}\left(\cos\delta-\sin\delta\right)}{\alpha}\sin\dfrac{\alpha}{2}e^{-\frac{\beta_{3}}{2}}\\[8pt]
						\dfrac{\beta_{R}\left(\sin\delta+\cos\delta\right)}{\alpha}\sin\dfrac{\alpha}{2}e^{\frac{\beta_{3}}{2}} & \cos\dfrac{\alpha}{2}e^{-\frac{\beta_{3}}{2}}
				\end{pmatrix},\label{eqn:T_R_matrix_form}\\[8pt]
				T_{L}&=
				\begin{pmatrix}
						\cos\dfrac{\alpha}{2}e^{-\frac{\beta_{3}}{2}} & -\dfrac{\beta_{L}\left(\sin\delta+\cos\delta\right)}{\alpha}\sin\dfrac{\alpha}{2}e^{\frac{\beta_{3}}{2}}\\[8pt]
						\dfrac{\beta_{R}\left(\cos\delta-\sin\delta\right)}{\alpha}\sin\dfrac{\alpha}{2}e^{-\frac{\beta_{3}}{2}} & \cos\dfrac{\alpha}{2}e^{\frac{\beta_{3}}{2}}
				\end{pmatrix},\label{eqn:T_L_matrix_form}
		\end{align}
\end{subequations}
and
\begin{subequations}
		\begin{align}
		  (B(\boldsymbol\eta)\mathcal{D}(\hat{\mathbf{m}},\theta_{s}))_{R}&=
			\begin{pmatrix}
			F\cos\dfrac{\theta_{s}}{2}+e^{i\phi_{s}}n_{L}\sinh\dfrac{\eta}{2}\sin\dfrac{\theta_{s}}{2} &\:\: -e^{-i\phi_{s}}F\sin\dfrac{\theta_{s}}{2}+n_{L}\sinh\dfrac{\eta}{2}\cos\dfrac{\theta_{s}}{2}\\[8pt]
			n_{R}\sinh\dfrac{\eta}{2}\cos\dfrac{\theta_{s}}{2}+e^{i\phi_{s}}G\sin\dfrac{\theta_{s}}{2} &\:\: -e^{-i\phi_{s}}n_{R}\sinh\dfrac{\eta}{2}\sin\dfrac{\theta_{s}}{2}+G\cos\dfrac{\theta_{s}}{2}
			\end{pmatrix},\label{eqn:BR_R}\\[8pt]
			(B(\boldsymbol\eta)\mathcal{D}(\hat{\mathbf{m}},\theta_{s}))_{L}&=
			\begin{pmatrix}
			G\cos\dfrac{\theta_{s}}{2}-e^{i\phi_{s}}n_{L}\sinh\dfrac{\eta}{2}\sin\dfrac{\theta_{s}}{2} &\:\: -e^{-i\phi_{s}}G\sin\dfrac{\theta_{s}}{2}-n_{L}\sinh\dfrac{\eta}{2}\cos\dfrac{\theta_{s}}{2}\\[8pt]
			-n_{R}\sinh\dfrac{\eta}{2}\cos\dfrac{\theta_{s}}{2}+e^{i\phi_{s}}F\sin\dfrac{\theta_{s}}{2} &\:\: e^{-i\phi_{s}}n_{R}\sinh\dfrac{\eta}{2}\sin\dfrac{\theta_{s}}{2}+F\cos\dfrac{\theta_{s}}{2}
			\end{pmatrix},
		\end{align}
\end{subequations}
with $n_{L}=n_{1}-in_{2}$, $n_{R}=n_{1}+in_{2}$, $(n_{1},n_{2},n_{3})=(\sin\theta\cos\phi,\sin\theta\sin\phi,\cos\theta)$, and $F\equiv \cosh\dfrac{\eta}{2}+n_{3}\sinh\dfrac{\eta}{2}$, $G\equiv \cosh\dfrac{\eta}{2}-n_{3}\sinh\dfrac{\eta}{2}$.

% section section_name (end)

\section{\texorpdfstring{$(0,\frac{1}{2})\oplus(\frac{1}{2},0)$}{(0,1/2)+(1/2,0)} helicity spinors for anti-particles}
\label{sec:helicity_spinors_for_anti_particles}
\begin{align}
  \vT^{(\sfrac{1}{2})}_{H}(P)=
  \begin{pmatrix}
  P^{L}\sqrt{\dfrac{\cos\delta-\sin\delta}{2 \Pp(\Pp+P_{\mT})}}
  \sqrt{P^{\pT}-\Pp}\\[16pt]
  -\sqrt{\dfrac{P_{\mT}+\Pp}{2 \Pp}} \sqrt{\dfrac{P^{\pT}-\Pp}{(\cos\delta-\sin\delta)}}\\[16pt]
  -P^{L}\sqrt{\dfrac{\sin\delta+\cos\delta}{2 \Pp(\Pp+P_{\mT})}}\sqrt{P^{\pT}+\Pp}\\[16pt]
  \sqrt{\dfrac{P_{\mT}+\Pp}{2 \Pp}}\sqrt{\dfrac{P^{\pT}+\Pp}{(\sin\delta+\cos\delta)}}
  \end{pmatrix},
  \quad
  \vT^{(-\sfrac{1}{2})}_{H}(P)=
  \begin{pmatrix}
    \sqrt{\dfrac{P_{\mT}+\Pp}{2 \Pp}}\sqrt{\dfrac{P^{\pT}+\Pp}{(\sin\delta+\cos\delta)}}\\[16pt]
    P^{R}\sqrt{\dfrac{\sin\delta+\cos\delta}{2 \Pp(\Pp+P_{\mT})}}\sqrt{P^{\pT}+\Pp}\\[16pt]
    -\sqrt{\dfrac{P_{\mT}+\Pp}{2 \Pp}}\sqrt{\dfrac{P^{\pT}-\Pp}{(\cos\delta-\sin\delta)}}\\[16pt]
    -P^{R}\sqrt{\dfrac{\cos\delta-\sin\delta}{2 \Pp(\Pp+P_{\mT})}}\sqrt{P^{\pT}-\Pp}
  \end{pmatrix},\label{eqn:Spinor_v1v2_P_Any_Interpolation_Angle}
 \end{align}

% section helicity_spinors_for_anti_particles (end)

\section{Dirac spinors in chiral representation}
\label{sec:dirac_spinor_in_chiral_representation}
The Dirac spinors $u^{(\sfrac{1}{2})}$ and $u^{(-\sfrac{1}{2})}$ in chiral representation are given by
% \begin{subequations}
  % \label{eqn:Dirac_spinors}
  \begin{alignat}{2}\label{eqn:Dirac_spinors}
    u^{(\sfrac{1}{2})}_{D}(P)&=\dfrac{1}{\sqrt{2(E+M)}}
    \begin{pmatrix}
      E+P^{3}+M\\
      P^{R}\\
      E-P^{3}+M\\
      -P^{R}
    \end{pmatrix}, &\quad\quad\quad
    u^{(-\sfrac{1}{2})}_{D}(P)&=\dfrac{1}{\sqrt{2(E+M)}}
    \begin{pmatrix}
      P^{L}\\
      E+P^{3}+M\\
      -P^{L}\\
      E+P^{3}+M
    \end{pmatrix}.
  \end{alignat}
% \end{subequations}

\section{Generalized \texorpdfstring{$(0,J)\oplus(J,0)$}{(0,J)+(J,0)} helicity spinors for arbitrary interpolation angle up to spin 2}
\label{sec:generalized_helicity_spinors_for_arbitrary_interpolation_angle_up_to_spin_2}
Following the notations in Ref.~\cite{Ji2001}, we will use $A=\cos\delta$, $B=-\sin\delta$ in the spinors listed below.
We also define $\Xx\equiv \dfrac{P^{\pT}-\Pp}{\Cc}=\dfrac{P^{\pT}-\sqrt{(P^{\pT})^{2}-M^{2}\Cc}}{\Cc}$ for convenience.
% \begin{align}
%   \Xx\equiv \dfrac{P^{\pT}-\Pp}{\Cc}=\dfrac{P^{\pT}-\sqrt{(P^{\pT})^{2}-M^{2}\Cc}}{\Cc},\label{eqn:xx}
% \end{align}
% for convenience.

Spin-1 helicity spinors for any interpolation angle in chiral representation:
\vspace{1em}
\begin{align}
  \label{eqn:spin_1}
  u_{H}^{(+1)}=\dfrac{1}{2\sqrt{M\Pp^{2}}}
  \begin{pmatrix}
    \dfrac{(P_{\hat{-}}+\Pp)(P^{\hat{+}}+\Pp)}{(A-B)} \\[10pt]
    \sqrt{2}P^{R}(P^{\hat{+}}+\Pp) \\[10pt]
    \dfrac{(A-B)(P^{R})^2(P^{\hat{+}}+\Pp)}{( P_{\hat{-}}+\Pp)} \\[10pt]
    (A-B)(P_{\hat{-}}+\Pp)\Xx \\[10pt]
    \sqrt{2}P^{R}(P^{\hat{+}}-\Pp) \\[10pt]
    \dfrac{(A+B)(P^{R})^2(P^{\hat{+}}- \Pp)}{(P_{\hat{-}}+\Pp)}
  \end{pmatrix},
  &\quad
  u_{H}^{(-1)}=\dfrac{1}{2\sqrt{M\Pp^{2}}}
  \begin{pmatrix}
    \dfrac{(A+B)(P^{L})^2(P^{\hat{+}}-\Pp)}{(P_{\hat{-}}+\Pp)}  \\[10pt]
    -\sqrt{2}P^{L}(P^{\hat{+}}-\Pp) \\[10pt]
    (A-B)(P_{\hat{-}}+\Pp)\Xx \\[10pt]
    \dfrac{(A-B)(P^{L})^2(P^{\hat{+}}+\Pp)}{(P_{\hat{-}}+\Pp)} \\[10pt]
    -\sqrt{2}P^{L}(P^{\hat{+}}+\Pp) \\[10pt]
    \dfrac{(P_{\hat{-}}+\Pp)(P^{\hat{+}}+\Pp)}{(A-B)}
  \end{pmatrix},%\nonumber\\
	% &\quad
	u_{H}^{(0)}=\sqrt{\dfrac{M}{2\Pp^{2}}}
  \begin{pmatrix}
    -(A+B)P^{L} \\[10pt]
    \sqrt{2}P_{\hat{-}} \\[10pt]
    (A-B)P^{R} \\[10pt]
    (-A+B)P^{L} \\[10pt]
    \sqrt{2}P_{\hat{-}} \\[10pt]
    (A+B)P^{R}
  \end{pmatrix}.
\end{align}

% \begin{align}
%   u_{H}^{(0)}(P)=\sqrt{\dfrac{M}{2\Pp^{2}}}
%   \begin{pmatrix}
%     -(A+B)P^{L} \\[5pt]
%     \sqrt{2}P_{\hat{-}} \\[5pt]
%     (A-B)P^{R} \\[5pt]
%     (-A+B)P^{L} \\[5pt]
%     \sqrt{2}P_{\hat{-}} \\[5pt]
%     (A+B)P^{R}
%   \end{pmatrix}.
% \end{align}

% In the light-front limit, they become
% \begin{equation}
% u(P,+1)=\dfrac{1}{\sqrt{2M}}\begin{pmatrix}
% 2P^+ \\[5pt] 2P^{R} \\[5pt] \dfrac{ (P^{R})^2}{P^+} \\[5pt]\dfrac{ M^2}{P^+} \\[5pt] 0 \\[5pt] 0
% \end{pmatrix}, \;
% u(P,0)=\dfrac{\sqrt{M}}{P^+} \begin{pmatrix}
%  0\\[5pt]P^+ \\[5pt] P^{R} \\[5pt] -P^{L} \\[5pt] P^+ \\[5pt] 0
% \end{pmatrix}, \;
% u(P,-1)=\dfrac{1}{\sqrt{2M}} \begin{pmatrix}
% 0 \\[5pt] 0 \\[5pt] \dfrac{M^2}{P^+} \\[5pt] \dfrac{(P^{L})^2}{P^+} \\[5pt] -2P^{L} \\[5pt] 2P^+
% \end{pmatrix}
% \end{equation}

Spin-$\frac{3}{2}$ helicity spinors for any interpolation angle in chiral representation:
\vspace{2em}
\begin{align}
  \label{eqn:spin_3/2}
  u_{H}^{(\sfrac{3}{2})}&=\dfrac{N}{M}
  \begin{pmatrix}
    \left(\dfrac{(P_{\mT}+\Pp)(P^{\pT}+\Pp)}{A-B} \right)^{3/2} \\[15pt]
    \dfrac{\sqrt{3(P_{\mT}+\Pp)} P^{R}(P^{\pT}+\Pp)^{3/2}}{\sqrt{A-B}} \\[15pt]
    \dfrac{\sqrt{3(A-B)}(P^{R})^2(P^{\pT}+\Pp)^{3/2}}{\sqrt{(P_{\mT}+\Pp)}} \\[15pt]
    \dfrac{(A-B)^{3/2}(P^{R})^3(P^{\pT}+\Pp)^{3/2}}{(P_{\mT}+\Pp)^{3/2}} \\[15pt]
    ( (A-B)\left(P_{\mT}+\Pp)\Xx\right)^{3/2} \\[15pt]
    \sqrt{3(A-B)}P^{R}(P^{\pT}-\Pp)\sqrt{(P_{\mT}+\Pp)\Xx} \\[15pt]
    \dfrac{\sqrt{3(A+B)}(P^{R})^2(P^{\pT}-\Pp)^{3/2}}{\sqrt{(P_{\mT}+\Pp)}} \\[15pt]
    \dfrac{(A+B)^{3/2}(P^{R})^3(P^{\pT}-\Pp)^{3/2}}{(P_{\mT}+\Pp)^{3/2}}
  \end{pmatrix},&
  u_{H}^{(-\sfrac{3}{2})}&=\dfrac{N}{M}
  \begin{pmatrix}
    -\dfrac{(A+B)^{3/2}(P^{L})^{3}(P^{\pT}-\Pp)^{3/2}}{(P_{\mT}+\Pp)^{3/2}} \\[15pt]
    \dfrac{\sqrt{3(A+B)}(P^{L})^2(P^{\pT}-\Pp)^{3/2}}{\sqrt{P_{\mT}+\Pp}} \\[15pt]
    -\sqrt{3(A-B)} P^{L}(P^{\pT}-\Pp)\sqrt{(P_{\mT}+\Pp)\Xx} \\[15pt]
    ((A-B)(P_{\mT}+\Pp)\Xx)^{3/2} \\[15pt]
    -\dfrac{(A-B)^{3/2}(P^{L})^3(P^{\pT}+\Pp)^{3/2}}{(P_{\mT}+\Pp)^{3/2}} \\[15pt]
    \dfrac{\sqrt{3(A-B)}(P^{L})^2(P^{\pT}+\Pp)^{3/2}}{\sqrt{(P_{\mT}+\Pp)}} \\[15pt]
    -\dfrac{\sqrt{3} P^{L}(P^{\pT}+\Pp)^{3/2}\sqrt{(P_{\mT}+\Pp)}}{\sqrt{A-B}} \\[15pt]
    \left(\dfrac{(P_{\mT}+\Pp)(P^{\pT}+\Pp)}{A-B}\right)^{3/2}
  \end{pmatrix},\nonumber\\[3em]
  u_{H}^{(\sfrac{1}{2})}&=N
  \begin{pmatrix}
    -\dfrac{\sqrt{3}(A+B)P^{L}\sqrt{(P_{\mT}+\Pp)(P^{\pT}+\Pp})}{\sqrt{A-B}} \\[15pt]
    \dfrac{(3P_{\mT}-\Pp)\sqrt{(P_{\mT}+\Pp)(P^{\pT}+\Pp)}}{\sqrt{A-B}} \\[15pt]
    \dfrac{(3P_{\mT}+\Pp)P^{R}\sqrt{(A-B)(P^{\pT}+\Pp)} }{\sqrt{P_{\mT}+\Pp}} \\[15pt]
    \dfrac{\sqrt{3}(A-B)^{3/2}(P^{R})^2\sqrt{P^{\pT}+\Pp}}{\sqrt{P_{\mT}+\Pp}} \\[15pt]
    -\sqrt{3}(A-B)^{3/2}P^{L}\sqrt{(P^{\pT}+\Pp)\Xx} \\[15pt]
    (3P_{\mT}-\Pp)\sqrt{(A-B)(P_{\mT}+\Pp)\Xx} \\[15pt]
    \dfrac{(3P_{\mT}+\Pp)P^{R} \sqrt{(A+B)(P^{\pT}-\Pp)}}{\sqrt{P_{\mT}+\Pp}} \\[15pt]
    \dfrac{\sqrt{3(P^{\pT}-\Pp)}(A+B)^{3/2} P_{R}^2 }{\sqrt{P_{\mT}+\Pp}}
  \end{pmatrix},&
	u_{H}^{(-\sfrac{1}{2})}&=N
  \begin{pmatrix}
    \dfrac{\sqrt{3(P^{\pT}-\Pp)}(A+B)^{3/2}(P^{L})^2}{\sqrt{P_{\mT}+\Pp}} \\[15pt]
    -\dfrac{(3P_{\mT}+\Pp)P^{L}\sqrt{(A+B)(P^{\pT}-\Pp)}}{\sqrt{P_{\mT}+\Pp}} \\[15pt]
    (3P_{\mT}-\Pp)\sqrt{(A-B)(P_{\mT}+\Pp)\Xx} \\[15pt]
    \sqrt{3}(A-B)^{3/2}P^{R}\sqrt{(P_{\mT}+\Pp)\Xx} \\[15pt]
    \dfrac{\sqrt{3}(A-B)^{3/2}(P^{L})^2\sqrt{(P^{\pT}+\Pp)}}{\sqrt{(P_{\mT}+\Pp)}} \\[15pt]
    -\dfrac{(3P_{\mT}+\Pp)P^{L}\sqrt{(A-B)(P^{\pT}+\Pp)}}{\sqrt{(P_{\mT}+\Pp)}} \\[15pt]
    \dfrac{(3P_{\mT}-\Pp) \sqrt{(P_{\mT}+\Pp)(P^{\pT}+\Pp)}}{\sqrt{A-B}} \\[15pt]
    \dfrac{\sqrt{3}(A+B)P^{R}\sqrt{(P_{\mT}+\Pp)(P^{\pT}+\Pp)}}{\sqrt{A-B}}
  \end{pmatrix},
\end{align}

% \begin{align}
%   u_{H}^{(-\sfrac{1}{2})}=N
%   \begin{pmatrix}
%     \dfrac{\sqrt{3(P^{\hat{+}}-\Pp)}(A+B)^{3/2}(P^{L})^2}{\sqrt{P_{\hat{-}}+\Pp}} \\[15pt]
%     \dfrac{-(3P_{\hat{-}}+\Pp)P_{L}\sqrt{(A+B)(P^{\hat{+}}-\Pp)}}{\sqrt{P_{\hat{-}}+\Pp}} \\[15pt]
%     (3P_{\mT}-\Pp)\sqrt{(A-B)(P_{\mT}+\Pp)\Xx} \\[15pt]
%     \sqrt{3}(A-B)^{3/2}P^{R}\sqrt{(P_{\hat{-}}+\Pp)\Xx} \\[15pt]
%     \dfrac{\sqrt{3}(A-B)^{3/2}(P^{L})^2\sqrt{(P^{\hat{+}}+\Pp)}}{\sqrt{(P_{\hat{-}}+\Pp)}} \\[15pt]
%     -\dfrac{(3P_{\mT}+\Pp)P^{L}\sqrt{(A-B)(P^{\hat{+}}+\Pp)}}{\sqrt{(P_{\hat{-}}+\Pp)}} \\[15pt]
%     \dfrac{(3P_{\mT}-\Pp) \sqrt{(P_{\mT}+\Pp)(P^{\hat{+}}+\Pp)}}{\sqrt{A-B}} \\[15pt]
%     \dfrac{\sqrt{3}(A+B)P^{R}\sqrt{(P_{\hat{-}}+\Pp)(P^{\hat{+}}+\Pp)}}{\sqrt{A-B}}
%   \end{pmatrix},
%   u_{H}^{(-\sfrac{3}{2})}=\dfrac{N}{M}
%   \begin{pmatrix}
%     -\dfrac{(A+B)^{3/2}(P^{L})^{3}(P^{\hat{+}}-\Pp)^{3/2}}{(P_{\mT}+\Pp)^{3/2}} \\[15pt]
%     \dfrac{\sqrt{3(A+B)}(P^{L})^2(P^{\hat{+}}-\Pp)^{3/2}}{\sqrt{P_{\hat{-}}+\Pp}} \\[15pt]
%     -\sqrt{3(A-B)} P^{L}(P^{\hat{+}}-\Pp)\sqrt{(P_{\hat{-}}+\Pp)\Xx} \\[15pt]
%     ((A-B)(P_{\hat{-}}+\Pp)\Xx)^{3/2} \\[15pt]
%     -\dfrac{(A-B)^{3/2}(P^{L})^3(P^{\hat{+}}+\Pp)^{3/2}}{(P_{\hat{-}}+\Pp)^{3/2}} \\[15pt]
%     \dfrac{\sqrt{3(A-B)}(P^{L})^2(P^{\hat{+}}+\Pp)^{3/2}}{\sqrt{(P_{\hat{-}}+\Pp)}} \\[15pt]
%     -\dfrac{\sqrt{3} P^{L}(P^{\hat{+}}+\Pp)^{3/2}\sqrt{(P_{\hat{-}}+\Pp)}}{\sqrt{A-B}} \\[15pt]
%     (\dfrac{\left(P_{\hat{-}}+\Pp)(P^{\hat{+}}+\Pp)}{A-B}\right)^{3/2}
%   \end{pmatrix},
% \end{align}
where $N=\dfrac{1}{2\sqrt{2\Pp^{3}}}$.

% In the light-front limit these become
%  \begin{equation*}
%  u(P,+3/2)=\dfrac{1}{\sqrt{\sqrt{2}P^+}M}\begin{pmatrix}
%  2(P^+)^2 \\[5pt] \sqrt{6}P^+P^{R} \\[5pt] \sqrt{3}(P^{R})^2 \\[5pt] \dfrac{(P^{R})^3}{(\sqrt{2}P^+)} \\[5pt] \dfrac{M^3}{(\sqrt{2})P^+} \\[5pt] 0 \\[5pt] 0  \\[5pt] 0
%  \end{pmatrix}, \; u(P,+\dfrac{1}{2})=\dfrac{1}{\sqrt{\sqrt{2}P^+}}\begin{pmatrix}
%  0 \\[5pt] \sqrt{2}P^+ \\[5pt] 2P^{R} \\[5pt] \sqrt{3/2}\dfrac{(P^{R})^2}{P^+} \\[5pt] -\sqrt{3/2}\dfrac{MP^{L}}{P^+} \\[5pt] M \\[5pt] 0 \\[5pt] 0
%  \end{pmatrix}
%  \end{equation*}
%  \begin{equation}
%  u(P,-\dfrac{1}{2})=\dfrac{1}{\sqrt{\sqrt{2}P^+}}\begin{pmatrix}
%  0 \\[5pt] 0 \\[5pt] M \\[5pt] \sqrt{3/2}\dfrac{MP^{R}}{P^+} \\[5pt] \sqrt{3/2}\dfrac{(P^{L})^2}{P^+} \\[5pt] -2P^{L} \\[5pt] \sqrt{2}P^+ \\[5pt] 0
%  \end{pmatrix}, \;
%   u(P,-3/2)=\dfrac{1}{\sqrt{\sqrt{2}P^+}M}\begin{pmatrix}
%  0 \\[5pt] 0 \\[5pt] 0 \\[5pt]\dfrac{M^3}{(\sqrt{2}P^+)} \\[5pt]-\dfrac{(P^{L})^3}{(\sqrt{2}P^+)} \\[5pt] \sqrt{3}(P^{L})^2 \\[5pt] -\sqrt{6}P^+P^{L} \\[5pt] 2 (P^+)^2
%  \end{pmatrix}
%  \end{equation}

Spin-2 helicity spinors for any interpolation angle in chiral representation:
\vspace{3em}
\begin{align}
  \label{eqn:spin_2}
  u_{H}^{(2)}&=\dfrac{1}{4M^{3/2}\Pp^{2}}
  \begin{pmatrix}
    \dfrac{(P_{\mT}+\Pp )^2(P^{\pT}+ \Pp)^2}{(A-B)^{2}} \\[12pt]
    \dfrac{2P^{R}(P_{\mT}+\Pp )(P^{\pT}+ \Pp)^2}{A-B} \\[12pt]
    \sqrt{6}(P^{R})^2(P^{\pT}+ \Pp)^2 \\[12pt]
    \dfrac{2(A-B)(P^{R})^3(P^{\pT}+ \Pp)^2 }{P_{\mT}+\Pp} \\[12pt]
    \dfrac{(A-B)^2(P^{R})^4(P^{\pT}+ \Pp)^2 }{(P_{\mT}+\Pp)^{2}} \\[12pt]
    (A-B)^{2}(P_{\mT}+\Pp )^2\Xx^{2} \\[12pt]
    2P_{R}(A-B)(P^{\pT}- \Pp)(P_{\mT}+\Pp )\Xx \\[12pt]
    \sqrt{6}(P^{R})^2(P^{\pT}- \Pp)^2 \\[12pt]
    \dfrac{2(A+B)(P^{R})^3(P^{\pT}- \Pp)^2}{P_{\mT}+ \Pp} \\[12pt]
    \dfrac{(A+B)^2(P^{R})^4(P^{\pT}- \Pp)^2}{(P_{\mT}+ \Pp)^2}
  \end{pmatrix},
	&u_{H}^{(-2)}&=\dfrac{1}{4M^{3/2}\Pp^{2}}
    \begin{pmatrix}
      \dfrac{(A+B)^{2}(P^{L})^{4}(P^{\pT}-\Pp)^{2}}{(P_{\mT}+\Pp)^{2}} \\[12pt]
      -\dfrac{2(A+B)(P^{\pT}-\Pp)^{2}(P^{L})^{3}}{P_{\mT}+\Pp } \\[12pt]
      \sqrt{6}(P^{L})^2 (P^{\pT}-\Pp)^{2}  \\[12pt]
      -2 P^{L}(A-B)(P_{\mT}+\Pp)(P^{\pT}-\Pp)\Xx \\[12pt]
      (A-B)^{2}(P_{\mT}+\Pp)^{2}\Xx^{2} \\[12pt]
      \dfrac{(A-B)^{2}(P^{L})^{4}(P^{\pT}+\Pp)^{2}}{(P_{\mT}+\Pp)^{2}} \\[12pt]
      \dfrac{2(A-B)(P^{L})^{3}(P^{\pT}+\Pp)^{2}}{P_{\mT}+\Pp} \\[12pt]
      \sqrt{6}(P^{L})^2(P_{\mT}+\Pp)^{2} \\[12pt]
      -\dfrac{2P^{L}(P_{\mT}+\Pp)(P^{\pT}+\Pp)^{2}}{A-B} \\[12pt]
      \dfrac{(P_{\mT}+\Pp)^{2}(P^{\pT}+\Pp)^{2}}{(A-B)^{2}}
    \end{pmatrix},\nonumber\\[3em]
	u_{H}^{(1)}&=\dfrac{1}{2\Pp^{2}\sqrt{M}}
  \begin{pmatrix}
    -\dfrac{(A+B)P^{L}(P^{\hat{+}}+ \Pp) (P_{\hat{-}}+ \Pp)}{A-B} \\[12pt]
    \dfrac{(2P_{\mT}-\Pp)(P^{\pT}+\Pp)(P_{\hat{-}}+ \Pp)}{A-B} \\[12pt]
    \sqrt{6}P^{R}P_{\hat{-}}(P^{\hat{+}}+ \Pp) \\[12pt]
    \dfrac{(A-B)(P^{R})^2(P^{\hat{+}}+ \Pp)(2P_{\mT}+\Pp)}{P_{\mT}+\Pp} \\[12pt]
    \dfrac{(A-B)^{2} (P^{R})^3(P^{\hat{+}}+ \Pp)}{P_{\hat{-}}+ \Pp}\\[12pt]
    -(A-B)^2P^{L}(P_{\mT}+\Pp)\Xx\\[12pt]
    (A-B)(2P_{\mT}-\Pp)(P_{\mT}+\Pp)\Xx  \\[12pt]
    \sqrt{6}P^{R} P_{\hat{-}}(P^{\hat{+}}- \Pp) \\[12pt]
    \dfrac{\Cc^{2}(2P_{\mT}+\Pp)(P^{R})^{2}\Xx}{(A-B)(P_{\mT}+\Pp)} \\[12pt]
    \dfrac{\Cc^{3}(P^{R})^{3}\Xx}{(A-B)^{2}(P_{\mT}+\Pp)}
  \end{pmatrix},
	&u_{H}^{(-1)}&=\dfrac{1}{2\Pp^{2}\sqrt{M}}
  \begin{pmatrix}
    -\dfrac{\Cc^{3}(P^{L})^3\Xx}{(A-B)^{2}(\Pp+P_{\mT})} \\[12pt]
    \dfrac{\Cc^{2}(2P_{\mT}+\Pp)(P^{L})^2\Xx}{(A-B)(P_{\mT}+\Pp)} \\[12pt]
    -\sqrt{6}P^{L}P_{\hat{-}}( P^{\pT}-\Pp) \\[12pt]
    (A-B)(2P_{\mT}-\Pp)(P_{\mT}+\Pp)\Xx \\[12pt]
    (A - B)^2 (\Pp + P_{\mT}) P^{R} \Xx \\[12pt]
    -\dfrac{(A - B)^2 (P^{L})^3 (P^{\pT}+\Pp)}{P_{\mT}+\Pp} \\[12pt]
    \dfrac{(A - B) (P^{L})^2 (P^{\pT} + \Pp)(2P_{\mT}+\Pp)}{P_{\mT}+\Pp} \\[12pt]
    -\sqrt{6} P^{L} P_{\mT} (P^{\pT} + \Pp) \\[12pt]
    \dfrac{(2 P_{\mT}-\Pp) (P_{\mT}+\Pp) (P^{\pT}+\Pp)}{A-B} \\[12pt]
    \dfrac{(A+B)(P_{\mT}+\Pp)(P^{\pT}+\Pp)P^{R}}{A-B}
  \end{pmatrix},\nonumber
\end{align}

% \begin{align}
%   u_{H}^{(0)}&=\dfrac{1}{2\Pp^{2}}
%     \begin{pmatrix}
%       \sqrt{3/2}(A+B)^2(P^{L})^2 \\[12pt]
%       -\sqrt{6}(A+B)P^{L}P_{\hat{-}} \\[12pt]
%       3{P_{\hat{-}}}^2-\Pp^{2} \\[12pt]
%       \sqrt{6}(A-B)P^{R}P_{\hat{-}} \\[12pt]
%       \sqrt{3/2}(A-B)^2(P^{R})^2 \\[12pt]
%       \sqrt{3/2}(A-B)^2(P^{L})^2 \\[12pt]
%       -\sqrt{6}(A-B)P^{L}P_{\hat{-}} \\[12pt]
%       3{P_{\hat{-}}}^2-\Pp^{2} \\[12pt]
%       \sqrt{6}(A+B)P^{R}P_{\hat{-}} \\[12pt]
%       \sqrt{3/2}(A+B)^2(P^{R})^2
%     \end{pmatrix},
%   &u_{H}^{(-1)}&=\dfrac{1}{2M\Pp^{2}}
%   \begin{pmatrix}
%     -\dfrac{\Cc^{3}(P^{L})^3\Xx}{(A-B)^{2}(\Pp+P_{\mT})} \\[12pt]
%     \dfrac{\Cc^{2}(2P_{\mT}+\Pp)(P^{L})^2\Xx}{(A-B)(P_{\mT}+\Pp)} \\[12pt]
%     -\sqrt{6}P^{L}P_{\hat{-}}( P^{\pT}-\Pp) \\[12pt]
%     (A-B)(2P_{\mT}-\Pp)(P_{\mT}+\Pp)\Xx \\[12pt]
%     (A - B)^2 (\Pp + P_{\mT}) P_{R} \Xx \\[12pt]
%     -\dfrac{(A - B)^2 P_{L}^3 (P^{\pT}+\Pp)}{(P_{\mT}+\Pp)} \\[12pt]
%     \dfrac{(A - B) P_{L}^2 (P^{\pT} + \Pp)(2P_{\mT}+\Pp)}{P_{\mT}+\Pp} \\[12pt]
%     -\sqrt{6} P_{L} P_{\mT} (P^{\pT} + \Pp) \\[12pt]
%     \dfrac{(2 P_{\mT}-\Pp) (P_{\mT}+\Pp) (P^{\pT}+\Pp)}{A-B} \\[12pt]
%     \dfrac{(A+B)(P_{\mT}+\Pp)(P^{\pT}+\Pp)P_{R}}{A-B}
%   \end{pmatrix},
% \end{align}

\begin{align}
    % u_{H}^{(-2)}&=\dfrac{1}{4M^2\Pp^{2}}
    % \begin{pmatrix}
    %   \dfrac{(A+B)^{2}P_{L}^{4}(P^{\pT}-\Pp)^{2}}{(P_{\mT}+\Pp)^{2}} \\[12pt]
    %   -\dfrac{2(A+B)(P^{\pT}-\Pp)^{2}P_{L}^{3}}{P_{\mT}+\Pp } \\[12pt]
    %   \sqrt{6}(P^{L})^2 (P^{\pT}-\Pp)^{2}  \\[12pt]
    %   -2 P_{L}(A-B)(P_{\mT}+\Pp)(P^{\pT}-\Pp)\Xx \\[12pt]
    %   (A-B)^{2}(P_{\mT}+\Pp)^{2}\Xx^{2} \\[12pt]
    %   \dfrac{(A-B)^{2}P_{L}^{4}(P^{\pT}+\Pp)^{2}}{(P_{\mT}+\Pp)^{2}} \\[12pt]
    %   \dfrac{2(A-B)P_{L}^{3}(P^{\pT}+\Pp)^{2}}{P_{\mT}+\Pp} \\[12pt]
    %   \sqrt{6}(P^{L})^2(P_{\mT}+\Pp)^{2} \\[12pt]
    %   -\dfrac{2P_{L}(P_{\mT}+\Pp)(P^{\pT}+\Pp)^{2}}{A-B} \\[12pt]
    %   \dfrac{(P_{\mT}+\Pp)^{2}(P^{\pT}+\Pp)^{2}}{(A-B)^{2}}
    % \end{pmatrix}.
		u_{H}^{(0)}&=\dfrac{\sqrt{M}}{2\Pp^{2}}
    \begin{pmatrix}
      \sqrt{3/2}(A+B)^2(P^{L})^2 \\[12pt]
      -\sqrt{6}(A+B)P^{L}P_{\hat{-}} \\[12pt]
      3{P_{\hat{-}}}^2-\Pp^{2} \\[12pt]
      \sqrt{6}(A-B)P^{R}P_{\hat{-}} \\[12pt]
      \sqrt{3/2}(A-B)^2(P^{R})^2 \\[12pt]
      \sqrt{3/2}(A-B)^2(P^{L})^2 \\[12pt]
      -\sqrt{6}(A-B)P^{L}P_{\hat{-}} \\[12pt]
      3{P_{\hat{-}}}^2-\Pp^{2} \\[12pt]
			\sqrt{6}(A+B)P^{R}P_{\hat{-}} \\[12pt]
      \sqrt{3/2}(A+B)^2(P^{R})^2
    \end{pmatrix},
\end{align}

% light-front limit in \Cchiral Representation
% \begin{equation*}
% u(P,+2)=\dfrac{1}{M^2}\begin{pmatrix}
% 2{P^+}^2 \\ 2\sqrt{2}P^+P_R \\ \sqrt{6}P_R^2 \\ \dfrac{\sqrt{2}P_R^3}{P^+} \\ \dfrac{P_R^4}{2{P^+}^2} \\ \dfrac{M^4}{2{P^+}^2} \\ 0 \\ 0 \\ 0 \\ 0
% \end{pmatrix} \;
% u(P,+1)=\dfrac{1}{M} \begin{pmatrix}
% 0 \\ \sqrt{2}P^+\\ \sqrt{6}P_R \\ \dfrac{3P_R^2}{\sqrt{2} P^+} \\ \dfrac{P_R^3}{{P^+}^2} \\ -\dfrac{P_LM^2}{{P^+}^2} \\ \dfrac{M^2}{\sqrt{2}P^+}  \\ 0 \\ 0 \\0
% \end{pmatrix} \; u(P,0)=\begin{pmatrix}
% 0 \\ 0 \\ 1 \\ -\dfrac{\sqrt{3}P_R}{P^+} \\ \sqrt{3/2}\dfrac{P_R^2}{{P^+}^2} \\ \sqrt{3/2} \dfrac{P_L^2}{{P^+}^2} \\ \dfrac{\sqrt{3}P_L}{P^+} \\ 1 \\ 0 \\ 0
% \end{pmatrix}
% \end{equation*}
% \begin{equation}
% u(P,-1)=\dfrac{1}{M} \begin{pmatrix}
% 0 \\ 0\\0 \\ \dfrac{M^2}{\sqrt{2}P^+} \\ \dfrac{M^2P_R}{{P^+}^2} \\ -\dfrac{P_L^3}{{P^+}^2} \\ \dfrac{3P_L^2}{\sqrt{2}P^+} \\ -\sqrt{6}P_L \\ \sqrt{2}P^+ \\ 0
% \end{pmatrix} \; u(P,-2)=\dfrac{1}{M^2}\begin{pmatrix}
% 0 \\ 0 \\ 0 \\ \dfrac{M^4}{2{P^+}^2} \\ \dfrac{P_L^4}{2{P^+}^2} \\ -\dfrac{\sqrt{2}P_L^3}{P^+} \\ \sqrt{6}P_L^2 \\ -2\sqrt{2}P_LP^+ \\2 {P^+}^2
% \end{pmatrix}
% \end{equation}

In the light-front limit ($\delta\rightarrow\pi/4$), $A-B\rightarrow\sqrt{2}$, $A+B\rightarrow0$, $\Xx\rightarrow\dfrac{M^{2}}{2P^{+}}$, and $P_{\mT}=P^{\pT}=\Pp\rightarrow P^{+}$, and one can verify that our spinors agree with the light-front spinors listed in Ref.~\cite{Ahluwalia1993}.
To compare our results to those in that paper, one should note that our spinors are written in the chiral representation while those in Ref.~\cite{Ahluwalia1993} are in the standard representation.
The standard representation (SR) and the chiral representation (CR) used in this paper are related by:
\begin{align}
  u_{SR}=
  \dfrac{1}{\sqrt{2}}
  \begin{pmatrix}
  1 & 1\\
  1 & -1
  \end{pmatrix}
  u=
  \begin{pmatrix}
    \chi_{R}\\
    \chi_{L}
  \end{pmatrix},
  \label{eqn:SR_CR}
\end{align}
where $\chi_{R}$ and $\chi_{L}$ are each 3, 4 and 5 components long for spin 1, 3/2, and 2.
In Ref.~\cite{Ahluwalia1993}, the notations of $P_{R}=P^{R}$ and $P_{L}=P^{L}$ were used.
One should also note that in our paper $P^{+}=(P^{0}+P^{3})/\sqrt{2}$, and the spinors are all normalized so that $\bar{u}u=2M$ while the spinors in Ref.~\cite{Ahluwalia1993} are normalized to $\bar{u}u=M^{2j}$ where $j=1/2,1,3/2,2$ is the total spin.
% section generalized_helicity_spinors_for_arbitrary_interpolation_angle_up_to_spin_2 (end)

% \section{Relation between spin orientation \texorpdfstring{$(\theta_s, \phi_s)$}{(theta_s,phi_s)} and the particle momentum for \texorpdfstring{$+$}{+} helicity spinor}
% \label{sec:relation_between_spin_orientation_theta_phi_and_the_particle_momentum_for_helicity_spinor}
% % \begin{subequations}
% % \label{eqn:theta/phi_relation_with_P}
%  \begin{align}
%   \cos\dfrac{\theta_{s}}{2}=\dfrac{1}{2}\sqrt{\dfrac{2M}{E+M}}\sqrt{\dfrac{P_{\mT}+\Pp}{2\Pp}}\left(\sqrt{\dfrac{P^{\pT}+\Pp}{M(\sin\delta+\cos\delta)}}+\sqrt{\dfrac{M(\sin\delta+\cos\delta)}{P^{\pT}+\Pp}}\right), \quad
%   \cos\phi_{s}=\dfrac{P^{1}}{\sqrt{\mathbf{P}_{\perp}^{2}}}, \quad \sin\phi_{s}=\dfrac{P^{2}}{\sqrt{\mathbf{P}_{\perp}^{2}}}. \label{eqn:theta_phi_relation_with_P}
%  \end{align}
% \end{subequations}

\section{Generalized Melosh transformation}
\label{Appendix:generalized_melosh_transformation}
For convenience, we write everything in terms of the angles $\theta_{s}$ and $\phi_{s}$ which are given by Eq.~(\ref{eqn:theta_s_phi_s_relation_with_alpha_and_beta}).
For spin i, the generalized Melosh transformation matrix $\Omega(i)$ which connect the Dirac spinors with our generalized helicity spinors is
\begin{align}
		\Omega(i)=
		\begin{pmatrix}
				\omega(i) & 0 \\
				0 & \omega(i)
		\end{pmatrix}.
		\label{eqn:Omega_i}
\end{align}
For spin 1,
\begin{align}
		\omega(1)=
		\begin{pmatrix}
      \cos ^2\dfrac{\theta_s}{2} & -\dfrac{e^{i \phi _s} \sin \theta_s}{\sqrt{2}} & e^{2 i \phi _s} \sin ^2\dfrac{\theta_s}{2} \\[8pt]
      \dfrac{e^{-i \phi _s} \sin \theta_s}{\sqrt{2}} & \cos \theta_s & -\dfrac{e^{i \phi _s} \sin \theta_s}{\sqrt{2}} \\[8pt]
      e^{-2 i \phi _s} \sin ^2\dfrac{\theta_s}{2} & \dfrac{e^{-i \phi _s} \sin \theta_s}{\sqrt{2}} & \cos ^2\dfrac{\theta_s}{2} \\[8pt]
		\end{pmatrix}.
		\label{eqn:omega_1}
\end{align}
For spin $3/2$,
\begin{align}
		\omega\left(\dfrac{3}{2}\right)=
		\begin{pmatrix}
      \cos ^3\dfrac{\theta_s}{2} & -\dfrac{1}{4} \sqrt{3} e^{i \phi _s} \csc \dfrac{\theta_s}{2} \sin ^2\theta_s & \dfrac{1}{2} \sqrt{3} e^{2 i \phi _s} \sin \dfrac{\theta_s}{2} \sin \theta_s & -e^{3 i \phi _s} \sin ^3\dfrac{\theta_s}{2} \\[8pt]
      \dfrac{1}{4} \sqrt{3} e^{-i \phi _s} \csc \dfrac{\theta_s}{2} \sin ^2\theta_s & \dfrac{1}{4} \left(\cos \dfrac{\theta_s}{2}+3 \cos \dfrac{3\theta_s}{2}\right) & \dfrac{1}{4} e^{i \phi _s} \left(\sin \dfrac{\theta_{s}}{2} - 3 \sin\dfrac{3\theta_{s}}{2}\right) \sin \dfrac{\theta_s}{2} & \dfrac{1}{2} \sqrt{3} e^{2 i \phi _s} \sin \dfrac{\theta_s}{2} \sin \theta_s \\[8pt]
     \dfrac{1}{2} \sqrt{3} e^{-2 i \phi _s} \sin \dfrac{\theta_s}{2} \sin \theta_s & -\dfrac{1}{4} e^{-i \phi _s} \left(\sin \dfrac{\theta_s}{2}-3 \sin \dfrac{3\theta_s}{2}\right) & \dfrac{1}{4} \left(\cos \dfrac{\theta_s}{2}+3 \cos \dfrac{3\theta_s}{2}\right) & -\dfrac{1}{4} \sqrt{3} e^{i \phi _s} \csc \dfrac{\theta_s}{2} \sin ^2\theta_s \\[8pt]
     e^{-3 i \phi _s} \sin ^3\dfrac{\theta_s}{2} & \dfrac{1}{2} \sqrt{3} e^{-2 i \phi _s} \sin \dfrac{\theta_s}{2} \sin \theta_s & \dfrac{1}{4} \sqrt{3} e^{-i \phi _s} \csc \dfrac{\theta_s}{2} \sin ^2\theta_s & \cos ^3\dfrac{\theta_s}{2}
		\end{pmatrix}.
		\label{eqn:omega_3/2}
\end{align}
For spin 2, $\omega(2)$ is defined via the five columns
\begin{align}
		&\omega(2)_{\alpha,1}=
		\begin{pmatrix}
	    \cos ^4\dfrac{\theta_s}{2}\\[8pt]
2 e^{-i \phi _s} \sin \dfrac{\theta_s}{2} \cos ^3\dfrac{\theta_s}{2}\\[8pt]
\dfrac{1}{2} \sqrt{\dfrac{3}{2}} e^{-2 i \phi _s} \sin ^2\theta_s\\[8pt]
e^{-3 i \phi _s} \sin ^2\dfrac{\theta_s}{2} \sin \theta_s\\[8pt]
e^{-4 i \phi _s} \sin ^4\dfrac{\theta_s}{2}
		\end{pmatrix},\;
		\omega(2)_{\alpha,2}=
		\begin{pmatrix}
		  -2 e^{i \phi _s} \sin \dfrac{\theta_s}{2} \cos ^3\dfrac{\theta_s}{2}\\[8pt]
\cos ^2\dfrac{\theta_s}{2} \left(2 \cos \theta_s-1\right)\\[8pt]
\sqrt{\dfrac{3}{2}} e^{-i \phi _s} \sin \theta_s \cos \theta_s\\[8pt]
e^{-2 i \phi _s} \sin ^2\dfrac{\theta_s}{2} \left(2 \cos \theta_s+1\right)\\[8pt]
e^{-3 i \phi _s} \sin ^2\dfrac{\theta_s}{2} \sin \theta_s
		\end{pmatrix},\;
		\omega(2)_{\alpha,3}=
		\begin{pmatrix}
		  \dfrac{1}{2} \sqrt{\dfrac{3}{2}} e^{2 i \phi _s} \sin ^2\theta_s\\[8pt]
-\dfrac{1}{2} \sqrt{\dfrac{3}{2}} e^{i \phi _s} \sin 2\theta_s\\[8pt]
\dfrac{1}{4} \left(3 \cos 2\theta_s+1\right)\\[8pt]
\sqrt{\dfrac{3}{2}} e^{-i \phi _s} \sin \theta_s \cos \theta_s\\[8pt]
\dfrac{1}{2} \sqrt{\dfrac{3}{2}} e^{-2 i \phi _s} \sin ^2\theta_s
		\end{pmatrix},
\\
		&\omega(2)_{\alpha,4}=
		\begin{pmatrix}
		  -e^{3 i \phi _s} \sin ^2\dfrac{\theta_s}{2} \sin \theta_s\\[8pt]
e^{2 i \phi _s} \sin ^2\dfrac{\theta_s}{2} \left(2 \cos \theta_s+1\right)\\[8pt]
-\dfrac{1}{2} \sqrt{\dfrac{3}{2}} e^{i \phi _s} \sin 2\theta_s\\[8pt]
\cos ^2\dfrac{\theta_s}{2} \left(2 \cos \theta_s-1\right)\\[8pt]
2 e^{-i \phi _s} \sin \dfrac{\theta_s}{2} \cos ^3\dfrac{\theta_s}{2}
		\end{pmatrix},\;
		\omega(2)_{\alpha,5}=
		\begin{pmatrix}
		  e^{4 i \phi _s} \sin ^4\dfrac{\theta_s}{2}\\[8pt]
-e^{3 i \phi _s} \sin ^2\dfrac{\theta_s}{2} \sin \theta_s\\[8pt]
\dfrac{1}{2} \sqrt{\dfrac{3}{2}} e^{2 i \phi _s} \sin ^2\theta_s\\[8pt]
-2 e^{i \phi _s} \sin \dfrac{\theta_s}{2} \cos ^3\dfrac{\theta_s}{2}\\[8pt]
\cos ^4\dfrac{\theta_s}{2}
		\end{pmatrix}.
		\label{eqn:omega_2}
\end{align}

% section generalized_melosh_transformation (end)

\section{Apparent Spin Orientation}
\label{sec:apparent-spin}
As explained in Sec.~\ref{sub:spin_orientation}, we apply the light-front boost in the $x$ direction to a spin 4-vector and figure out the resulted ``apparent'' spin direction.
With the boost and rotation generators written in the following 4-vector representation:
\begin{alignat}{3}
  K_{1}=
  \begin{pmatrix}
  0 & i & 0 & 0 \\
  i & 0 & 0 & 0 \\
  0 & 0 & 0 & 0\\
  0 & 0 & 0 & 0
  \end{pmatrix},
  K_{2}=
  \begin{pmatrix}
  0 & 0 & i & 0 \\
  0 & 0 & 0 & 0 \\
  i & 0 & 0 & 0 \\
  0 & 0 & 0 & 0
  \end{pmatrix},
  K_{3}=
  \begin{pmatrix}
  0 & 0 & 0 & i \\
  0 & 0 & 0 & 0 \\
  0 & 0 & 0 & 0 \\
  i & 0 & 0 & 0
  \end{pmatrix},
  % \nonumber\\
  J_{1}=
  \begin{pmatrix}
  0 & 0 & 0 & 0 \\
  0 & 0 & 0 & 0 \\
  0 & 0 & 0 & -i\\
  0 & 0 & i & 0
  \end{pmatrix},
  J_{2}=
  \begin{pmatrix}
  0 & 0 & 0 & 0 \\
  0 & 0 & 0 & i \\
  0 & 0 & 0 & 0 \\
  0 & -i & 0 & 0
  \end{pmatrix},
  J_{3}=
  \begin{pmatrix}
  0 & 0 & 0 & 0 \\
  0 & 0 & -i & 0 \\
  0 & i & 0 & 0 \\
  0 & 0 & 0 & 0
  \end{pmatrix},
  \label{eqn:J_K}
\end{alignat}
and $\beta_{3}=0$, $\beta_{1}=0$ in Eq.~(\ref{eqn:T_transformation_for_any_interpolation_angle}), the $T_{12}$ transformation can be applied to the momentum 4-vector $P^{\mu}_{0}=(M,0,0,0)$ and the spin 4-vector $S^{\mu}_{0}=(0,0,0,1)$ defined in the rest frame.
The resulted momentum 4-vector is $P^{\mu}=(M+M\beta_{1}^{2}/2, M\beta_{1}, 0, -M\beta_{1}^{2}/2)$, where $\beta_{1}=P^{x}/M$ and the relation $P^{z}=-\frac{(P^{x})^{2}}{2M}$ is satisfied~\cite{Ji2013}.
The resulted spin 4-vector is $S^{\mu}=(\beta_{1}^{2}/2,\beta_{1},0,1-\beta_{1}^{2}/2)$, or written in terms of the final momentum
\begin{align}
  S^{\mu}=\left( \dfrac{(P^{x})^{2}}{2M^{2}}, \dfrac{P^{x}}{M}, 0, \dfrac{2M^{2}-(P^{x})^{2}}{2M^{2}} \right).
	\label{eqn:S_mu_P}
\end{align}
Therefore, after this transverse light-front boost, the momentum direction and the ``apparent'' spin direction are given by
\begin{align}
		\theta=-\arctan \left( \dfrac{2M}{P^{x}} \right),\quad\quad \theta_{a}=\arccos\left( \dfrac{2M^{2}-(P^{x})^{2}}{\sqrt{4M^{4}+(P^{x})^{4}}} \right).
		\label{eqn:theta_and_theta_apparent}
\end{align}
Finally, the ``apparent'' spin orientation angle $\theta_{a}$ as a function of the momentum direction $\theta$ is
\begin{align}
		\theta_{a}=\arccos \left( \dfrac{\tan^{2}\theta-2}{\sqrt{\tan^{4}\theta+4}} \right),\label{eqn:theta_apparent_in_terms_of_theta}
\end{align}
which is plotted as the magenta dot-dashed line in Fig.~\ref{fig:LF_profile1}. From Eq.~(\ref{eqn:theta_apparent_in_terms_of_theta}),
we note that $\theta_{a} \approx 0$ as $\theta \approx \pi/2$ and $\theta_{a} \approx \pi$ as $\theta \approx \pi$. This behavior can be seen
in Fig.~\ref{fig:LF_profile1}.

\section{Interpolated helicity scattering amplitudes}
\label{sec:Interpolated_scattering_amplitudes}
In this Appendix, we plot the helicity amplitudes of two particle scattering, as given by Eq.~(\ref{eqn:scattering_M}), in terms of both the interpolation angle $\delta$ and the total momentum $P^{z}$.
The same parameters in Sec.~\ref{sec:interpolating_helicity_probabilities_for_fermion_scattering} are used.
All the 16 helicity amplitudes are shown in Fig.~\ref{fig:Helicity_Scattering_Amplitudes}.
% , where the annihilation angle $\theta=\pi/3$, the center of mass energy $M=2\epsilon=4\;\text{GeV}$, and $m_{\text{ini}}=m_{1}=m_{2}=1$ GeV and $m_{\text{final}}=m_{3}=m_{4}=1.5$ GeV for the mass of the initial and final particles.

\begin{figure*}[!hbp]
 \begin{align}
    &\subfloat{
    \includegraphics[width=0.23\textwidth]{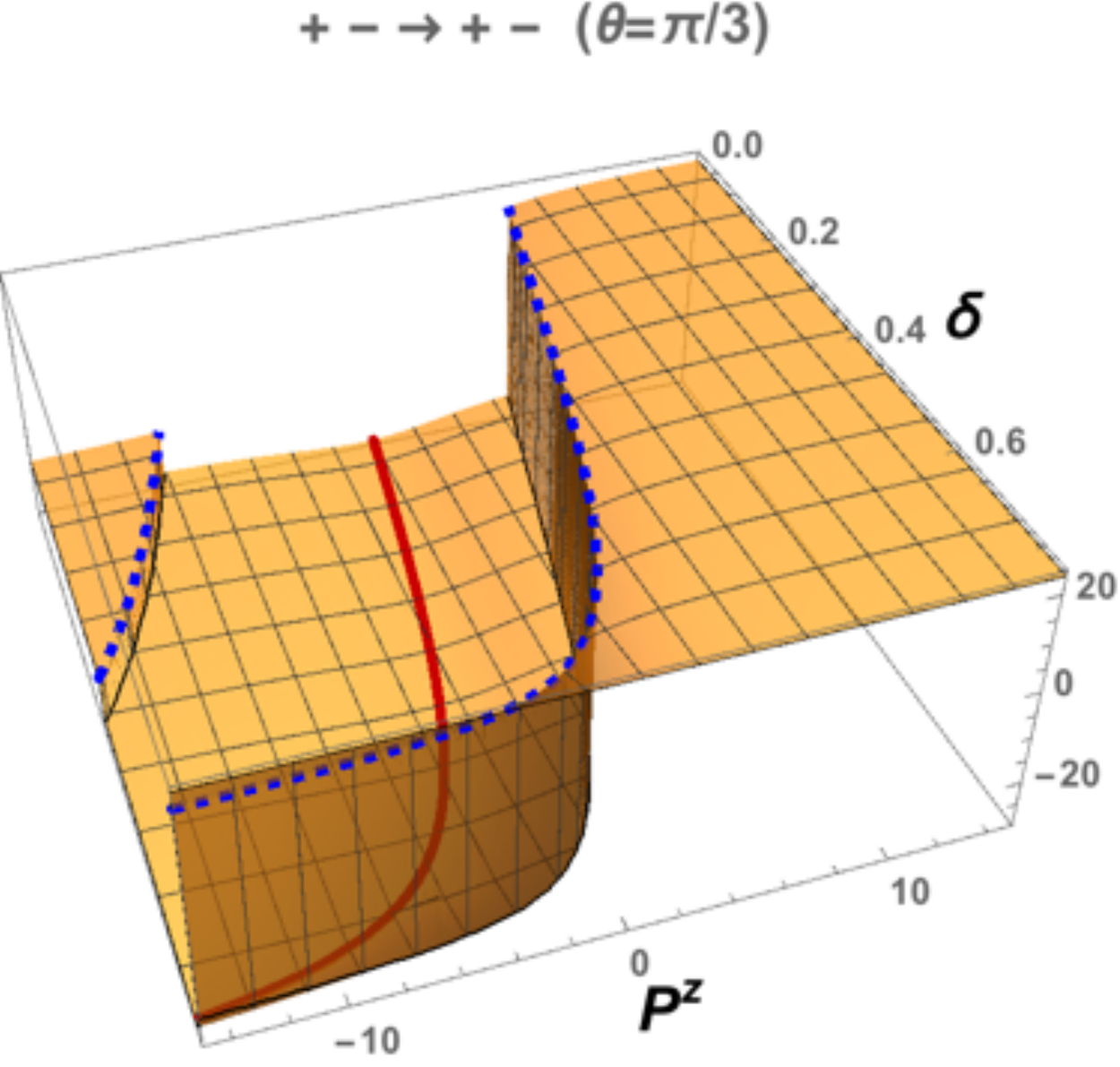}
    }
    %\quad
    \subfloat{
    \includegraphics[width=0.23\textwidth]{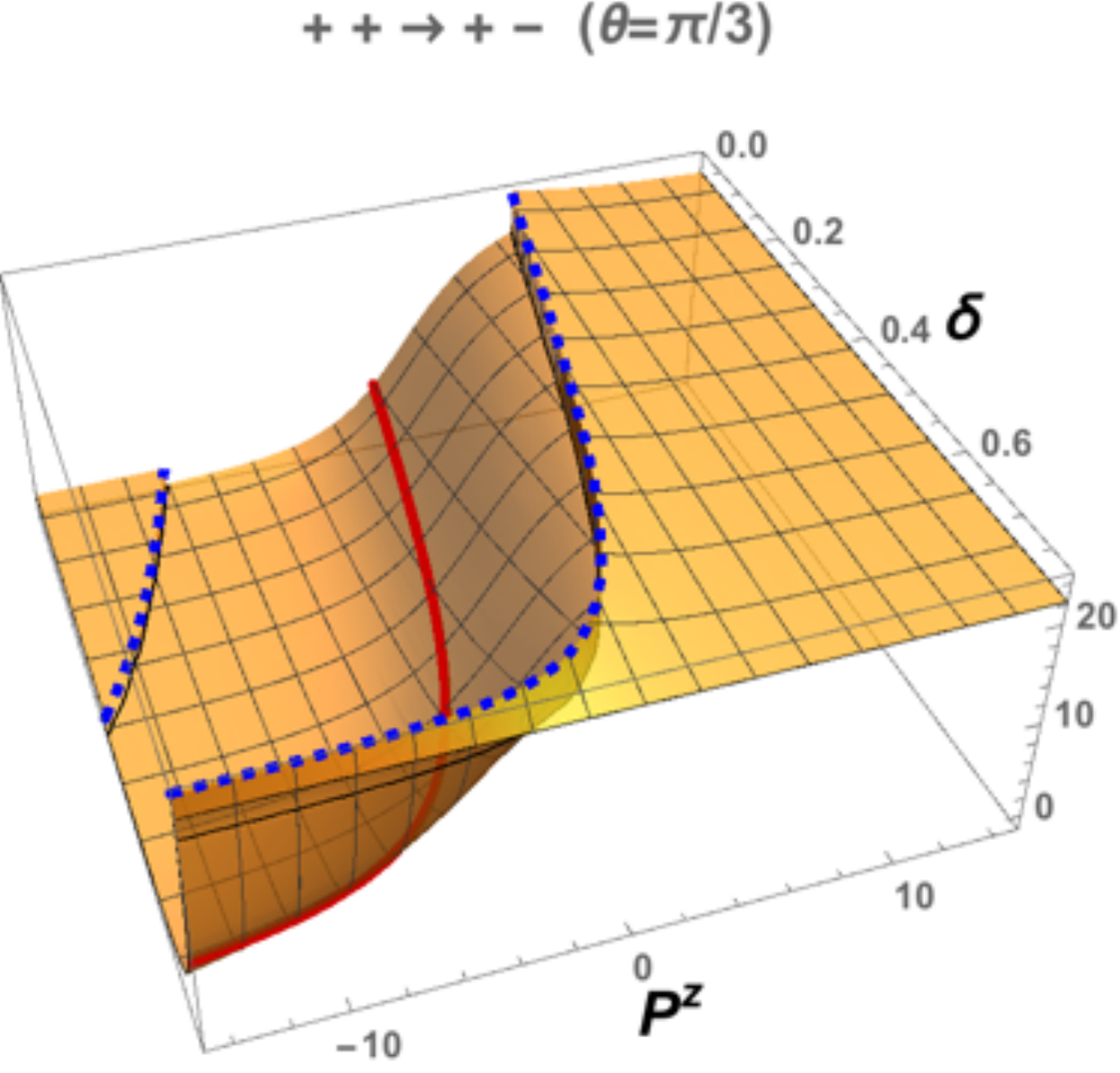}
    }
    %\quad
    \subfloat{
    \includegraphics[width=0.23\textwidth]{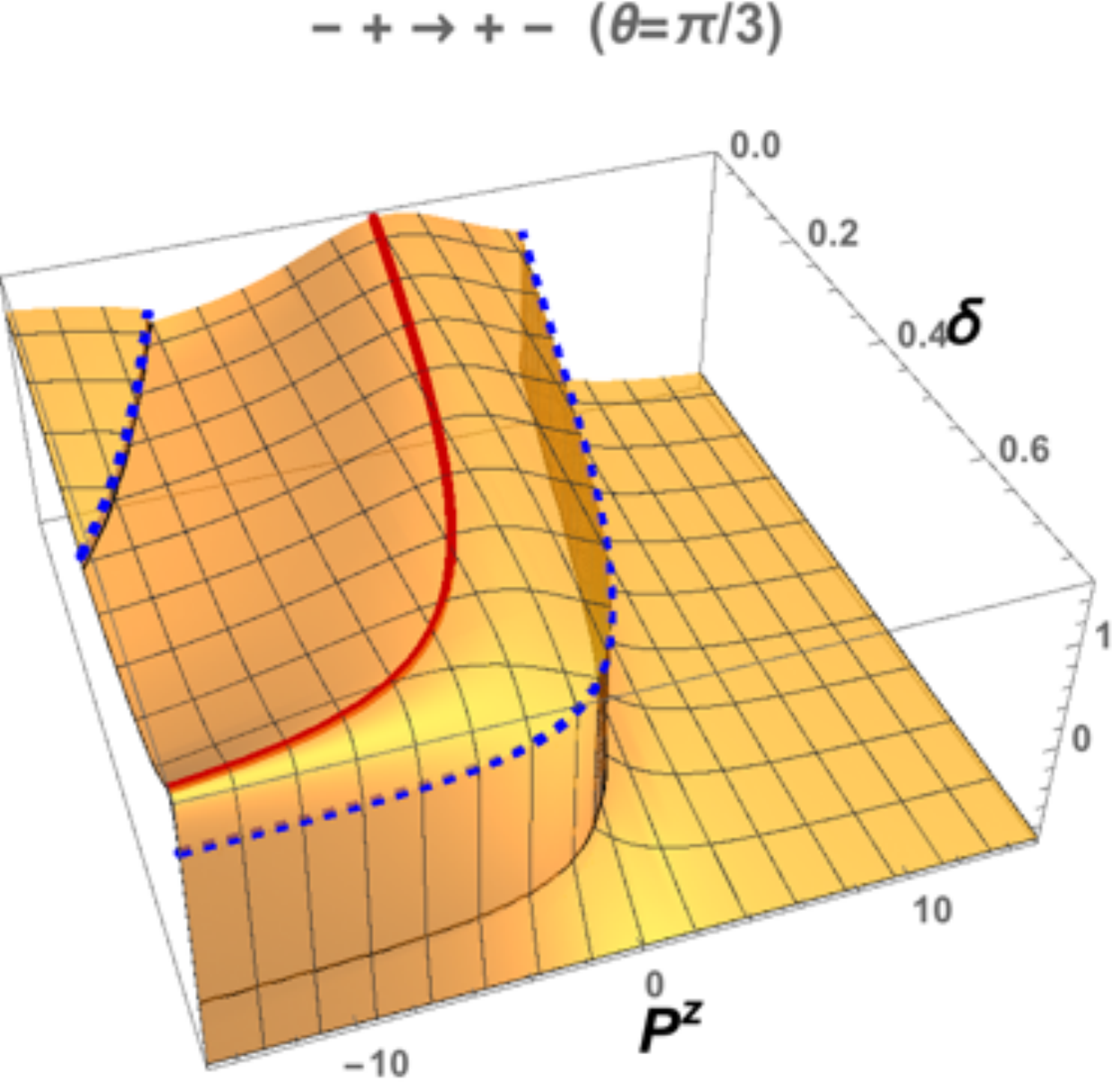}
    }
		%\quad
    \subfloat{
    \includegraphics[width=0.23\textwidth]{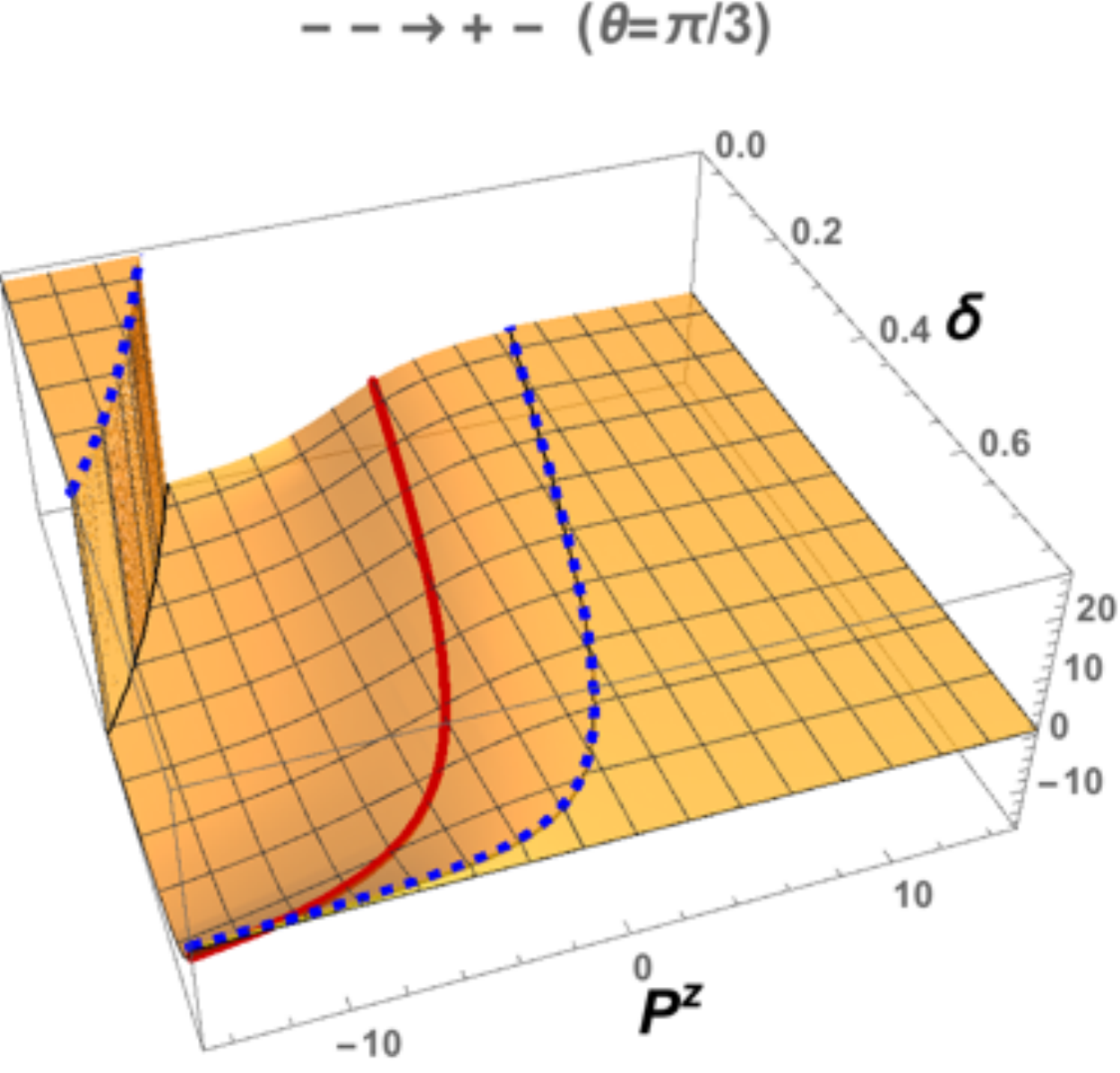}
    }
    \nonumber\\
    &\subfloat{
    \includegraphics[width=0.23\textwidth]{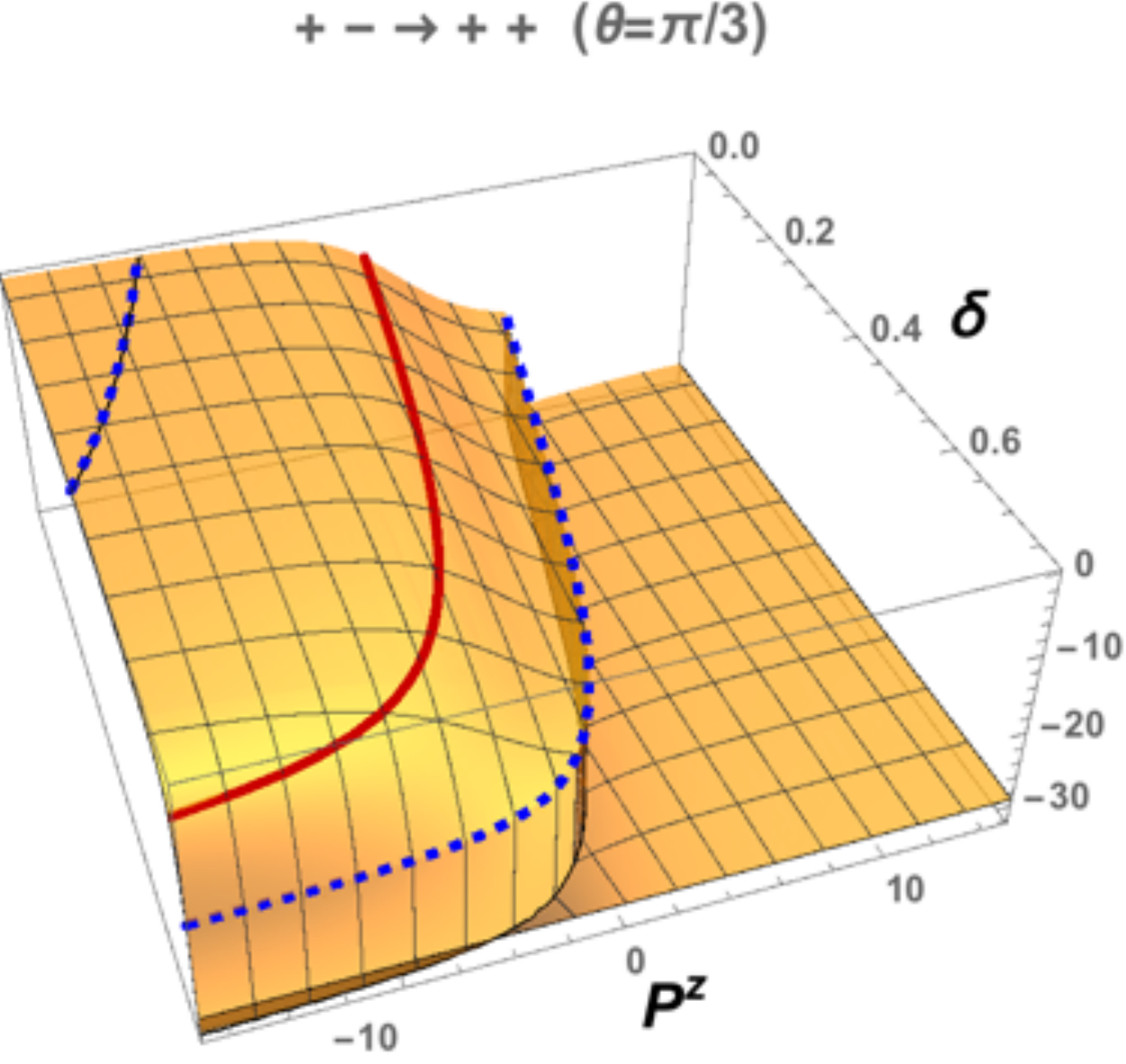}
    }
    %\quad
    \subfloat{
    \includegraphics[width=0.23\textwidth]{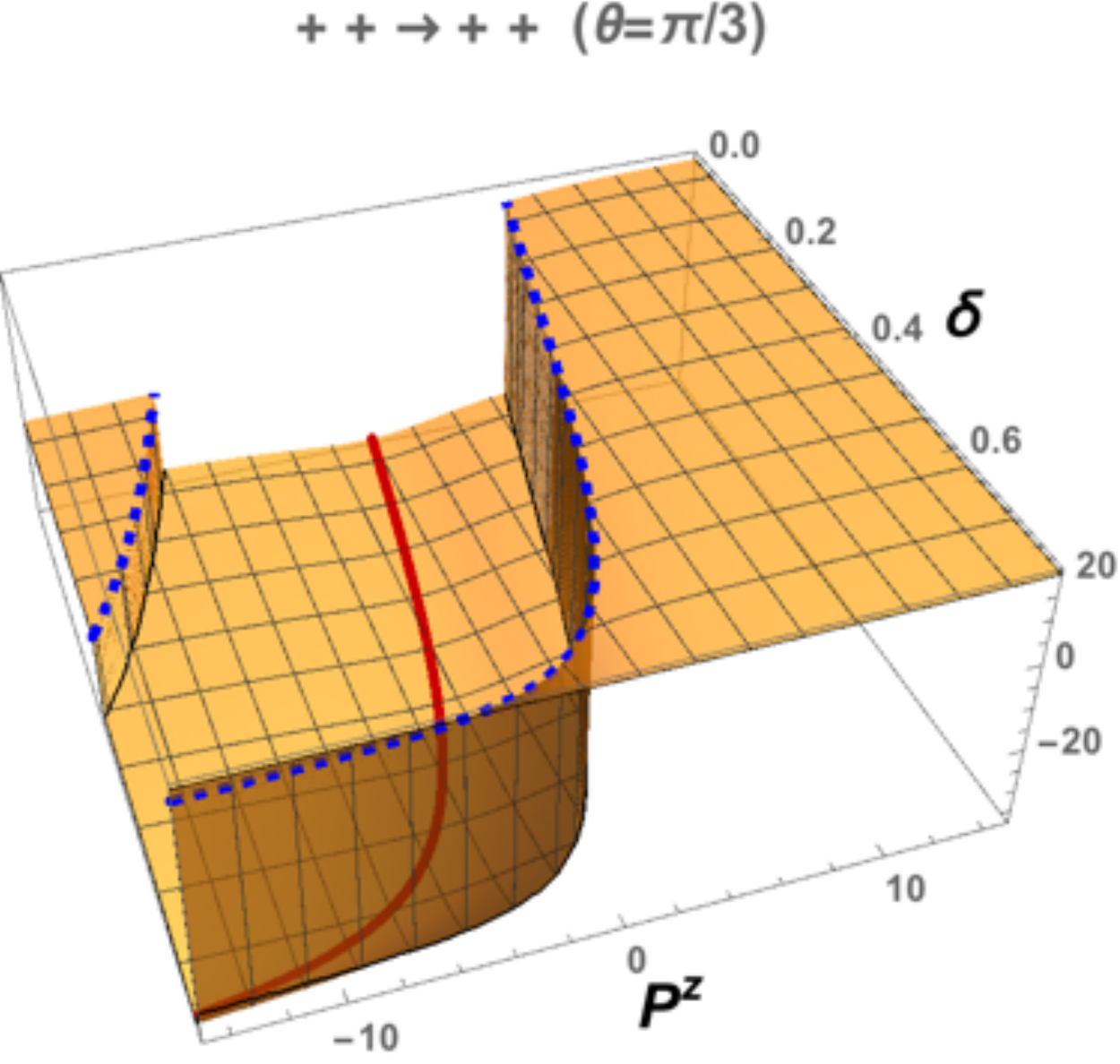}
    }
    %\quad
    \subfloat{
    \includegraphics[width=0.23\textwidth]{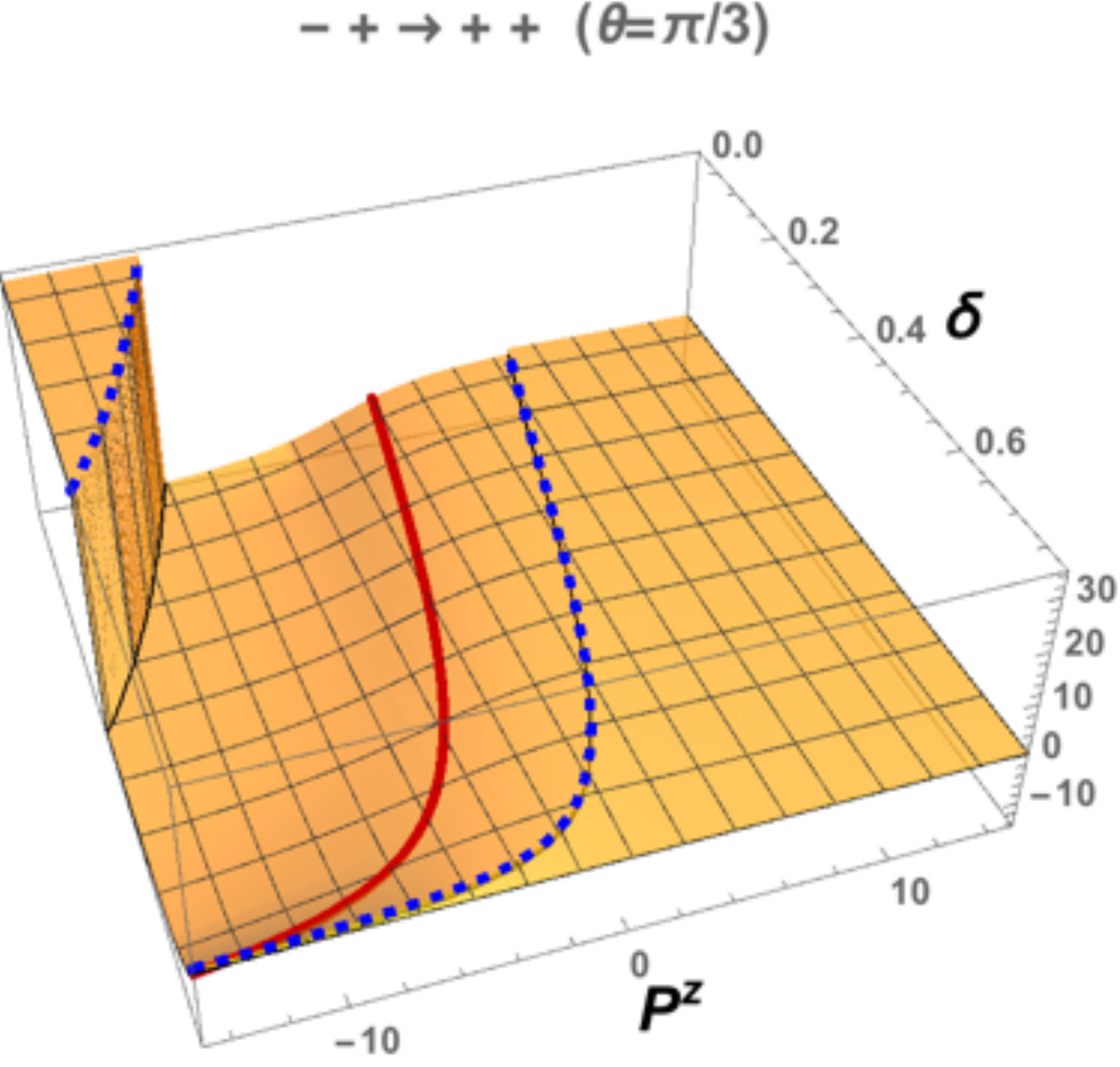}
    }
    %\quad
    \subfloat{
    \includegraphics[width=0.23\textwidth]{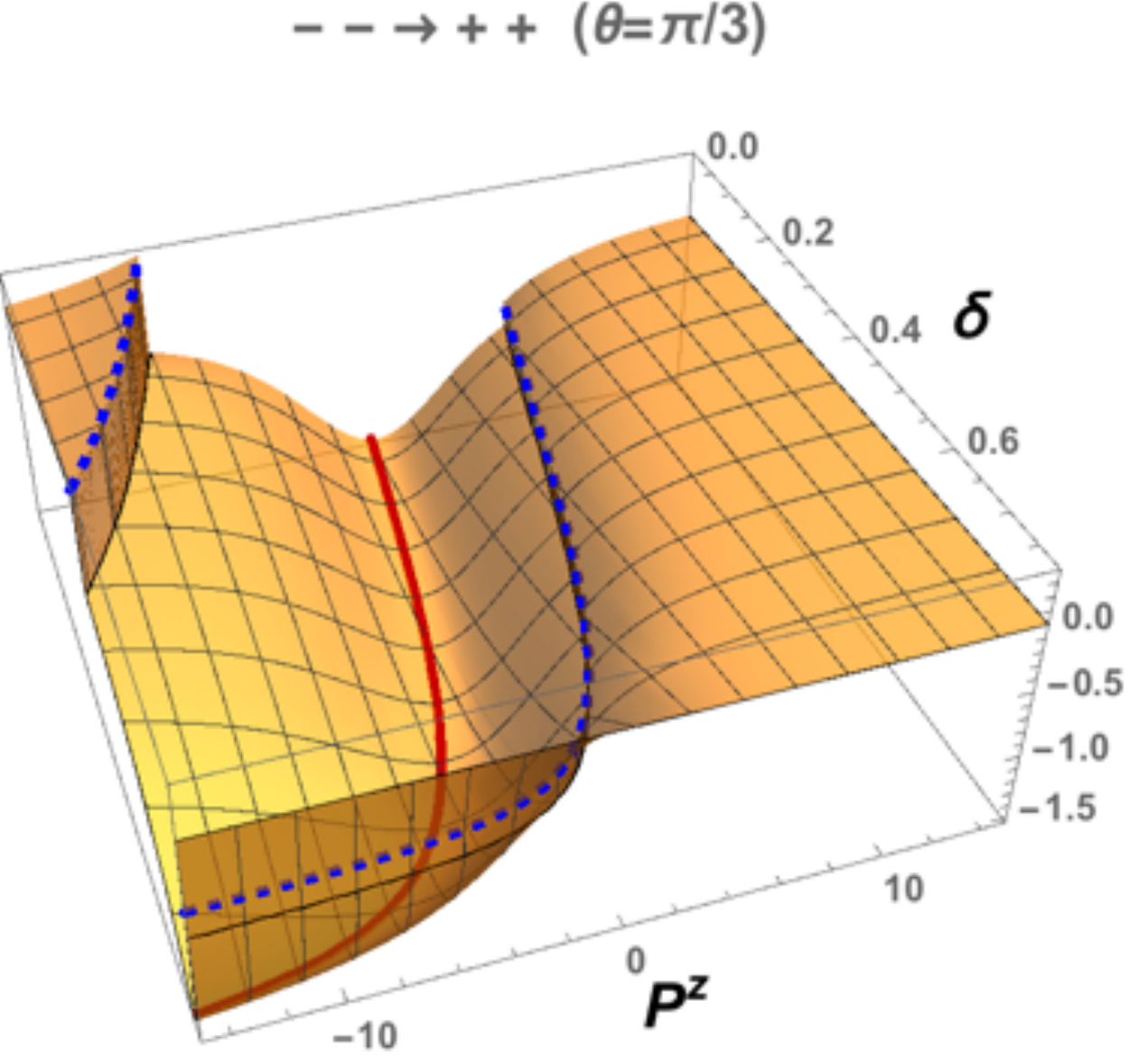}
    }
    \nonumber\\
    &\subfloat{
    \includegraphics[width=0.23\textwidth]{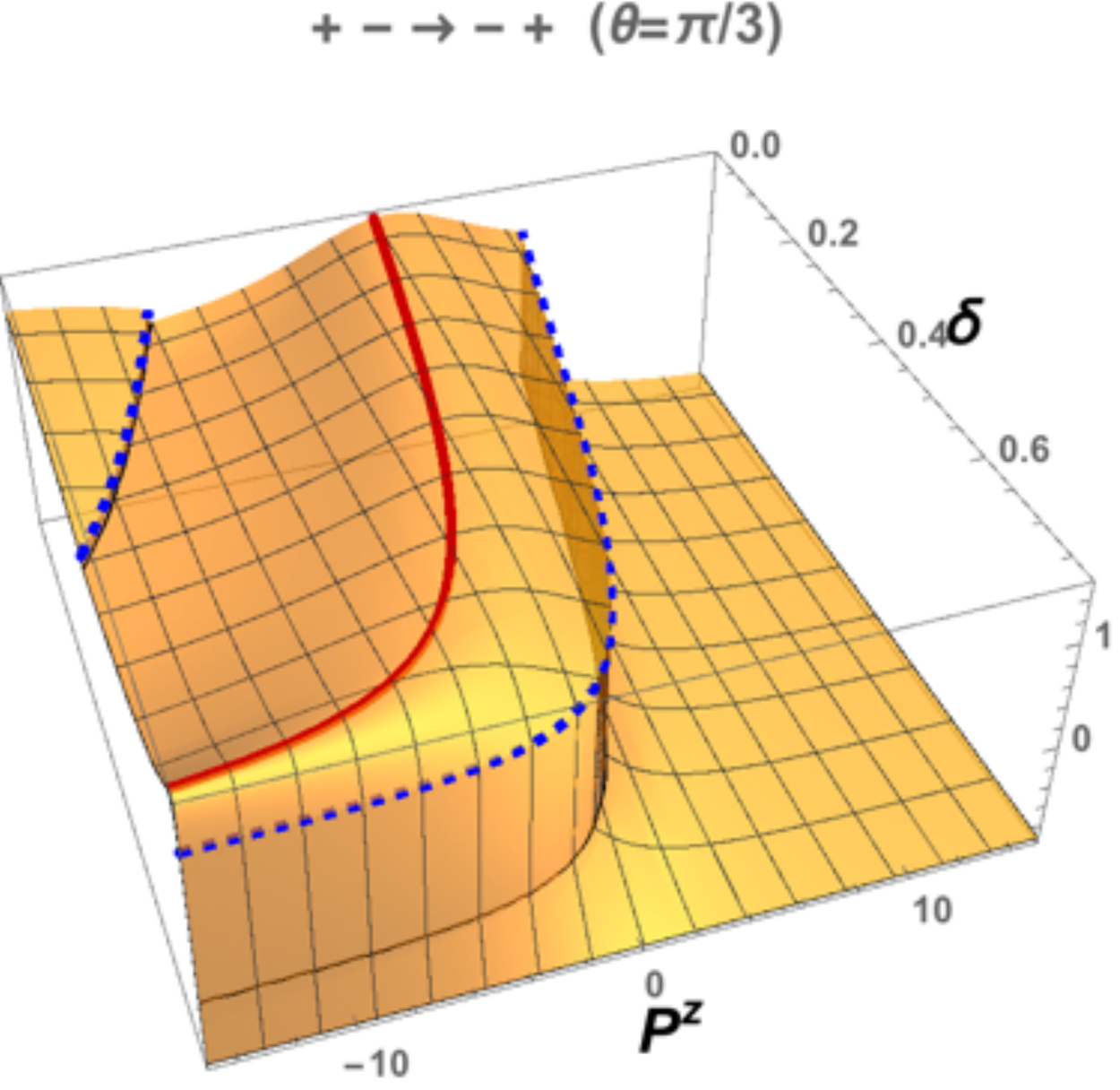}
    }
    %\quad
    \subfloat{
    \includegraphics[width=0.23\textwidth]{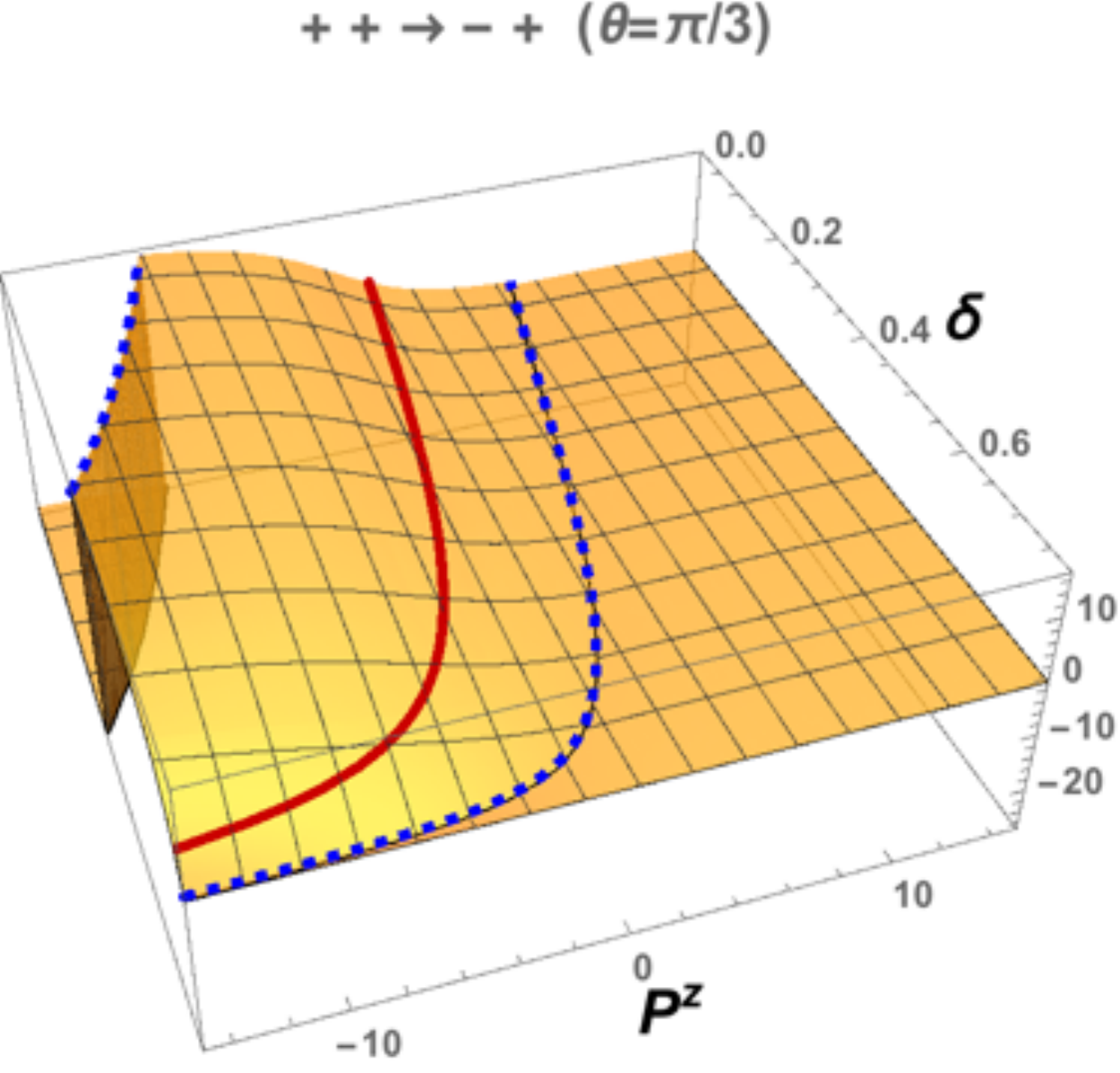}
    }
    %\quad
    \subfloat{
    \includegraphics[width=0.23\textwidth]{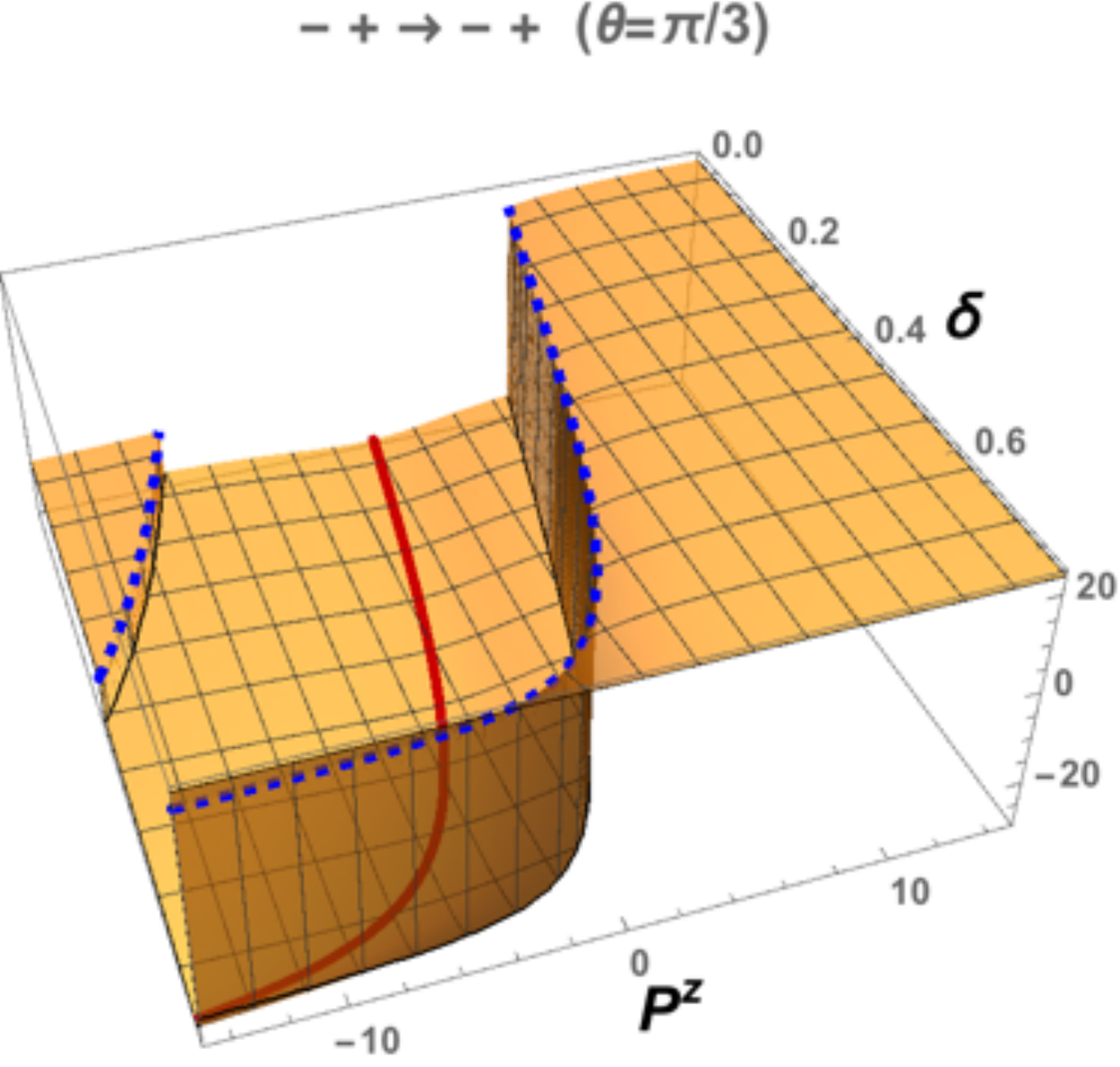}
    }
    %\quad
    \subfloat{
    \includegraphics[width=0.23\textwidth]{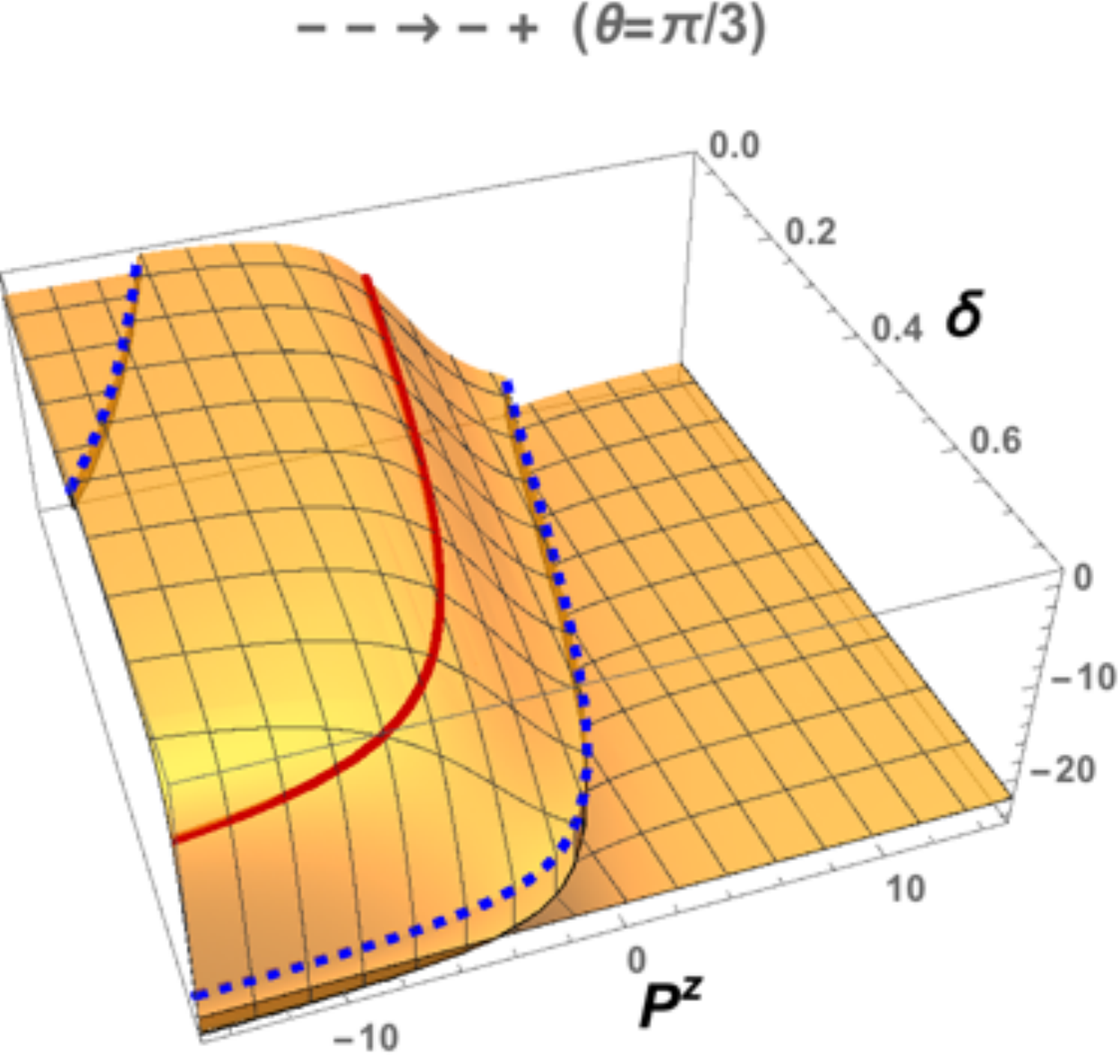}
    }
    \nonumber\\
    &\subfloat{
    \includegraphics[width=0.23\textwidth]{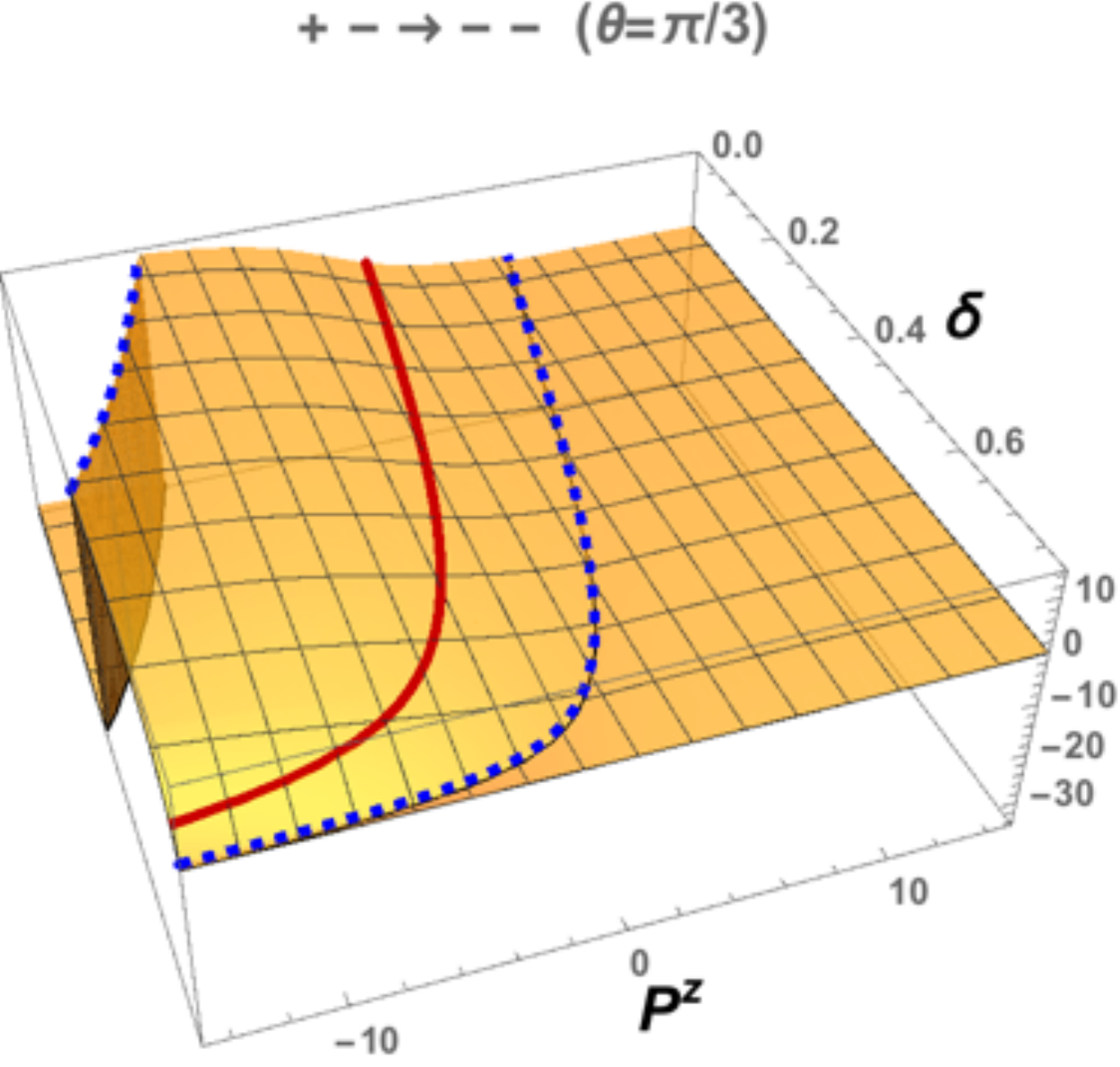}
    }
    %\quad
    \subfloat{
    \includegraphics[width=0.23\textwidth]{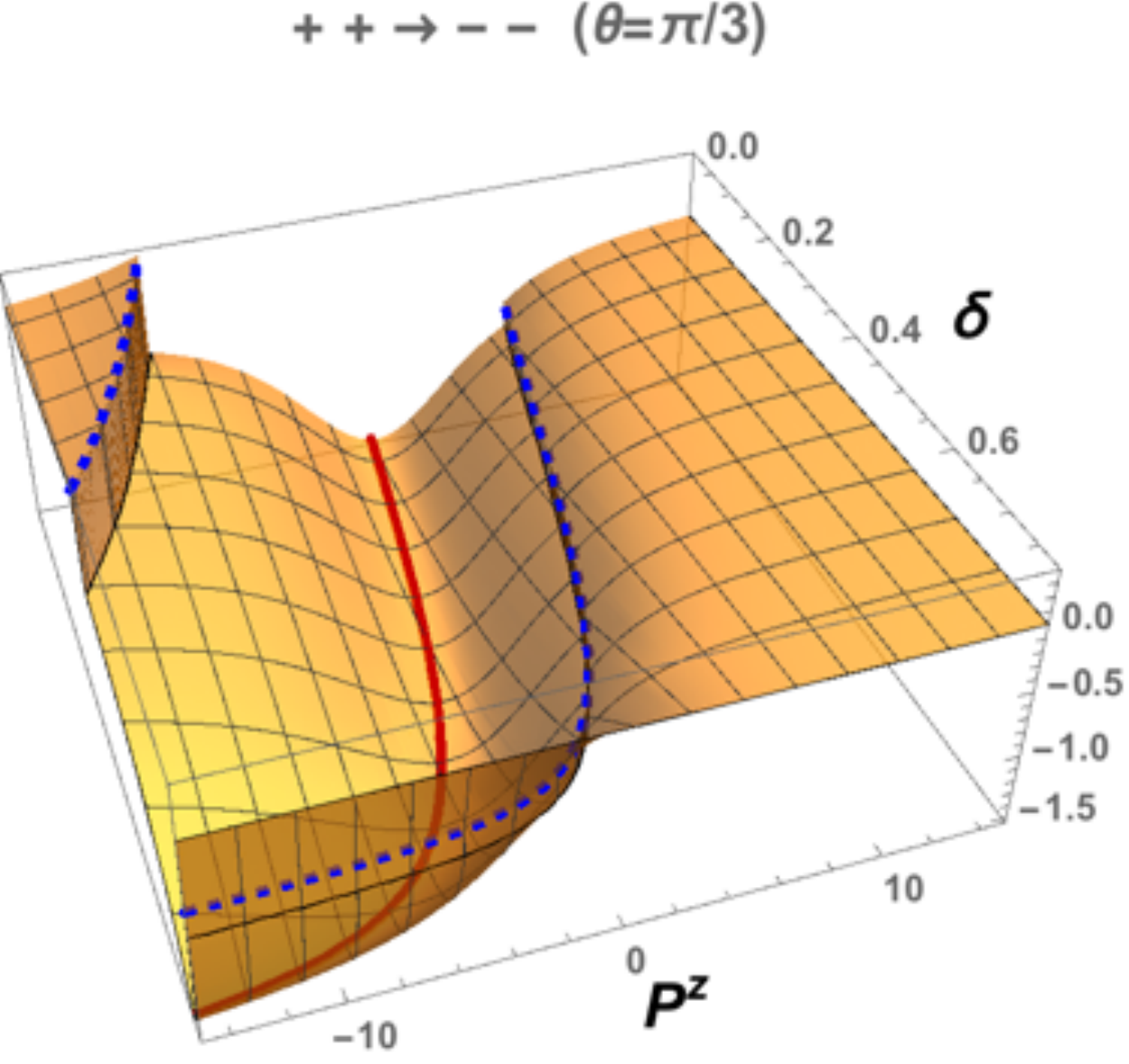}
    }
    %\quad
    \subfloat{
    \includegraphics[width=0.23\textwidth]{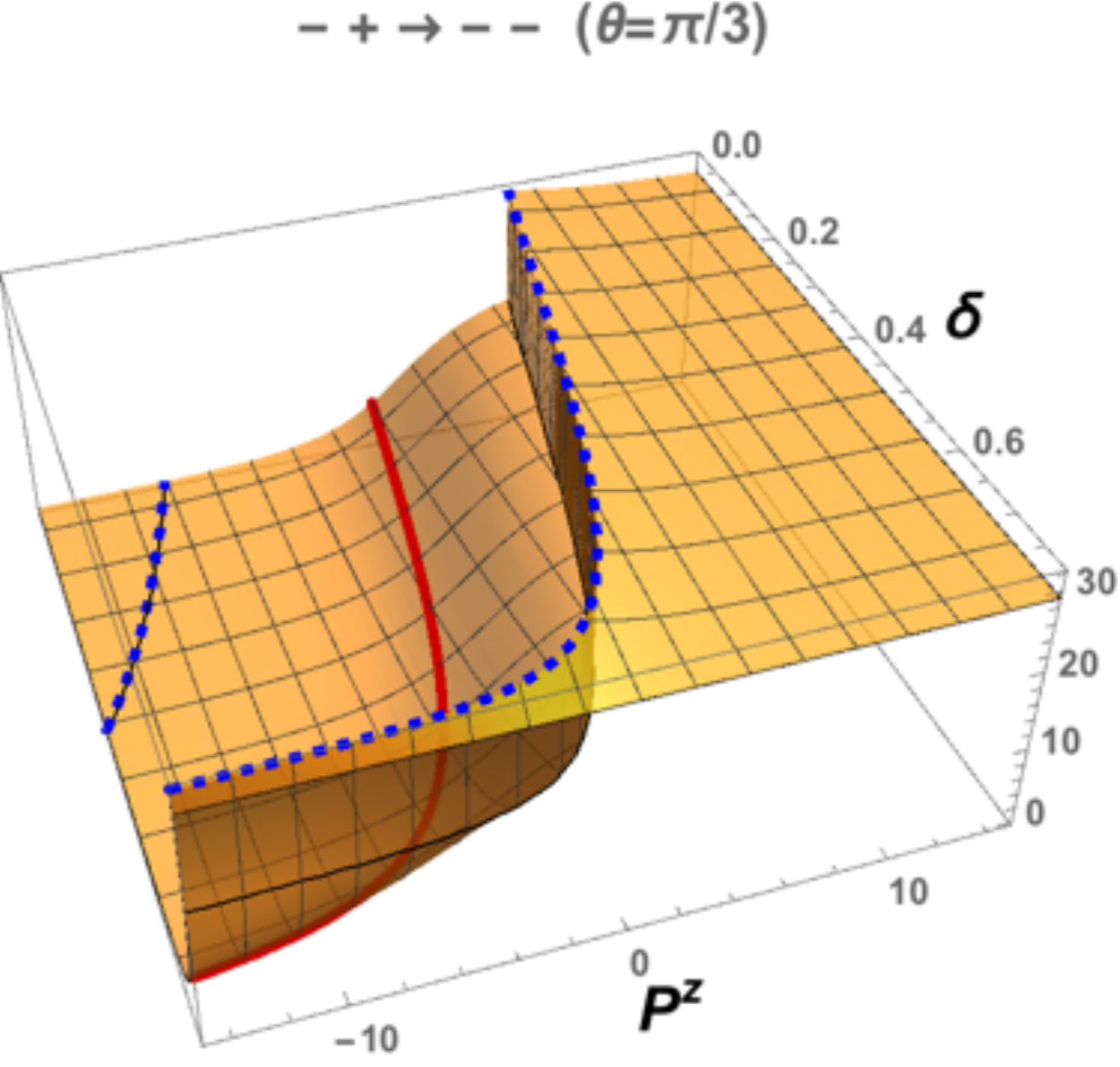}
    }
    %\quad
    \subfloat{
    \includegraphics[width=0.23\textwidth]{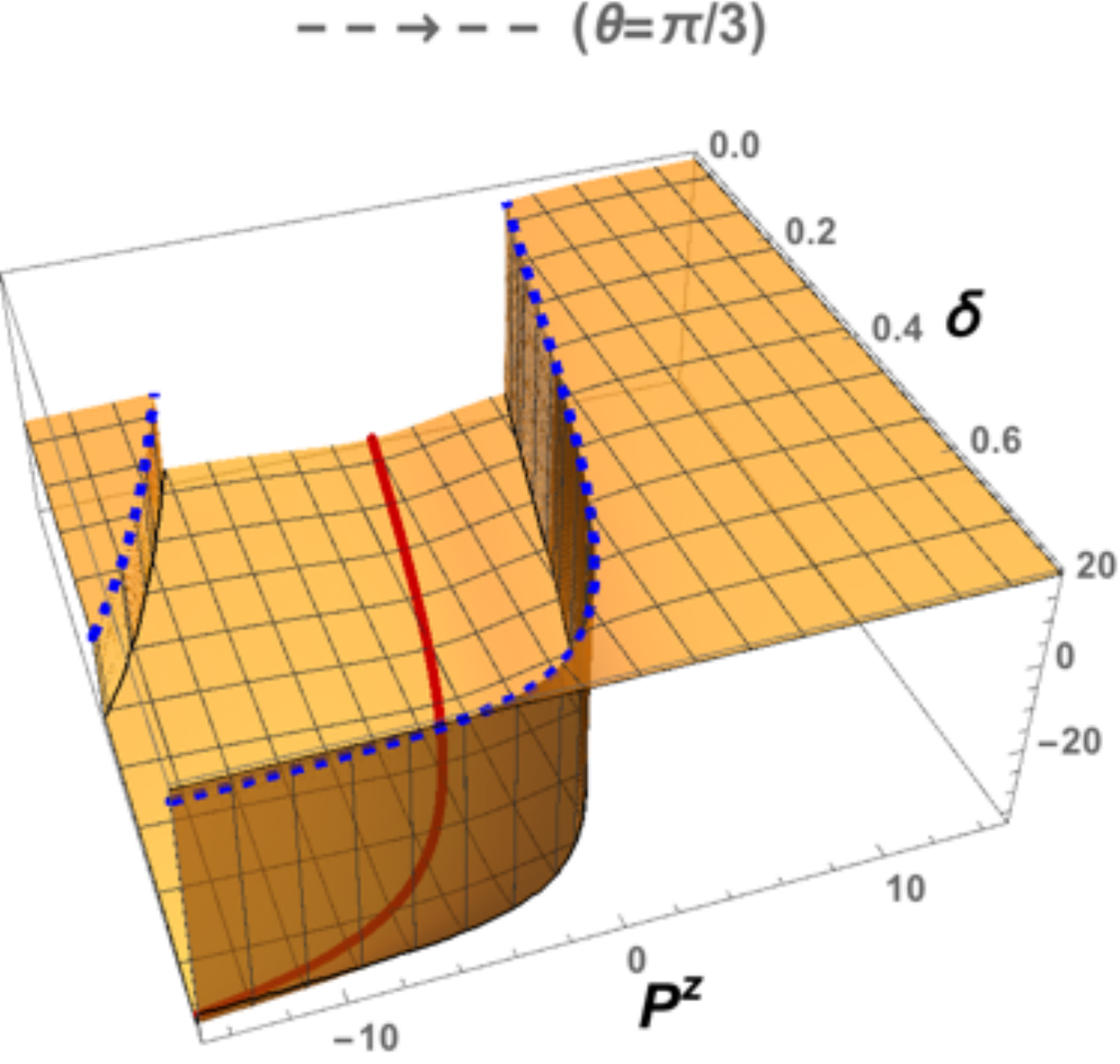}
    }
    % \nonumber\\
    % &\subfloat{
    % \includegraphics[width=0.23\textwidth]{MRL.pdf}
    % }
    % %\quad
    % \subfloat{
    % \includegraphics[width=0.23\textwidth]{MRR.pdf}
    % }
    % %\quad
    % \subfloat{
    % \includegraphics[width=0.23\textwidth]{MLR.pdf}
    % }
    % %\quad
    % \subfloat{
    % \includegraphics[width=0.23\textwidth]{MLL.pdf}
    % }
    \nonumber
 \end{align}
  \caption{\label{fig:Helicity_Scattering_Amplitudes} Scattering Amplitudes (with the factor $-e^{2}/q^{2}$ dropped) for 16 different helicity configurations, where the masses of the two colliding particles are $m_{1}=1$ GeV and $m_{2}=1.5$ GeV, the center of mass momentum for each particle is $p=2$ GeV and the scattering angle $\theta=\pi/3$.}
% 	The two blue dashed lines are the boundary lines across which the values of helicity amplitudes suddenly change and the red solid line is again the same universal J-curve.}
\end{figure*}

% section scattering_amplitudes (end)

\end{widetext}

\section{Interpolated helicity annihilation amplitudes and probabilities}
\label{sec:Interpolated_annihilation_amplitudes_and_probabilities}
\begin{figure}[H]
  \includegraphics[width=\columnwidth]{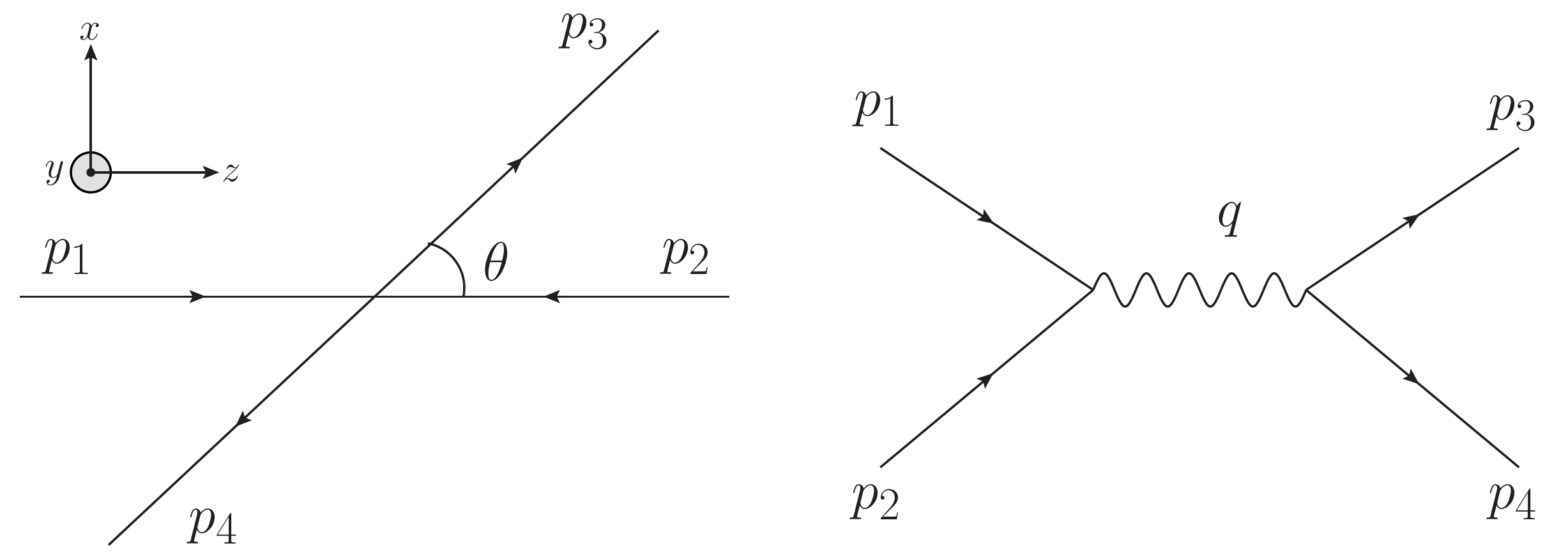}
  \caption{\label{fig:annihilation} Two particle annihilation at angle $\theta$, and its corresponding Feynman diagram at the lowest tree level.}
\end{figure}

For a annihilation process depicted in Fig.~\ref{fig:annihilation}, the helicity-dependent amplitude $\mathcal{M}(\lambda_{1},\lambda_{2},\lambda_{3},\lambda_{4 })$ is given by
\begin{align}
  \bar{v}^{(\lambda_{2})}(p_{2})\gamma^{\mu}u^{(\lambda_{1})}(p_{1}) \bar{u}^{(\lambda_{3})}(p_{3})\gamma_{\mu}v^{(\lambda_{4})}(p_{4}),\label{eqn:annihilation_M}
\end{align}
where $\lambda$ denotes the helicity of the particle.
Again, we dropped the coupling constant factor $(-ie)^{2}$ and the Lorentz invariant part of the propagator $-1/q^{2}$, since they are irrelevant to our discussion.

The 4-momentum of these four particles in the center of mass frame are given by Eq.~(\ref{eqn:p_for_four_particles}), with $m_{1}=m_{2}\equiv m_{\rm ini}$, $m_{3}=m_{4}=m_{\rm final}$ and $\epsilon_{1}=\epsilon_{2}=\sqrt{m_{\text{ini}}^{2}+p_{\text{ini}}^{2}}=\sqrt{m_{\text{final}}^{2}+p_{\text{final}}^{2}}=\epsilon_{3}=\epsilon_{4}\equiv \epsilon$.
The boosted 4-momentum for each particle is given by Eq.~(\ref{eqn:p_for_four_particles_boosted}), with the center of mass energy $M=2\epsilon$ and the total energy in the boosted frame $E=\sqrt{(P^{z})^{2}+M^{2}}$.
Just as in the scattering case, we use these frame dependent helicity spinors to calculate the helicity amplitudes, as well as the probabilities which are given by the square of Eq.~(\ref{eqn:annihilation_M}).
In our calculation, we choose $m_{\text{ini}}=1$ GeV and $m_{\text{final}}=1.5$ GeV, $\theta=\pi/3$ in the center of mass frame, and $M=2\epsilon=4\text{ GeV}$.

The annihilation amplitudes and probabilities for all 16 different spin configurations are plotted in terms of both the interpolation angle $\delta$ and the total momentum $P^{z}$ in Fig.~\ref{fig:Annihilation_Helicity_Amplitudes} and Fig.~\ref{fig:Annihilation_Helicity_Probabilities}.
In the last row of Fig.~\ref{fig:Annihilation_Helicity_Probabilities}, we also plot the probabilities of a certain helicity configuration going into all possible helicity configurations.

The boundaries lines and the J-curve are still described by Eq.~(\ref{eqn:Boundary_lines}) and Eq.~(\ref{eqn:J_curve}), except here $\epsilon_{1}=\epsilon_{2}=\epsilon=2$ GeV, $p=\sqrt{\epsilon^{2}-m_{\rm ini}^{2}}$, $M=2\epsilon$, and $E=\sqrt{(P^{z})^{2}+4\epsilon^{2}}$.

\begin{widetext}

\begin{figure*}[!hbp]
 \begin{align}
    &\subfloat{
    \includegraphics[width=0.22\textwidth]{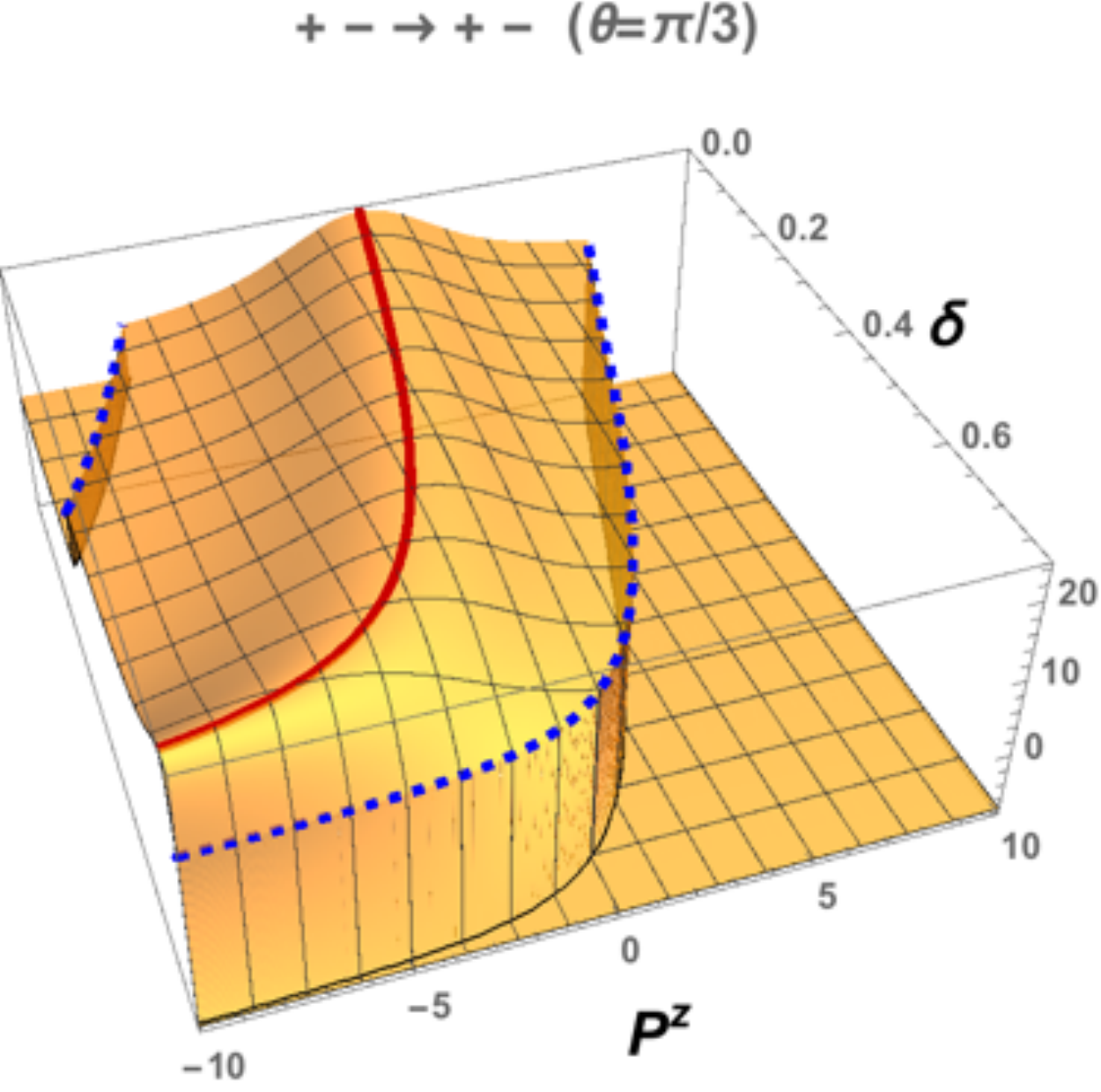}
    }
    %\quad
    \subfloat{
    \includegraphics[width=0.22\textwidth]{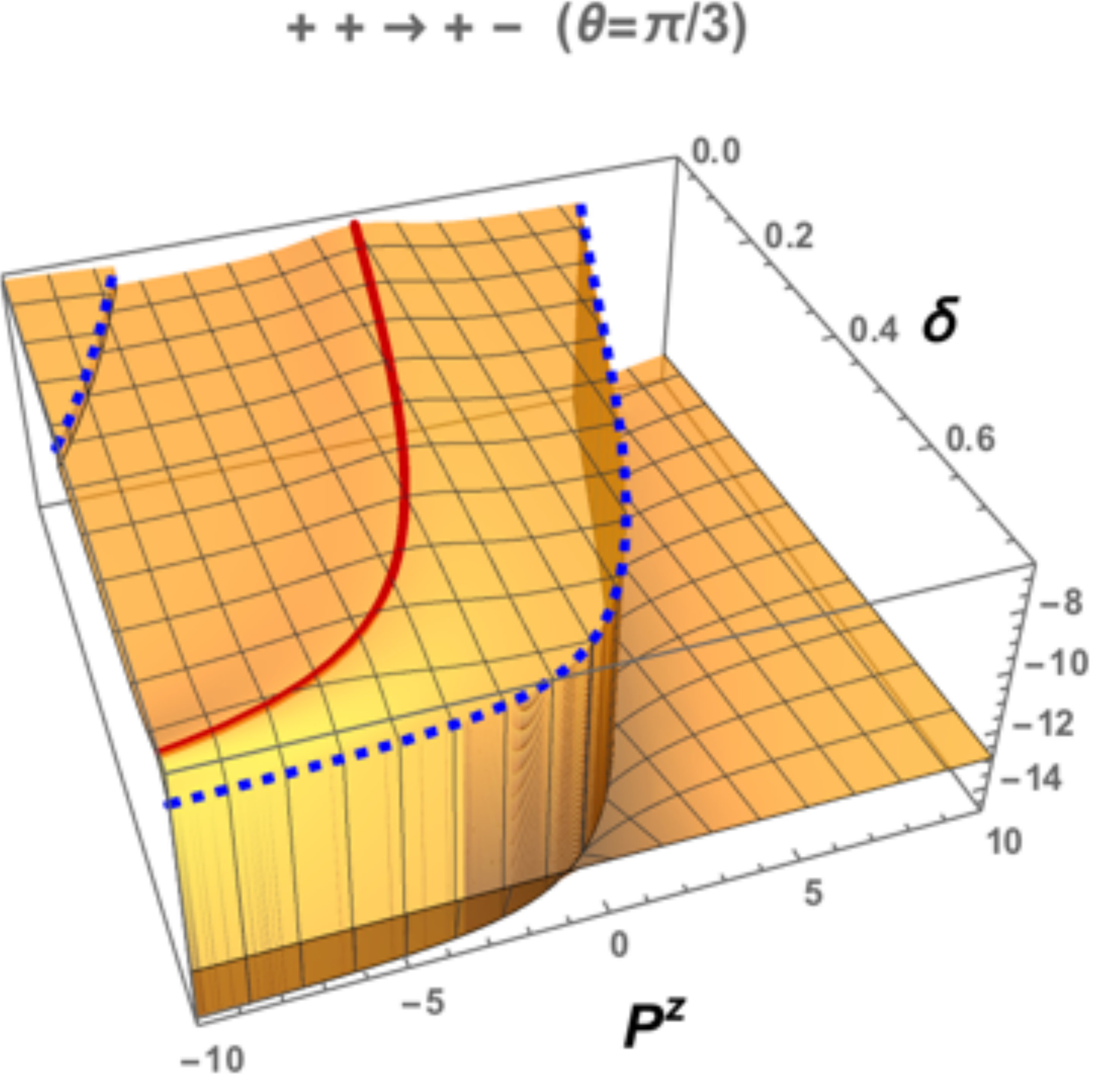}
    }
    %\quad
    \subfloat{
    \includegraphics[width=0.22\textwidth]{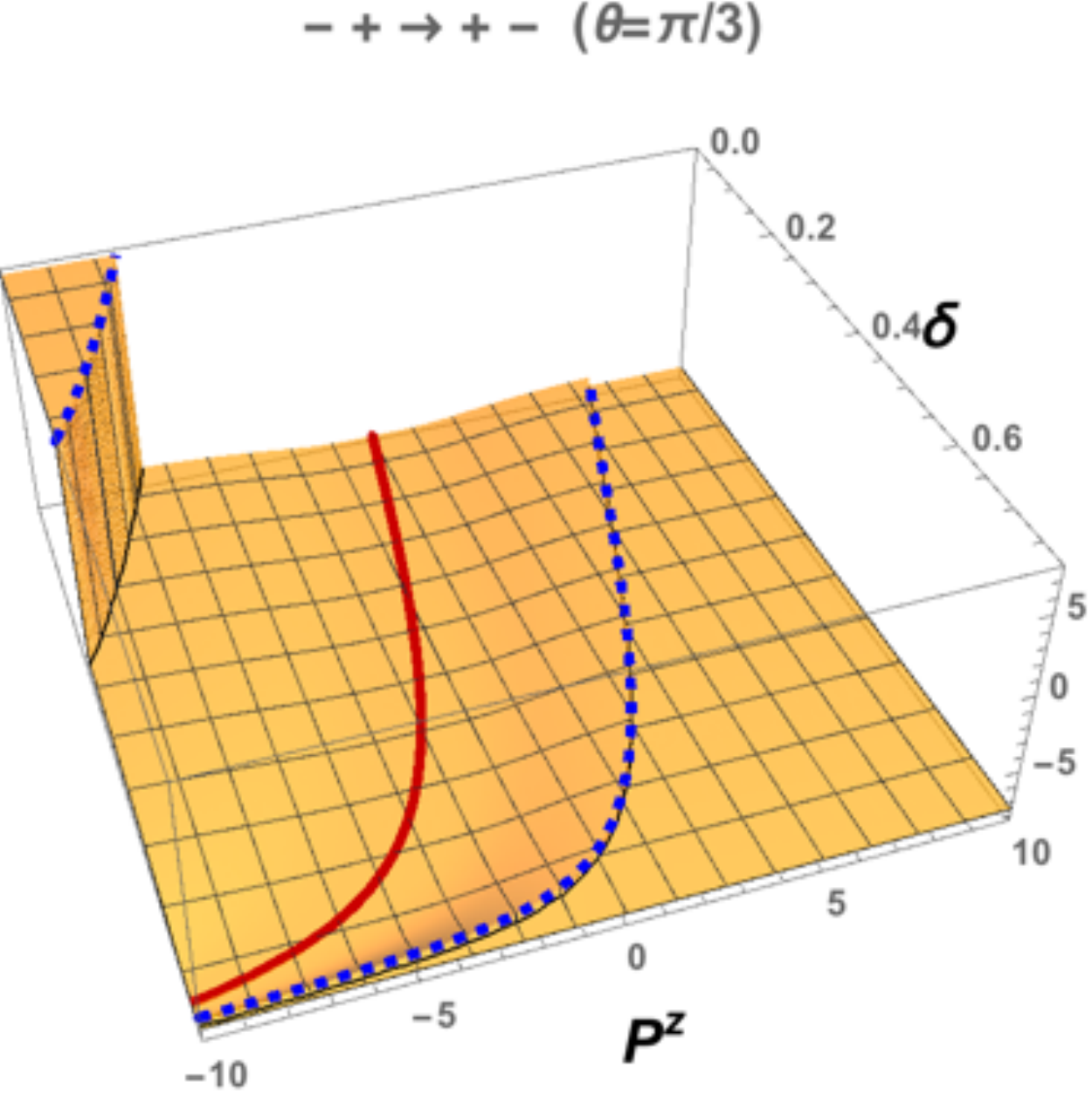}
    }
		%\quad
    \subfloat{
    \includegraphics[width=0.22\textwidth]{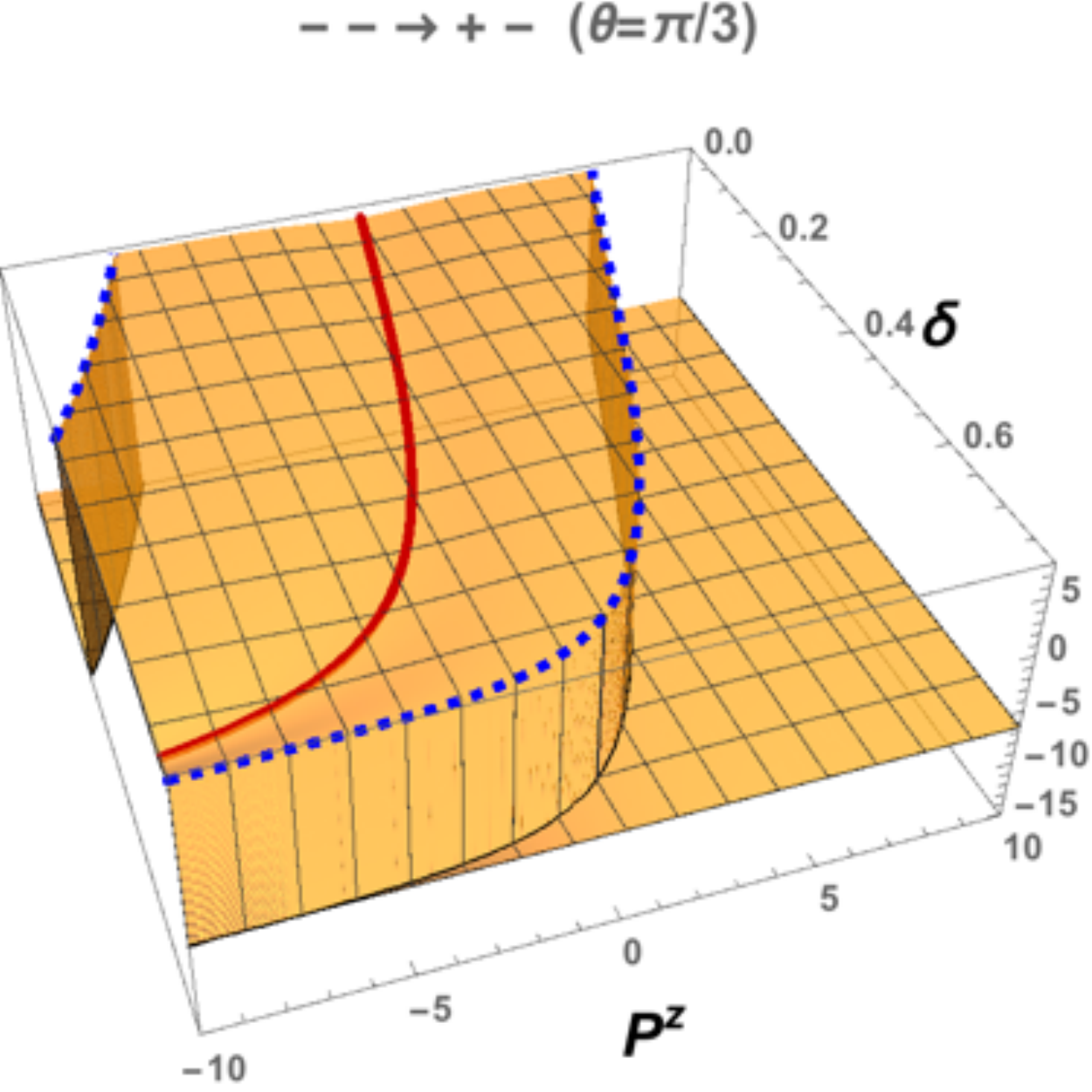}
    }
    \nonumber\\
    &\subfloat{
    \includegraphics[width=0.22\textwidth]{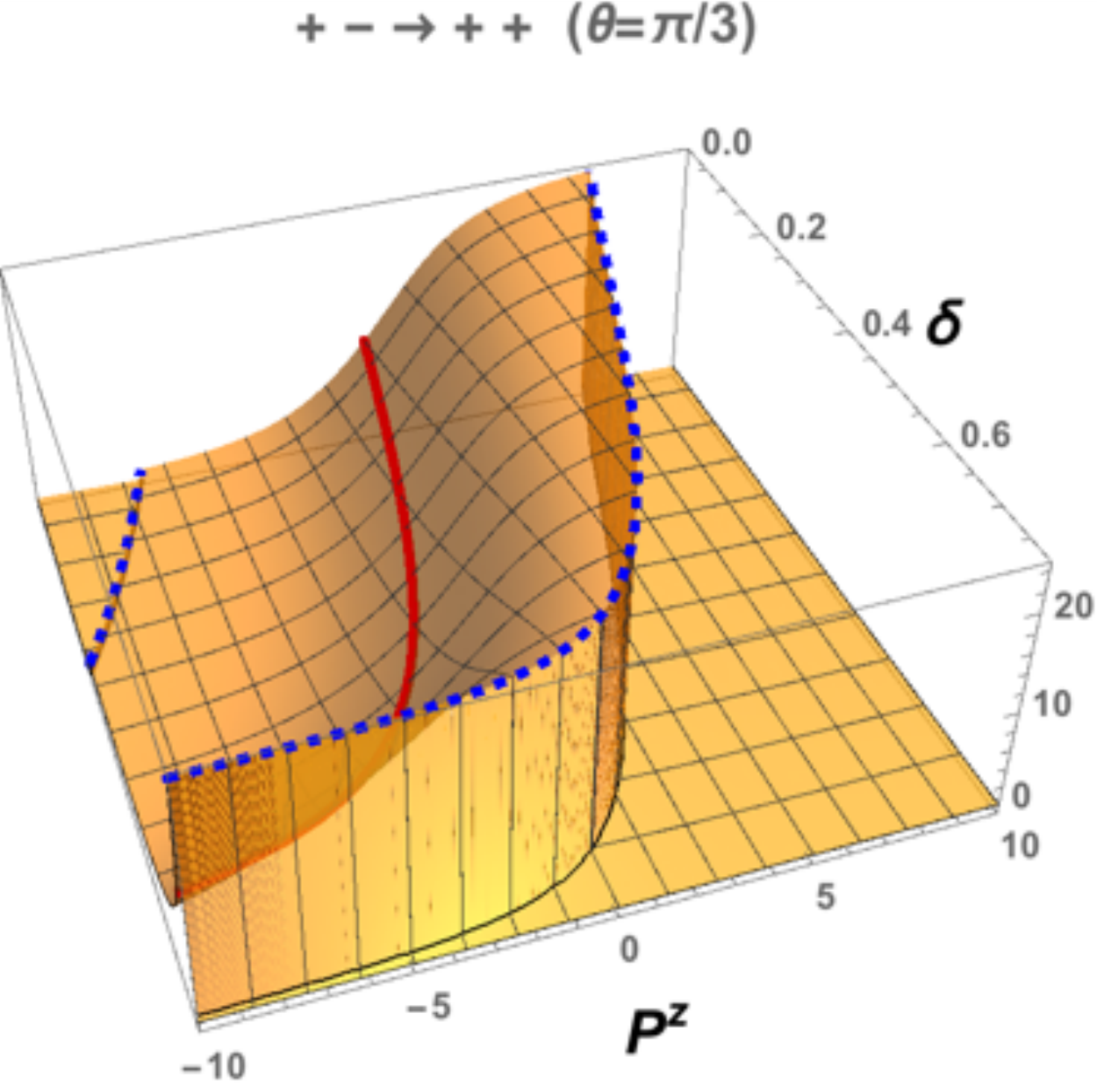}
    }
    %\quad
    \subfloat{
    \includegraphics[width=0.22\textwidth]{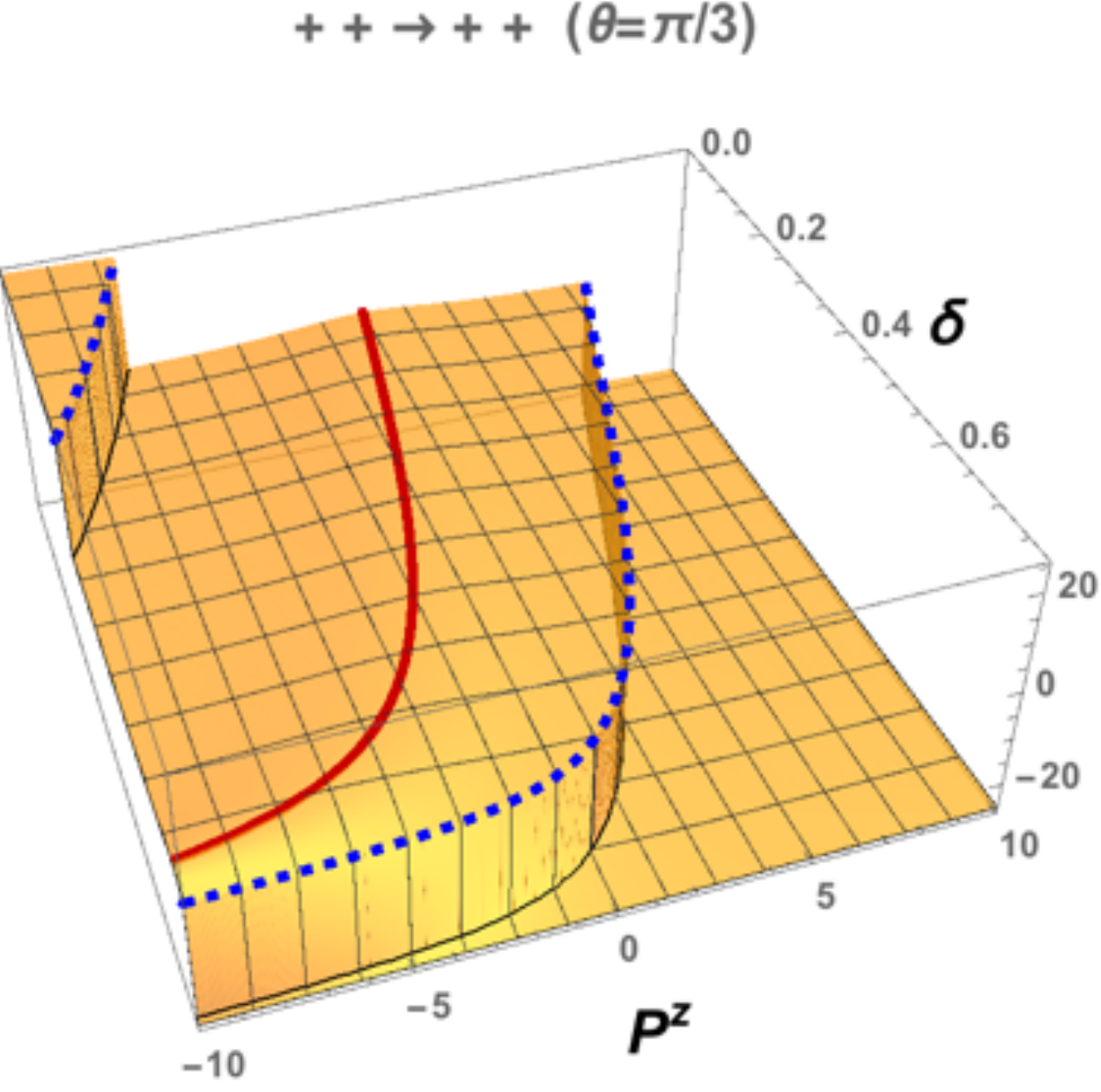}
    }
    %\quad
    \subfloat{
    \includegraphics[width=0.22\textwidth]{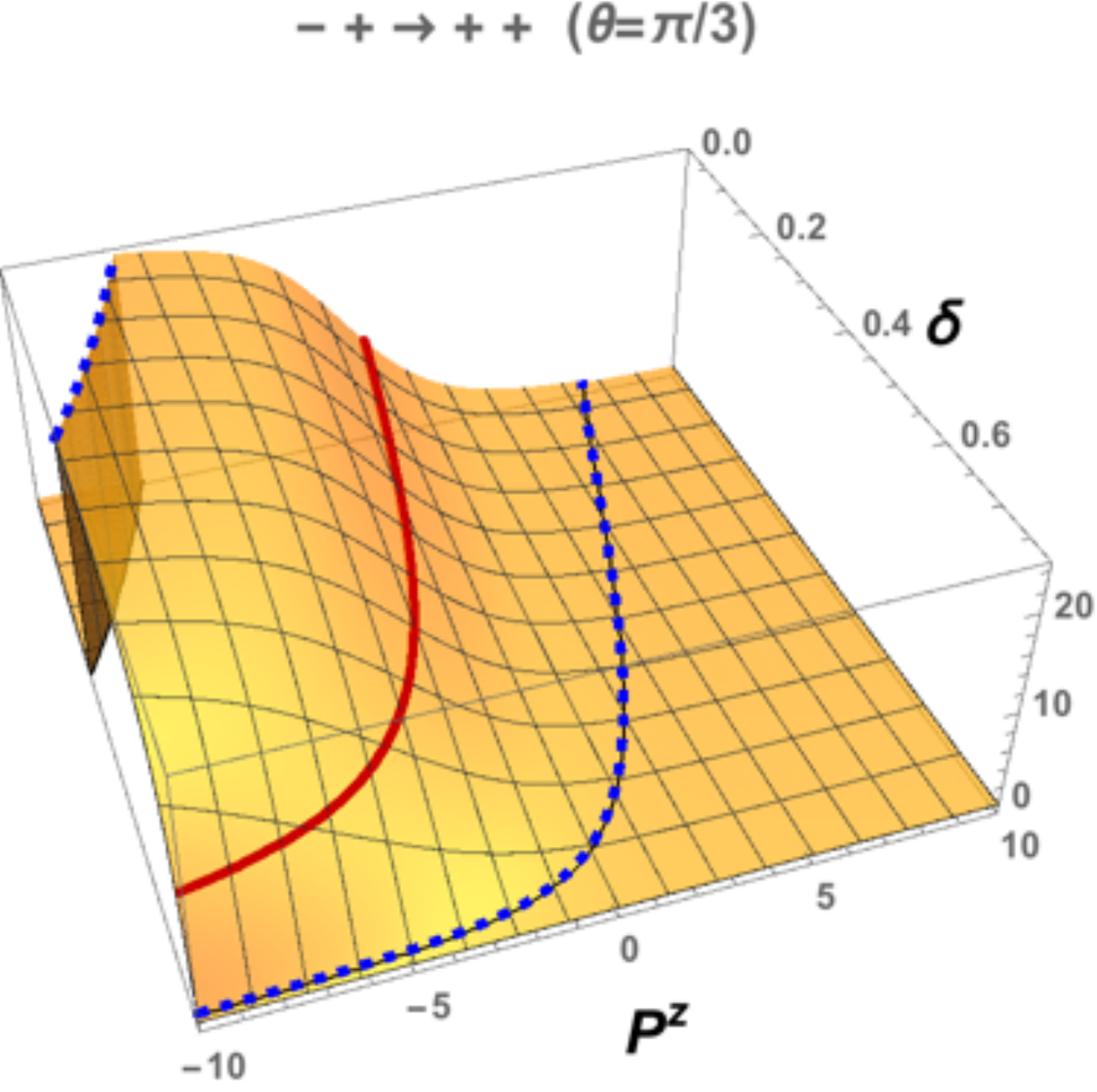}
    }
    %\quad
    \subfloat{
    \includegraphics[width=0.22\textwidth]{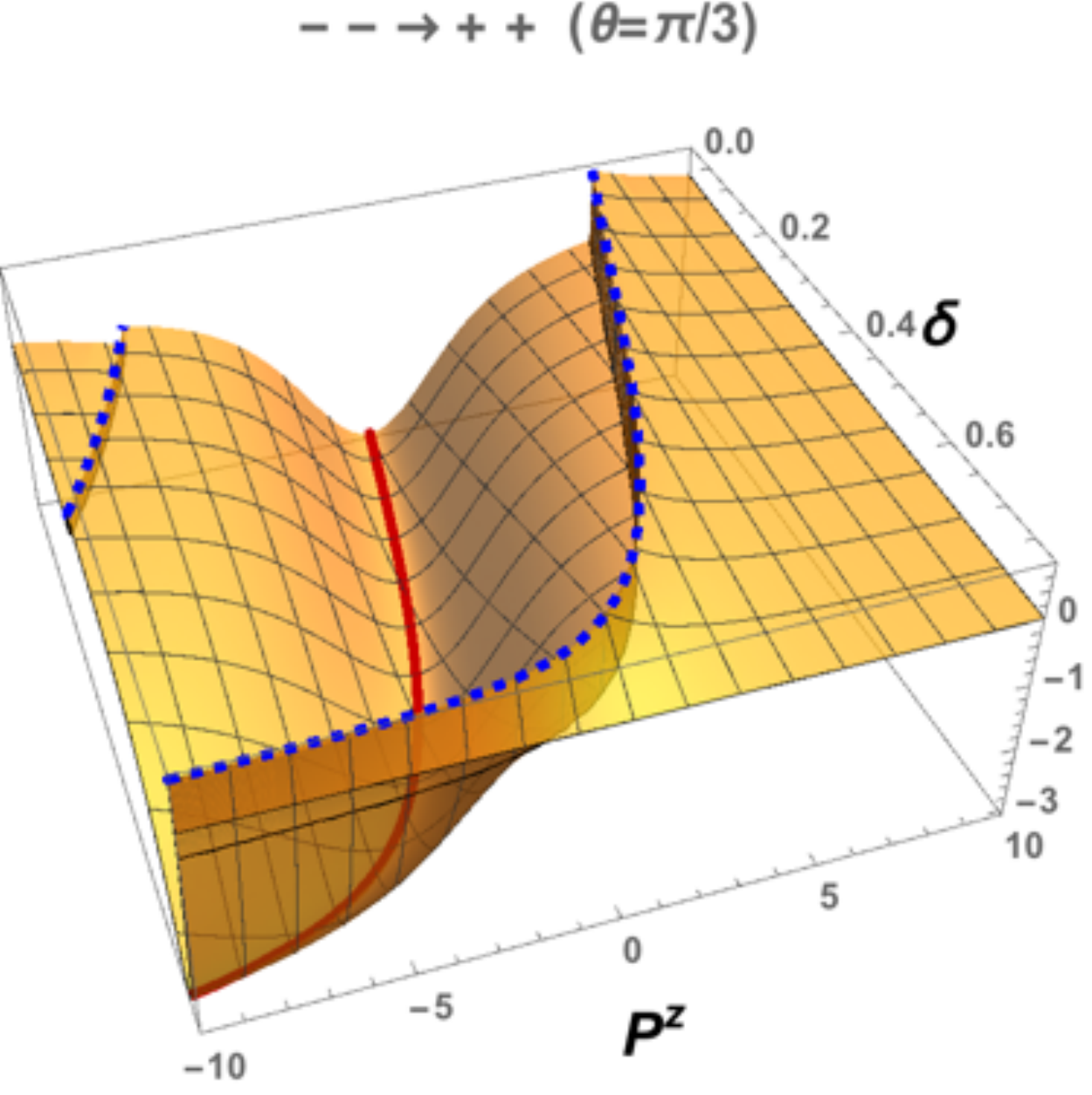}
    }
    \nonumber\\
    &\subfloat{
    \includegraphics[width=0.22\textwidth]{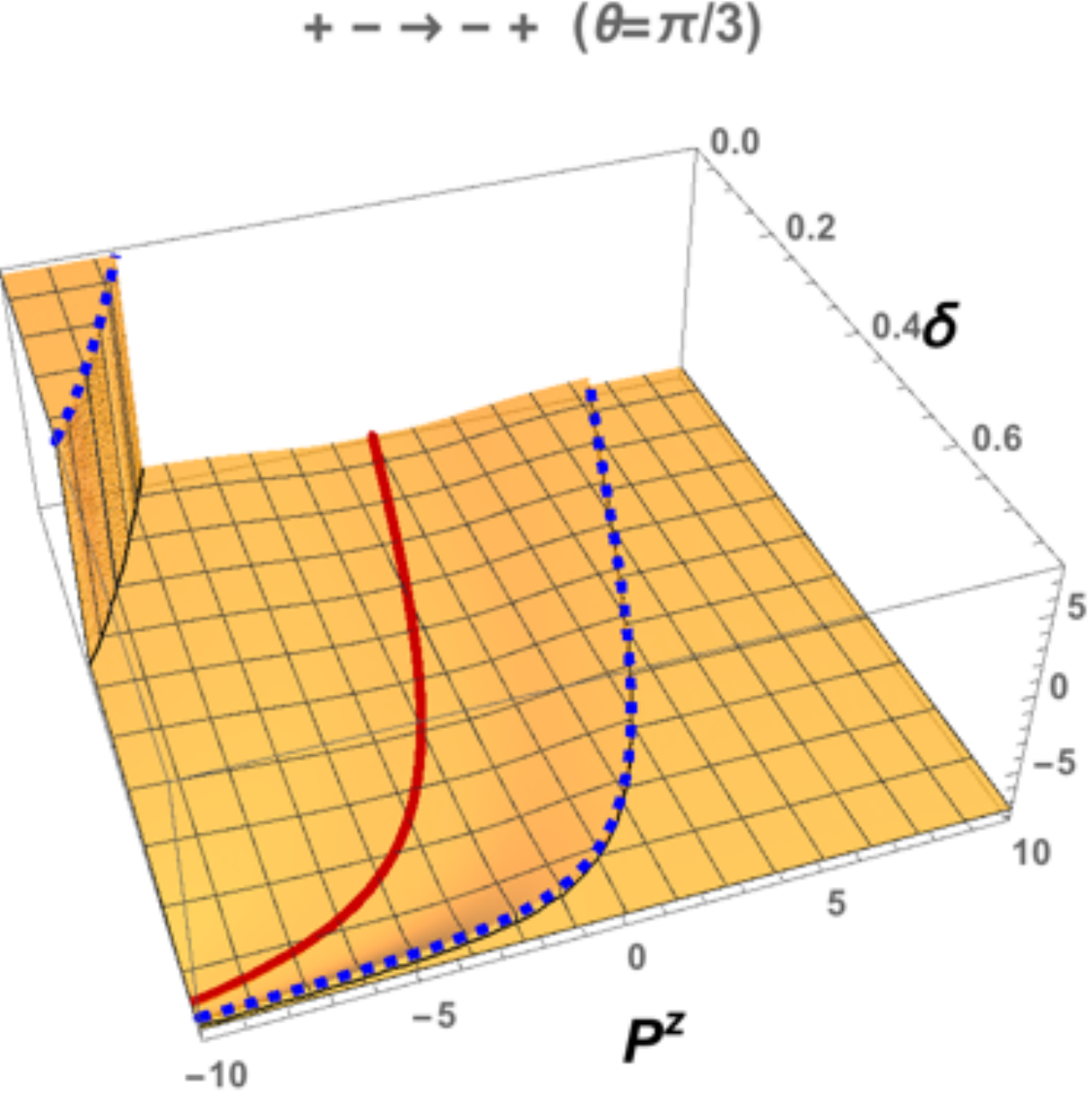}
    }
    %\quad
    \subfloat{
    \includegraphics[width=0.22\textwidth]{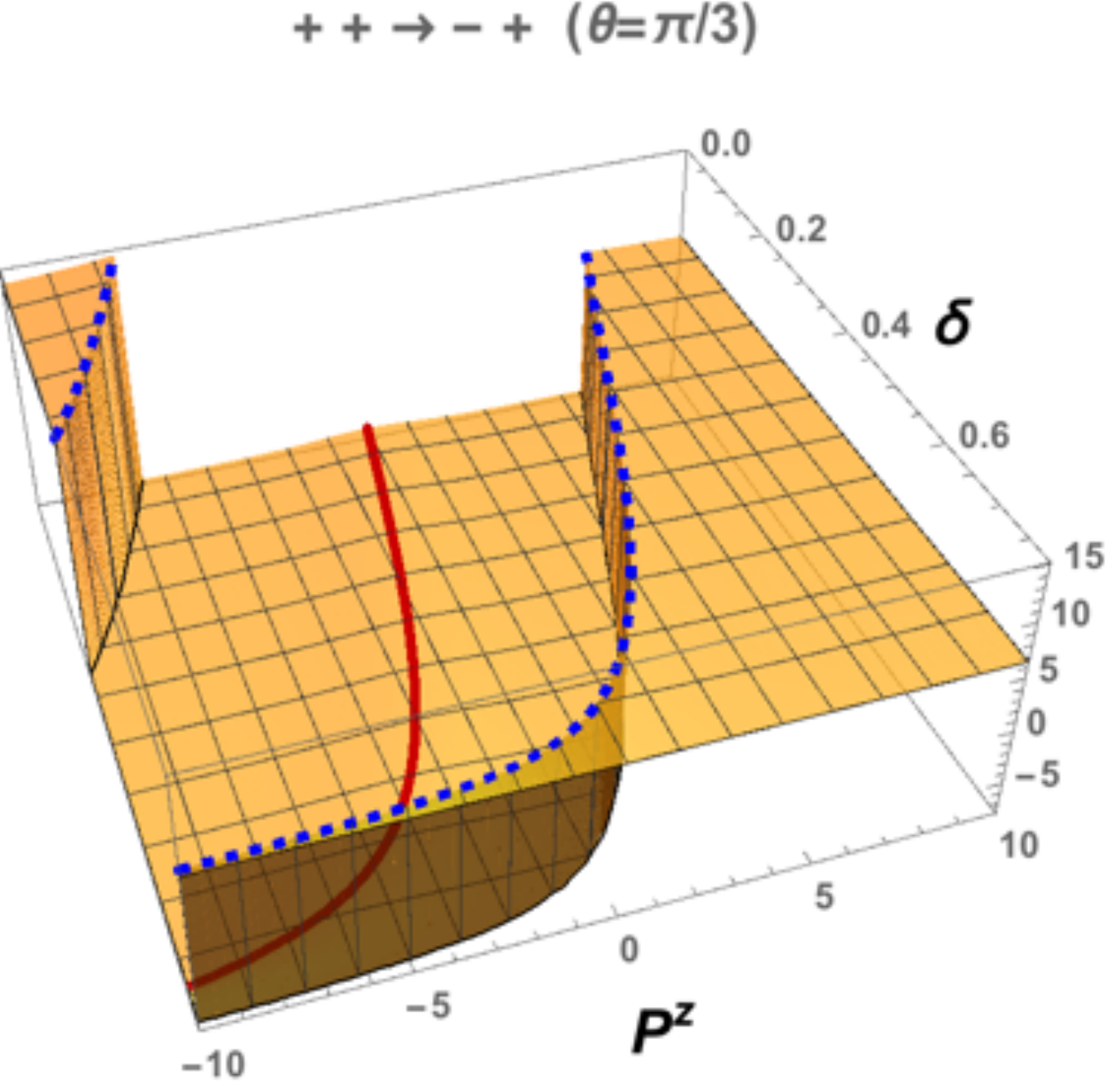}
    }
    %\quad
    \subfloat{
    \includegraphics[width=0.22\textwidth]{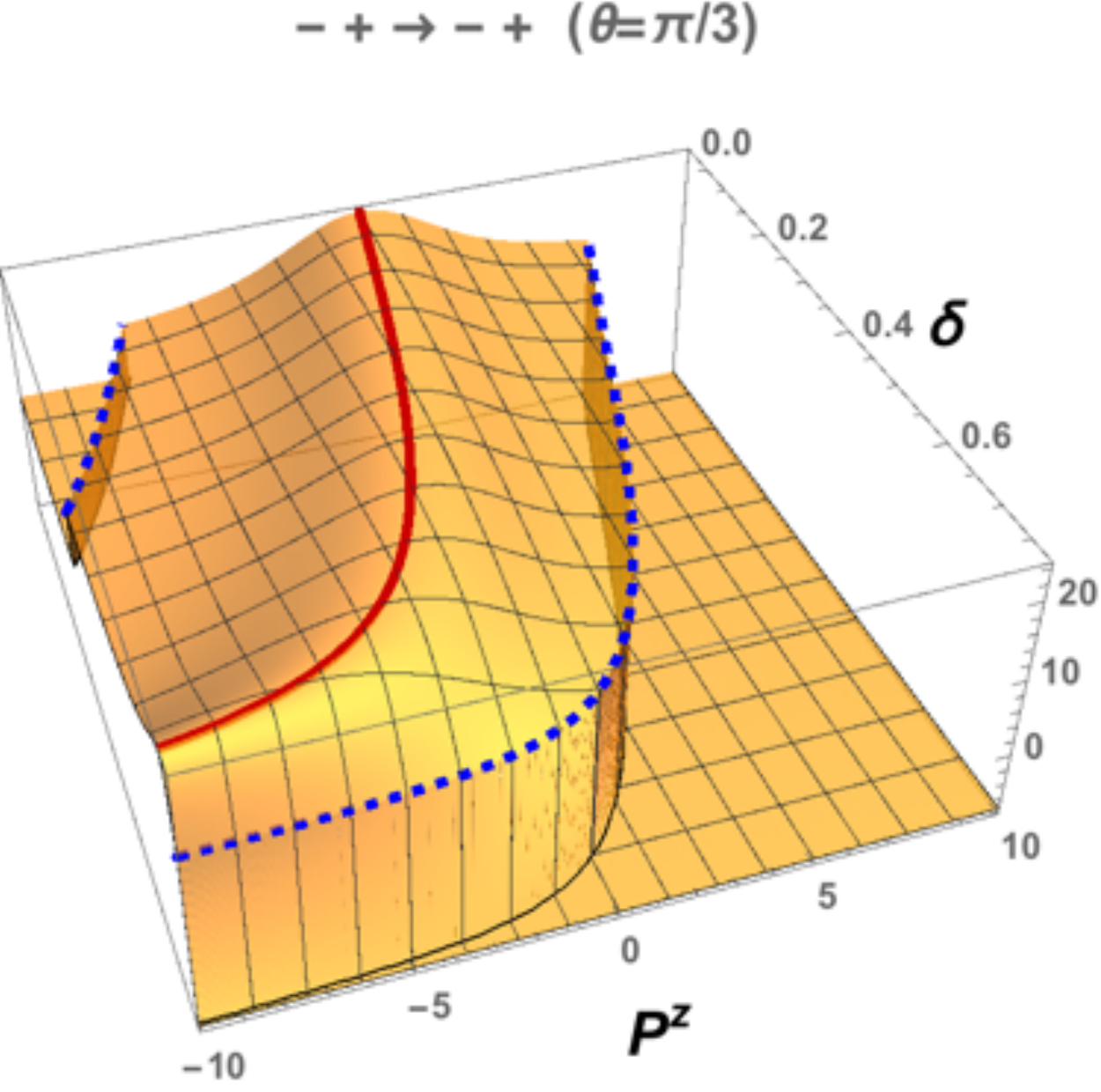}
    }
    %\quad
    \subfloat{
    \includegraphics[width=0.22\textwidth]{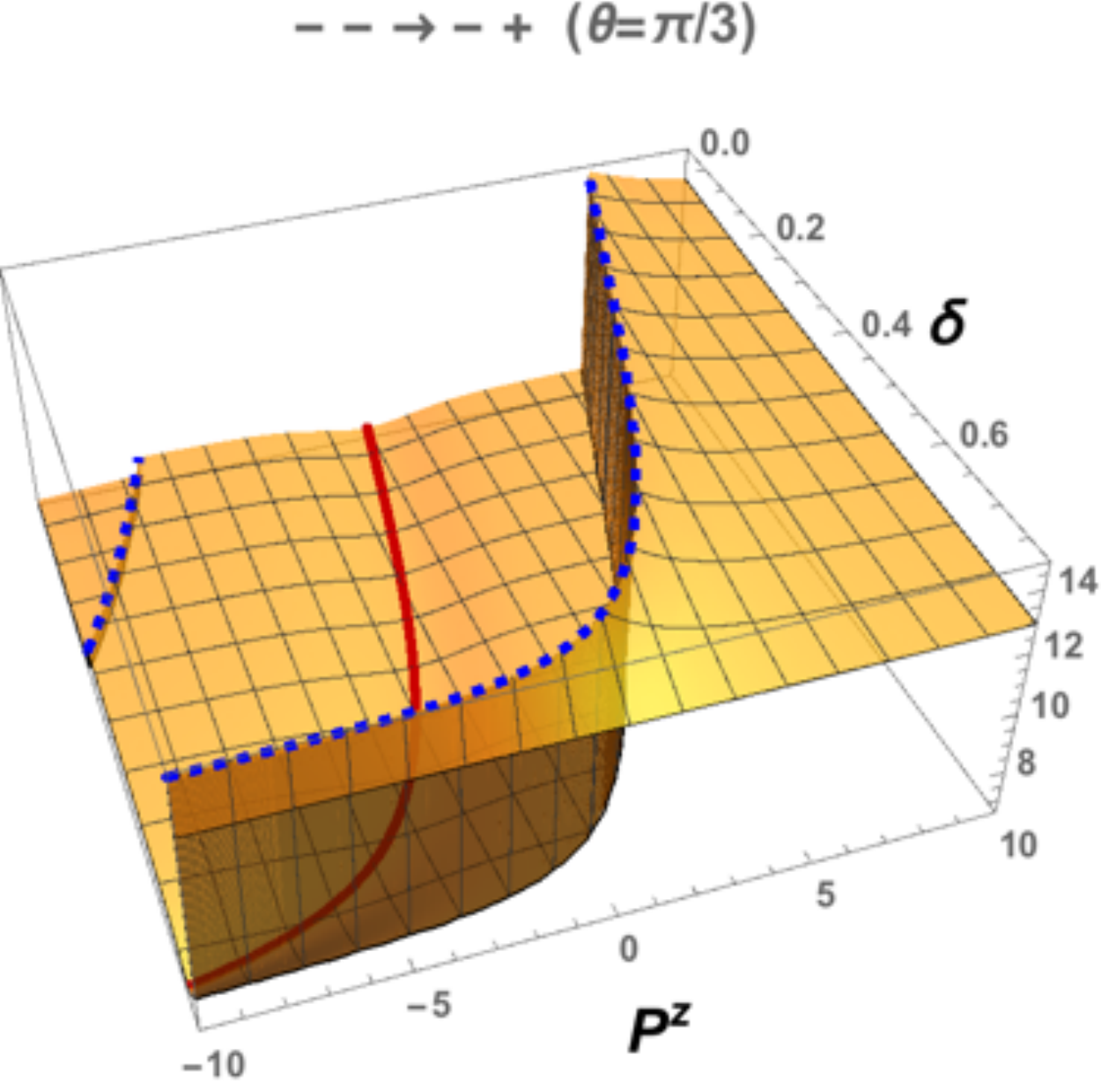}
    }
    \nonumber\\
    &\subfloat{
    \includegraphics[width=0.22\textwidth]{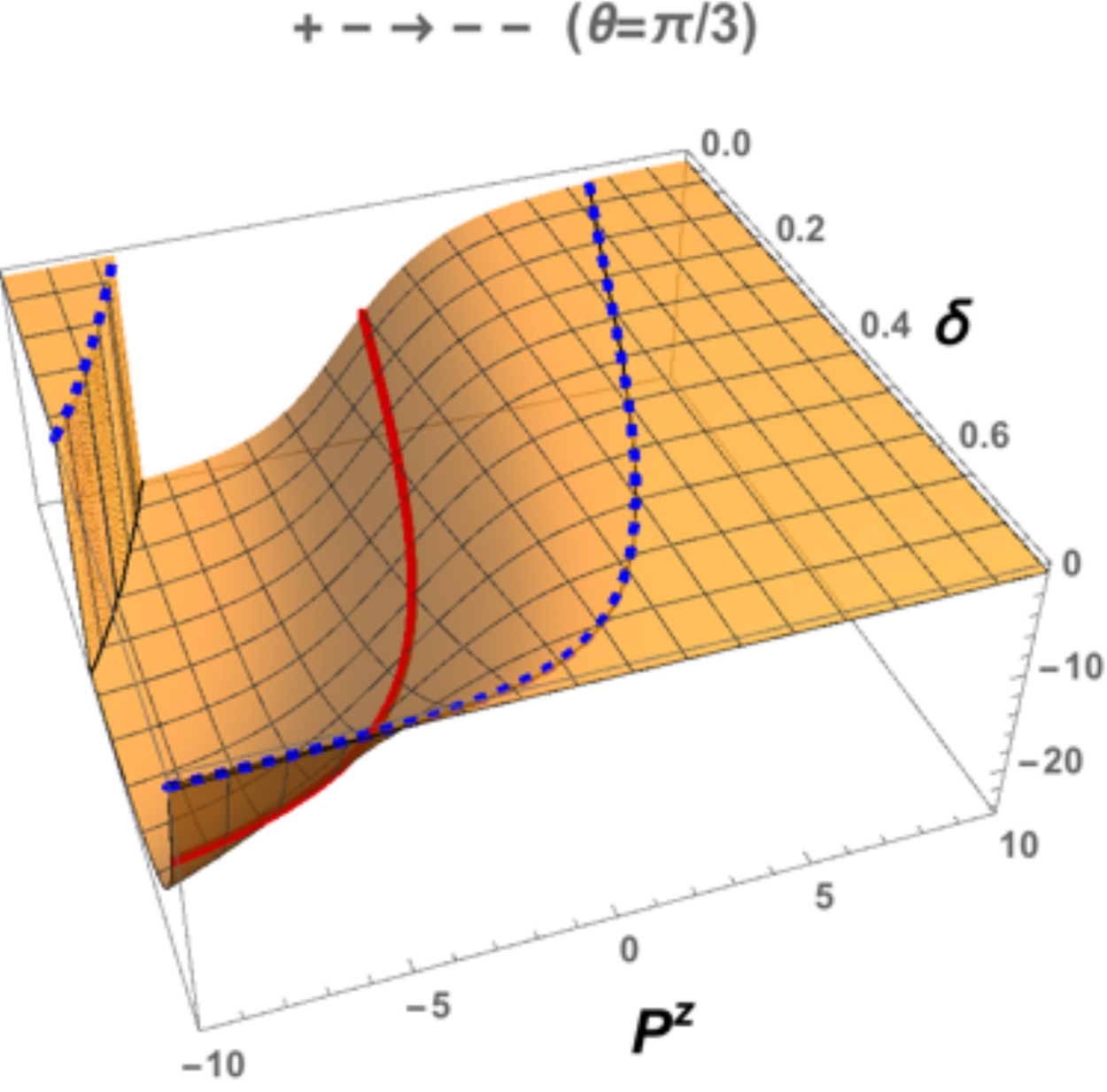}
    }
    %\quad
    \subfloat{
    \includegraphics[width=0.22\textwidth]{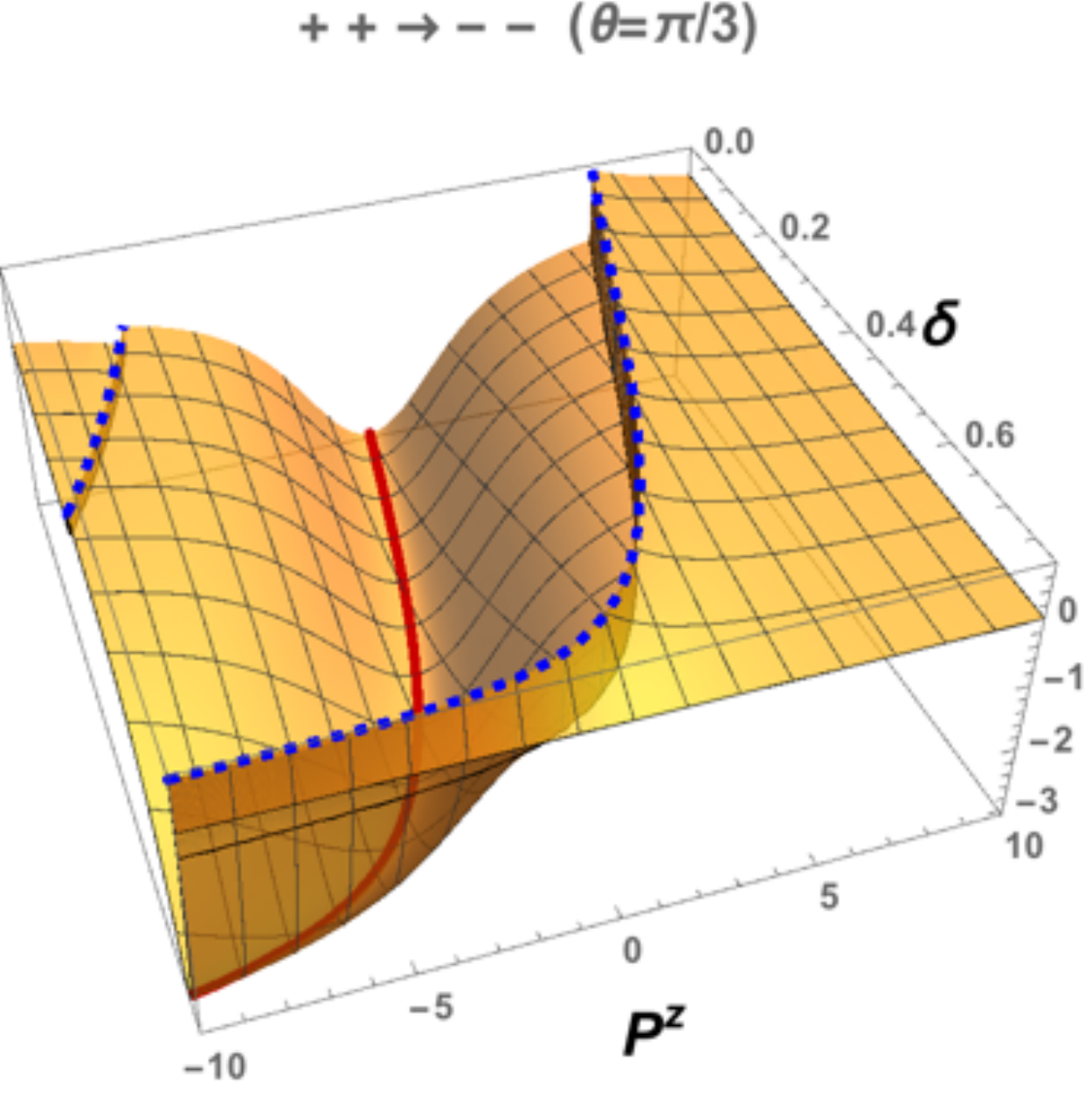}
    }
    %\quad
    \subfloat{
    \includegraphics[width=0.22\textwidth]{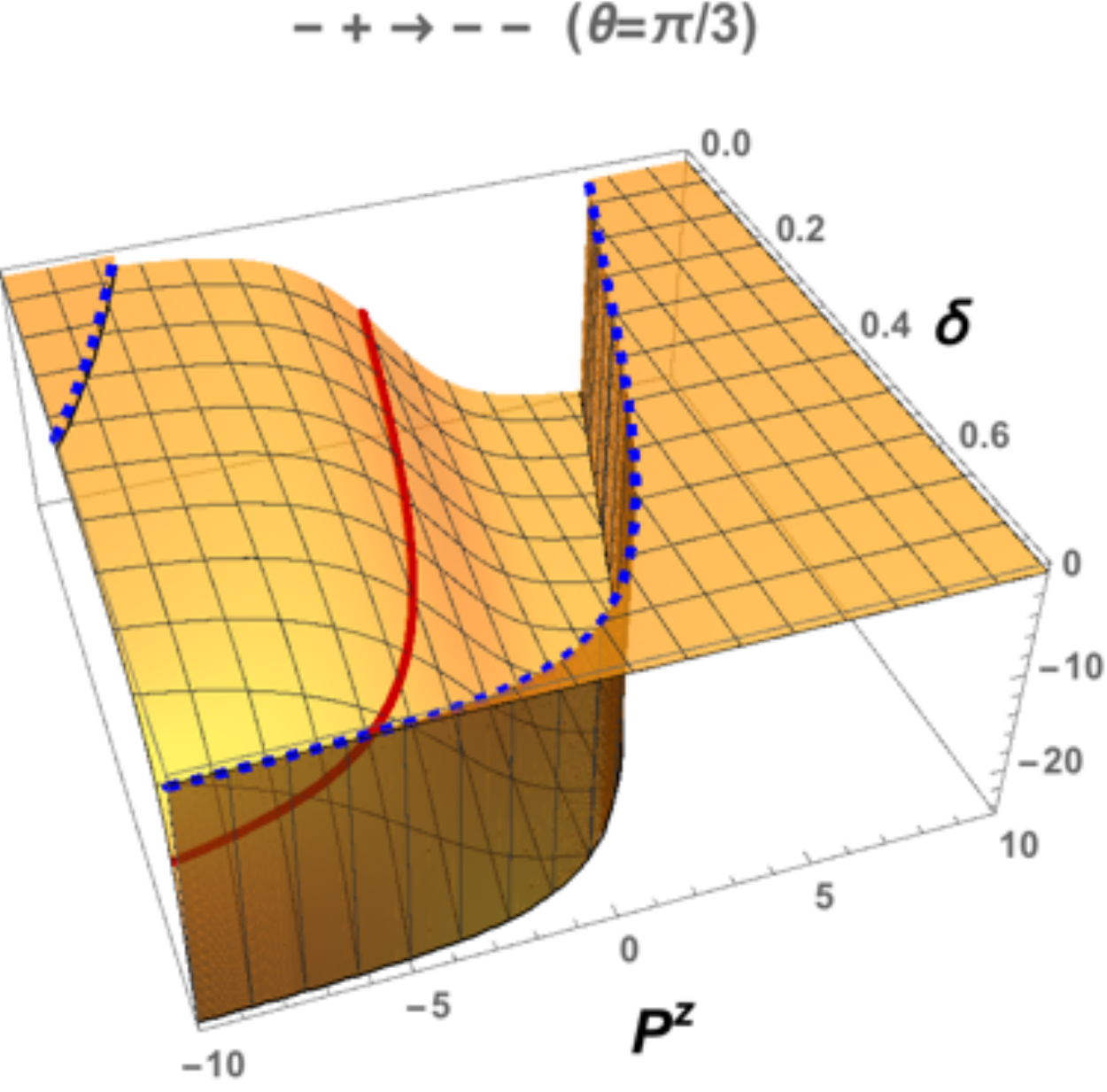}
    }
    %\quad
    \subfloat{
    \includegraphics[width=0.22\textwidth]{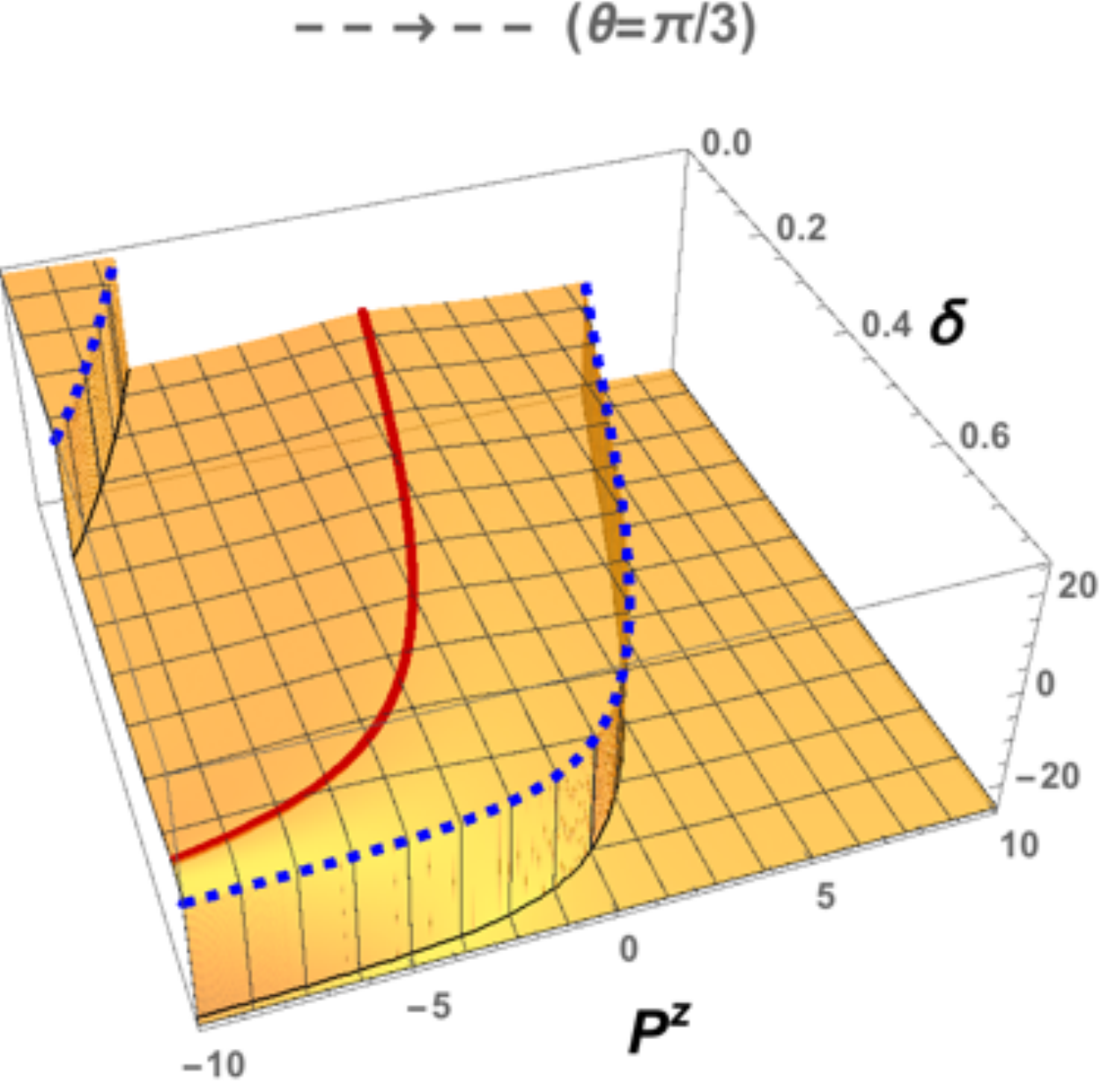}
    }
    % \nonumber\\
    % &\subfloat{
    % \includegraphics[width=0.22\textwidth]{MRL.pdf}
    % }
    % %\quad
    % \subfloat{
    % \includegraphics[width=0.22\textwidth]{MRR.pdf}
    % }
    % %\quad
    % \subfloat{
    % \includegraphics[width=0.22\textwidth]{MLR.pdf}
    % }
    % %\quad
    % \subfloat{
    % \includegraphics[width=0.22\textwidth]{MLL.pdf}
    % }
    \nonumber
 \end{align}
  \caption{\label{fig:Annihilation_Helicity_Amplitudes} (Color online) Fermion annihilation amplitudes (with the factor $-e^{2}/q^{2}$ dropped) for 16 different spin configurations with the center of mass energy $M=4$ GeV and the annihilation angle $\theta=\pi/3$.
  The masses of the initial and final particles are $m_{\text{ini}}=1$ GeV and $m_{\text{final}}=1.5$ GeV.}
\end{figure*}

\begin{figure*}[!hbp]
 \begin{align}
    &\subfloat{
    \includegraphics[width=0.22\textwidth]{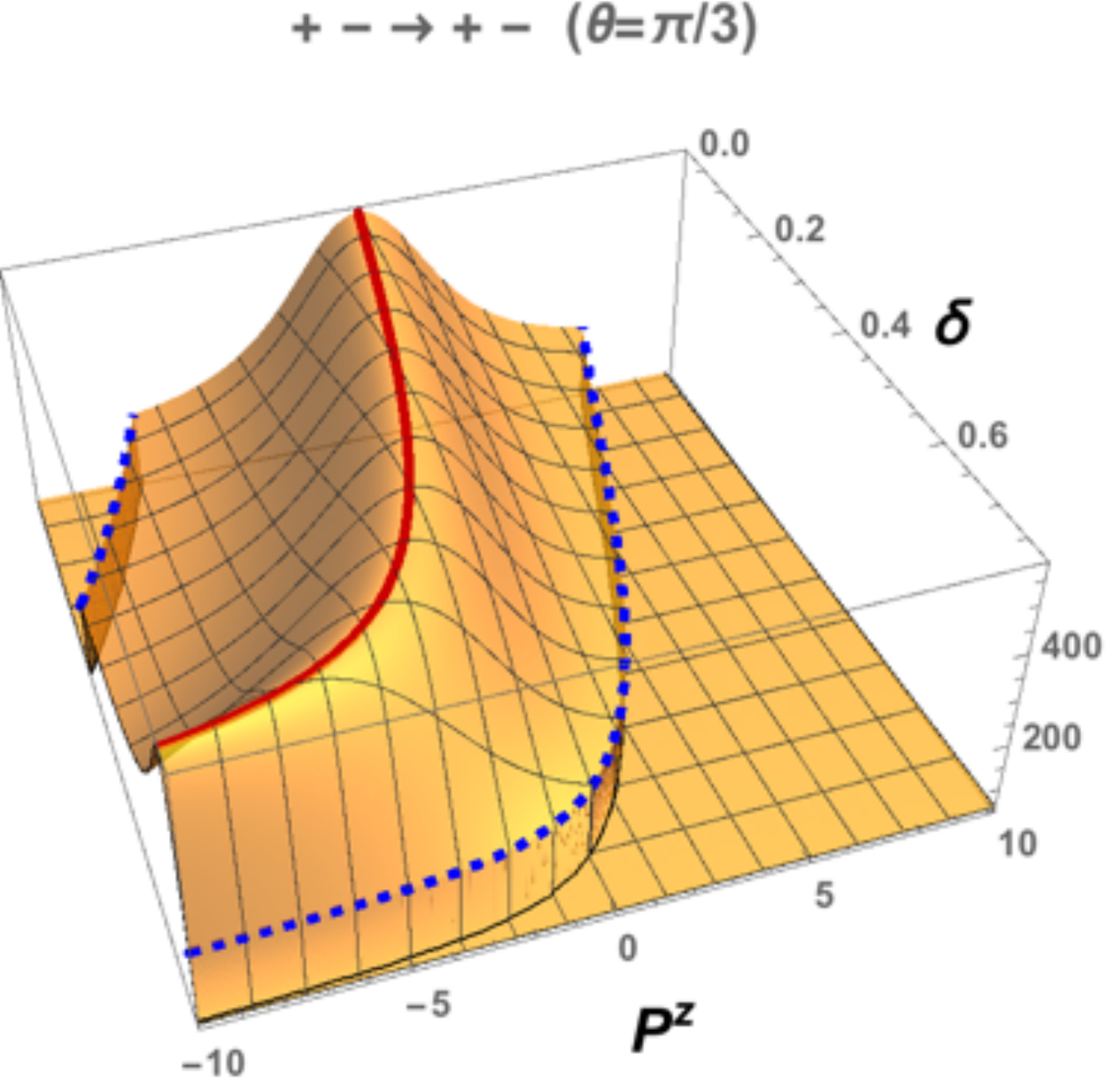}
    }
    %\quad
    \subfloat{
    \includegraphics[width=0.22\textwidth]{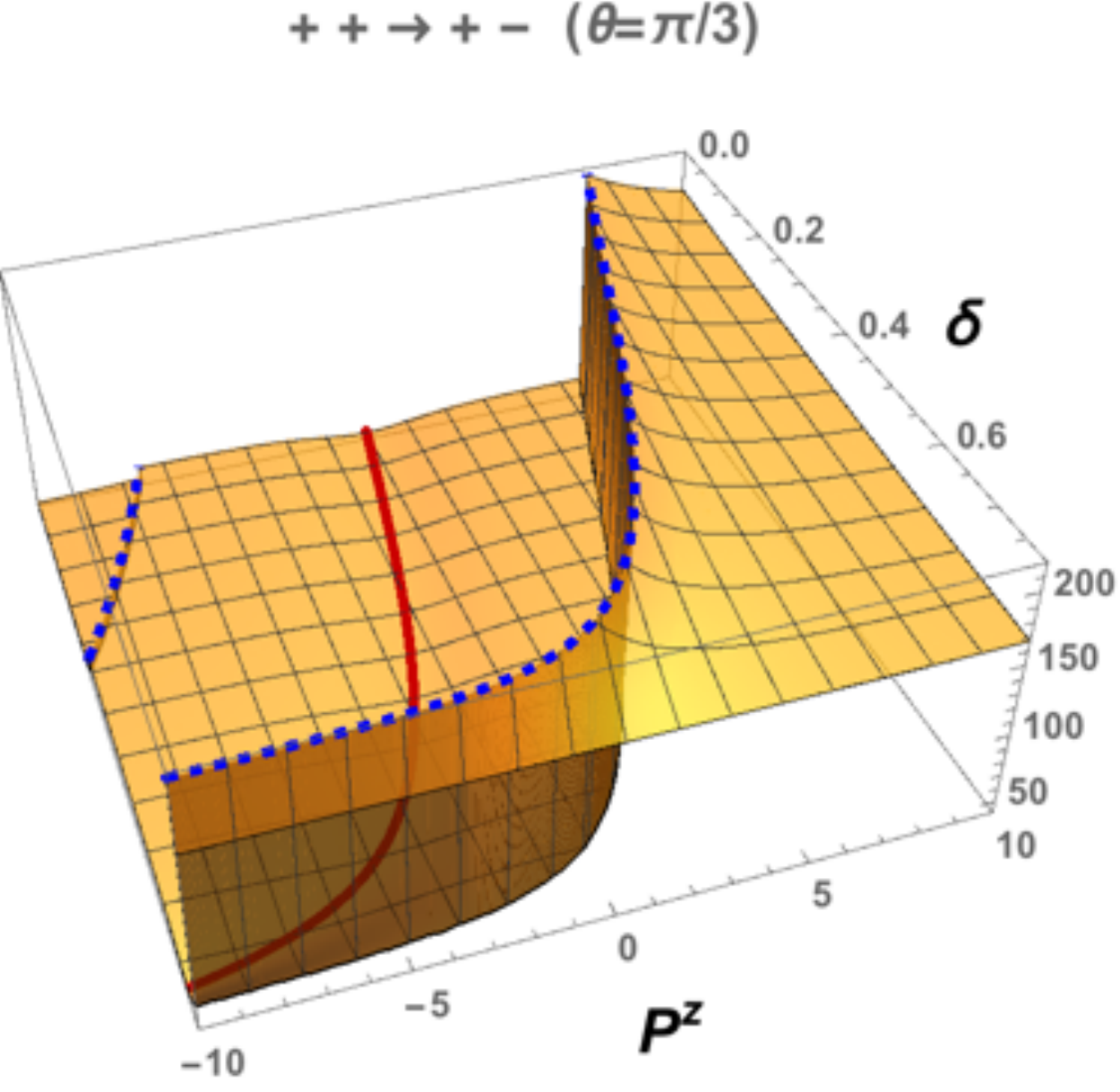}
    }
    %\quad
    \subfloat{
    \includegraphics[width=0.22\textwidth]{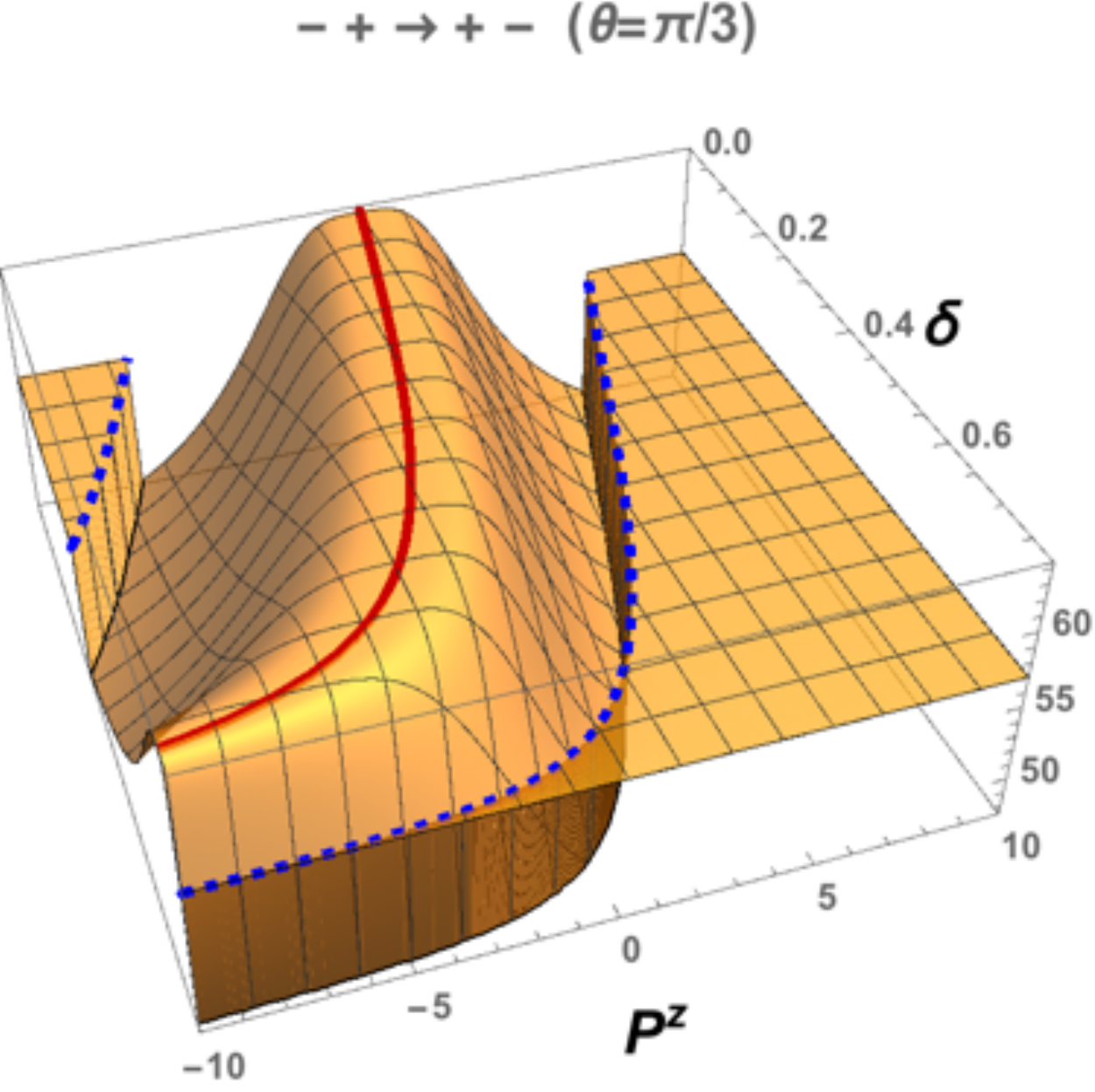}
    }
		%\quad
    \subfloat{
    \includegraphics[width=0.22\textwidth]{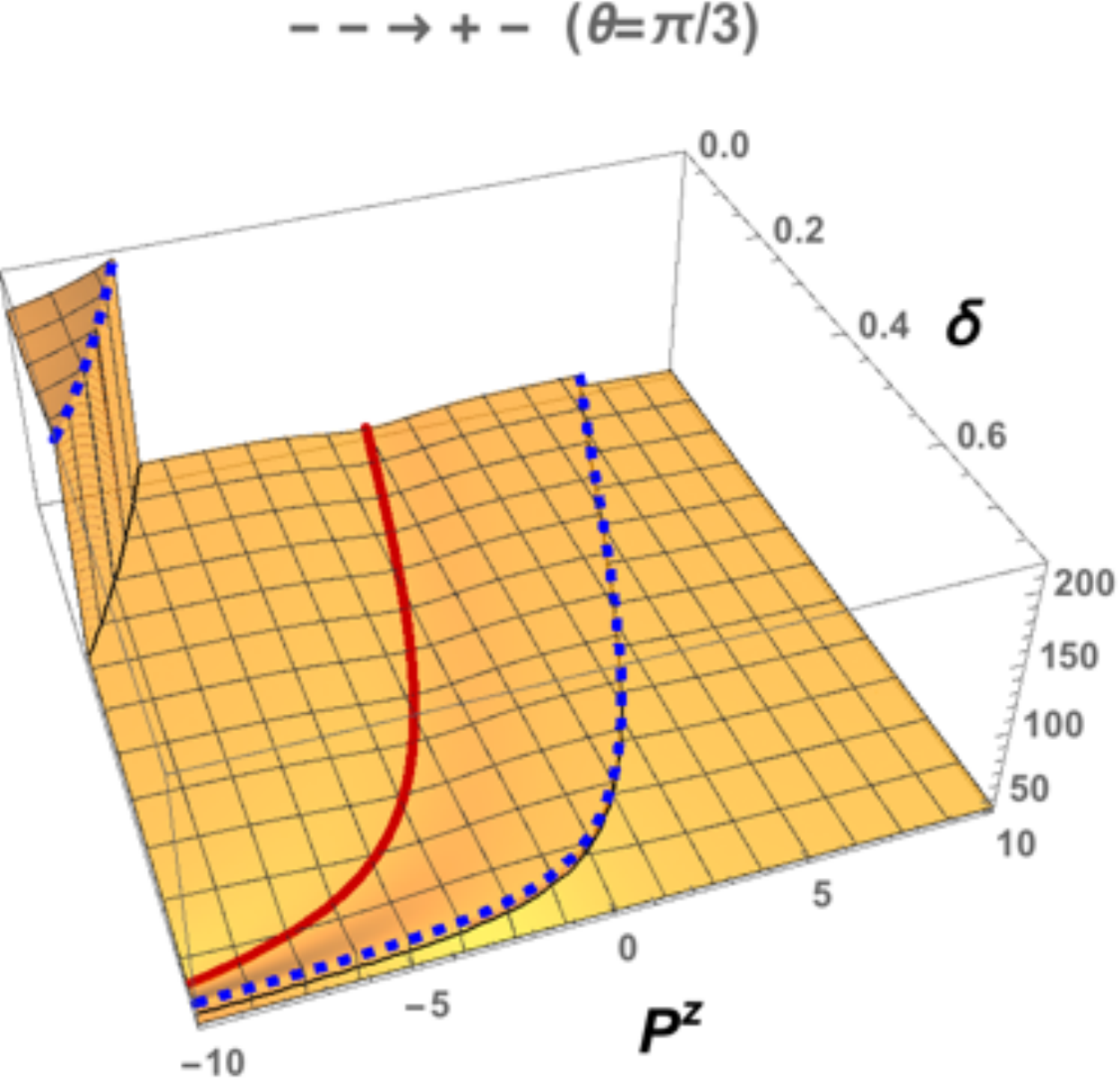}
    }
    \nonumber\\
    &\subfloat{
    \includegraphics[width=0.22\textwidth]{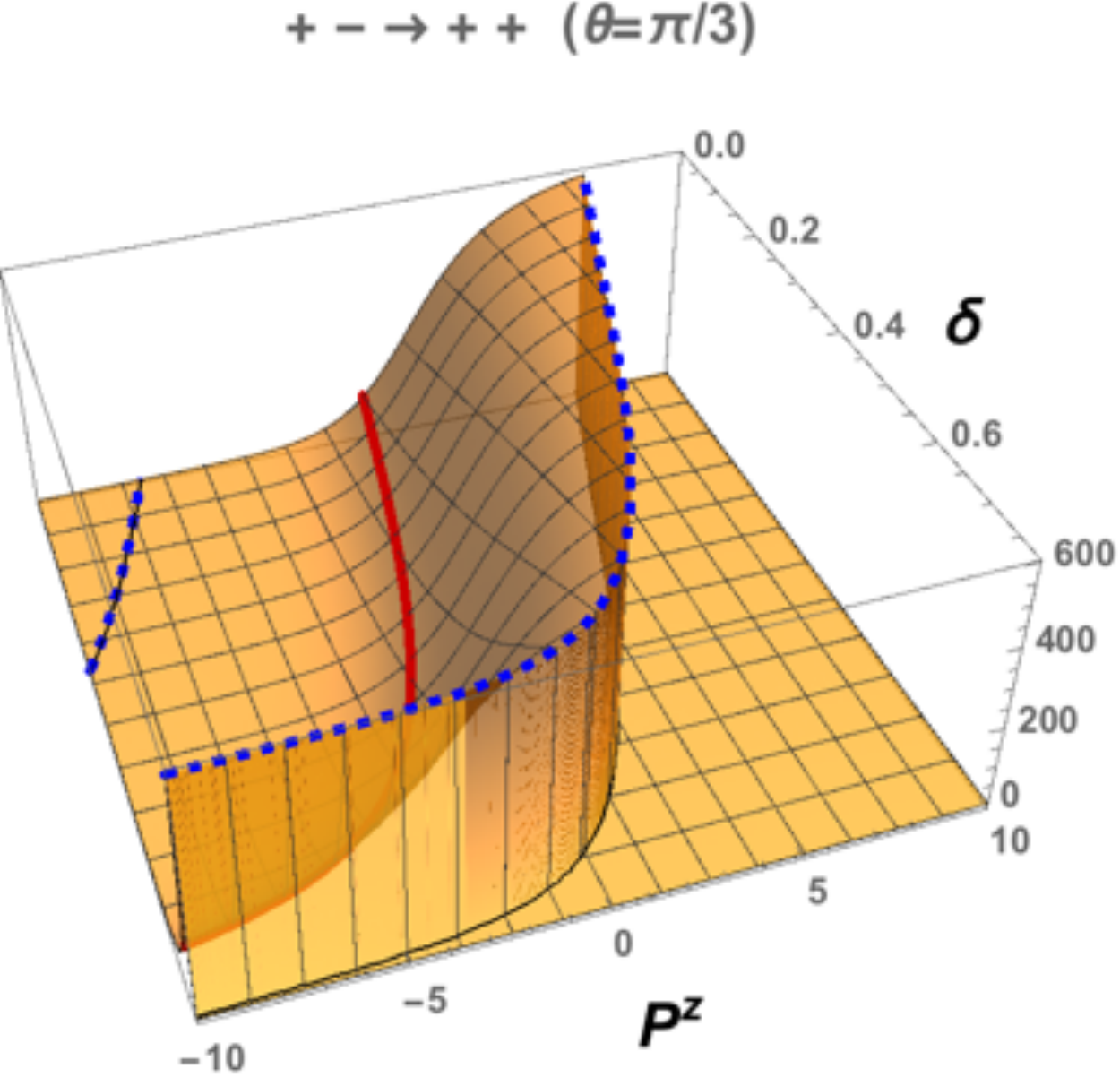}
    }
    %\quad
    \subfloat{
    \includegraphics[width=0.22\textwidth]{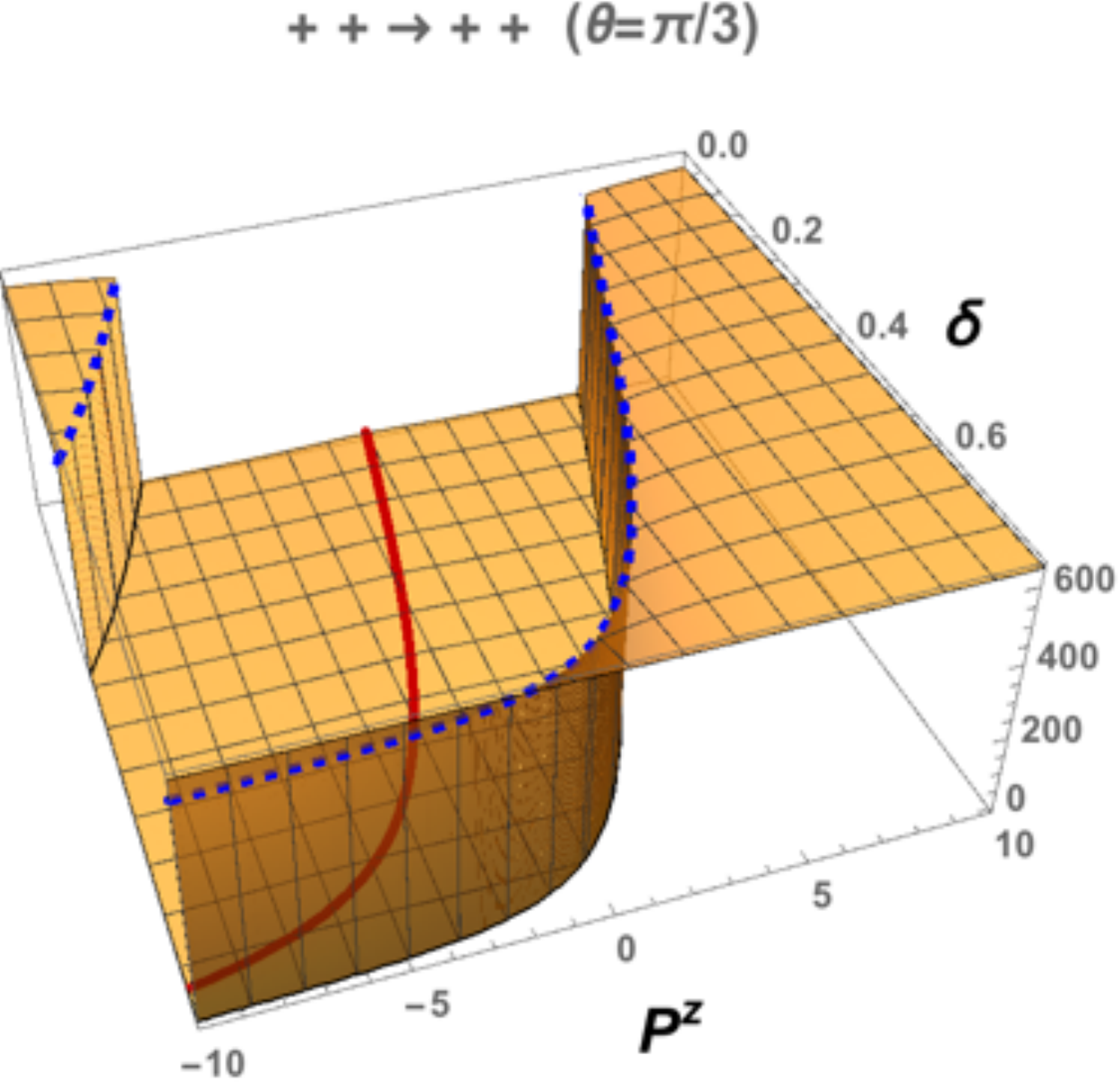}
    }
    %\quad
    \subfloat{
    \includegraphics[width=0.22\textwidth]{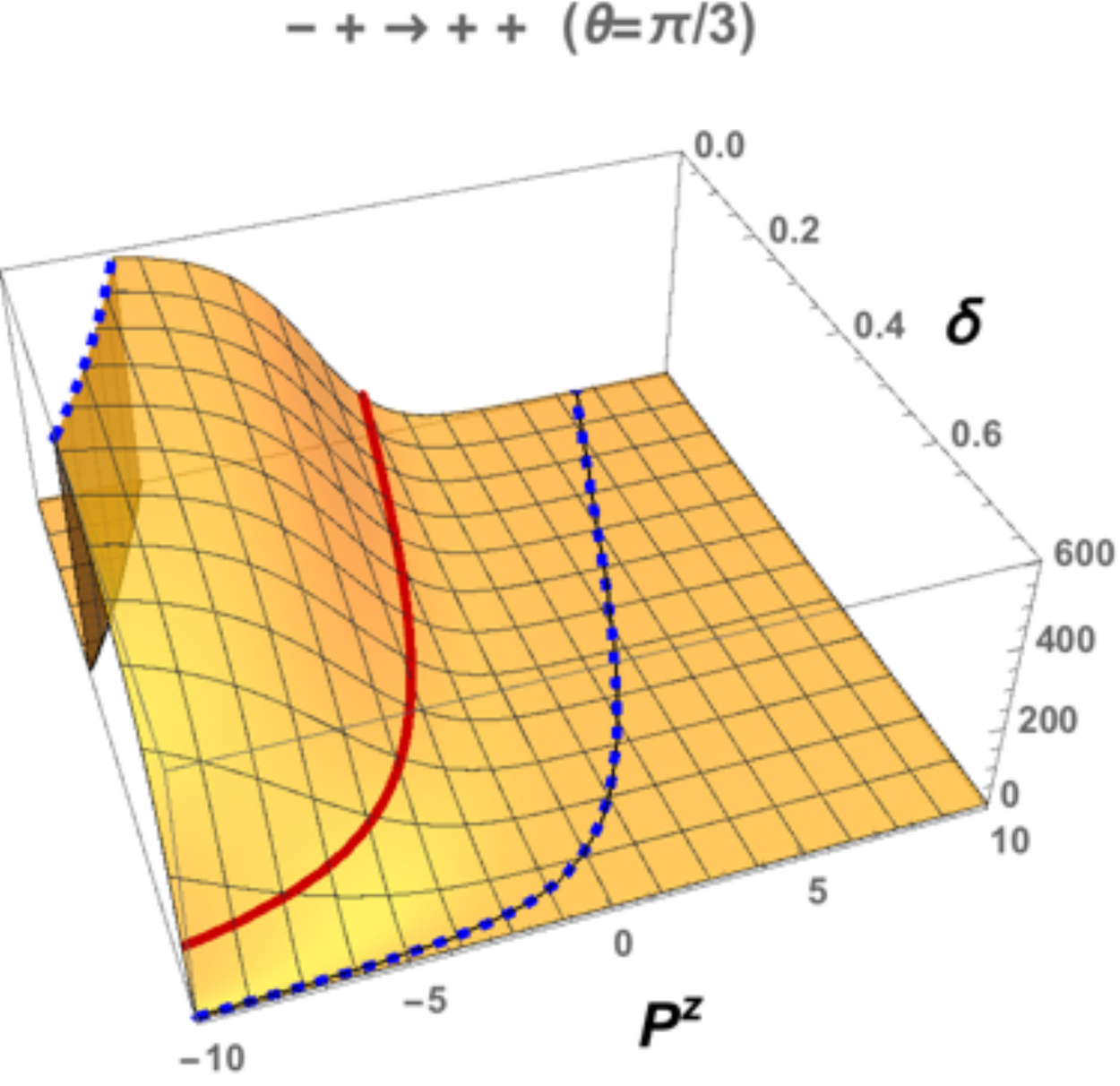}
    }
    %\quad
    \subfloat{
    \includegraphics[width=0.22\textwidth]{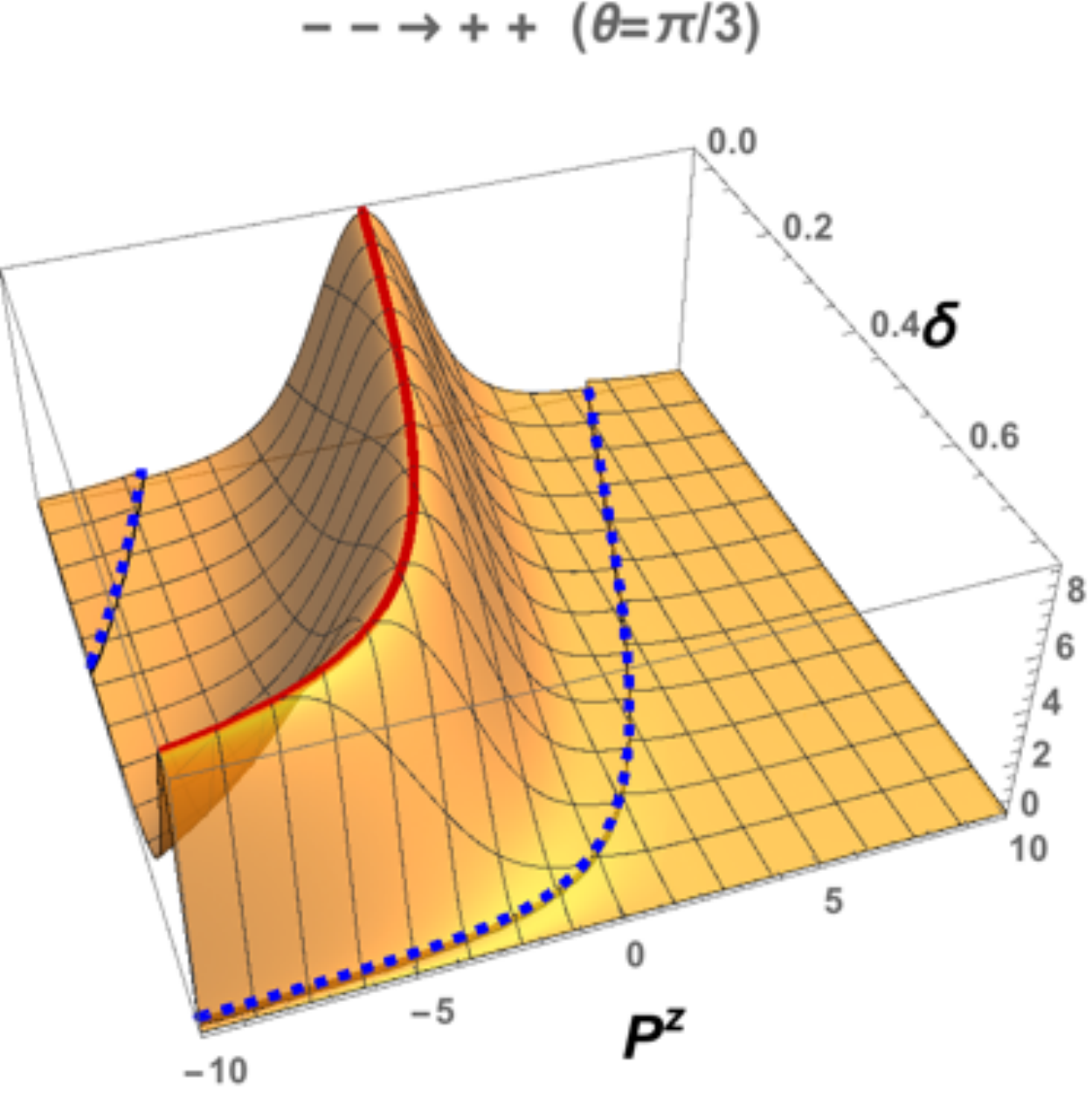}
    }
    \nonumber\\
    &\subfloat{
    \includegraphics[width=0.22\textwidth]{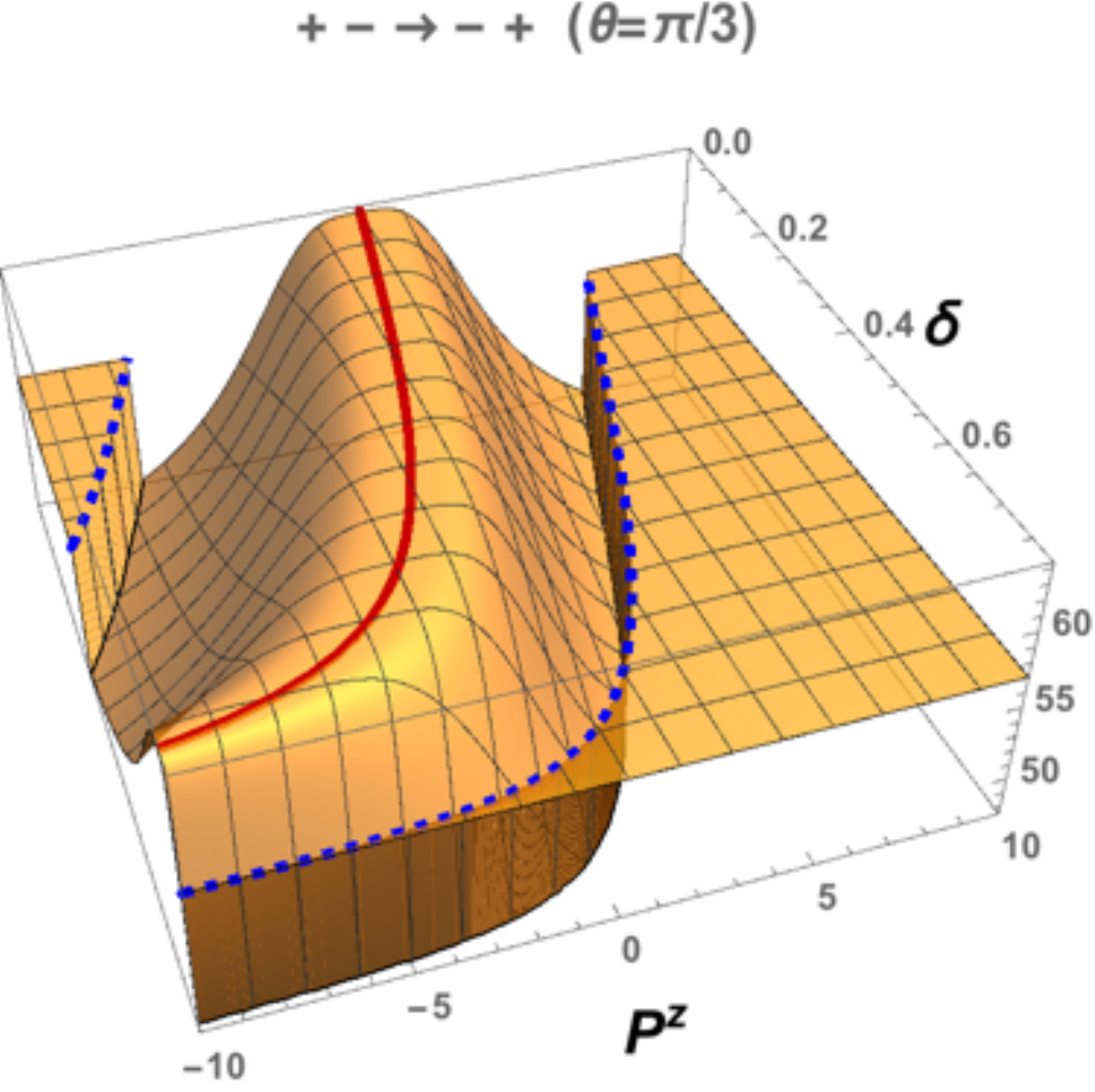}
    }
    %\quad
    \subfloat{
    \includegraphics[width=0.22\textwidth]{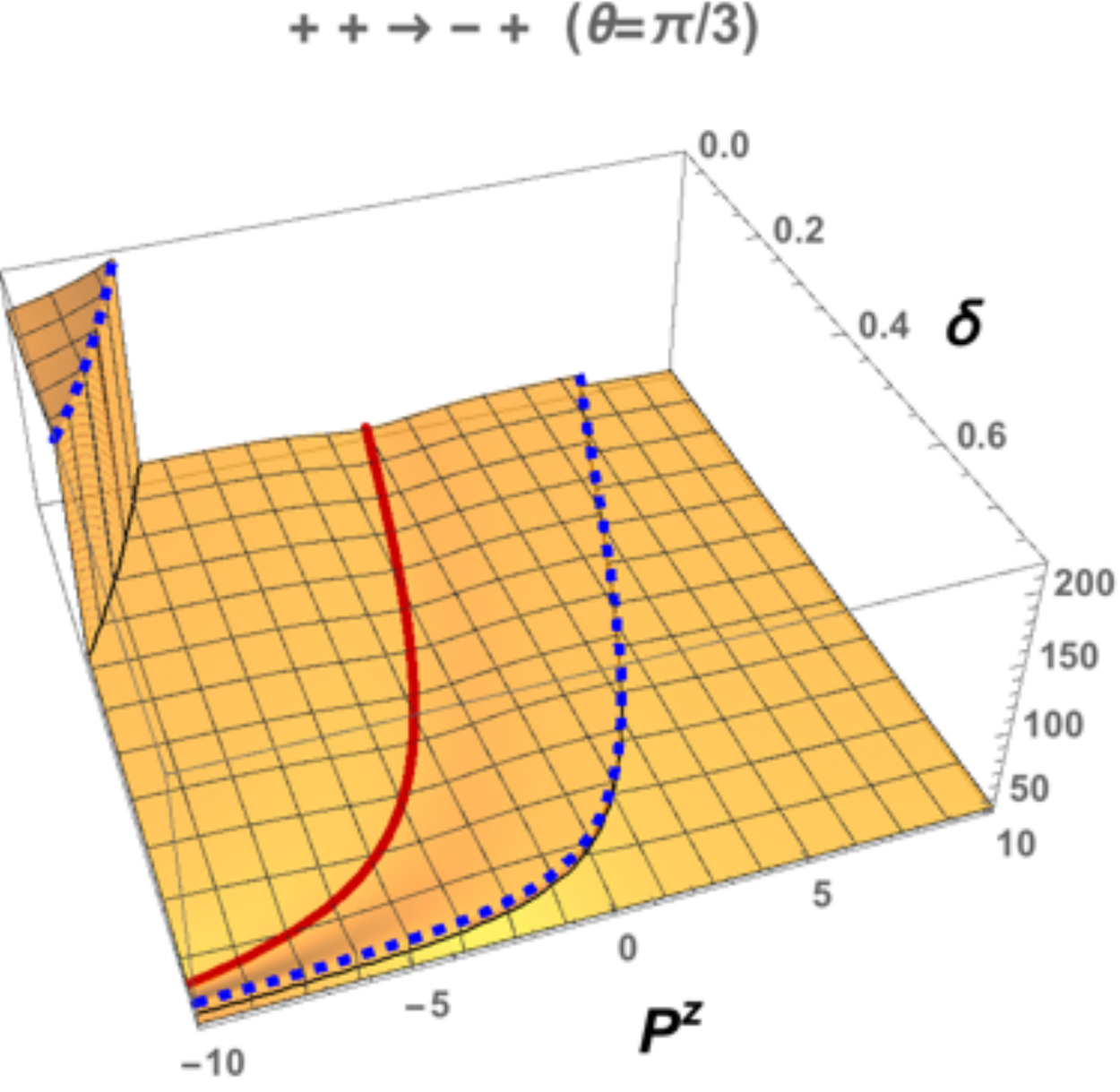}
    }
    %\quad
    \subfloat{
    \includegraphics[width=0.22\textwidth]{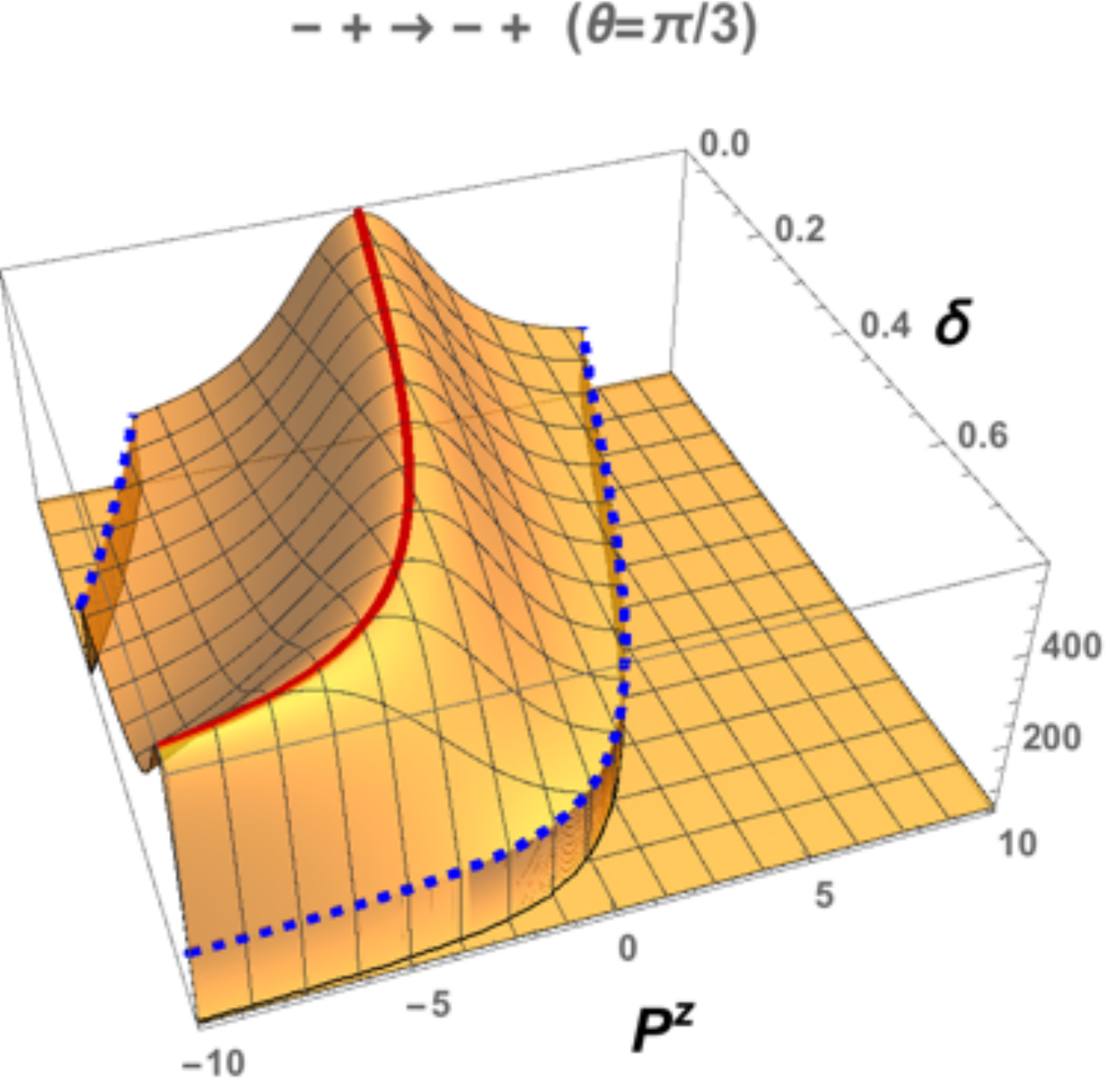}
    }
    %\quad
    \subfloat{
    \includegraphics[width=0.22\textwidth]{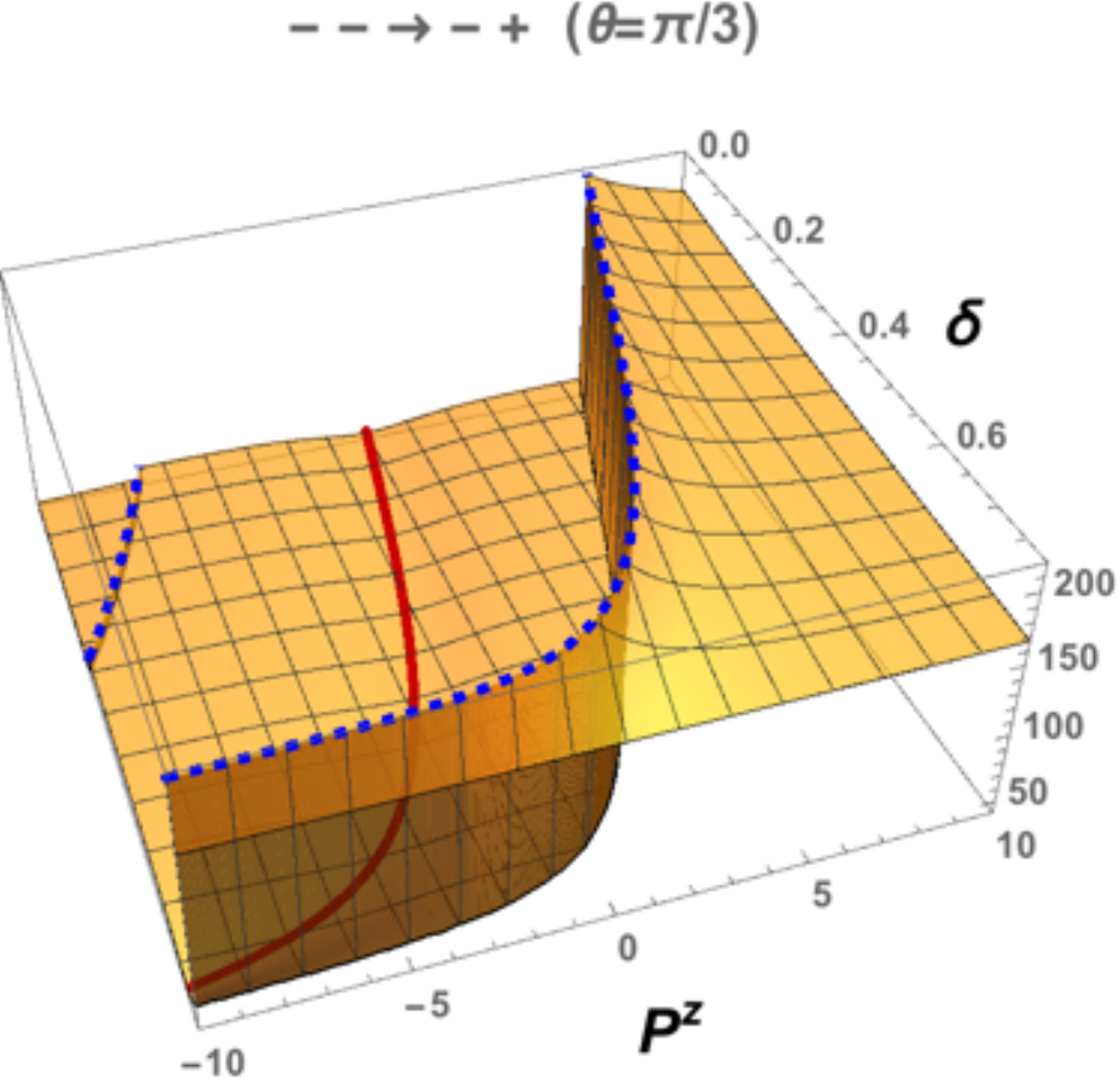}
    }
    \nonumber\\
    &\subfloat{
    \includegraphics[width=0.22\textwidth]{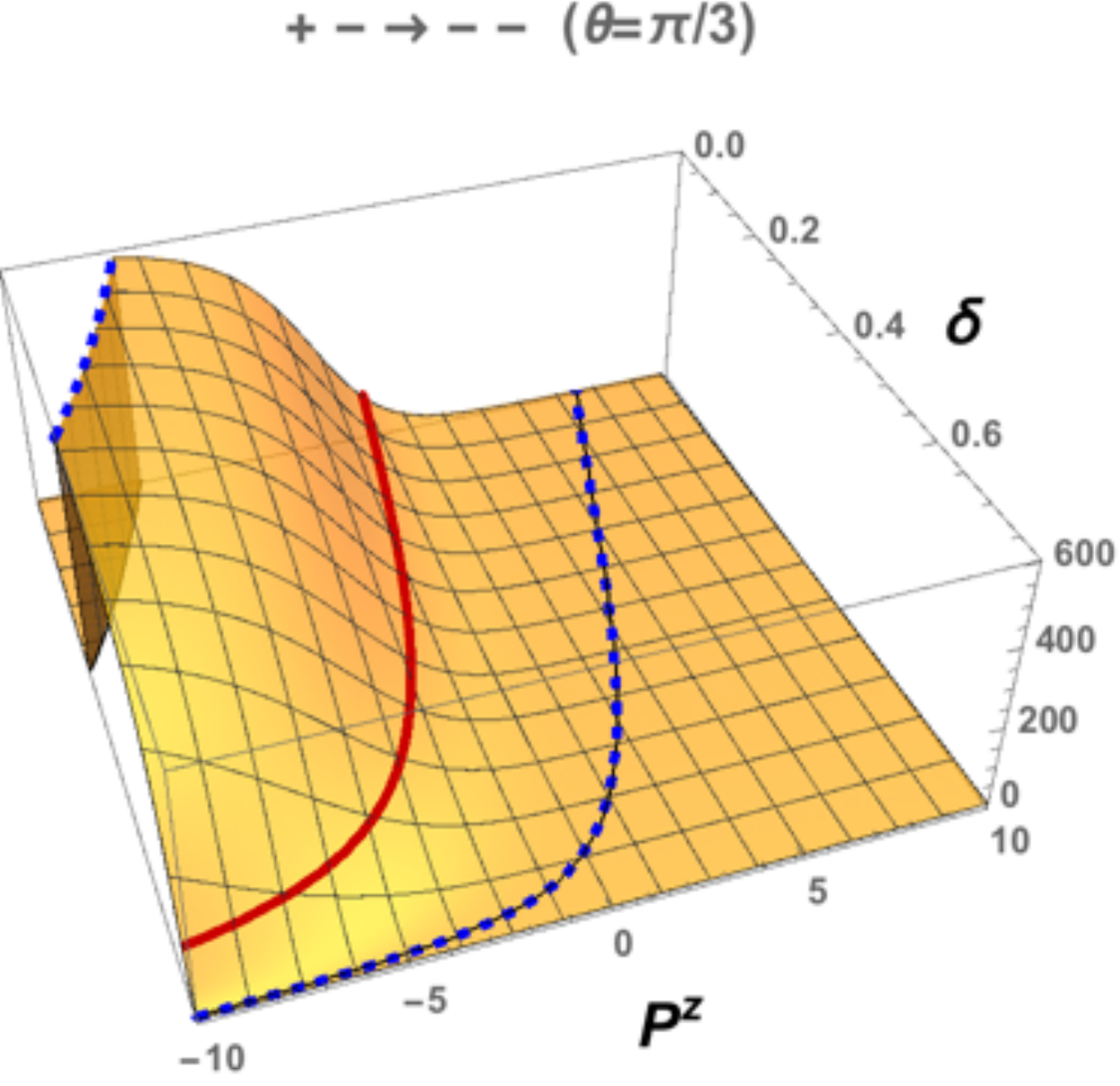}
    }
    %\quad
    \subfloat{
    \includegraphics[width=0.22\textwidth]{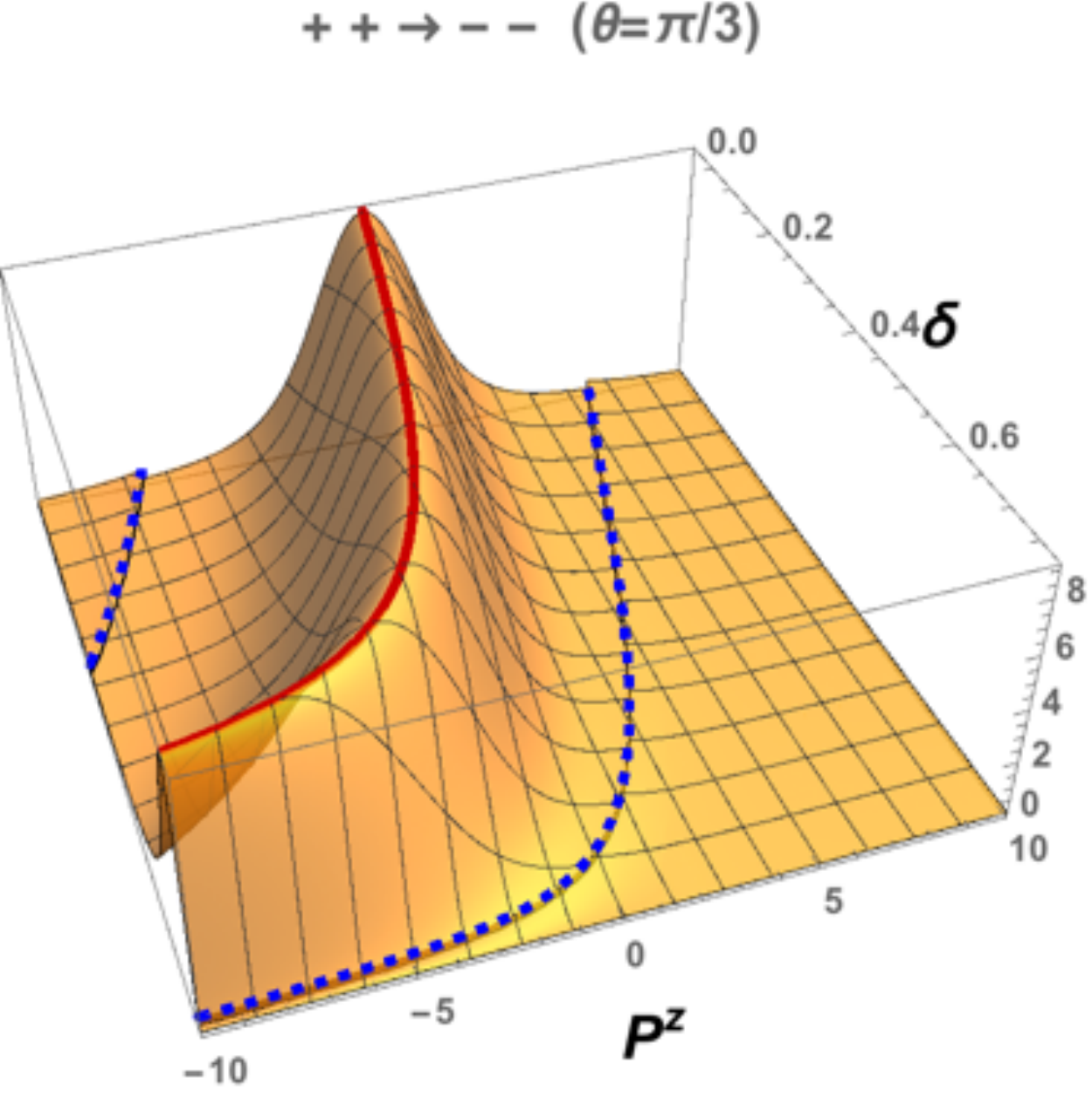}
    }
    %\quad
    \subfloat{
    \includegraphics[width=0.22\textwidth]{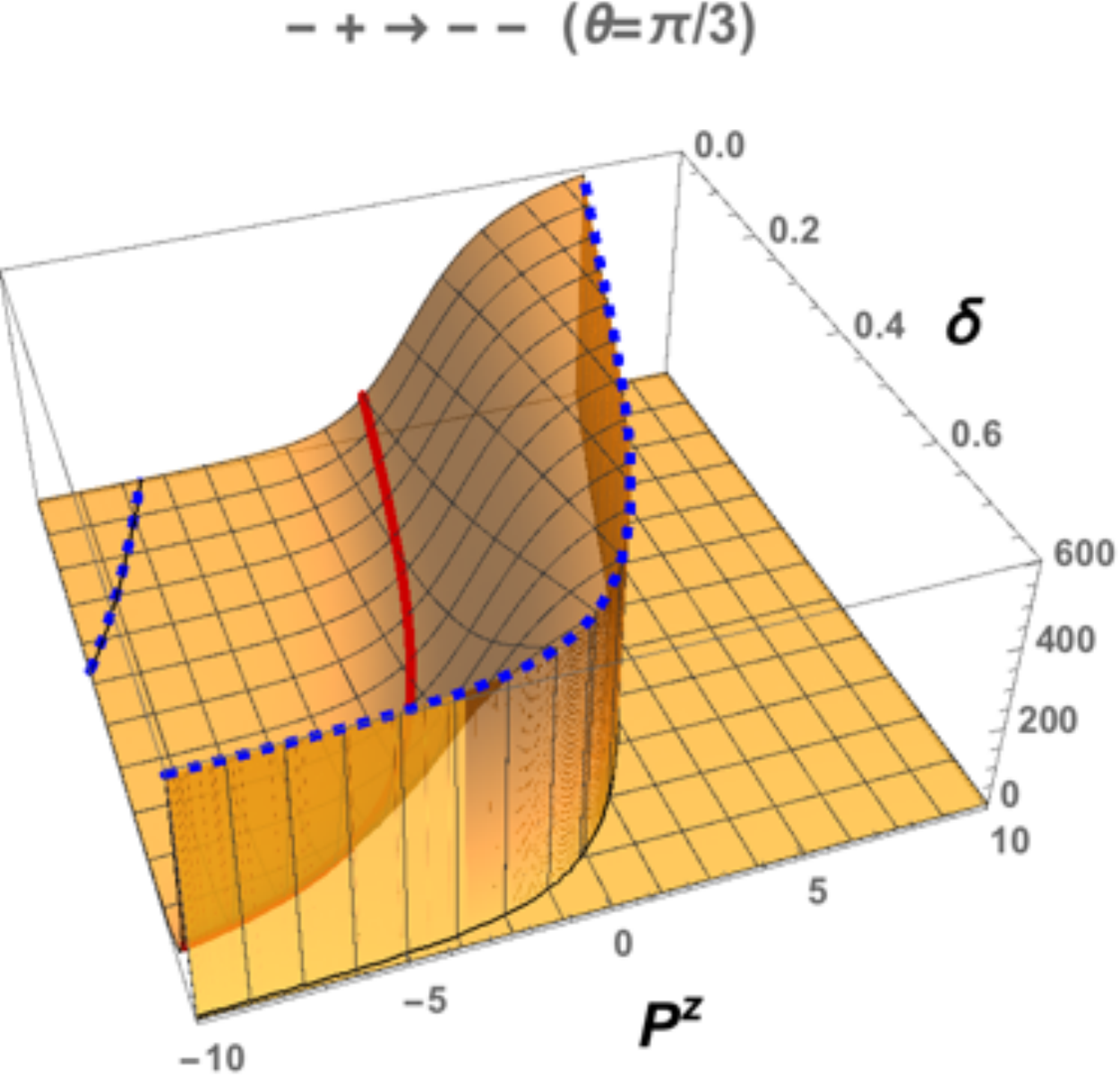}
    }
    %\quad
    \subfloat{
    \includegraphics[width=0.22\textwidth]{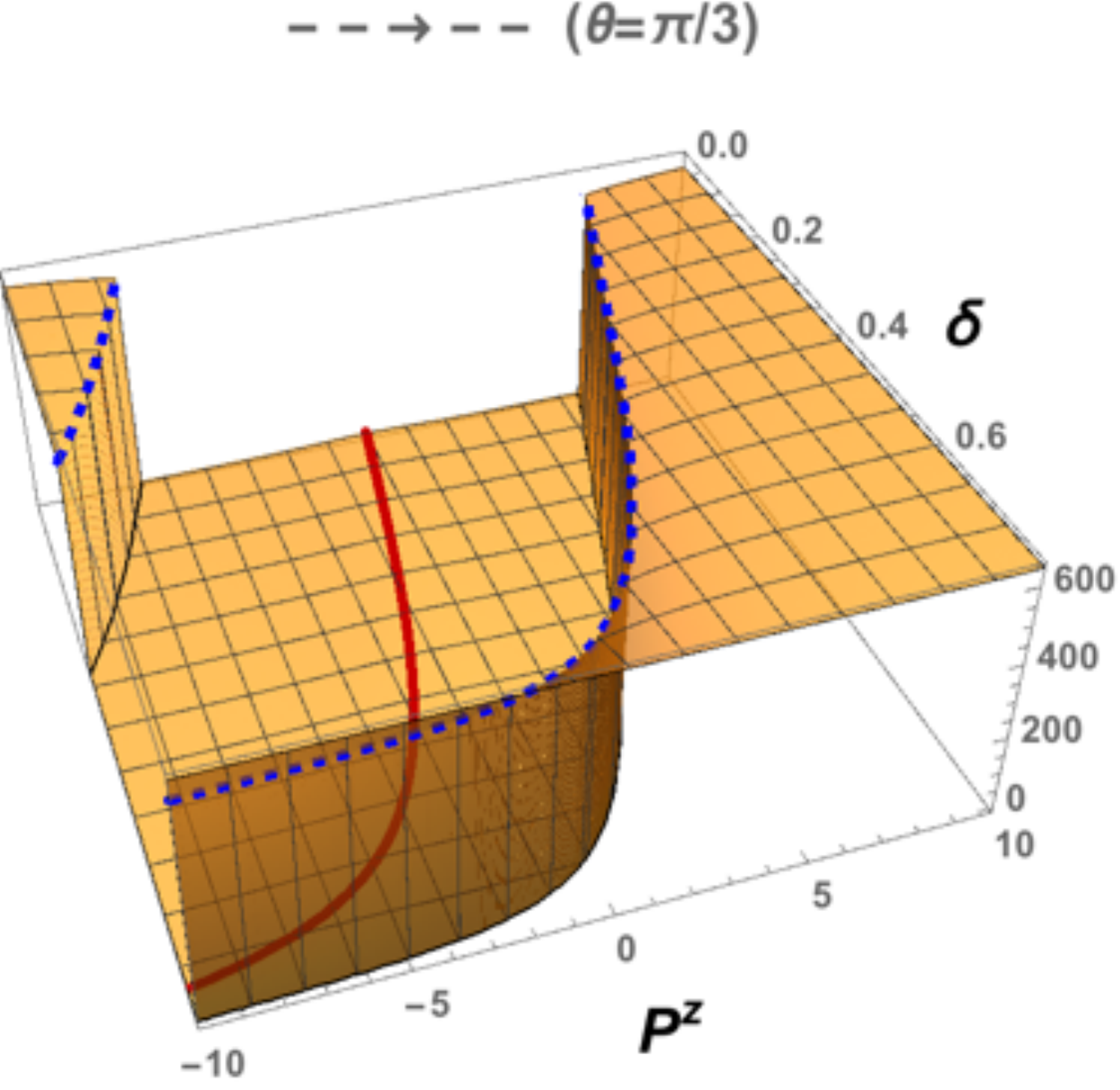}
    }
		\nonumber\\
		% \hline
		&\subfloat{
    \includegraphics[width=0.22\textwidth]{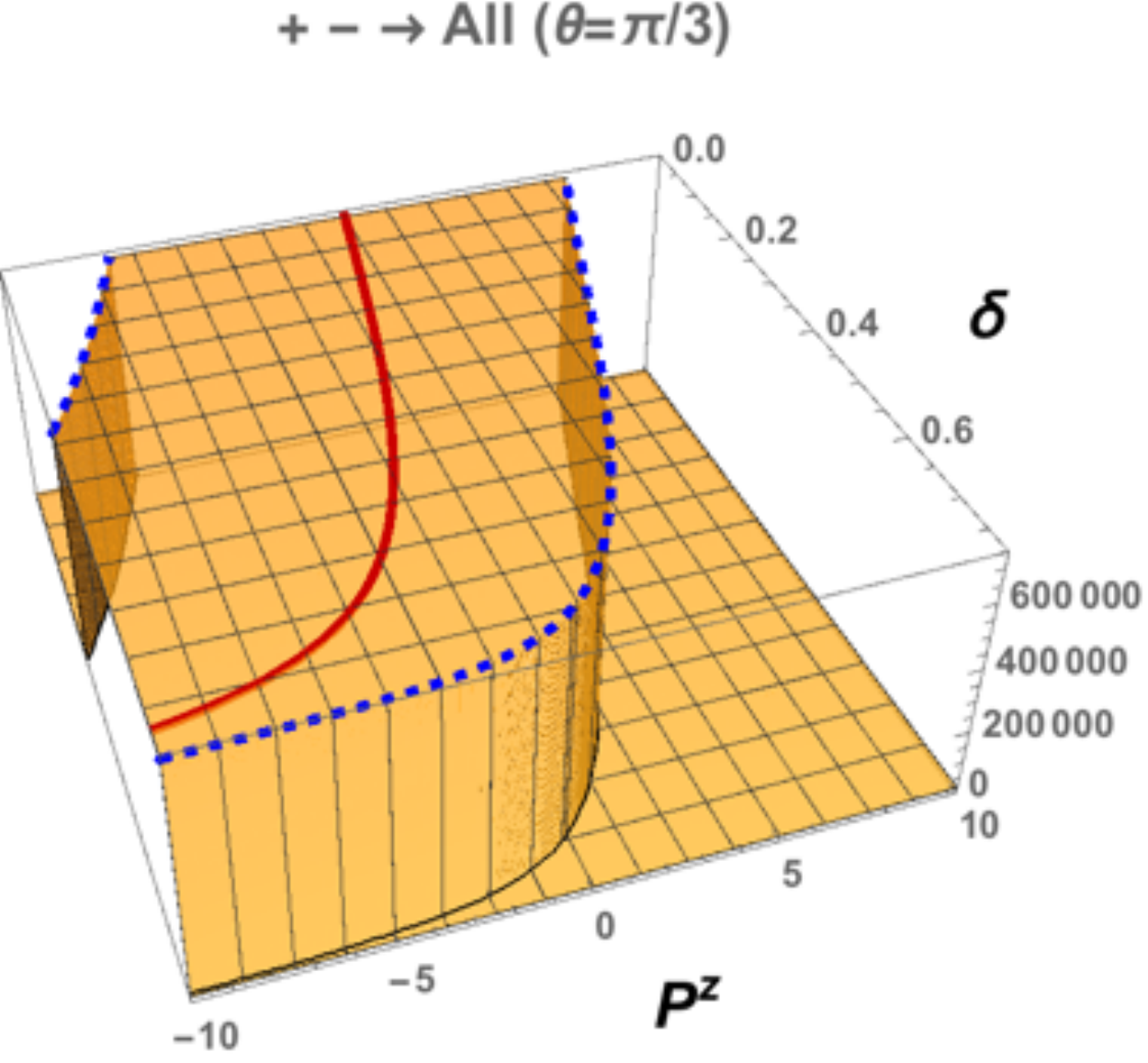}
    }
		%\quad
		\subfloat{
    \includegraphics[width=0.22\textwidth]{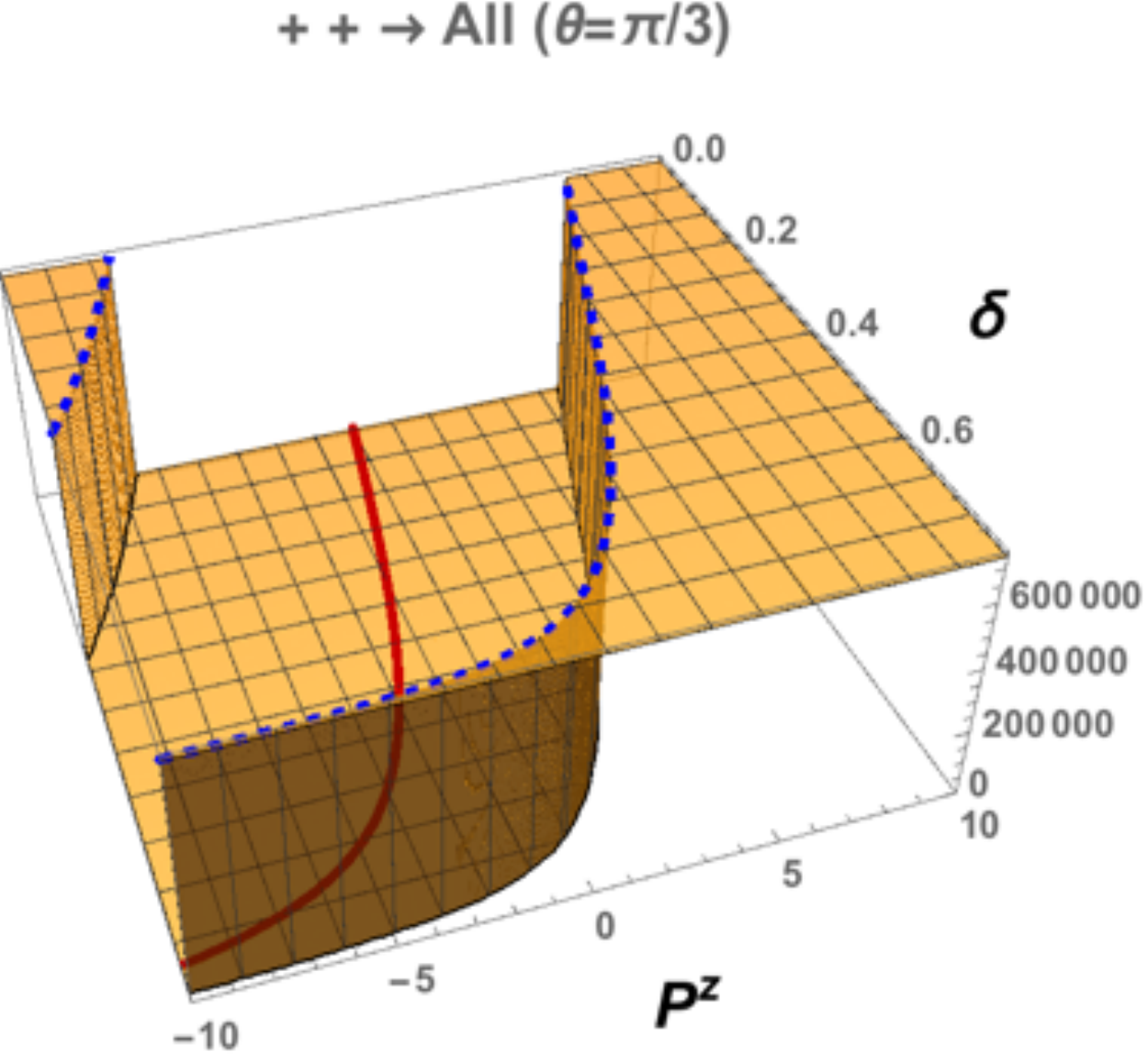}
    }
		%\quad
		\subfloat{
    \includegraphics[width=0.22\textwidth]{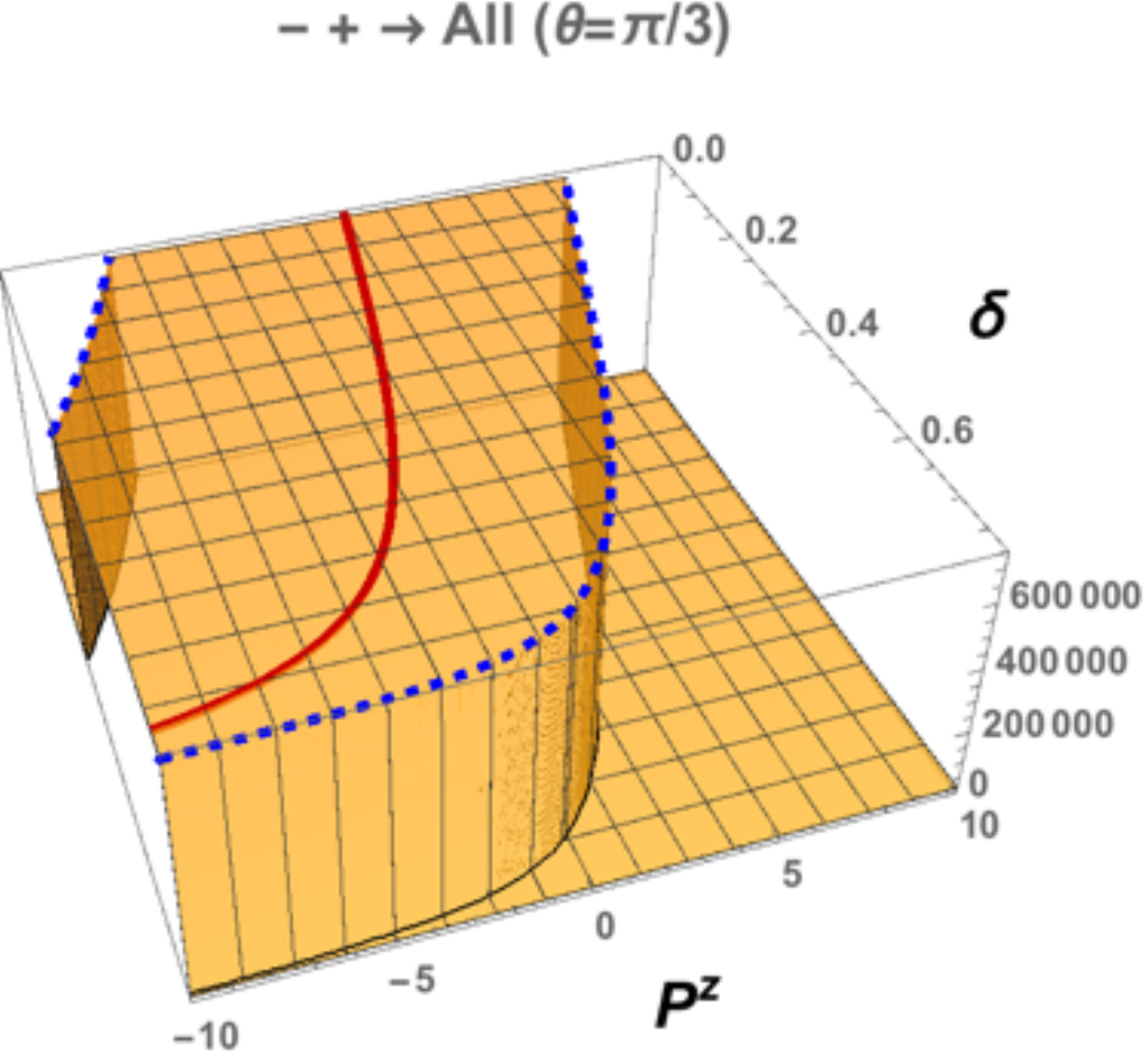}
    }
		%\quad
		\subfloat{
    \includegraphics[width=0.22\textwidth]{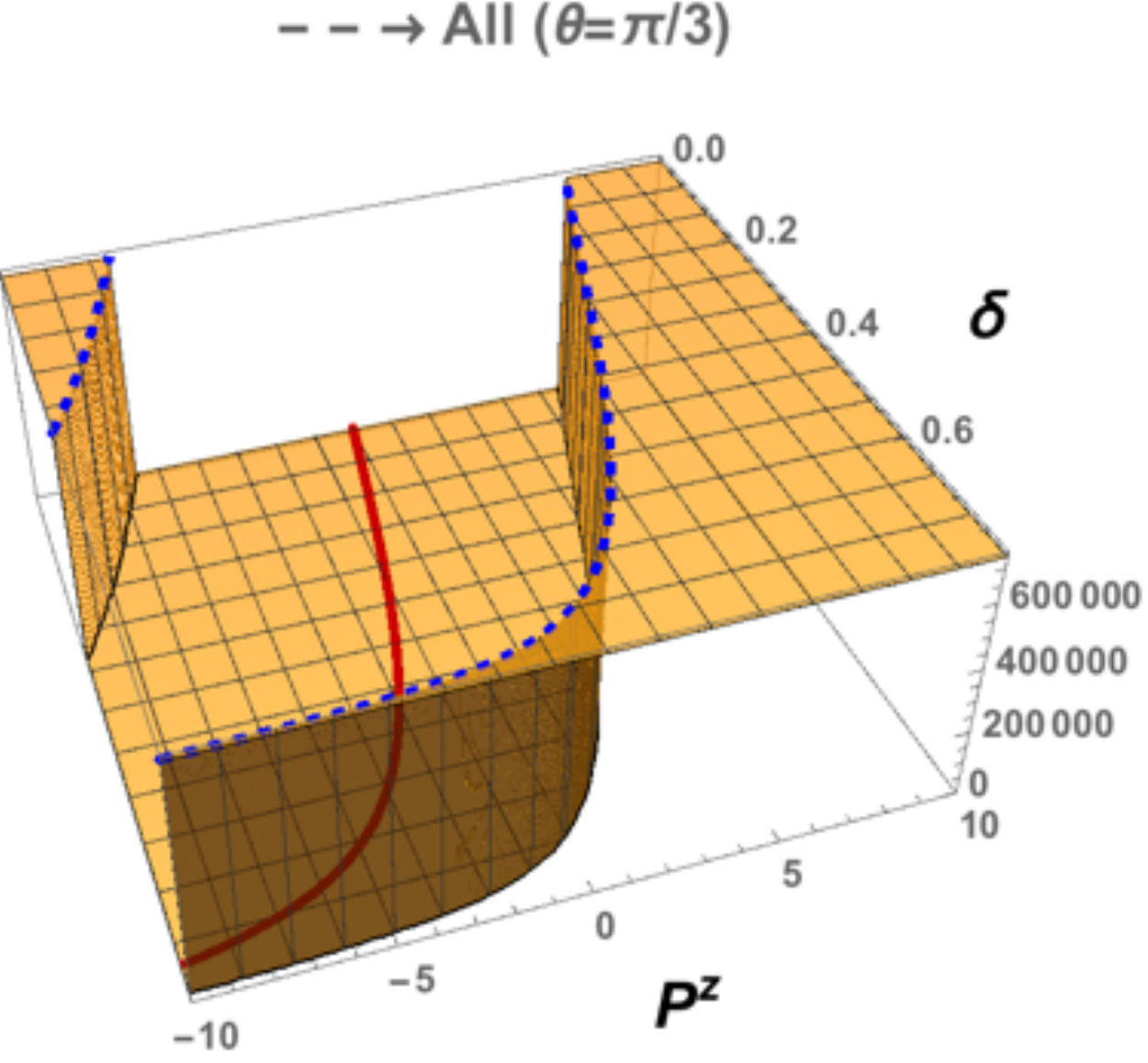}
    }
    \nonumber
 \end{align}
  \caption{\label{fig:Annihilation_Helicity_Probabilities} (Color online) fermion annihilation probabilities (with the factor $e^{4}/q^{4}$ dropped) for 16 different spin configurations with the center of mass energy $M=4$ GeV and the annihilation angle $\theta=\pi/3$.
  The masses of the initial and final particles are $m_{\text{ini}}=1$ GeV and $m_{\text{final}}=1.5$ GeV.
  The 4 plots in the last row are each a summation of the 4 plots above, representing the probability of a certain helicity configuration going into all possible final helicity states.
	% The two blue dashed lines are the boundary lines across which the values of helicity probabilities suddenly change and the red solid line is the universal J-curve mentioned in the paper.
	}
\end{figure*}

% section _interpolated_annihilation_amplitudes (end)

\end{widetext}

\clearpage
\bibliography{Interpolation_spinor_updated}
\bibliographystyle{apsrev4-1}

\end{document}